\documentclass[aps,preprint,showpacs,preprintnumbers,amsmath,amssymb,floatfix,superscriptaddress,nofootinbib,showkeys]{revtex4-1}

\usepackage{amssymb}
\usepackage{amsmath}
\usepackage{graphicx}
\usepackage{xcolor}
\usepackage{times}
\usepackage{upgreek}
\usepackage{longtable}
\usepackage{soul}

\renewcommand{\arraystretch}{2}

\makeatletter
\renewcommand{\p@section}{}
\renewcommand{\p@subsection}{}
\renewcommand{\p@subsubsection}{}
\renewcommand{\p@paragraph}{}
\renewcommand{\p@subparagraph}{}
\makeatother

\begin{document}

\title{
  {Charge and spin dynamics  driven by  ultrashort extreme broadband
 pulses: a theory perspective}}

\author{Andrey S. Moskalenko}
\email{andrey.moskalenko@uni-konstanz.de}
\affiliation{Institut f\"{u}r Physik,
Martin-Luther-Universit\"{a}t Halle-Wittenberg,
 06099 Halle, Germany}
\affiliation{Department of Physics and Center for Applied
Photonics, University of Konstanz, 78457 Konstanz, Germany}
\author{Zhen-Gang Zhu}
\email{zgzhu@ucas.ac.cn}
 \affiliation{Institut f\"{u}r Physik,
Martin-Luther-Universit\"{a}t Halle-Wittenberg,
 06099 Halle, Germany}
 \affiliation{School of Electronic, Electrical and Communication Engineering, University of Chinese Academy of Sciences, Beijing 100049, China}

\author{Jamal Berakdar}
\thanks{Corresponding author}
\email{jamal.berakdar@physik.uni-halle.de}
\affiliation{Institut f\"{u}r Physik,
Martin-Luther-Universit\"{a}t Halle-Wittenberg,
 06099 Halle, Germany}

\date{\today}

\begin{abstract}
This article gives an overview on recent theoretical progress in controlling  the
charge and spin dynamics in  {low-dimensional electronic systems}  by means of ultrashort  {and ultra}broadband
electromagnetic pulses. A particular focus is put on  {sub-cycle}
and single-cycle pulses  { and their utilization} for coherent control.
The discussion is mostly limited to cases where the
pulse  duration is shorter than the characteristic time
scales  { associated with the involved  spectral features of the excitations.
The relevant current theoretical
knowledge is}  {presented}  {in a coherent, pedagogic
manner. We work out that the pulse action amounts  in essence to a quantum map
between the}  {quantum states of the system}  {at an appropriately
chosen time moment during the pulse.} The influence of a particular
pulse shape on the post-pulse dynamics is reduced to several
integral parameters entering the expression for the quantum map.
The validity range of this reduction scheme for different strengths of
the driving  fields is established and discussed for
particular nanostructures.  {
 Acting   with  a periodic pulse sequence, it is shown how the system can be steered to and largely maintained in predefined
states.  The conditions for this}  {nonequilibrium}  {sustainability  are worked out by means of geometric phases,  which are identified as
 the appropriate  quantities to
indicate  quasistationarity}  {of}  {periodically driven
quantum systems.
 Demonstrations are presented for the control of the charge, spin, and valley degrees
of freedom in nanostructures on picosecond and subpicosecond time
scales.}  The theory is illustrated with several applications to
one-dimensional semiconductor quantum wires and superlattices,
double quantum dots, semiconductor and graphene quantum rings.  { In
the case of a periodic pulsed driving the influence of the
relaxation and decoherence processes}  {is}  {included by utilizing the density matrix approach.}
The integrated and
time-dependent spectra of the light emitted from the driven
system deliver information on its spin-dependent dynamics. We review
examples of such spectra of photons emitted from  pulse-driven
nanostructures as well as a possibility to characterize and
control the light polarization on an ultrafast time scale. Furthermore, we consider
the response of strongly correlated systems to short broadband pulses and show that this case
bears a great potential to unveil high order correlations while they build up upon excitations.

\end{abstract}


\maketitle

\tableofcontents

\vfill \pagebreak

%

\section{Introduction}
Electromagnetic waves are omnipresent  in modern society with a vast variety  of applications ranging from  TV, radio,
 and cell phones to high power lasers and ultra precision metrology.
In scientific research,  newly invented methods offer a wide  range of  pulse durations
from nanoseconds, through picoseconds, femtoseconds to currently
attoseconds  \cite{Krausz2009,French1995,Walmsley2009} opening so
 new avenues for research  to explore the time evolution
in a desired  spectral regime  which has lead
to landmark discoveries in physics and chemistry.
The key point thereby is the exploitation of the
 light-matter interaction to steer the system in a controlled manner
 out of the equilibrium or to stabilize it
 in target states by irradiation with shaped electromagnetic waves.
The study of the behavior of nonequilibrium quantum systems driven by
short light pulses has evolved so, depending on
the goals and applications,  to diverse sub-branches  such as  photovoltaics
\cite{Timmerman2008,Shabaev2006,Schaller2005,Schaller2006},
optical, electro- and magnetooptical devices
\cite{Huber2001,Ropers2007a,Ropers2007,Alex_PRL2005,Kimel2007,Hsia2008}
as well as  efficient  schemes for the control of chemical processes
\cite{Zewail1988,DeSchryver_book,Brixner2003a,Marquetand2005,Wollenhaupt2005}.
Particularly, the  studies of nonequilibrium processes in nanostructures are
fueled by the equally impressive progress in nanoscience allowing to fabricate
and  engineer
structures with desired geometric and electronic properties and bringing them to
real applications, e.g. as  an efficient radiation emitter in a broad frequency
 range or parts in electronic circuits.
 From a theoretical point of view, the currently available nanostructures with
well-defined and simple topology like quantum wells
\cite{Bastard_book,Harrison_book}, quantum rings
\cite{mailly,Nitta1999,Lorke2000,Rabaud2001,Fuhrer2001,Yu2007,Russo2008,Recher2007,Hoffmann2007,Barth2006a,Barth2006b},
quantum dots
\cite{Harrison_book,Bimberg_book,Delerue_book,Klimov_book,Rogach_book},
and quantum spheres \cite{Dresselhaus_book,Fullerenes_book} are particularly appealing,
as they allow for a clear
understanding of their static  and nonequilibrium behavior. Hence, our main focus will
 be on these structures.
As for the driving electromagnetic fields, emphasis  is put on the utilization of
 broadband  ultrashort pulses because
they offer efficient schemes for steering the nonequilibrium states of matter.
 There has been an enormous  progress in the generation and
design of ultrashort pulses allowing to control the  duration, the shape, the strength,
the polarization properties, the
 focusing,   the repetition rates, as well as the spectral bandwidth
\cite{Diels_book,You1993,You1997,Weinacht1999,Brabec2000,Huber2000,Brixner2001,Polachek2006,Carr2002,Keller2003,Kubler2005,Aeschlimann2007,Corkum2007,Bucksbaum2007,Persson2006,Sell2008,Junginger2010,Krauss2010,Goulielmakis2008,Krausz2009,Popmintchev2010,Wu2012,Zhao2006,Shalaev2007,Zhang2008,Lee2009,Zhao2012,Hassan2016,Seifert2016}.
%
 {The pulses which are in the focus of this review are briefly introduced and discussed in  {Section}} {\ref{sec:exp}.}
Excitations  by   short electromagnetic pulses may
 proceed resonantly or non-resonantly. In the first
case, the light frequency is selected as to match a certain quantum
transitions in the system. A paradigm of resonant excitations are
 driven two-level
systems \cite{CohenTannoudji_book1}. For instance, the application of
 resonant circular polarized $\pi$-pulses [$\int \Omega_{\rm
R}(t) \mbox{d} t=\pi$, where $\Omega_{\rm R}(t)$ is the Rabi
frequency] to quantum rings leads to a population transfer
between the  ring quantum states, provided the pulse duration is much shorter
than the typical time scales of  dissipative processes in the
system. It was theoretically demonstrated how to generate
nonequilibrium charge currents in semiconductor and molecular
quantum rings with the help of an appropriate resonant excitation
by light pulses \cite{Pershin2005,Barth2006b}.
In another
characteristic case of the $\pi/2$-pulses applied to the same
system, a rotating charge density is generated in the rings,
additionally to the current, which is in this case smaller by a
factor of two \cite{Barth2006a}. For quantum dots, the resonant
excitation with short light pulses can lead to population
inversion of confined exciton states, as it was demonstrated
experimentally using $\pi$-pulses \cite{Stievater2001}. The
reduction of the light-matter interaction to transitions in driven
two-level systems is based on the so-called ``rotating wave''
approximation. It is effective only if the pulse duration is long
enough, on the order of ten wave cycles or longer, and the central
frequency of the pulse exactly matches the frequency of
the induced transition. The required number of wave cycles can be
slightly reduced if the optimal control theory is implemented for the
driving pulse \cite{Rasanen2007}. The resonant excitation with few
wave cycles seems to be inappropriate if the desired result of
the excitation requires transitions between many levels of the
driven system, which are generally not equidistantly spaced in
energy. A predictable result may require application of a pulse
sequence with different central frequencies \cite{Pershin2005}, at
the cost of much longer duration of such an excitation.

To stay with the  example of a phase-coherent ring,
if the driving field is non-resonant,  and if its
 strength is sufficiently large, the states of the ring become dressed by the photon
 field \cite{Faisal_book,CohenTannoudji_book2,Scully_book}.
If the field is circularly  polarized, the degeneracy between
the field-counter and anti-counter propagating
ring states is lifted and a finite current emerges in the
 ring (in the presence of the field) \cite{Kibis2011}. The phase change associated
 with this break of symmetry goes, as usual for non-resonant effects, at least quadratically
 with the field strength and hence becomes important at higher intensities. On the other hand, at high intensities
 multiphoton processes or tunnelling in the electric field of the laser
 may also contribute substantially   depending   on the frequencies \cite{Yamanouchi_book}.
We deal in this work with a
 further kind of processes which are
  not really resonant but still may occur
to the first order in the
driving field.
This is the case of a broadband pulse covering a large number of the system
excitations \cite{Yamashita_book}.
An example of an ultrabroadband pulse is an
 asymmetric monocycle electromagnetic pulse,  also called
\textit{half-cycle pulse} (HCP)
\cite{You1993,You1997,Jones1993,tiekling,Jones1996,Dion2001,Zeibel2003,Persson2006,Wu2012}.
The electric field of a linear polarized HCP performs a short and
strong oscillation half-cycle followed by a long but much weaker
tail of an opposite polarity. If the duration of the tail is much
longer than the characteristic time scales of the excited system
then its effect can be neglected.  Such a pulse  contains
a broad band of frequencies, particularly with a decreasing pulse duration.
If  the pulse duration becomes
significantly  smaller than the characteristic time scales of the
 system under study, then the  action of the HCP subsumes
to  an appropriate matching
of the wave functions (or the density matrix if a
many-body consideration of the system is required) before and just after
 the pulse application. This does not mean that the  state after the pulse is
 in general an eigenstate, but usually a coherent state.
  Classically,  the matching condition corresponds in fact to an instantaneous
  transfer
of a momentum $\Delta p$ (a \textit{kick}) to the system
\cite{Dhar1983,Carnegie1984,Reinhold1997,Frey1999,Henriksen1999,Yoshida2000,Tannian2000}.
The transferred momentum is proportional to the pulse strength and
its duration. For confined electrons, usually the momentum operator does not commute
with the field-free Hamiltonian and hence
the pulse-induced momentum shift generates a coherent state.
Quantum mechanically, the wave
function $\Psi(x,t)$ of a one-dimensional system subjected at the time moment $t=0$ to the
action of a HCP  obeys  the matching
condition
 $\Psi(x,t=0^+)
    =\exp(i\Delta p x/\hslash)\Psi(x,t=0^-)$. Here  $t=0^-$ is the time moment
     just before the pulse
 and $t=0^+$ is right  after it.
This matching condition is the essence of the \textit{impulsive
(or sudden) approximation} (IA).  The
pulse-generated coherent state develops in the time after the pulse   according to the original
Hamiltonian. Below we work out the validity range of this stroboscopic evolution scenario.
Terahertz (THz) HCPs  and trains of   HCPs were considered in the impulsive regime  to orient polar molecules
\cite{Henriksen1999,Dion2001,Alex_Molecules2003}, to manipulate the
populations and control the orbital motion of electrons in
Rydberg states
\cite{Jones1993,Reinhold1993,tiekling,Tiekling1995,Bugacov1995,Jones1996,Reinhold1997,Frey1999,Tannian2000,Zhao2006,Mestayer2007},
and to steer the electronic density of ionized atoms and
molecules on the  attosecond time scale
\cite{Dimitrovski2005,Briggs2008,Emmanouilidou2008,Persson2009}.

 {
Generally, the area of driven quantum systems is huge with a number of sub-branches depending on the type
of driving, the system under consideration, and the intended goals. The focus of this review is on the theory of
quantum dynamics driven by
ultra broadband short pulses. To be more specific we discuss briefly in
Section \ref{sec:exp} the type of the appropriate experimental pulses and mention
some methods of generating them.}
 In Section~\ref{Sec:Theory} we discuss a general
perturbation theory for the unitary evolution operator of a
quantum system driven by ultrashort external pulsed fields, where
the small parameter is the pulse duration. Such a development
is important for the understanding of the approximation steps
leading to the IA in the case of HCPs and determining its limits
of validity. Apart from this, we discuss cases when a theory
beyond the IA should be applied. The corresponding theoretical
considerations can be found in literature
\cite{Dykhne1978,Henriksen1999,Fernandez1990,Krstic1994,Daems2004,Dimitrovski2005,Dimitrovski2006}
but a development of a consistent perturbation theory with the
pulse duration as a small parameter was absent until
recently when it was formulated for atoms excited by light pulses
confined to a small and finite time range
\cite{Klaiber2008,Klaiber2009}. We present here an alternative
derivation which is suitable also for pulse-driven nanostructures
and includes the natural case of short light pulses with decaying
tails which are however not necessarily strictly confined to a
finite time range. With this approach we get an approximative
description of the action of ultrashort pulses of a general shape,
e.g. also in the cases of single-cycle pulses and few-cycle
pulses, as a map between the states of the driven system before
and after the pulse. Further in Section 2, we concern ourselves with the limits
of validity of such a treatment of the excitation process. Interestingly,
the IA may remain
valid in some range of parameters even when the entire unitary perturbation expansion
in the pulse duration breaks down
due to the increase of the pulse amplitude. In this regime of short but very strong
(SVS)
interactions the next order correction to the unitary evolution operator beyond the
 IA can be also found.
We discuss implications of the IA, unitary perturbation theory and SVS result for
general one-dimensional
geometries and two-level systems. In last part of Section 2 these findings are used
 to describe
driving by periodic trains of the pulses and characterize the resulting quantum
dynamics. We describe conditions for the controlled periodicity and quasistationarity
of the evolution.


 {
  Sections~\ref{Sec:Appl} and \ref{Chapter:Zhengang} introduce }
various applications of the developed theoretical methods for particular
 nanostructures.
Here we start with the pulse-driven dynamics of electrons moving along a
spatially-periodic potential energy landscape (mimicking semiconductor superlattices,
 or generally
crystal lattices and superlattices). Indirect transitions and charge currents
can be induced in unbiased structures on extremely short time scales
 \cite{Moskalenko_PLA2006}. These results are especially appealing in view of
  an impressive ongoing progress on ultrafast
 control of the electron dynamics in solids by strong light pulses \cite{Schultze2013,Schultze2014,Schiffrin2013,Yakovlev2016,Hohenleutner2015,Maag2015,Mayer2015}. The reviewed approach provides access to this dynamics in a different, complementary
and so far unexplored regime with distinct and unique features.
Further in Section~\ref{Sec:Appl}, we discuss how the charge
polarization can be induced in double quantum wells and controlled
 by periodic pulse trains
\cite{Alex_QW2004,Alex_Europhysics2005a}. Then we switch our
attention to the light-driven semiconductor quantum rings, where
apart from the charge polarization dynamics also nonequilibrium
charge currents can be induced by an appropriate sequence of two
light pulses
\cite{Alex_PRB2004,Alex_Europhysics2005,Alex_PRL2005}. This
dynamics can be influenced by a perpendicular magnetic flux
piercing the semiconductor ring \cite{Moskalenko_EPL2007}. The
induced polarization dynamics and current are subjects to
decoherence and relaxation processes
\cite{Moskalenko_PRB2006,Moskalenko_PRA2008}. The capability to
model these processes allows to create schemes for the charge
current switching and generation of local magnetic fields with a
tunable time structure \cite{Moskalenko_PRB2006}. We show that if
transferred to graphene quantum rings, these ideas suggest a way
for an ultrafast generation of pure valley currents
\cite{Moskalenko_PRB2009}. In Section~\ref{Chapter:Zhengang}, we
concentrate our attention on the spin dynamics triggered by
ultrashort light pulses in semiconductor quantum structures
and discuss how the spin dynamics can be steered to generate spin dependent
polarization, spin-polarized current, and pure spin currents in nanostructures \cite{zgzhu2008,zgzhu2008,zgzhu2009,zgzhu2010,Schueler2014}.
Proposals for an optically driven spin field-effect transistor
 \cite{zgzhu2010qw} and ultrafast spin filtering \cite{Waetzel2011} are described.

Section~\ref{Sec:emission} is devoted to the emission properties and their control associated with the dynamics of the pulse-driven nonequilibrium dynamics. Finally, in Section \ref{Sec:many_body}
 we discuss briefly how short, broadband  pulses can be utilized to explore many-body effects in correlated systems \cite{Pavlyukh2014}, finishing
with   a summary and concluding remarks.

\section{Generation of short broadband pulses}\label{sec:exp}
\begin{figure*}
  \includegraphics[width=14cm]{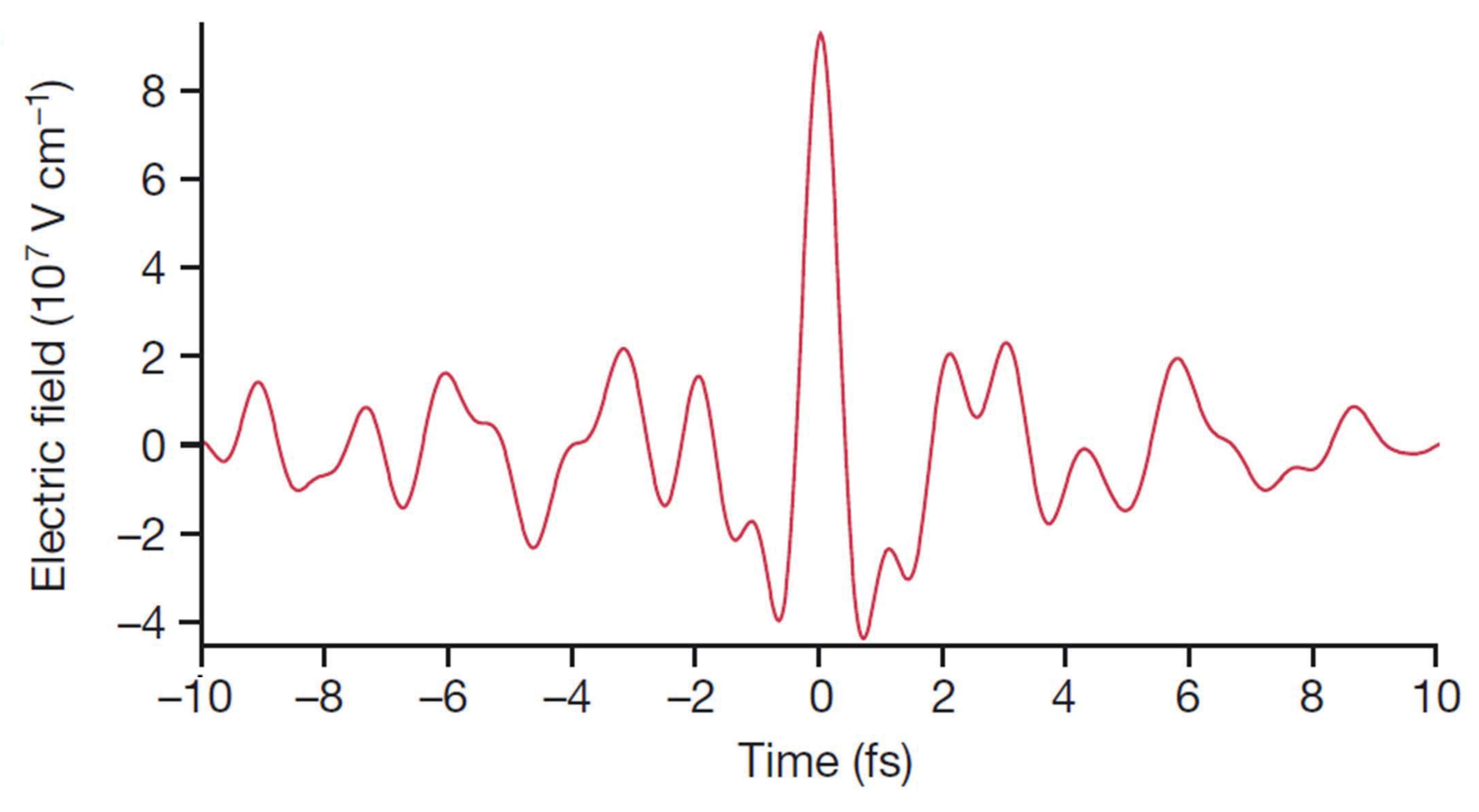}
  \caption{ {Electric} field of a femtosecond unipolar pulse produced via  an attosecond light-field synthesizer. Reproduced from Ref.~\cite{Hassan2016} with permission from NPG group.
   \label{fig:pulse1}}
\end{figure*}

 {While we mainly aim in this report at the theoretical aspects of the short-time dynamics triggered by} {broadband}  {pulses, it is useful to briefly discuss the appropriate experimentally available pulses.
In this review we}  {consider a pulse to be short and call it  also ``ultrashort''}  {if its duration is on the scale or smaller than the generic times of the involved transitions}  {that is}  {is reflected in a respective frequency range of the pulse. Thus, depending on the problem at hand a picosecond pulse might be short} {enough,}  {as for instance for the case of intra conduction band excitations in micron-size, semiconductor-based quantum rings. Other processes may require}  {femtosecond or sub-femtosecond} {pulses.} {Even the latter}  {pulses became recently available. An example is shown in Fig.~\ref{fig:pulse1}.}  {The field transient}   {was produced by synthesizing
intense optical attosecond pulses
 in the visible and nearby spectral ranges \cite{Hassan2016}. In this scheme
 1.1  to 4.6 eV  wide-band pulses  are divided by dichroic}  {beam splitters}  {into  spectral bands and then  each band is compressed  and  spatiotemporally superimposed to yield a pulse such as the one  {in} Fig.~\ref{fig:pulse1}. The intensity profile duration is approximately 380 as at FWHM.}  {Moreover, the carrier-envelope phase of such field transients can be adjusted to produce
``near-cosine'' and ``near-sine'' waveforms \cite{Garg2016}.}
  {Synthesized sub-cycle pulses in the  mid-infrared  which are suitable for our purposes were reported also in Ref.~\cite{Liang2016} and further references therein.}

\begin{figure*}
  \includegraphics[width=0.95\textwidth]{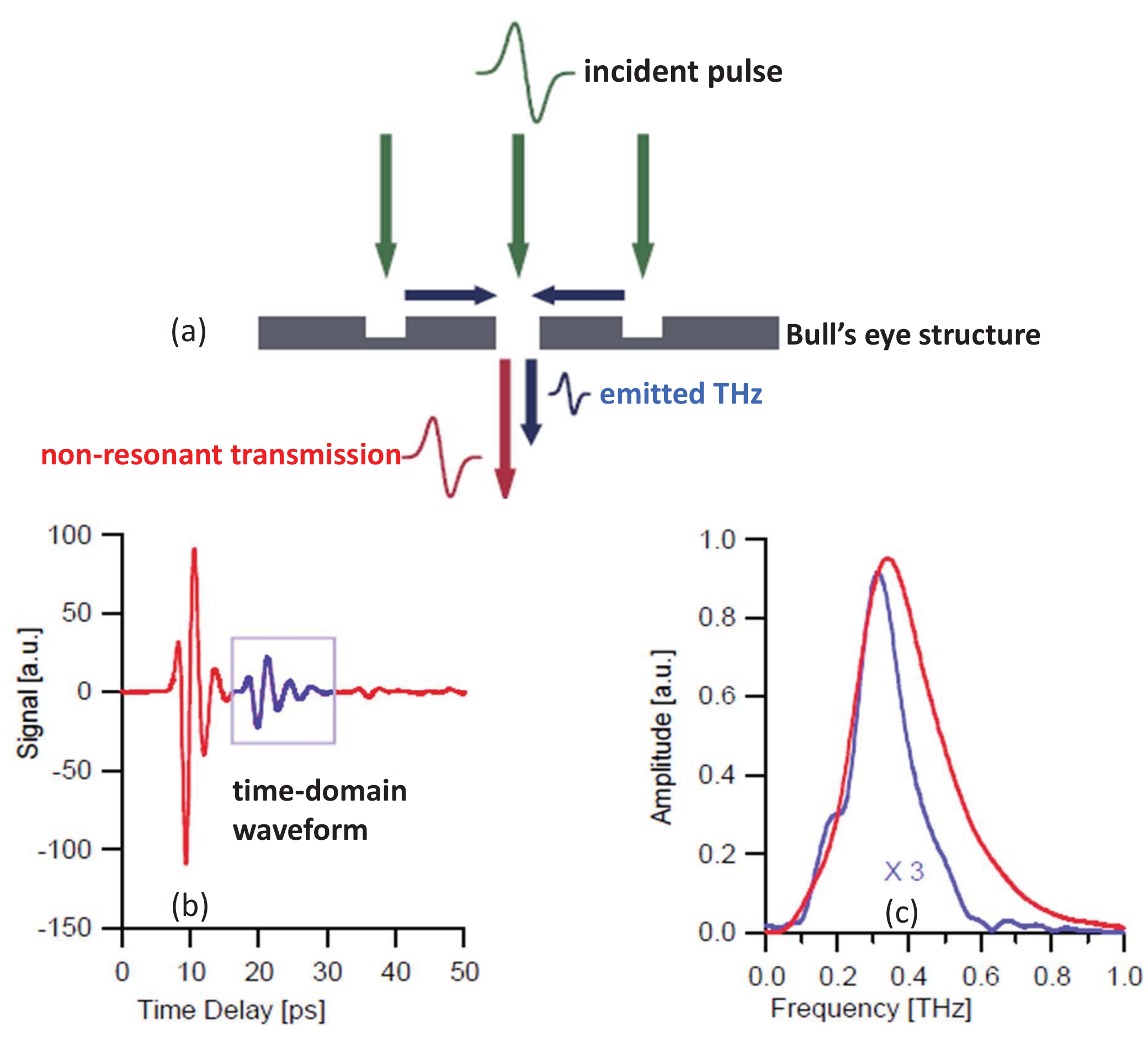}
  \caption{A cross-sectional line diagram of a bullseye structure with a  {sub-wavelength aperture which is irradiated by} a THz pulse. The non-resonant
transmitted part  of the incident THz pulse (red curve) adds  to the
 {waveform} (blue curve) emitted with a time delay  upon the  pulse interaction  with the bullseye structure. Red and blue parts in the  temporal waveform (b) and the  amplitude spectra (c) correspond to the wave portion with  respectively the same color in (a).
Reproduced  from  {Ref.~}\cite{Agrawal2005} with permission from OSA.
   \label{fig:pulse2}}
\end{figure*}

 {There is a}  {possibility to generate strong}  {near-field}  {pulses that may drive impulsively charge and spin dynamics in the THz regime}  {by using}  {plasmonic structures such}  {as  bullseye} {structures consisting of  annular grooves \cite{Agrawal2005}. A cross-sectional line diagram}  {illustrating the setup}  {is shown in   Fig.~\ref{fig:pulse2}} {,}  {which also includes the time-domain waveform and the amplitude spectra. In recent years there has been an enormous progress in designing and applying plasmonic structures for}  {near-field}  {THz generation} {;}  {we refer to Refs.~\cite{Mittleman2013_2,Kampfrath2013} and the references therein for further details on this topic.}
 {Theory illustrations presented in the following sections}  {employ mostly
strong}  {single-}  {or sub-cycle THz pulses applied to systems with spectral
features in the}  {far-infrared}  {range.}  {In this context we will discuss in detail the
 ballistic charge and spin manipulation in mesoscopic rings and quantum wires.
 Although the dynamics of the molecular rotations is outside of the scope of the present review, it is important to note that the same excitation regime can be successfully realized in the case of orientation of polar molecules and alignment of
nonpolar molecules (cf. the review article \cite{Stapelfeldt2003} and references therein). Apart from this, the pulses of this type were used to study the acceleration of carriers and postpulse dynamics in semiconductor heterostructures \cite{Hebling2010,Gaal2007}.}

\begin{figure*}
   \includegraphics[width=0.45\textwidth]{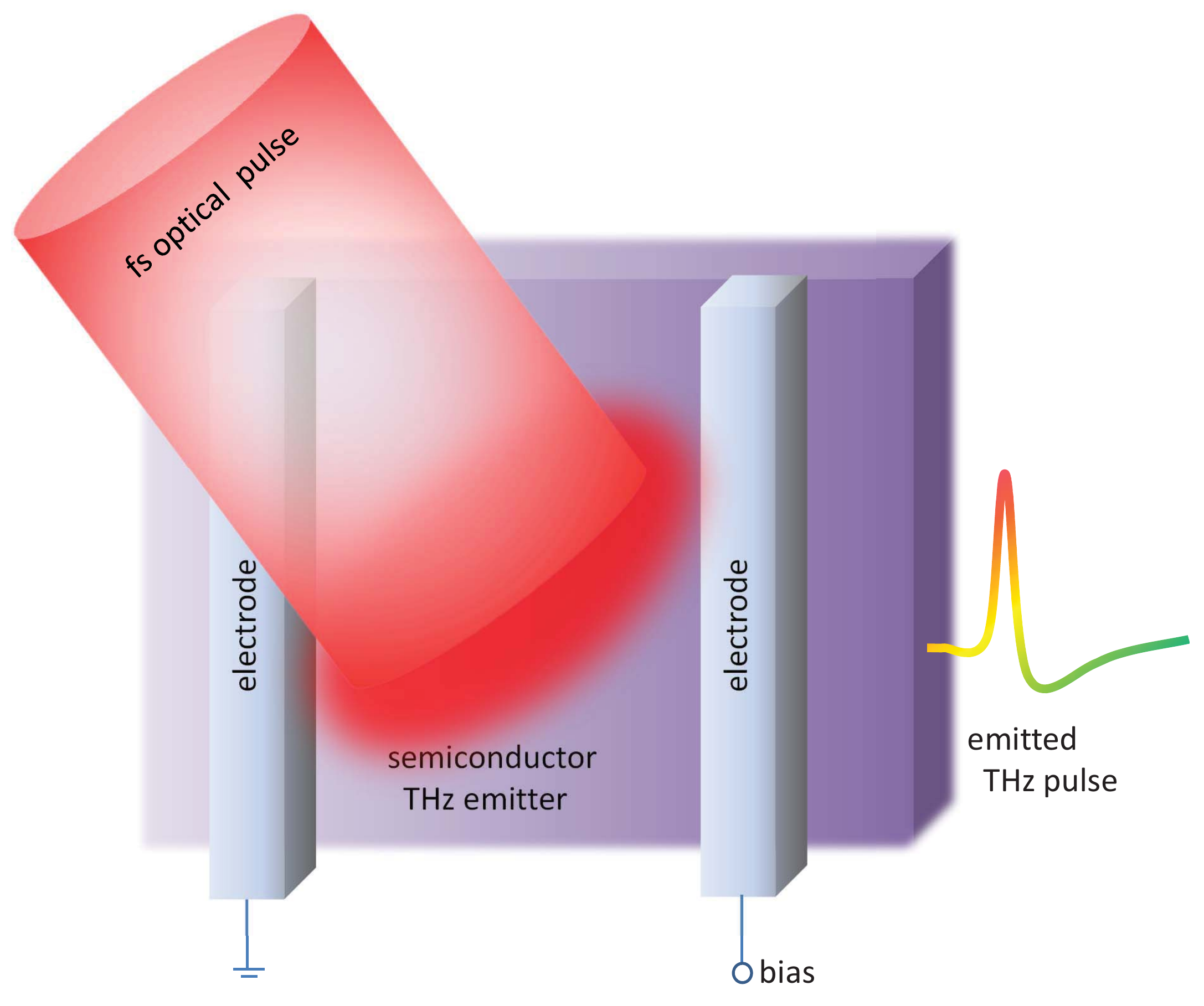}
  \caption{Schematics for THz pulse generation via a conventional photoconductive Auston switch. The electrodes on the semiconducting sample are separated by few tens of micrometers and are  biased by tens of volts generating an electric field of  several  kV/cm  across the sample. The switch is electrically
  shortened by a femtosecond laser pulse with a frequency above the  {band gap of the biased} sample leading to a swift change in the polarization and hence the emission of a THz pulse. In Ref.~\cite{You1993}  {an almost unipolar pulse was produced, with duration of 1 ps and peak} amplitude of $\approx 150~$kV/cm.\label{fig:pulse3} }
\end{figure*}

 {Another}  {method to generate the appropriate pulses is to use
 photoconductive}  {(Auston)}  {switches \cite{Auston1983,Auston1984_2}.}  {The}  {schematics}  {is}  {shown in Fig.~\ref{fig:pulse3}:
  A}   {semiconductor-based}  {structure with short carrier lifetime, for instance GaAs or silicon on sapphire, is biased
with tens of}  {volts}  {amounting to  an electric field of few kV/cm acting across the photoconductive area (cf. Fig.~\ref{fig:pulse3}).
  The switch is then
 electrically shortened by a
femtosecond laser pulse with a frequency above the band gap  {of the biased semiconductor}, resulting in {the generation of} free carriers and}  {their following votage-induced}  {acceleration. This process leads to
 an abrupt  polarization change which goes along with the emission of
 a sub-picosecond, single-cycle coherent electromagnetic pulse that propagates  along the electrodes and in free
space with a  polarization being predominantly along the bias field. The free-space pulses are time-asymmetric {, as} evident from the way they are generated (cf. Fig.\ref{fig:pulse3}). Yet, the integral}  {of}  {the amplitude of the electric field that propagates
 in free-space}  {over its full duration}   {vanishes. The}  {temporal asymmetry of these pulses} {is essential for
  a number}  {of phenomena discussed in this review from the theory point of view,}  {such as the impulsive driving of charge and spin.}
    {Relevant}  {experimental demonstrations}  {exist in the field of atomic physics and}  {include the impulsive  ionization  and the controllable steering of  wave packets in Rydberg states of atoms
\cite{Jones1993,You1993,bensky,Jones1996,Dunning2005,Mestayer2008,Mestayer2007}. The formal theory behind this type of dynamics is reviewed in this work.
 For further discussions of the Auston-switch-type technique for generating  pulses
 as well as  for antenna geometries other than the one shown in Fig. \ref{fig:pulse3}  (such as  interdigitated structures,  bow-tie,
and spiral antennas) we refer to the dedicated literature, for instance
\cite{Sakai2010,Mittleman2013,Shen2003,Sartorius2008}.}

 {Here we}  {will be also concerned with}  {high-field}  {pulses triggering}  {excitation} {processes} {which are strongly nonlinear in the field strength.}  {Such pulses were accomplished
by using miniaturized interdigitated metal-semiconductor-metal structures  \cite{Dreyhaupt2005,Beck2010}  or by enlarging the photoconductive antenna area
(up to  cm) and increasing the bias voltage (up to several kV) leading to pulse energies in the range of $\mu$J \cite{Jones1993,You1993}} {, which}  {were
demonstrated to cause field ionization of Rydberg states.}

 {A further way to generate intense single-cycle THz pulses relies on nonlinear processes in gas plasmas subjected to}  {an}
 {intense femtosecond
laser \cite{Hamster1993,Roskos2007,Karpowicz2009}.
As a gas   ambient air, nitrogen or a noble gas  were utilized.
By this method
 THz pulses with frequency band}  {extending}  {up to 100 THz}  {and supporting a sub-20-fs duration} {were reported \cite{Thomson2010}.
Also optical rectification of}  {conventional (relatively narrow-band) picosecond or}  {femtosecond
laser pulses \cite{Bass1962,Yang1971},}  {in electro-optical materials like}  {LiNbO$_3$} {, ZnTe, GaP, or GaSe, under appropriate conditions can lead}  {to ultrashort}  {broadband}  {THz pulses}  {suitable for the aims of}  {this work.}   {A further promising route relies on novel metallic spintronic emitters, which produce femtosecond pulses  covering the 1--30 THz range with flat spectral amplitude and phase \cite{Seifert2016}.}
%
%
%
%

\section{Theoretical description of the unitary evolution}\label{Sec:Theory}
In this section we develop a systematic description and
approximation schemes  for the  dynamics of an electronic quantum
system driven by ultrashort pulses of electromagnetic radiation.
 Let us consider a
general system described by the Hamiltonian $H_0$ which is
subjected at $t=t_{1}$ to an electromagnetic pulse. The
 pulse duration is $\tau_{\mathrm{d}}$.
   The system evolves without time-dependent external forces from
a time moment $t_0$ before the pulse application.
For brevity we may choose $t_0=0$, i.e., the evolution involving external driving  is
prescribed by the  operator $U(t,0)$ that satisfies the
equation of motion
\begin{equation}\label{Eq:evol}
    i\hslash \frac{\partial U(t,0)}{\partial
    t}=[H_0+V(t)]U(t,0)\;,
\end{equation}
where $V(t)$ describes the  coupling of the  pulse  to
the system. For clarity of notation we do not explicitly indicate
spatial coordinates, unless deemed necessary.
To separate the field-free propagation before
and after the pulse we write the evolution operator in the form
($t>t_1$)  \cite{Henriksen1999,Daems2004,Matos_Indian_paper}
\begin{equation}\label{Eq:erel}
    U(t,0)=U_{0}(t,t_{1})\mathcal{U}(t,t_{1},0)U_{0}(t_{1},0)\;,
\end{equation}
where $U_{0}(t,t')\equiv U_{0}(t-t')=\exp\left[-iH_0(t-t')/\hslash\right]$
is  the evolution operator of the unperturbed system in the time
interval from $t'$ to $t$ and $\mathcal{U}$ is yet to be determined. The unitarity of $U_{0}$ dictates  that
\begin{equation}\label{Eq:erel_recipr}
\mathcal{U}(t,t_{1},0)=U_{0}^{\dag}(t,t_{1})U(t,0)U_{0}^{\dag}(t_{1},0)
\end{equation}
applies. Inserting
Eq.~\eqref{Eq:erel_recipr} into Eq.~\eqref{Eq:evol} we infer  a
 relation  for $\mathcal{U}(t,t_{1},t_0)$ that
can be written   formally as
\begin{equation}\label{Eq:U_time_ordered}
    \mathcal{U}(t,t_{1},0)=\hat{\textrm{T}}\exp\left[
    \int_{-t_{1}}^{t-t_{1}} A(t',t_{1},0) dt'\right],
\end{equation}
where
\begin{equation}\label{Eq:A}
    A(t,t_{1},0)=-\frac{i}{\hslash}
    e^{iH_0t/\hslash}V(t+t_{1})e^{-iH_0t/\hslash}
\end{equation}
and $\hat{\textrm{T}}$ is the time-ordering operator. It was
shown by Magnus \cite{Magnus1954} and thoroughly discussed and
illustrated in  following works
\cite{Pechukas1966,Wilcox1967,Blanes2009} that
Eq.~\eqref{Eq:U_time_ordered} can be expressed in a form that
does not involve  time ordering by writing
\begin{equation}\label{Eq:U_exp}
    \mathcal{U}(t,t_{1},0)=\exp\left[\Omega(t,t_{1},0)\right],
\end{equation}
where
\begin{equation}\label{Eq:Omega}
    \Omega(t,t_{1},0)=\sum_{k=1}^\infty \Omega_k(t,t_{1},0)\;
\end{equation}
with the first two terms of this \textit{Magnus expansion} given
by
\begin{eqnarray}
    \Omega_1(t,t_{1},0)&=&\int_{-t_{1}}^{t-t_{1}}\!\!dt'\: A(t',t_{1},0)\;, \label{Eq:Omega1}\\
    \Omega_2(t,t_{1},0)&=&\frac{1}{2}\int_{-t_{1}}^{t-t_{1}}\!\!dt'\int_{-t_{1}}^{t'}\!\!dt''\: [A(t',t_{1},0),A(t'',t_{1},0)]\;.
    \label{Eq:Omega2}
\end{eqnarray}
In many studies  considerations are limited to  only the first
term in this expansion
\cite{Henriksen1999,Matos_Indian_paper,Dimitrovski2005,Dimitrovski2006}
which  is just equivalent to  neglecting the time ordering in
Eq.~\eqref{Eq:U_time_ordered}. However, such an approach alone does not
allow for a consistent expansion in the
pulse duration which is necessary to unveil the range of validity of this doing.

\subsection{Unitary perturbation expansion in powers of the pulse
duration}\label{Sec:Unitary_expansion}

To proceed further we  use the Baker-Hausdorff operator
identity
\begin{equation}\label{ident}
    e^{-X}Ye^{X}=Y+[Y,X]+\frac{1}{2!}\big[[Y,X],X\big]+\ldots
\end{equation}
for $A(t,t_{1},0)$ in Eq.~\eqref{Eq:A} and write
\begin{equation}\label{Eq:A_expand}
    A(t,t_{1},0)=-\frac{i}{\hslash}V(t+t_{1})-\frac{1}{\hslash^2}t
    [V(t+t_{1}),H_0]+\frac{i}{2\hslash^3}t^2
    \big[[V(t+t_{1}),H_0],H_0\big]+\ldots\;,
\end{equation}
\begin{equation}\label{Eq:AA_expand}
   \begin{split}
    [A(t,t_{1},0),A(t',t_{1},0)]=&-\frac{1}{\hslash^2}[V(t+t_{1}),V(t'+t_{1})]\\
    &+\frac{i}{\hslash^3}\Big\{V(t\!+\!t_{1})t'[V(t'\!+\!t_{1}),H_0]-t[V(t\!+\!t_{1}),H_0]V(t'\!+\!t_{1})\Big\}
    +\ldots\;.
   \end{split}
\end{equation}
The coupling of the external field to the system  $V(t)$ can be usually factorized as
\begin{equation}\label{Eq:V}
   V(t)=V_0 f(t)\;,
\end{equation}
where $V_0$ is a time-independent Hermitian operator and $f(t)$ is
a dimensionless time-dependent function determining the temporal
profile of the excitation. We can rewrite so
Eqs.~\eqref{Eq:A_expand} and \eqref{Eq:AA_expand} as
\begin{equation}\label{Eq:A_expand_case}
    A(t,t_{1},0)=f(t+t_1)\left\{-\frac{i}{\hslash}V_0-\frac{1}{\hslash^2}t
    [V_0,H_0]+\frac{i}{2\hslash^3}t^2
    \big[[V_0,H_0],H_0\big]+\ldots\right\}\;,
\end{equation}
\begin{equation}\label{Eq:AA_expand_case}
    [A(t,t_{1},0),A(t',t_{1},0)]=\frac{i}{\hslash^3}f(t+t_{1})f(t'+t_{1})(t'-t)\big[V_0,[V_0,H_0]\big]
    +\ldots\;.
\end{equation}

Using these equations in Eqs.~\eqref{Eq:Omega1} and
\eqref{Eq:Omega2} and including all terms up to
$\mathcal{O}(\tau^3_{\rm d})$ in Eq.~\eqref{Eq:Omega} we find
\begin{equation}\label{Eq:U_result}
\begin{split}
    \mathcal{U}(t,t_{1},0)=\exp\bigg[&-\frac{i\tau_{\rm d}}{\hslash}s_1V_{0}-\frac{\tau_{\rm
    d}^2}{\hslash^{2}}s_2[V_0,H_0]\\
    &+\frac{i\tau_{\rm d}^3}{2\hslash^{3}}s_3\big[[V_0,H_0],H_0\big]-\frac{i\tau_{\rm d}^3}{4\hslash^{3}}s'_3\big[V_0,[V_0,H_0]\big]+\ldots
    \bigg]\;\;,
\end{split}
\end{equation}
where
\begin{equation}\label{Eq:s_n}
    s_n=\frac{1}{\tau_{\rm d}^n}\int_{0}^{t}\!\! dt'\: (t'-t_1)^{n-1} f(t'),\ \ \
    n=1,2,3,
\end{equation}
\begin{equation}\label{Eq:s_33}
    s'_3=\frac{1}{\tau_{\rm d}^3}\int_{0}^{t}\!\! dt'\int_{0}^{t}\!\! dt''\: |t'-t''|
    f(t')f(t'')\;.
\end{equation}
If there is no electric field (or it is sufficiently low) at the
initial time moment and the final time moment is selected late
enough, we can write
\begin{equation}\label{Eq:s_n_infinity}
    s_n=\frac{1}{\tau_{\rm d}^n}\int_{-\infty}^{\infty}\!\! dt'\: (t'-t_1)^{n-1} f(t'),\ \ \
    n=1,2,3,
\end{equation}
\begin{equation}\label{Eq:s_33_infinity}
    s'_3=\frac{1}{\tau_{\rm d}^3}\int_{-\infty}^{\infty}\!\! dt'\int_{-\infty}^{\infty}\!\! dt''\: |t'-t''|
    f(t')f(t'')\;.
\end{equation}
The coefficients $s_n$ and $s'_3$ are dimensionless factors.
Due to the hermiticity of $V_0$, the evolution operator given by
Eq.~\eqref{Eq:U_result} is unitary up to the selected order in
$\tau_{\rm d}$. In this description, when we are not interested in
the dynamics of the system in the short time range during the
pulse, the total evolution of the system can be summarized as a
free evolution before the time moment $t_1$, the momentary action
of the pulse, and the free evolution afterwards. The momentary
action of the pulse is given by mapping the wave function of the
system at the time moment just before $t=t_1$ to the wave function
at the time moment just after $t=t_1$:
\begin{equation}\label{Eq:IA_mapping_psi}
    \Psi(t=t_1^+)=\mathcal{U}(t_{1})\Psi(t=t_1^-)\; .
\end{equation}
Equivalently, we can describe  the system by a density matrix
$\rho(t)$ and the corresponding mapping reads then
\begin{equation}\label{Eq:IA_mapping_rho}
    \rho(t=t_1^+)=\mathcal{U}(t_{1})\rho(t=t_1^-)\mathcal{U}^\dagger(t_{1})\;,
\end{equation}
where $\mathcal{U}(t_{1})\equiv \mathcal{U}(\infty,t_{1},0)$ is
given by Eq.~\eqref{Eq:U_result} with Eqs.~\eqref{Eq:s_n_infinity}
and ~\eqref{Eq:s_33_infinity}.

In such a treatment the question arises as  how  to select in theory the
time moment $t_1$  to achieve   simplicity in
 description while maintaining accuracy. This depends generally on
the shape of the applied light pulse. The first order term in the
exponent of Eq.~\eqref{Eq:U_result} is determined by $s_1$ as given
 by Eq.~\eqref{Eq:s_n_infinity} for $n=1$ and hence it is
independent of $t_1$. If  $s_1$  is finite then the time moment
$t_1$ should be selected such that the second order term
governed by $s_2$ vanishes. This is always possible by
picking the value of $t_1$ at the center of gravity of the experimentally  applied
pulse. If the first order term is zero, the
second order term does not depend on $t_1$.  A reasonable choice
would be to select $t_1$ such that the absolute value of $s_3$ is
minimized. The third order term determined by $s_3'$ does not
depend on $t_1$.

Notice that  pulses with nonzero value of $s_1$ are not
possible for freely propagating light beams in the far field
\cite{Kaplan1998}. However they can be generated in the near field, close
to the emitter or to a proper nonlinear optical element
transforming the incident wave, as well as in a waveguide
configuration \cite{Guertler2000,Kozlov2011,Yamashita_book}.

With the help of the Zassenhaus formula for disentanglement of
exponential operators \cite{Wilcox1967}
\begin{equation}\label{Eq:Zassenhaus}
    e^{\tau(X+Y)}=e^{\tau X}e^{\tau Y}e^{-\frac{\tau^2}{2}[X,Y]}
    e^{\frac{\tau^3}{6}\left\{2[Y,[X,Y]]+[X,[X,Y]]\right\}}e^{\mathcal{O}(\tau^4)}\;,
\end{equation}
where $\tau$ is a small number, it is possible to rewrite the
exponential of the sum of the operators in Eq.~\eqref{Eq:U_result}
as a product of exponentials, e.g., as
\begin{equation}\label{Eq:U_result_Zassenhaus}
\begin{split}
    \mathcal{U}(t_{1})=&\exp\!\bigg[-\frac{i\tau_{\rm d}}{\hslash}s_1V_{0}\bigg]\exp\!\bigg[-\frac{\tau_{\rm
    d}^2}{\hslash^{2}}s_2[V_0,H_0]\bigg]
    \exp\!\bigg[\frac{i\tau_{\rm d}^3}{2\hslash^{3}}s_3\big[[V_0,H_0],H_0\big]\bigg]\\
    &
    \times\exp\!\bigg[-\frac{i\tau_{\rm d}^3}{4\hslash^{3}}(s'_3-2s_1s_2)\big[V_0,[V_0,H_0]\big]
    \bigg]\exp[\mathcal{O}(\tau^4_{\rm d})]\;.
\end{split}
\end{equation}
In practice, as mentioned above, we either have $s_1$ equal to
zero, or if it is not true then $t_1$ is selected such that $s_2$
is zero, or they are both independent of $t_1$ and equal to zero.
In all these cases the product $s_1s_2$ vanishes and the third
order terms in $\tau_{\rm d}$ are determined by the pulse shape
parameters $s_3$ and $s'_3$ only.  {Equation \eqref{Eq:U_result_Zassenhaus} may be also obtained based on the Wilcox product expansion  \cite{Wilcox1967,Blanes2009} in place of the Magnus expansion. Both expansions are, of course, closely related.}

Let $H_0$ be the atomic Hamiltonian. We employ the
light-electron interaction within the dipole approximation and in the
length gauge \cite{Yamanouchi_book,Faisal_book}.   Assuming the
light pulses to be confined to a finite time range and setting $t_1$ to
be the initial time moment of this range, we find that in such a
case our result given by Eq.~\eqref{Eq:U_result}  coincides exactly
with the result of Refs.~\cite{Klaiber2008,Klaiber2009}.
Also all conclusions of Refs.~\cite{Klaiber2008,Klaiber2009}
concerning the unitarity of the derived expression for the
evolution operator, the equivalence between the length and the
velocity gauges \cite{Yamanouchi_book,Faisal_book} of the
interaction Hamiltonian, its convergence to the result of the
time-dependent perturbation theory for the case of small field
amplitudes, and its relation to the sudden-perturbation expansion
of Ref.~\cite{Dykhne1978}, are valid in the more general case
considered here.

The approximation when only the first factor on the right hand side (rhs) of Eq.~\eqref{Eq:U_result_Zassenhaus} [or equivalently only the first term in the exponent of Eq.~\eqref{Eq:U_result}] is taken into account (assuming $s_1\neq 0$) is called impulsive approximation (IA), generalizing the generic case of a light-driven electronic motion mentioned in the Introduction.
Generally, we will inspect the appropriate choice of $t_1$   which removes the second order correction.
In the next sections at various places we will detail the physical aspects associated with the IA.
  {The IA is of a fundamental character providing the appropriate approximate description for an important limit case of the excitation of quantum systems.
One can draw an analogy to the conventional time-dependent perturbation theory (TDPT). Whereas for the TDPT the strength of the perturbation is used as a small parameter and to  the first order the Fermi's golden rule results, it is the vanishing pulse duration that is essential in this respect for the reviewed method and the first order gives the IA. }

 {An important aspect underlying the ansatz \eqref{Eq:erel} is the assumption that the action of the evolution operator of the unperturbed system $U_0(t,t')$ on a given state is  easy to evaluate analytically or to compute numerically. The same should be valid also for the operator $\mathcal{U}(t_{1})$ determining the effective instantaneous action of the excitation pulse. In general, especially for complex many-body systems, already the computation of the free evolution of an excited system might be demanding. Even more complicated might be numerically exact simulations of the dynamics governed by the full time-dependent Hamiltonian which includes the pulsed driving. This task arises, e.g., if a comparison between the approximate and numerically exact solutions is required. For advanced numerical methods, which can be used in this context, such as the split-operator and other (higher-order) splitting (Suzuki-Trotter) schemes, non-equilibrium Green's function approach, and Magnus integrators, we refer the reader to the corresponding specialized literature \cite{Tannor_book,Kosloff1988,Gaury2014,Blanes2009,Castro2004} and references therein. Some of these methods require approximate evaluation of the operator exponential at each propagation step\footnote{For $U_0(t,t')$ it may correspond to the whole time interval.}. Typically it is much easier to evaluate the result of the application of the operator exponential to a particular given state than to find an appropriate approximation for the exponential itself. Here polynomial expansions with fixed coefficients, e.g. based on Chebychev polynomials, or the iterative Lanczos method generating an orthogonal basis in the corresponding Krylov subspace \cite{Hochbruck1997} are commonly used to approximate the action of the exponential operator on the state \cite{Tannor_book,Blanes2009,Castro2004}. In principle, in case of the reviewed approach these methods would be relevant for the quantum mapping, Eqs.~\eqref{Eq:IA_mapping_psi} and \eqref{Eq:IA_mapping_rho},  determined by $\mathcal{U}(t_{1})$.
However, the first order in $\tau_{\rm d}$ in the exponent of Eq.~\eqref{Eq:U_result}, i.e. the result of the IA, depends solely on the interaction part $V(t)$ of the total Hamiltonian. In many cases, e.g., when we deal with the light-matter interaction in the dipole approximation, $\mathcal{U}(t_1)$ is then just a multiplication operator. Therefore, its action on any state has a simple analytical description, which is a pronounced strength of the considered approach. For more sophisticated interactions or higher orders in $\tau_{\rm d}$  one might encounter a situation when computing the corresponding exponential term would in fact represent a certain numerical problem (see, e.g., Section~\ref{Sec:1d_ring}).}

\begin{figure*}
  \includegraphics[width=0.95\textwidth]{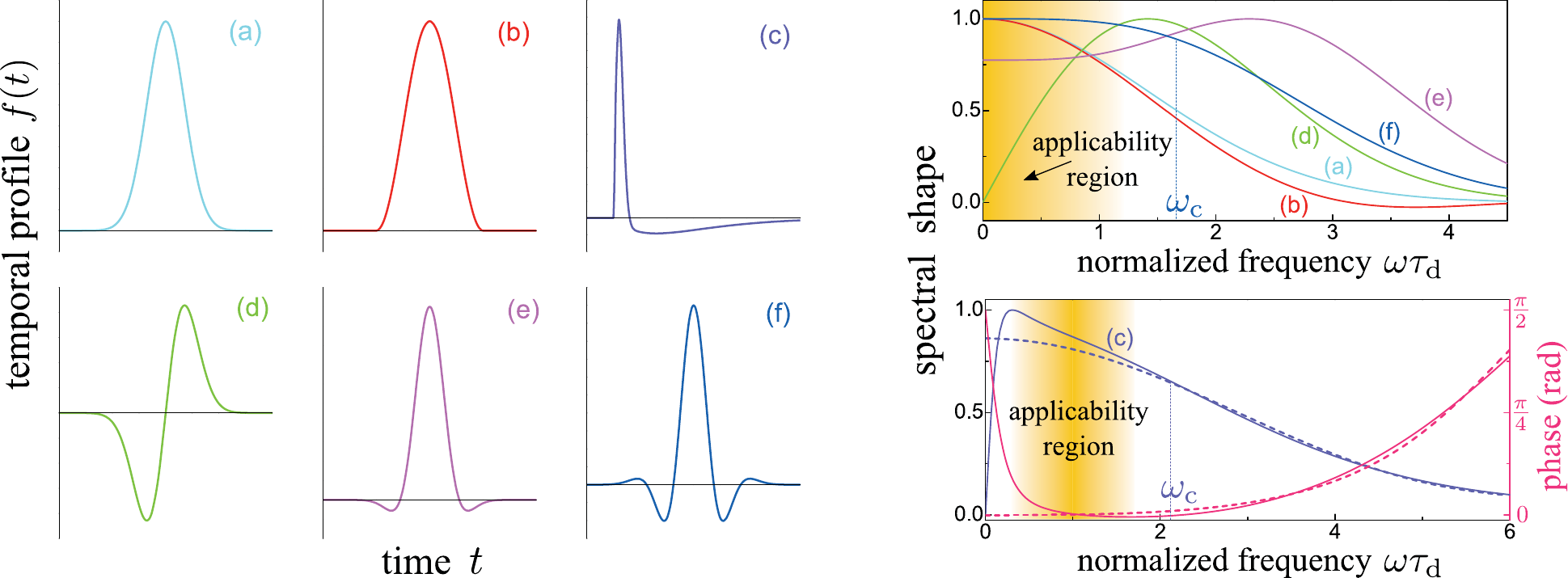}
  \caption{Left panel shows model temporal shapes of broadband ultrashort light pulses $f(t)$: (a) Gaussian profile,
  (b) sine-square profile, (c) strongly asymmetric HCP as discussed in Section~\ref{Sec:Strongly_as_HCPs},
  (d) single-cycle pulse as discussed
  in Section~\ref{Sec:single_cycle}, (e) harmonic few-cycle
  pulse with a Gaussian envelope as discussed in Section~\ref{Sec:Harmonic_Gauss},
  and (f) polynomial few-cycle
  pulse with a Gaussian envelope as discussed in Section~\ref{Sec:Polynom_Gauss}.
  Right panel illustrates the spectral properties of these pulses. Top plot displays the normalized Fourier transform $f_\omega$
  for the cases (a), (b), (e) and (f), whereas for the case (d) the normalized spectral amplitude $|f_\omega|$ is shown (the corresponding spectral phase is constant and equals to $-\pi/2$). Position of the central frequency $\omega_\mathrm{c}$ is marked for (f).
  Bottom plot: spectral amplitude $|f_\omega|$ (blue solid line, central frequency $\omega_\mathrm{c}$) and spectral phase (pink solid line) for the case (c), spectral amplitude $|\tilde{f}_\omega|$ (blue dashed line) and spectral phase (pink dashed line) for the positive half-cycle of (c).
  Yellow color in both plots of the left panel indicates the interval where the relevant transition frequencies of the driven system should be situated for the applicability of the IA and expansion \eqref{Eq:U_result}.\label{Fig:time_profiles}}
\end{figure*}

Next, for  illustrations  let us   discuss the pulse shape parameters for
several typical model broadband ultrashort pulses.
We note from the outset that the profiles shown in Fig.~\ref{Fig:time_profiles} and Fig.~\ref{Fig:time_profile_fc} are the most generic ones for our purpose:
The duration of the pulse should be below the characteristic time scales of the system. So consider, for instance, the electric field of a moderate intensity pulse
in Fig.~\ref{Fig:time_profiles}f consisting of few oscillation cycles limited by a quickly decaying envelope.
In the case relevant for our study the frequency of these oscillations is comparable with the bandwidth of the pulse and is located beyond
the relevant frequency spectrum of the system  (cf.  Fig.~\ref{Fig:time_profiles}).
 On the other hand, if the oscillation frequency is lowered entering the  characteristic frequency spectrum  of the system whereas the number of oscillation cycles is kept the same, then the envelope can be considered adiabatic and this type of envelope has only a marginal
effect on the physics. This effectively continuous time-periodic wave  case is well captured either numerically or by means of the well-documented Floquet approach. We are interested in  (ultra) broadband short-pulse excitations. Examples of such pulses are illustrated in Fig.~\ref{Fig:time_profiles}.


\subsection{Half-cycle pulses (HCPs)}\label{Sec:HCPs}
An excitation by HCPs is the most widely used form of ultrafast
broadband excitations, both in theory and experiment.
Theoretically, the simplest consideration is based on their
description by just a delta function in time, i.e., a pulse with a
zero duration, imposing a kick to the excited system. Below we
consider some more realistic model temporal profiles beyond this
simplification.

\subsubsection{Gaussian temporal profile}\label{Sec:Guassian_HCP}
One example for the temporal profile of a HCP is a Gaussian
shape given by
\begin{equation}\label{Eq:Gaussian_pulse}
    f(t)=\exp[-t^{2}/\tau_{\rm d}^2]\;,
\end{equation}
where the time parameter $\tau_{\rm d}$ characterizes the pulse
width. For such a pulse we calculate using Eq.~\eqref{Eq:s_n_infinity} that
\begin{equation*}\label{Eq:Gaussian_pulse_s_1_s2}
    s_1=\sqrt{\pi}\ \ \ {\rm and}\ \ \ s_2=-\sqrt{\pi}\frac{t_1}{\tau_{\rm
    d}}\;.
\end{equation*}
In order to minimize the second order contribution, we select then
$t_1=0$ leading to $s_2=0$. For this selection we calculate
\begin{equation*}\label{Eq:Gaussian_pulse_s_3}
    s_3=\frac{\sqrt{\pi}}{2} \ \ \ {\rm and}\ \ \
    s'_3=\sqrt{2\pi}\;.
\end{equation*}
The Gaussian temporal profile is shown in
Fig.~\ref{Fig:time_profiles}a.

\subsubsection{Sine-square temporal profile}\label{Sec:sine_square_HCP}
Another frequently used shape representing HCPs is the sine-square
temporal profile:
\begin{equation}\label{Eq:sine_pulse}
    f(t)=\sin^2(\pi t/\tau_{\rm d}) \ \ \ {\rm for}\ 0<t<\tau_{\rm d},\ \ f(t)=0 \ \ \ {\rm else}.
\end{equation}
For this type of pulses we get
\begin{equation*}\label{Eq:sine_pulse_s_1_s2}
    s_1=\frac{1}{2}, \ \ \ s_2=\frac{1}{2}\left(\frac{1}{2}-\frac{t_1}{\tau_{\rm d}}\right).
\end{equation*}
Selection of $\;t_1\!=\!\tau_{\rm d}/2\;$ leads to $s_2=0$ and
\begin{equation*}\label{Eq:sine_pulse_s_3}
    s_3=\frac{\pi^2-6}{24\pi^2}\:, \ \ s'_3=\frac{4\pi^2-15}{48\pi^2}\:.
\end{equation*}
These pulses belong to the type (i) pulses following the
classification of Ref.~\cite{Klaiber2009}. The sine-square
temporal profile is shown in Fig.~\ref{Fig:time_profiles}b.

\subsubsection{Strongly asymmetric HCPs}\label{Sec:Strongly_as_HCPs}
Experimentally generated HCPs can be highly asymmetric like in
Refs.~\cite{You1993,Jones1993}. The electric field of such a
HCP performs the first strong, asymmetric oscillation half-cycle
that is followed by a much longer, much weaker and also asymmetric
second half-cycle of the opposite polarity. The integral over the
whole temporal profile of the field vanishes, as it should be at
large distances from the light emitter for a freely propagating
light pulse \cite{Kaplan1998,Guertler2000,Kozlov2011}. The case relevant for the interaction of such broadband, short
pulses with a quantum system is that when the first half-cycle is much shorter than
the characteristic transition time scales (reciprocal transition
frequencies) of the system whereas the second half-cycle is considerably longer than them. Both the long duration and the
weakness of the latter half-cycle lead to the smallness of the
spectral components of the field at the transition frequencies
of the driven system.
As a consequence, the action of the second half-cycle on the
system can be neglected with respect to the impact of the short and
strong first half-cycle. Therefore the $s$-factors should be calculated by integrating merely over the
first half-cycle. This justifies the name
``half-cycle'' for the asymmetric light pulses of this type. A
numerical example demonstrating the limits of the applicability of
the sketched approach is provided in Section~\ref{Sec:1d_ring} for
the case of  light-driven quantum rings. In order to model
realistic temporal profiles of the experimentally generated
pulses~\cite{You1993,Jones1993} the following function was suggested in Ref.~\cite{Moskalenko_PRB2006}:
\begin{equation}\label{Eq:pulse_shape}
   f(t)=\frac{t}{\tau_0}\left[\exp\left(-\frac{t^2}{2\tau_0^2}\right)
   -\frac{1}{b^2}\exp\left(-\frac{t}{b\tau_0}\right)\right] \ \ \ {\rm for}\ t>0,\ \ f(t)=0 \ \ \ {\rm else}.
\end{equation}
The parameters $\tau_0$ and $b$ determine respectively the
duration and the asymmetry of the HCP. The duration $\tau_{\rm d}$
of the positive half-cycle is calculated to $\tau_{\rm
d}=\tau_0(1+\sqrt{1+2b^2\ln b^2})/b$. By choosing $b=8$ we ensure
that the pulse determined by Eq.~\eqref{Eq:pulse_shape} has
approximately the experimentally observed ratio 13:1 between the
maximum field values of the positive and negative parts
\cite{Jones1993} and therefore the chosen shape of the pulse
reproduces the main features of the experimentally generated HCPs.
For the selected value of $b$ we obtain $\tau_{\rm d}=3.012\tau_0$. The
maximum value $\max\!\left[ f(t)\right]=0.593$ is achieved at
$t_{\rm max}=0.329\tau_{\rm d}$. One should use
Eqs.~\eqref{Eq:s_n} and \eqref{Eq:s_33}
with $t=\tau_{\rm d}$ for the calculation of
the $s$-factors.
We get $s_1=0.934$. Setting
$t_1=0.395\tau_{\rm d}$ results in $s_2=0$, $s_3=0.323$ and
$s'_3=0.583$. The temporal profile of a light pulse corresponding to
this choice of parameters is depicted in
Fig.~\ref{Fig:time_profiles}c. Its spectral properties as well as those of its short and strong positive half-cycle $\tilde{f}(t)$
are illustrated in the right bottom plot of Fig.~\ref{Fig:time_profiles}, whereby the origin of the time axis for the corresponding Fourier transforms has been shifted to $t_1$.
Note that  the selected value of $t_1$ does not coincide
with $t_{\rm max}$ because of the asymmetry of the
pulse shape. A marginal drawback of the function
\eqref{Eq:pulse_shape} for  modelling of
temporal profiles of realistic HCPs is  the non-smoothness at $t=0$, which however,
 has practically no effect on the resulting $s$-factors given by Eqs.~\eqref{Eq:s_n_infinity} and
\eqref{Eq:s_33_infinity}.


\subsection{Single-cycle pulses}\label{Sec:single_cycle}
Another characteristic case is that of a light pulse with
an electric field  performing exactly one oscillation cycle.
As an example let us  consider a pulse having the Gaussian temporal profile
\eqref{Eq:Gaussian_pulse} in the near field. In the far field the
on-axis electric field replicates the time derivative of the
original pulse \cite{Kaplan1998}. Therefore, the initially
Gaussian temporal profile of the field transforms to
\begin{equation}\label{Eq:single_cycle_pulse}
    f(t)=t/\tau_{\rm d}\exp[-t^{2}/\tau_{\rm d}^2]
\end{equation}
with a vanishing electric field area: $s_1=0$.
Consequently, the parameter $s_2$ does not depend on $t_1$ and we
find
\begin{equation*}\label{Eq:single_cycle_pulse_s_2t1}
    s_2=\frac{\sqrt{\pi}}{2}\;.
\end{equation*}
The third order parameters are
\begin{equation*}\label{Eq:single_cycle_pulse_s_3}
    s_3=-\sqrt{\pi}\frac{t_1}{\tau_{\rm d}}\:, \ \ s'_3=-\frac{\sqrt{2\pi}}{4}\:.
\end{equation*}
Here it is natural to select $t_1=0$ so that $s_3$ is zero. These
pulses belong to the type (ii) pulses discussed in
Ref.~\cite{Klaiber2009}. They are illustrated in
Fig.~\ref{Fig:time_profiles}d.

\subsection{Few-cycle pulses}\label{Sec:few_cycle}
In many cases ultrashort light pulses are generated experimentally
in the form of few-cycle pulses. Here, we consider three theoretical
model functions for their temporal profile.
\subsubsection{Harmonic with a Gaussian envelope}\label{Sec:Harmonic_Gauss}
The temporal profile of such a pulse is given by
\begin{equation}\label{Eq:few_cycle_pulse}
    f(t)=\exp[-t^{2}/\tau_{\rm d}^2]\cos(\Omega t+\Phi),
\end{equation}
where $\Omega$ is the central frequency of the pulse, $\tau_{\rm}$
determines the temporal width of its envelope, and $\Phi$ is the carrier-envelope phase. In this
case we calculate
\begin{equation*}\label{Eq:few_cycle_pulse_s_1}
    s_1=\sqrt{\pi}\exp\left(-\frac{\Omega\tau_{\rm
    d}}{4}\right)\cos\Phi\;,
\end{equation*}
\begin{equation*}\label{Eq:few_cycle_pulse_s_2t1}
    s_2=-\sqrt{\pi}\left[\frac{t_1}{\tau_{\rm d}}\cos\Phi
             +\frac{1}{2}\Omega\tau_{\rm d}\sin\Phi\right]\exp\left(-\frac{\Omega\tau_{\rm d}}{4}\right).
\end{equation*}
Two cases should be differentiated: $\cos\Phi=0$ and $\cos\Phi\neq
0$.

If light pulses with $\cos\Phi=0$ are applied we get $s_1=0$. In
this case $s_2$ is independent of $t_1$ and is given by
\begin{equation*}
    s_2=-\sin\Phi\frac{1}{2}\sqrt{\pi}\Omega\tau_{\rm d}\exp\left(-\frac{\Omega\tau_{\rm
             d}}{4}\right),
\end{equation*}
where $\sin\Phi$ is just 1 or -1. Further, we get
\begin{equation*}
    s_3=-\sin\Phi\sqrt{\pi}\Omega t_1\exp\left(-\frac{\Omega\tau_{\rm
             d}}{4}\right),
\end{equation*}
which can be made exactly zero by setting $t_1=0$. The factor
$s'_3$ can be calculated numerically as a function of the
parameter $\Omega\tau_{\rm d}$. We found that $s'_3$ is always
negative for all possible values of the pulse parameters so that
the corresponding third order term is always present for the
pulses of the considered type.

For light pulses with $\cos\Phi\neq 0$, implicating that
$s_1\neq 0$, we can select
\begin{equation*}
    t_1=-\frac{\tau_{\rm d}}{2}\Omega\tau_{\rm d}\tan\Phi
\end{equation*}
and get $s_2=0$. Then we also have
\begin{equation*}
    s_3=\frac{1}{4}\sqrt{\pi}\frac{2\cos^2\Phi-(\Omega\tau_{\rm d})^2}{\cos\Phi}\exp\left(-\frac{\Omega\tau_{\rm d}}{4}\right).
\end{equation*}
In this case the factor $s'_3$ can be again calculated numerically
but now it depends on two parameters: $\Phi$ and $\Omega\tau_{\rm
d}$. We see that $s_3=0$ for $\Omega\tau_{\rm
d}=\sqrt{2}|\cos\Phi|$. In particular, with $\cos\Phi=1$ the
electric field of the light pulse behaves in time as shown in
Fig.~\ref{Fig:time_profiles}e. It is also possible to achieve
$s'_3=0$ for another choice of parameters. However, it happens that we
can not get $s_3=0$ and $s'_3=0$ simultaneously for this pulse
type.

\subsubsection{Polynomial with a Gaussian envelope}\label{Sec:Polynom_Gauss}
Alternatively, we can model few-cycle pulses by
\begin{equation}\label{Eq:few_cycle_pulse_polynomial}
    f(t)=\exp[-t^{2}/\tau_{\rm d}^2]P(t/\tau_{\rm d})\:,
\end{equation}
where $P(x)$ is a polynomial \cite{Parali2010}. Selection of
appropriate polynomials allows for the engineering of the action
of the ultrashort pulse on the system as the coefficients
$s_1,s_2,s_3,$ and $s'_3$ are varied.

For example, it might be desirable  to generate a pulse with a
non-zero parameter $s_1$ and vanishing parameters $s_2$,$ s_3$, and
$s'_3$. In such a case the IA would give a
correct result up to the third order in $\tau_{\rm d}$,
inclusively. As an illustration, let us analyze
 the following fourth order polynomial
\begin{equation*}\label{Eq:Polynomial_IA}
   P(x)=1+ax^2+bx^4
\end{equation*}
depending on two parameters, $a$ and $b$. Selecting here
\begin{equation*}\label{Eq:Polynomial_ab}
   a={\textstyle -\frac{502}{123}+\frac{20\sqrt{154}}{123}}\approx
   -2.06, \ \ \ b={\textstyle \frac{56}{41}-\frac{8\sqrt{154}}{123}}\approx 0.559
\end{equation*}
we get
\begin{equation*}\label{Eq:Polynomial_s1}
   s_1={\textstyle -\frac{2\sqrt{\pi}}{123}+\frac{4\sqrt{\pi}\sqrt{154}}{123}}\approx 0.686
\end{equation*}
and $s_2=s_3=s_3'=0$. Thus, such a pulse would deliver a ``perfect
kick'' (i.e. in the simplest case provide just a transfer of momentum) to the excited quantum system at $t=0$ while all contributions
up to the third order in $\tau_{\rm d}$ are taken into account. The temporal profile of the
corresponding light pulse is shown in
Fig.~\ref{Fig:time_profiles}f.

\subsubsection{Frequency-domain model}\label{Sec:freq_dom_model}
\begin{figure*}
  \includegraphics[width=0.95\textwidth]{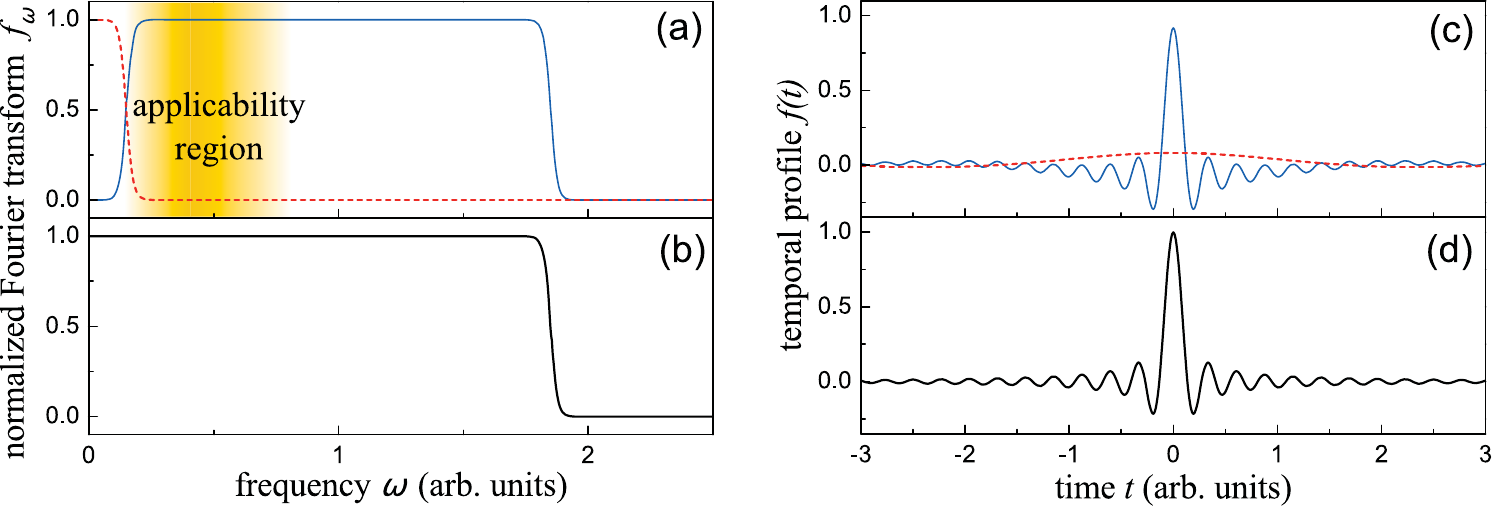}
  \caption{Normalized spectrum $f_\omega$ [blue solid line in (a)] of a model few-cycle pulse $f(t)$  [blue solid line in (c)].
  It can be split into auxiliary spectra $\tilde{f}_\omega$ (b) and $\bar{f}_\omega$ [red dashed line in (a)]. Correspondingly, $f(t)$  is constituted by a superposition of  quickly-varying $\tilde{f}(t)$ (d) and weak, slowly-varying  $\bar{f}(t)$ [red dashed line in (a)] waveforms.
  The applicability range in (a) is indicated analogously to Fig.~\ref{Fig:time_profiles}. \label{Fig:time_profile_fc}}
\end{figure*}
Some realistic few-cycle pulses, e.g., generated with the Er:fiber technology \cite{Brida2014}, can be appropriately modelled starting from their shape in the frequency domain \cite{Moskalenko2015}. Let us consider pulses with an almost rectangular shape of the
spectrum and a constant vanishing phase. The frequency- and time-domain properties for a typical case are illustrated in Fig.~\ref{Fig:time_profile_fc}. The Fourier transform $f_\omega$ (blue solid line in Fig.~\ref{Fig:time_profile_fc}a) can be viewed as a superposition
of two auxiliary spectra remaining flat in the long wavelength limit: $f_\omega=\tilde{f}_\omega-\bar{f}_\omega$. The main component $\tilde{f}_\omega$ (Fig.~\ref{Fig:time_profile_fc}b) coincides with $f_\omega$,  excluding the low frequency region where $f_\omega$ vanishes and the complimentary component $\bar{f}_\omega$ (red dashed line in Fig.~\ref{Fig:time_profile_fc}a) shows up, compensating the difference to $f_\omega$. Transforming into the time domain, one sees that the temporal profile of the pulse $f(t)$ (blue solid line in Fig.~\ref{Fig:time_profile_fc}c) is contributed by two waveforms: a quickly oscillating one with a central dominating half-cycle $\tilde{f}(t)$  (Fig.~\ref{Fig:time_profile_fc}d) and a slowly oscillating complimentary wave  $\bar{f}(t)$ (red dashed line in Fig.~\ref{Fig:time_profile_fc}c). The integral over the oscillating tails of $\tilde{f}(t)$ vanishes. Notice that even though $\tilde{f}(t)$  and $\bar{f}(t)$ do not exist separately as forms of propagating waves in the far field, their difference does describe a temporal profile allowed in this region. This situation has exciting connections  to the issue of the symmetry breaking and mass of gauge bosons in the near-field zone \cite{Keller_book}. The corresponding discussion is outside of the scope of the present report.

If the relevant transition frequencies of the driven system are situated substantially above the range of $\bar{f}_\omega$ (see Fig.~\ref{Fig:time_profile_fc}a) the dynamics of $\bar{f}(t)$ is essentially adiabatic.
Hence its impact averages to zero and the whole effect of the interaction is well determined solely by $\tilde{f}(t)$. The consideration of Section~\ref{Sec:Unitary_expansion} applies when $\bar{f}_\omega$ is broad enough (see Fig.~\ref{Fig:time_profile_fc}a). Then we can calculate
$$s_1=\tilde{f}_{\omega=0}\;, \quad s_2=s_3=0.$$
The parameter $s_3'$ is generally different from zero.

\subsection{Short broadband but very strong interaction case}\label{Sec:ultrastrong_theory}
The accuracy and even the validity of the presented approximation scheme based on the Magnus
expansion and leading to
Eq.~\eqref{Eq:U_result} depend in general  on the strength of the pulse.
 For instance, if more than one term of the series in
Eq.~\eqref{Eq:A_expand_case} is included then the second term of
the Magnus expansion \eqref{Eq:Omega2} contains the second order
of $\tau_{\rm d}V_0/\hslash$ [see Eq.~\eqref{Eq:AA_expand_case}], the third term
of the Magnus expansion contains the third order corrections in
$\tau_{\rm d}V_0/\hslash$ and so on.
For ultrashort pulses we demand \footnote{This condition is formulated  more precisely below for several specific examples.} that $\tau_{\rm d}$ is very short with respect to the characteristic time scales of the driven quantum system, implying that the energies of all involved quantum transitions are much smaller than $\hslash/\tau_{\rm d}$.
Generally, even if this condition is fulfilled the interaction determined by $V_0$ can be strong enough to invalidate the neglect
of the higher order terms in $\tau_{\rm d}V_0/\hslash$ and furthermore the convergence of the corresponding series.
In such a situation Eq.~\eqref{Eq:U_result} can not be considered as a good
approximation. Resorting to another approach is then more appropriate for a
correct unitary perturbation expansion of the evolution operator.
On the other hand,  the lowest possible
approximation, when only the first term in the expansion
\eqref{Eq:A_expand_case} is retained, can still deal with short but very
strong (SVS) interactions. In this case all terms in the Magnus
expansion except for the first term \eqref{Eq:Omega1} vanish.
Therefore, no higher order terms in $\tau_{\rm d}V_0/\hslash$ appear and the
result is determined by Eq.~\eqref{Eq:U_result} where only the
first term in the exponent is retained. This can be also
understood in a simple way just by considering that during the pulse
the Hamiltonian of the undriven system $H_0$ in
Eq.~\eqref{Eq:evol} can be viewed as a perturbation in comparison
with the interaction part {$V(t)=V_0f(t)$}.
This is the case when the energies
of all induced  quantum transitions  are much smaller than $V_0$
(if some relevant transition energy is comparable with $V_0$ then in the considered limit of ultrashort
pulses both of them must be much smaller than $\hslash/\tau_{\rm d}$ --- in such a case, in turn, the consideration
of Section \ref{Sec:Unitary_expansion} is applicable). The evolution of the system
during the pulse from the time moment $t'$ to the time moment $t$
in the lowest approximation would be determined by
$U(t,t')=\exp\left[-\frac{i}{\hslash}V_0\int_{t'}^t f(t'') {\rm d}
t'' \right]$. Outside of the time interval taken by the pulse the evolution is
dictated by $U_0$, whereas any time shift on the order of
$\tau_{\rm d}$ leads just to a negligible correction as far as all the relevant transition energies are much smaller than  $\hslash/\tau_{\rm d}$. Combining these facts we see again that
the dynamics will be determined by Eq.~\eqref{Eq:erel} and
Eq.~\eqref{Eq:U_result} accounting only for the first term in the exponent:
\begin{equation}\label{Eq:U_1}
\begin{split}
    \mathcal{U}(t,t_{1},0)\equiv U_1(t)=\exp\bigg[-\frac{i\tau_{\rm
    d}}{\hslash}s_1(t)V_{0}\bigg],
\end{split}
\end{equation}
where $s_1(t)$ is given by Eq.~\eqref{Eq:s_n}. This result corresponds to the IA.

The question remains is how to determine the next order correction in $\tau_{\rm d}H_0/\hslash$ beyond the IA. To this end we
 make the  ansatz for the evolution operator
\begin{equation}\label{Eq:Ansatz_V_strong}
    U(t,0)= U_0(t,t_1)U_1(t)U_2(t)U_0(t_1,0).
\end{equation}
 The operator $U_2(t)$ is supposed to encapsulate  the correction. For brevity we have omitted here the dependence of $U_2$ on $t=0$  and $t_1$.
 We insert this equation into
Eq.~\eqref{Eq:evol} and find after some transformations
\begin{equation}\label{Eq:U_2_evolution}
    \frac{\partial}{\partial
    t}U_2(t)=-\frac{i}{\hslash}U_1^{-1}(t)\left[U_0^{-1}(t,t_1)V(t)U_0(t,t_1)-V(t)\right]U_1(t)\;.
\end{equation}
Expanding the first term in the brackets on the rhs
using the Baker-Hausdorff operator identity \eqref{ident} and
keeping only the lowest order term in $(t-t_1)H_0/\hslash$ we arrive at
\begin{equation*}
    \frac{\partial}{\partial
    t}U_2(t)=-\frac{i}{\hslash}U_1^{-1}(t)\left[V(t),-\frac{i}{\hslash}H_0(t-t_1)\right]U_1(t)\;.
\end{equation*}
Under the assumption \eqref{Eq:V}  solution of this equation
reads
\begin{equation}\label{Eq:U_2_solution}
    U_2(t)=\exp\left[-\frac{1}{\hslash^2}\int_0^t (t'-t_1)f(t') U_1^{-1}(t')\left[V_0,H_0\right]U_1(t'){\rm d}t'\right].
\end{equation}
Notice that the operator $U_1^{-1}(t)\left[V_0,H_0\right]U_1(t)$
commutes with itself at different time moments. This fact allows
us to write Eq.~\eqref{Eq:U_2_solution} without the time-ordering
operator, meaning that only the first term of the Magnus expansion
is required. All other terms of this expansion vanish. This is a consequence of Eq.~\eqref{Eq:V} and of  keeping
only the first, leading term on the rhs of
Eq.~\eqref{Eq:U_2_evolution}. For a time moment after
the light pulse, the limits of the integration in
Eq.~\eqref{Eq:U_2_solution} can be extended to  minus and plus
infinity which renders  $U_2$ independent of $t$.
For a particular system,
this solution might be more difficult to use practically in
comparison with Eq.~\eqref{Eq:U_result}. We will discuss it below
for several model cases. For strong fields we limit our
consideration to the lowest order correction of the IA,
as far as it is sufficient for the applications of the theory
which we are going to present as well as for the understanding of
its limits of validity in these cases. Higher order corrections are
obtainable along the same line developed here.  {The corresponding treatment is to a certain extent similar to the Fer product expansion of the propagator \cite{Fer1958,Wilcox1967,Blanes2009} but is not the same. In our case each consecutive term contains only the respective order in $\tau_{\rm d}$.}


Let us go back to Eq.~\eqref{Eq:U_1}: For sufficiently  small $\tau_{\rm d}V_0/\hslash$ (as to allow for fast convergence)
 we can expand its rhs in this small parameter  and insert
into Eq.~\eqref{Eq:U_2_solution}. This  leads to an expression that coincides
with Eq.~\eqref{Eq:U_result} with regard to the second order of $\tau_{\rm
d}$ in the exponent.
 Generally, the solution given by Eq.~\eqref{Eq:U_2_solution} is to be adopted to
  correct the IA in the case of very high  peak  fields
 in the range where the unitary perturbation expansion
leading to Eq.~\eqref{Eq:U_result} fails.
This situation occurs if the driven system gains so much energy \textit{during} the pulse
 that the energies of the involved quantum transitions become comparable or exceed $\hslash/\tau_{\rm d}$. Hence
 the crucial assumption of ultrashort pulses would be invalidated.
We illustrate this
restriction below for a specific  example of a single-channel
semiconductor quantum ring in Section \ref{Sec:1d_ring}. As expected  this condition is irrelevant
 for a two-level system (see Section \ref{Sec:driven_TLS}).

\subsection{One-dimensional motion}
Firstly, we are going to detail the concepts presented in the foregoing sections for the case of a one-dimensional motion
of an electron driven by linearly-polarized ultrashort light
pulses.
\subsubsection{Unbound electrons driven by broadband pulses}\label{Sec:1d_free_driven}
Assume there is an electron with a mass $m^*$ in a free
space subjected to the action of the electric field of ultrashort
light pulses linearly-polarized along the direction of motion.
Choosing the $x$-axis along the light polarization direction, in the dipole approximation we
have $V_0=-exE_0$. Here $E_0$ is the electric field
amplitude and $e$ is the electron charge. In this case $H_0=p^2/(2m^*)$ with $p=-i\hslash\partial/\partial x$. The terms in the
exponent of Eq.~\eqref{Eq:U_result} read
\begin{equation}\label{Eq:free_commut1}
    -\frac{i\tau_{\rm d}}{\hslash}s_1V_0=\frac{i}{\hslash}s_1(eE_0\tau_{\rm d})x\;,
\end{equation}
\begin{equation}\label{Eq:free_commut2}
    -\frac{\tau_{\rm d}^2}{\hslash^2}s_2[V_0,H_0]=\frac{i}{\hslash}s_2\frac{eE_0\tau_{\rm d}^2}{m^*}p\;,
\end{equation}
\begin{equation}\label{Eq:free_commut3}
    \frac{i\tau_{\rm
    d}^3}{2\hslash^{3}}s_3\big[[V_0,H_0],H_0\big]=0\;,
\end{equation}
and
\begin{equation}\label{Eq:free_commut3_prime}
    -\frac{i\tau_{\rm d}^3}{4\hslash^{3}}s'_3\big[V_0,[V_0,H_0]\big]=\frac{i}{\hslash}s'_3\frac{e^2E_0^2\tau_{\rm
    d}^3}{4m^*}\;.
\end{equation}
Equation \eqref{Eq:free_commut3} follows straightforwardly from
Eq.~\eqref{Eq:free_commut2} due to the vanishing commutator
between $p$ and $p^2$. These results allow us to write
Eq.~\eqref{Eq:U_result} in the considered case as
\begin{equation}\label{Eq:U_result_free_space}
    \mathcal{U}(t_{1})=\exp\left[\frac{i}{\hslash}\Delta p x+\frac{i}{\hslash}\Delta x p +i\Delta\phi\right],
\end{equation}
where the quantity
\begin{equation}\label{Eq:Delta_p_free_space}
    \Delta p=s_1(eE_0\tau_{\rm d})\;
\end{equation}
is the momentum transferred to the electron by the light pulse  and
\begin{equation}\label{Eq:Delta_x_free_space}
    \Delta x=s_2\frac{eE_0\tau_{\rm d}^2}{m^*}\;
\end{equation}
has the meaning of a position shift induced by the action of the
light pulse.
The same momentum transfer and position shifts are found in the
case of the classical consideration of the electron motion under the
influence of the light pulse. The last term in the exponent of
Eq.~\eqref{Eq:U_result_free_space} induces just a phase shift
\begin{equation}\label{Eq:Delta_phi_free_space}
    \Delta \phi=s'_3\frac{1}{\hslash}\frac{e^2E_0^2\tau_{\rm
    d}^3}{4m^*}\;
\end{equation}
and has no classical counterpart. It has also no physical meaning
in the considered simple case because of the absent coupling to
any measurable physical quantity. One can show that in the
velocity gauge of the light-matter interaction and in the dipole
approximation it is related to the phase change induced by the
term in the interaction Hamiltonian proportional to the square of
the vector potential $A^2(t)$ \cite{Klaiber2008,Klaiber2009}. The
situation may change, e.g., when such a phase becomes
spin-dependent in the case of a spin-dependent light-matter
interaction \cite{Waetzel2011,zgzhu2010qw} or for a more complex
system topology as in the case of quantum rings presented  below.

\subsubsection{Driven electron in a one-dimensional
confinement}\label{Sec:motion_1D} If the
electron experiences initially the time-independent  potential $U(x)$, i.e. for
$H_0=p^2/(2m^*)+U(x)$, the first and the second order terms in
$\tau_{\rm d}$ in the exponent of Eq.~\eqref{Eq:U_result} [given by
Eqs.~\eqref{Eq:free_commut1} and \eqref{Eq:free_commut2}] remain
the same as for the case of the driven potential-free electron. The
difference with  respect to latter  is in the third order of
$\tau_{\rm d}$. The third order term in $\tau_{\rm d}$ given by
Eq.~\eqref{Eq:free_commut3_prime} and leading to the phase shift
is also unchanged but the other third order term given by
Eq.~\eqref{Eq:free_commut3}, which is zero for the driven unbound
electron, does not vanish now and can be expressed as
\begin{equation}\label{Eq:1d_potential_commut3}
    \frac{i\tau_{\rm d}^3}{2\hslash^{3}}s_3\big[[V_0,H_0],H_0\big]=
    -\frac{i}{\hslash}\frac{s_3}{2}  (eE_0\tau_{\rm d})\frac{\tau_{\rm d}^2
    U'(x)}{m^*}\;.
\end{equation}
With this expression we obtain from Eq.~\eqref{Eq:U_result} the
following result:
\begin{equation}\label{Eq:U_result_1d_potential}
     \mathcal{U}(t_{1})=\exp\left[\frac{i}{\hslash}\Delta p x+\frac{i}{\hslash}\Delta x p +i\Delta\phi
     -\frac{i}{\hslash}\frac{s_3}{2}(eE_0\tau_{\rm d})\frac{\tau_{\rm d}^2 U'(x)}{m^*} \right].
\end{equation}

As a simple illustration let us consider the electron motion
confined to the potential of a harmonic oscillator
$U(x)=m^*\omega_0^2x^2/2$, where $\omega_0$ is the characteristic
frequency. This leads to
\begin{equation}\label{Eq:U_result_1d_harmonic osillator}
     \mathcal{U}(t_{1})=\exp\left[\frac{i}{\hslash}\Delta p x+\frac{i}{\hslash}\Delta x p +i\Delta\phi
     +\frac{i}{\hslash}\tilde{\Delta} p x \right],
\end{equation}
where the additional momentum shift $\tilde{\Delta}p$ is given by
\begin{equation}\label{Eq:tilde_Delta_p_free_space}
    \tilde{\Delta} p=-\frac{s_3}{2}(eE_0\tau_{\rm d})(\tau_{\rm
    d}\omega_0)^2\;.
\end{equation}
Comparing this equation with Eq.~\eqref{Eq:Delta_p_free_space} we
see that the momentum shift induced by the first order term
dominates over the momentum shift induced by the third order term
only if the condition $\tau_{\rm d}\ll 1/\omega_0$ is fulfilled.
This condition for the duration of the light pulse can be
also written as
\begin{equation}\label{Eq:condition_IA_validity}
    \tau_{\rm d}\ll \frac{\hslash}{\Delta E}\;,
\end{equation}
where $\Delta E=\hslash\omega_0$ is the difference between the
neighboring energy levels. It is a necessary requirement for the validity of the
IA including its generalization to the second order in $\tau_{\rm
d}$ with the corresponding choice of $t_1$.


In the SVS case we can not immediately
apply Eq.~\eqref{Eq:U_result} and should use
Eq.~\eqref{Eq:Ansatz_V_strong} supplemented by Eqs.~\eqref{Eq:U_1} and
\eqref{Eq:U_2_solution}. Calculating
\begin{equation*}
    U_1^{-1}(t)[V_0,H_0]U_1(t)=-i\hslash\frac{eE_0}{m^*}p-i\hslash\frac{e^2E_0^2\tau_{\rm d}}{m^*}s_1(t)\;,
\end{equation*}
and inserting this into Eq.~\eqref{Eq:U_2_solution} we get
\begin{equation}\label{Eq:U_2_1d}
    U_2(t)=\exp\left[\frac{i}{\hslash} s_2(t)\frac{eE_0\tau_{\rm d}^2}{m^*}p
    +i\Delta\phi_2(t)\right],
\end{equation}
where the time-dependent phase $\Delta\phi_2(t)$ is given by
\begin{equation}\label{Eq:phi_2}
    \Delta\phi_2(t)=\tilde{s}_2(t)\frac{1}{\hslash}\frac{e^2E_0^2\tau_{\rm d}^3}{m^*}\;,
\end{equation}
with $\tilde{s}_2(t)$ defined as
\begin{equation}\label{Eq:s_2_tilde}
    \tilde{s}_2(t)=\frac{1}{\tau_{\rm d}^3}
\int_0^t\!\!{\rm d}t'\; (t'-t_1)f(t')\int_0^{t'}\!\! {\rm d}t''\,
f(t'')\;.
\end{equation}
 After the pulse is gone, the phase
factor in the exponent becomes a constant and has no physical
importance. We see that the result determined by
Eq.~\eqref{Eq:U_2_1d} is physically equivalent to that of
Eq.~\eqref{Eq:U_result_1d_potential} where in the latter only the two first
terms in the exponent are taken into account. Thus, for the driven one-dimensional
systems  Eq.~\eqref{Eq:U_result} limited to the two lowest orders in $\tau_{\rm d}$ in the exponent may be applied also in the
case of SVS interactions. It is interesting that the same
situation happens for the atomic potentials as in
Ref.~\cite{Klaiber2009}, explaining a good agreement of the
theoretical results   with the numerical
simulations observed also for the SVS case
up to the second order in $\tau_{\rm d}$. Note that the
simulations of Ref.~\cite{Klaiber2009} under the conditions when the third order terms should be
most important do not show a good correspondence to the
theoretical result following from Eq.~\eqref{Eq:U_result} in the
case of SVS interactions (for $|e| a_{_{\rm
B}}E_0\tau_{\rm d}/\hslash \gtrsim 1$, where $a_{_{\rm B}}$ is the
Bohr radius).

It should be mentioned that another restriction of the considered
approach might arise if a high energy is transferred through the
applied field to the excited system leading to the population of
the distant energy levels. In order to fulfill the condition
\eqref{Eq:condition_IA_validity} for all pairs of levels such that
the excitation leads to transitions between them, the applied
 pulses should have a  relatively small duration. This means
that the applied field contains high  frequencies, and hence
 at some point the dipole approximation has to be revisited.

\subsubsection{Electrons in a single-channel quantum
ring}\label{Sec:1d_ring} A free electron with an (effective) mass
$m^*$ in a one-dimensional quantum ring (QR) with a radius $r_0$
has the Hamiltonian
\begin{equation}\label{Eq:Hamiltonian_ring}
    H_0=-\frac{\hslash}{2m^*r_0^2}\frac{\partial^2}{\partial\varphi^2}\;.
\end{equation}
Its eigenstates are characterized by the angular quantum number
$m=0,\pm 1, \pm 2, \ldots$ with the stationary wave functions
$\psi_m(\varphi)=\frac{1}{\sqrt{2\pi}}\exp(im\varphi)$ and the
energies
\begin{equation}\label{Eq:energies_ring}
    \varepsilon_m=\frac{\hslash^{2}m^{2}}{2m^{\ast}r_{0}^{2}}\;.
\end{equation}
We consider the case when the electrons are driven by
linearly-polarized light pulses. The polarization
vector $\mathbf{\hat{e}_x}$ of the electric field $\mathbf{E}(t)$ is in the plane of
the ring.  Let us write $\mathbf{E}(t)=\mathbf{\hat{e}_x} E_0 f(t)$, where $f(t)$ describes
solely the time structure of the pulse with the amplitude strength $E_0$.
 In the dipole approximation the coupling to the electronic system reads
 $V(t)=-e\mathbf{r}\cdot\mathbf{E}(t)$. Taking into
account the ring geometry, this leads to $V(t)=V_0f(t)$ with  the
spatial part $V_0$  given by
\begin{equation}\label{Eq:interaction_ring}
    V_0=-eE_0r_0\cos\varphi\;.
\end{equation}
Here $E_0$ is the electric field amplitude and $\varphi$ is the
angle between the electron position vector $\mathbf{r}$ and $\mathbf{\hat{e}_x}$.
For the considered case we find for the terms in the exponent of
Eq.~\eqref{Eq:U_result}:
\begin{equation}\label{Eq:ring_commut1}
    -\frac{i\tau_{\rm d}}{\hslash}s_1V_0=is_1a_{_{\!E}}\cos\varphi\;,
\end{equation}
\begin{equation}\label{Eq:ring_commut2}
    -\frac{\tau_{\rm d}^2}{\hslash^2}s_2[V_0,H_0]=i s_2a_{_{\!E}}b_0  \left[i\cos\varphi+2i\sin\varphi\frac{\partial}{\partial \varphi}\right],
\end{equation}
\begin{equation}\label{Eq:ring_commut3}
    \frac{i\tau_{\rm
    d}^3}{2\hslash^{3}}s_3\big[[V_0,H_0],H_0\big]=i\frac{s_3}{2}a_{_{\!E}}b_0^2\left[\cos\varphi+4\sin\varphi\frac{\partial}{\partial \varphi}
    -4\cos\varphi\frac{\partial^2}{\partial \varphi^2}\right],
\end{equation}
\begin{equation}\label{Eq:ring_commut3_prime}
    -\frac{i\tau_{\rm d}^3}{4\hslash^{3}}s'_3\big[V_0,[V_0,H_0]\big]=is'_3
    a_{_{\!E}}^2b_0\sin^2\varphi\;,
\end{equation}
where
\begin{eqnarray}
    a_{_{\!E}}&=&\frac{er_0E_0\tau_{\rm d}}{\hslash}\;, \label{Eq:ring_a}\\
    b_0&=&\frac{\hslash\tau_{\rm d}}{2m^*r_0^2}\;. \label{Eq:ring_b}
\end{eqnarray}
The operators in the square brackets of
Eqs.~\eqref{Eq:ring_commut2} and \eqref{Eq:ring_commut3} are
self-adjoint which ensures the required unitarity of the
approximate evolution operator.

In the framework of the IA, the impact of the whole pulse is
characterized by a single dimensionless parameter
\begin{eqnarray}\label{Eq:alpha_def}
    \alpha=s_1 a_{_{\!E}}\;,
\end{eqnarray}
which has the meaning of an action (in units of $\hslash$)
transferred by the pulse and is referred to as \textit{kick
strength}.
Within the IA $\mathcal{U}(t_{1})$ does not
depend on $t_1$. Denoting it as\ $\mathcal{U}_1$ we can write
\begin{equation}\label{Eq:U_1_ring_result}
   \mathcal{U}_1=e^{i\alpha\cos\varphi} = \sum_{n=0}^\infty \frac{(i\alpha)^n}{n!} \cos^n\varphi= 1 + i\frac{\alpha}{r_0}\mathbf{\hat{e}}_x\mathbf r -\frac{1}{2r_0^2}\alpha^2 (\mathbf{\hat{e}}_x\mathbf r)  (\mathbf{\hat{e}}_x\mathbf r) +\dots\;.
\end{equation}

The terms in
Eqs.~\eqref{Eq:ring_commut2} and \eqref{Eq:ring_commut3} contain
products of angular functions and angular derivatives. It is thus
cumbersome  in practice to calculate the corresponding operator
exponentials beyond the IA. For HCPs, to assure the validity of the approximation to a second order in $\tau_{\rm d}$
one  needs to select $t_1$ appropriately. If it is required,
the third order term given by Eq.~\eqref{Eq:ring_commut3} can be
eliminated by an appropriate pulse shape engineering as discussed
in Section~\ref{Sec:few_cycle}. In the case of single-cycle pulses
the leading term is given by Eq.~\eqref{Eq:ring_commut2} and the
calculation of the operator exponentials containing the angular
derivatives can not be avoided.  {Numerically, the evaluation of the action of these operators on a given state can be performed effectively with the methods mentioned in Section~\ref{Sec:Unitary_expansion} \cite{Tannor_book,Blanes2009,Castro2004,Hochbruck1997}.}

\subsubsection{Range of validity of the impulsive approximation for the case of quantum rings}
Let us discuss the limits of validity of the IA for QRs.
In a realistic semiconductor QR there are many electrons, which in
equilibrium obey the Fermi-Dirac distribution. At low temperatures
the electrons states located not far from the Fermi level
participate in the excitation process if the excitation strength
is not too high. For such states the angular derivatives in
Eqs.~\eqref{Eq:ring_commut2} and \eqref{Eq:ring_commut3} lead to
the appearance of additional factors on the order of the angular
quantum number at the Fermi level $m=m_{_\mathrm{F}}$. Therefore, we have to fulfil the
condition  $b_0m_{_\mathrm{F}}\ll 1$
 to justify neglecting  the second order term \eqref{Eq:ring_commut2} and the
third order term \eqref{Eq:ring_commut3} as
compared to the first order term \eqref{Eq:ring_commut1}. Given that the
neighboring energy level spacing $\Delta E$
close to the Fermi level is equal to $\hslash^2m_{_\mathrm{F}}/(m^*r_0^2)$, we infer again  the condition
\eqref{Eq:condition_IA_validity}. This restriction for the IA can
be easily understood considering a classical electron moving with
the Fermi velocity $v_{_\mathrm{F}}\approx m_{_\mathrm{F}}\hslash/(r_0m^*)$
around the ring. The impulsive approximation obviously breaks down
if the pulse duration $\tau_{\rm d}$ is longer than the ballistic
time $\tau_{_\mathrm{F}}=2\pi r_0/v_{_\mathrm{F}}$, i.e. the time for a free
electron at the Fermi level  to perform a turn around the
ring. Therefore, the condition $\tau_{\rm d}\ll \tau_{_\mathrm{F}}$ is
equivalent to Eq.~\eqref{Eq:condition_IA_validity}, up to the
factor of $2\pi$ that is just a question of conventions in this
case.

Comparing another third order term, which is given by Eq.~\eqref{Eq:ring_commut3_prime}, with
the lower order terms \eqref{Eq:ring_commut1} and
\eqref{Eq:ring_commut2}, we see that two additional
conditions, $|a_{_{\!E}}|b_0\ll 1$ and $|a_{_{\!E}}|\ll m_{_\mathrm{F}}$, must hold for the convergence of the presented scheme and
therefore for the validity of  the IA. When  $b_0m_{_\mathrm{F}}\ll 1$ is fulfilled, the condition
$|a_{_{\!E}}|\ll m_{_\mathrm{F}}$ is more restrictive than
$|a_{_{\!E}}|b_0\ll 1$. Thus, to justify the IA two
conditions are required: Eq.~\eqref{Eq:condition_IA_validity} and
\begin{eqnarray}\label{Eq:second_condition_IA_ring}
    |a_{_{\!E}}|\ll m_{_\mathrm{F}}\;.
\end{eqnarray}
For a HCP characterized by a kick strength $\alpha$ Eq.~\eqref{Eq:second_condition_IA_ring} implies
$|\alpha|\ll m_{_\mathrm{F}}$.

\begin{figure*}
\includegraphics[width=0.75\textwidth]{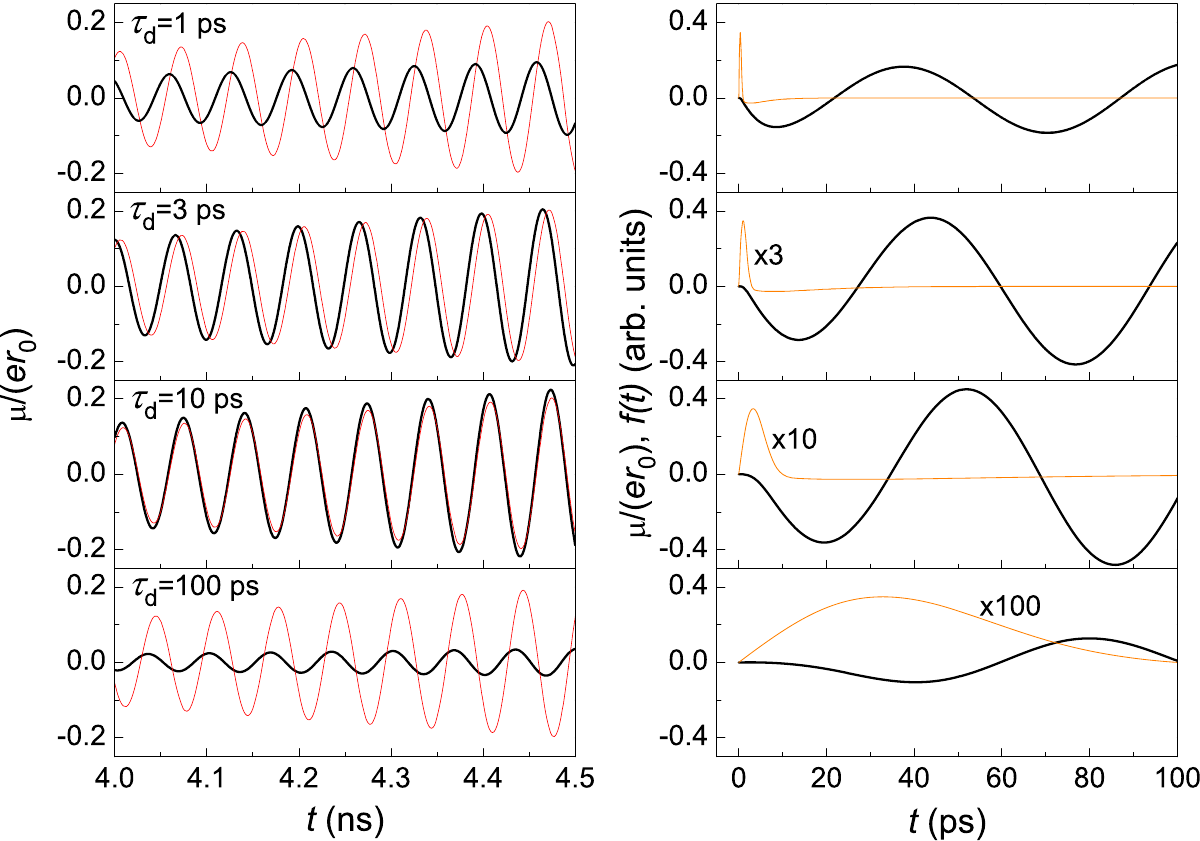}
\caption{\label{Fig:IA_pulses} Validity of the IA is tested for
different durations of strongly asymmetric HCPs
\eqref{Eq:pulse_shape} initiating the dipole moment dynamics
$\mu(t)$  in a single-channel GaAs QR, while keeping constant the
kick strength ($\alpha=0.2$). The duration $\tau_{\rm d}$ of the
HCPs for the figures of the right and left panels is 1~ps, 3~ps,
10~ps, and 100~ps (from top to bottom). The corresponding values of
the peak field strength is  1.86~V/cm, 0.62~V/cm, 0.186~V/cm, and
0.0186~V/cm, respectively. Thick black lines in the figures of
both panels correspond to full numerical simulations of the
density matrix dynamics, whereas thin red lines in the left
panel show the result of the IA with $t_1$ selected at the center
of gravity of the positive half-cycle. The left panel shows the
dipole moment dynamics long after the HCP. The right panel
illustrates its early stage as well as the time profile of the
applied pulses (thin orange lines) on the same time scale.
Parameters of the strongly asymmetric HCPs are selected as
discussed after Eq.~\eqref{Eq:pulse_shape}. The numbers near the
orange lines indicate the scaling factors for the field strength
required to obtain the same peak value as the HCP with $\tau_{\rm
d}=1~\mbox{ps}$ (top figure). The QR radius is $r_0=1.35$~$\mu$m
and the number of electrons is $N=400$.}
\end{figure*}

In Fig.~\ref{Fig:IA_pulses} we compare the dynamics of the dipole
moment $\mu(t)$ calculated using the IA with the result of full
numerical calculations based on the equation of motion for the
density matrix at zero temperature. For this comparison we use
strongly asymmetric HCPs \eqref{Eq:pulse_shape} varying their
duration $\tau_{\rm d}$ while keeping constant the kick strength
$\alpha=0.2$. Such a kick strength is well below the limit set by
Eq.~\eqref{Eq:second_condition_IA_ring}. Parameters of the pulse
shape are selected in a way as it is described in the text after
Eq.~\eqref{Eq:pulse_shape}. In the left panel
we show the dipole moment dynamics after 4~ns of the evolution past
from the time moment of the pulse application. In the right panel
the temporal profile of the pulse and the corresponding initial stage of
the dipole moment evolution are shown. The calculations illustrate
that the IA is well justified for the strongly asymmetric HCPs
when the duration of the positive half-cycle $\tau_{\rm d}$ is
smaller than, roughly, a quarter of the period of the dipole
moment oscillations, i.e., shorter than $\tau_{_\mathrm{F}}/4$. However, the
pulse should not be too short because  the duration of the
long negative half-cycle should be considerably longer than $\tau_{_\mathrm{F}}/4$.

In the case of SVS light pulses, i.e. when
Eq.~\eqref{Eq:condition_IA_validity} is valid but we have
$|a_{_{\!E}}|\gtrsim m_{_\mathrm{F}}$, Eqs.~\eqref{Eq:U_1} and
\eqref{Eq:U_2_solution} for the considered pulse-driven QR take the form
\begin{equation}\label{Eq:U_1_ring}
   U_1(t)=\exp\left[is_1(t)a_{_{\!E}}\cos\varphi\right]
\end{equation}
and
\begin{equation}\label{Eq:U_2_ring}
    U_2(t)=\exp\left[is_2(t)a_{_{\!E}}b_0\left(i\cos\varphi+2i\sin\varphi\frac{\partial}{\partial
    \varphi}\right)+i\tilde{s}_2(t)a_{_{\!E}}^2b_0\sin^2\varphi\right],
\end{equation}
where $\tilde{s}_2(t)$ is given by Eq.~\eqref{Eq:s_2_tilde} and
$s_1(t),s_2(t)$ are determined by Eq.~\eqref{Eq:s_n}. Notice that
the second term in the exponent of Eq.~\eqref{Eq:U_2_ring} is absent in
Eq.~\eqref{Eq:ring_commut2}. For a very strong HCP it still makes
sense to select $t_1$ at the center of the gravity of the pulse,
so that the first term in the exponent of Eq.~\eqref{Eq:U_2_ring}
is eliminated. In such a case, if a precision beyond the IA is
demanded then the correction to the evolution operator given by
Eq.~\eqref{Eq:U_2_ring} can be calculated in a similar way to
Eq.~\eqref{Eq:U_1_ring} in the basis of the stationary
eigenfunctions of the unperturbed ring.

Comparing Eq.~\eqref{Eq:U_2_ring} with Eq.~\eqref{Eq:U_1_ring} we
see that $U_2(t)$ represents a correction to $U_1(t)$ only if
Eq.~\eqref{Eq:condition_IA_validity} and additionally
$|a_{_{\!E}}|b_0\ll 1$, i.e.
\begin{eqnarray}\label{Eq:second_condition_ring_strong}
    |a_{_{\!E}}|\ll 2 m_{_\mathrm{F}}\left(\frac{\hslash}{\Delta
    E \tau_{\rm d}}\right),
\end{eqnarray}
hold. As it has been already mentioned above, if
Eq.~\eqref{Eq:condition_IA_validity} is fulfilled the condition
\eqref{Eq:second_condition_ring_strong} limiting the field
strength is less restrictive in comparison with the condition
\eqref{Eq:second_condition_IA_ring}, required for the validity of
the perturbation theory leading to Eq.~\eqref{Eq:U_result} with
Eqs.~\eqref{Eq:ring_commut1}-\eqref{Eq:ring_commut3_prime}. Thus
there is a range of pulse strengths beyond
Eq.~\eqref{Eq:second_condition_IA_ring} where
Eqs.~\eqref{Eq:U_1_ring} and \eqref{Eq:U_2_ring} can be used. Even
stronger fields would generally lead to the breakdown of the IA
and both approximation schemes. Then a full numerical solution is
required to describe the system evolution.

\subsubsection{Optical transitions via broadband ultrashort asymmetric pulses}
\begin{figure*}
\includegraphics[width=0.35\textwidth,angle=90]{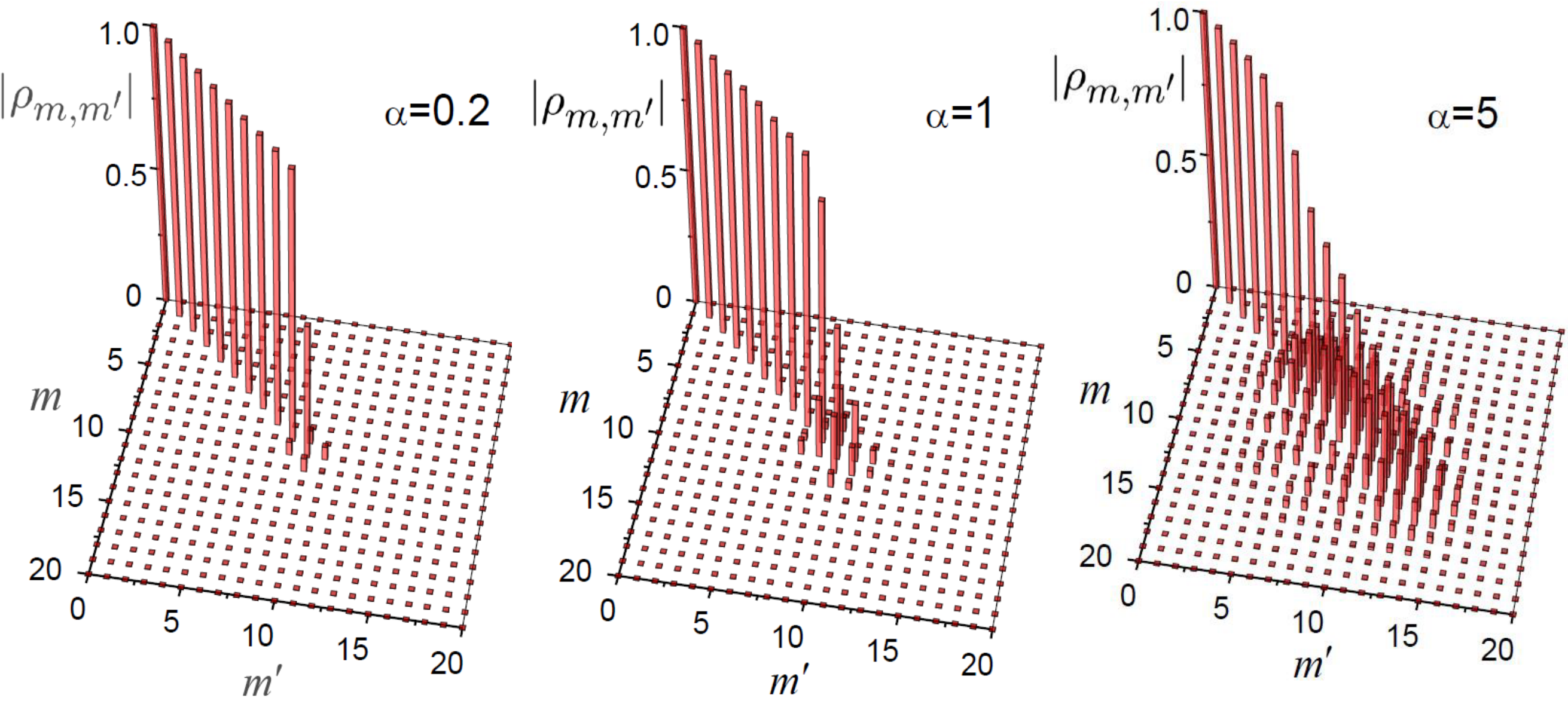}
\caption{\label{Fig:denisty_matrix}
Typical structure of the density matrix expressed in the eigenstates of the one channel QR $|m\rangle$ right after the pulse
for different values of the kick strength $\alpha$. There are $N=40$ electrons in the QR. Initially they are at equilibrium (i.e.
the density matrix has a diagonal structure) at low temperature ($k_\mathrm{B} T=\varepsilon_1$, where $\varepsilon_1$ is determined by Eq.~\eqref{Eq:energies_ring} with $m=1$ and $k_\mathrm{B}$ is the Boltzmann's constant).}
%
\end{figure*}

%
%
%
The structure of the propagator \eqref{Eq:U_1_ring_result} allows for some general statements on the nature of optical transitions via ultrashort, broadband pulses: A broadband pulse may mediate  quasi instantaneously  a multiple of  coherent multipolar,  highly nonlinear (in the field strength)
 transitions. The resulting coherent state contains thus contributions from excited states
 that would not have been reached via vertical transitions if we had employed  a harmonic pulse.
  The weight of these contributions  enhances
with the effective pulse strength $\alpha$, as evident from the way $\alpha$ enters the terms on the rhs of Eq.~\eqref{Eq:U_1_ring_result}. It is this feature of asymmetric  ultrabroadband pulses which
offers exciting new possibilities such as non-vertical transitions or optical bulk-type plasmon excitations.
An illustration is presented  by Fig.~\ref{Fig:denisty_matrix}  where the density matrix directly after the excitation of a QR with a HCP  gives insight into the modified population and the induced coherence. For $\alpha=5$ a huge
 angular momentum of up to $\Delta m= 8$ is transferred while for $\alpha=0.2$ transitions remain mainly  dipolar and of a linear  character.
 This hints on the relevance of this type of excitations  for  high harmonics  emission
 which will be discussed at length  in Section~\ref{Sec:HHG}.
It is worthwhile noting that, physically, the pulse offers a broad frequency spectrum.
Only certain frequencies are admitted however by the quantized system, depending on the pulse field strength. For instance, a pulse with a small $\alpha$  mediates solely
  intraband transitions between the angular momentum  states {$|m\rangle$ of the QR} [cf. $\alpha$-term of Eq.~\eqref{Eq:U_1_ring_result}]  which satisfy the dipolar selection rules, $\Delta m = \pm 1$.
  Transitions with $\Delta m =\pm 1$ correspond to certain energy level spacings,
whereas only the initial states close to the Fermi level play a role in the  low temperature case presented in Fig.~\ref{Fig:denisty_matrix}, due to the phase space filling. The pulse must be short enough  as to contain
 the required transition frequencies.
Pulses with large   strengths lead quasi  instantaneously to highly nonlinear processes.
The total transition energy may be set in relation to the sum  energy of  absorbed photons, which allows to connect the
effective pulse strength to the highest excited energy states (see Fig.~\ref{Fig:denisty_matrix}).

Let us look at the transitions caused by broadband ultrashort asymmetric pulses from the perspective of the time-dependent perturbation
theory (TDPT) and contrast with harmonic fields (cf. also Ref.~\cite{Klaiber2009}). In the first order of the TDPT and in the dipole approximation there is the following relation for the transition amplitude $A_{fi}$
between states $m_i$ and $m_f$:
$$A_{fi}\propto \langle m_f| V_0|m_i\rangle \int_{-\infty}^\infty \!\!\mathrm{d}t\; f(t) e^{i\omega_{fi} t} =
\langle m_f| V_0|m_i\rangle \int_{-\infty}^\infty   \!\!\mathrm{d}\omega\; f_\omega \frac{1}{2\pi} \int_{-\infty}^\infty  \!\!\mathrm{d}t\;  e^{-i(\omega -\omega_{fi}) t}.$$
 In the case of a harmonic driving field the Fourier transform $f_\omega$ of its temporal profile $f(t)$ has discrete finite number of contributing frequencies $\bar\omega_j$ \big[i.e., $f_\omega(\omega)=\sum_j
\delta (\omega -\bar\omega_j)$\big] which leads further to the Fermi's golden rule.
For an ideal unipolar HCP (cf. Sections \ref{Sec:Guassian_HCP} and \ref{Sec:sine_square_HCP}) with a wide and in the range from 0 to above $\omega_{fi}$ flat spectrum (so that $f_{\omega_{fi}} \approx  f_{\omega=0}$, see top right plot of Fig.~\ref{Fig:time_profiles})  we simplify to
 $$A_{fi}\propto \langle m_f| V_0|m_i\rangle \int_{-\infty}^\infty \!\!\mathrm{d}\omega\;  f_\omega(\omega) \delta(\omega -\omega_{fi}) \propto
\langle m_f| V_0|m_i\rangle   f_{\omega_{fi}}   \propto
\langle m_f| V_0|m_i\rangle  \int_{-\infty}^\infty \!\!\mathrm{d}t\; f(t) $$
which is the result we find from the IA to the first order in the field strength (or equivalently the first order in $\alpha$). If the pulse spectrum does not contain $\omega_{fi}$ then such a transition does not take place. The same arguments apply for few-cycle pulses designed to deliver a momentum kick (cf. Section~\ref{Sec:few_cycle}). For strongly asymetric HCPs (Section~\ref{Sec:Strongly_as_HCPs}) we have $f_{\omega_{fi}} \approx  \tilde{f}_{\omega=0}$ (see bottom right plot of Fig.~\ref{Fig:time_profiles}) where $\tilde{f}(t)$ reflects only the short and strong half-cycle of the pulse and otherwise vanishes (analogously for the few-cycle pulses considered in Section~\ref{Sec:freq_dom_model}).
In a distinct case of single-cycle pulses (Section~\ref{Sec:single_cycle}) with $ f_{\omega=0}=0$ we may use $f_{\omega_{fi}} \approx  \omega_{fi}\partial f/\partial \omega\big| _{\omega=0}=-i\omega_{fi} \int_{-\infty}^\infty \!\!\mathrm{d}t\; t f(t)$, when $\omega_{fi}$ is again situated inside the applicability region indicated in the top right plot of Fig.~\ref{Fig:time_profiles}. Taking also the relation $\langle m_f| [V_0,H]|m_i\rangle=-\hslash\omega_{fi}\langle m_f| V_0|m_i\rangle $ into account we get
$$A_{fi}  \propto
\frac{i}{\hslash}\left\langle m_f\big| [V_0,H]\big|m_i\right\rangle  \int_{-\infty}^\infty \!\!\mathrm{d}t\; t f(t). $$
This result coincides with the outcome of Eq.~\eqref{Eq:U_result} when the second term in the exponent, correcting the IA, is taken into account and again only the first order in the field strength is considered. Notice that in this case the first term in the exponent of Eq.~\eqref{Eq:U_result} actually vanishes, i.e. within the IA the pulse does not perturb the system ($\alpha=0$).

Repeating the above steps for the second order of the TDPT describing two-photon transitions we recover the quadratic term in $\alpha$ in the expansion of
Eq.~\eqref{Eq:U_1_ring_result}.  In fact  Eq.~\eqref{Eq:U_1_ring_result} accounts  for all perturbative orders of the pulse-system interaction.
Roughly speaking a strong kick strength $\alpha\gg 1$ (cf. Fig.~\ref{Fig:denisty_matrix}) selects  preferentially  nonlinear terms responsible for ``kicked'' multiphoton transitions  \big(if allowed by matrix elements $\langle m_f| V_0|m_i\rangle$\big). From this perspective and
as the term ``instantaneous kick'' is rather theoretical (each pulse will have a finite duration), it
would be interesting to inspect the weight of these multiphonton processes as a function of the pulse duration
but for a fixed $\alpha$ (one has then to vary the field strength). This may give access to the time
on which such multiphoton processes  usually (i.e. for harmonic driving) unfold.

 {Here it is important to contrast the considered excitation regime with the schemes of coherent control by shaped non-resonant optical pulses involving multiphoton transitions such as reported in Ref.~\cite{Meshulach1999}. Although the light pulses in those studies are also called ``ultrashort'' and their frequency bandwidth together with the internal spectral phase behavior  are essential ingredients in the control design, such pulses are very long and narrowband from the viewpoint of the present report, which is defined in Chapter~\ref{sec:exp} (see also Fig.~\ref{Fig:time_profiles}). Such many-cycle pulse driving a transition between certain quantum states,
 photons of various energies, which belong to the narrow frequency band of the pulse, participate in a particular multiphoton process whose order is determined by the relation of  the energy of the driven transition $\Delta E$
  to the energy set by the central frequency of the pulse $\hbar\omega_{\rm c}$. The energies of the participating photons sum up to $\hbar\omega_{\rm c}$. In contrast, the pulses considered in this review are shorter than $\hbar/\Delta E$, i.e. the interaction takes place on a subcycle time scale where the energy conservation breaks  down.  As one consequence in the frequency domain picture, the sum of the energies of the involved photons may vary significantly with respect to $\Delta E$ in a range determined by a broad band of our pulse $\hbar\Delta\omega>\Delta E$. Another consequence is that multiphoton transitions of different orders may contribute simultaneously and coherently to the transition amplitude. In the time domain picture, we deal here with a fundamentally different approach operating on subcycle time scales, i.e. allowing for drastically faster control schemes, which are illustrated in the next Chapters for various target quantum systems. The issue of operational speed is of crucial importance for a broad range of applications of light-driven nanostructures, and low-dimensional electronic systems in general, ranging from optoelectronics to quantum information.}

The interrelation between the perturbation expansion and Eq.~\eqref{Eq:U_1_ring_result} viewed from the perspective of
charged particle impact  makes clear how the asymmetric electromagnetic  pulses deliver effectively a momentum kick to the system, even though
the pulse field is treated within the dipole approximation. Namely, let us consider a swift (with respect to the Fermi velocity) charged particle
impinging with a well-defined (sharp) momentum  onto  an electronic system in the state $|i\rangle$ and transferring so a small amount of momentum $\mathbf q=
q \mathbf{\hat{e}_q}$ to the system which then goes over into the excited final state  $|f\rangle$. The transition amplitude for this
process within the first Born approximation, valid for our setup here (potentials are assumed short-ranged) \cite{Inokuti1971}, is
 $A_{if}^{1B}=\langle f|e^{iq (\mathbf{\hat{e}_q}\cdot\mathbf r)}|i\rangle = \delta_{ij} + iq  \langle f|\mathbf{\hat{e}_q}\cdot\mathbf r|i\rangle
 - \frac{q^2}{2}  \langle f|(\mathbf{\hat{e}_q}\cdot\mathbf r)(\mathbf{\hat{e}_q}\cdot\mathbf r)|i\rangle + \dots$. In the optical limit, i.e., for $q\to 0$, transitions caused by charged particle impact have the nature of  dipole optical transitions with the linear
  polarization vector being along $\mathbf{\hat{e}_q}$.
  With increasing $q$ higher multipoles, as in the case of strong HCPs, contribute  subsuming to the momentum transfer $\mathbf q$.
  The fundamental difference in dispersions of photons and particles is circumvented by varying for a HCP the two independent parameters: the pulse
  duration (offers the frequency range) and the pulse peak amplitude (multiplied by the pulse duration is proportional to the momentum kick). In practice, electron beams with
  well defined energy and momentum transfer are routinely employed, while the temporal control on the electron pulse
  duration is still a challenge. For HCPs the situation is   opposite. In both cases impressive
  advances have been made recently \cite{Sciaini2011,Kampfrath2013}.

\subsection{Two-level systems driven by short broadband pulses}\label{Sec:driven_TLS}
\begin{figure*}
  \includegraphics[width=8cm]{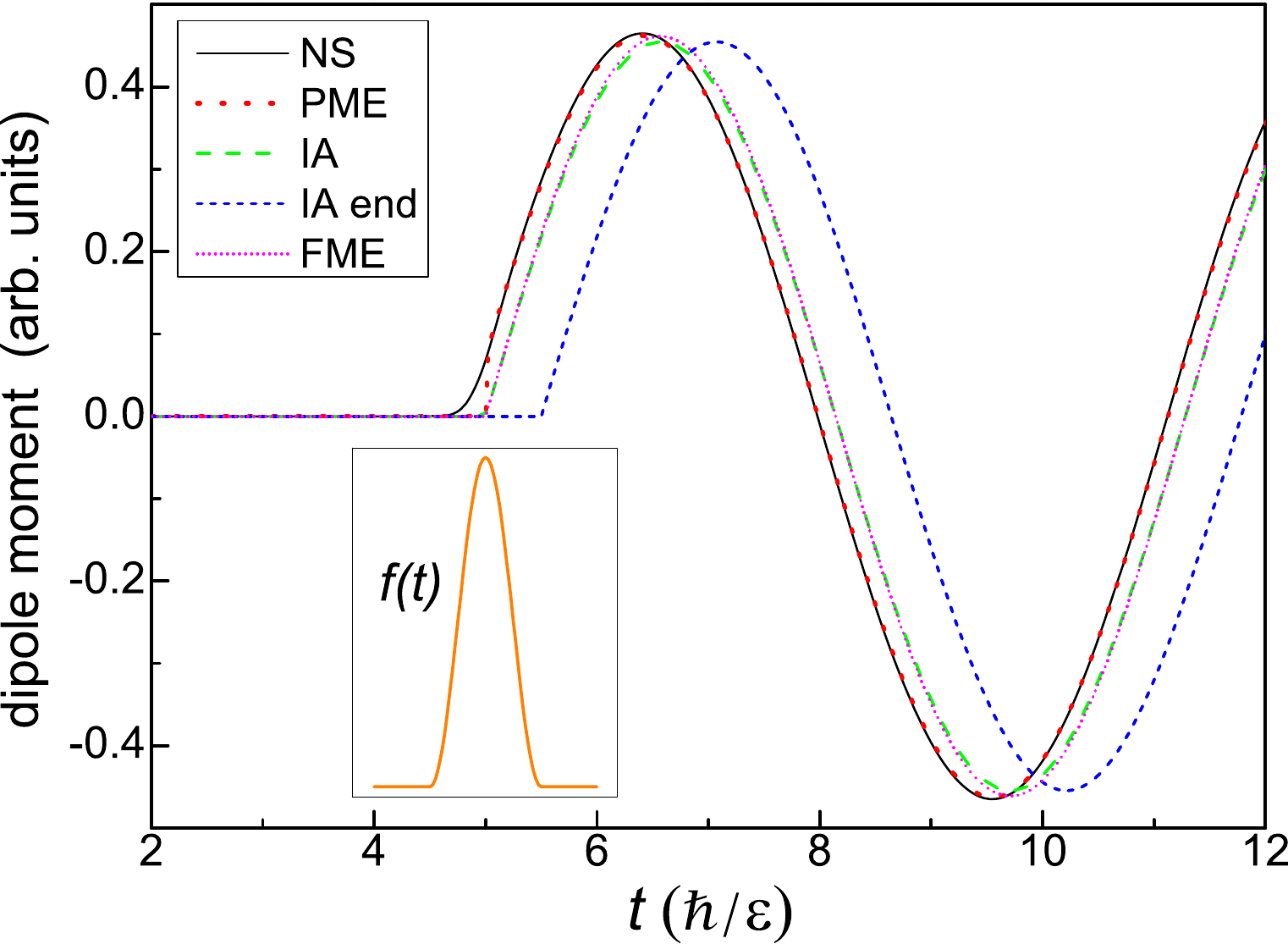}
  \caption{Dynamics of the induced dipole moment of a TLS being in the ground state at $t=0$.
  Inset illustrates the temporal profile of the applied sine-square pulse
  \eqref{Eq:sine_pulse} with $\tau_{\rm d}=\hslash/\varepsilon$ and $v=2\varepsilon$. The time
  scale of the inset is the same as for the main figure.
  In the main figure several approximations are compared with the full numerical solution (NS, full black line).
  Red dotted line shows the result of the unitary
  perturbation theory based on the Magnus expansion (PME) which is determined by Eq.~\eqref{Eq:U_result}  with Eqs.~\eqref{Eq:s_n_infinity}
  and \eqref{Eq:s_33_infinity}. Green dashed line corresponds to
  the IA [first term in the exponent of Eq.~\eqref{Eq:U_result}] with the parameter $t_1$ selected at
  the center of gravity of the pulse. Blue short-dashed line shows
  the same but $t_1$ is selected just at the end of the pulse (IA end).
  Magenta short-dotted line shows the result of the approximation
  based on the exact inclusion of the first term \eqref{Eq:Omega1} in the Magnus
  expansion (FME).
  \label{Fig:dip_TLS_weak}}
\end{figure*}

The dynamics of a two-level system (TLS) driven by an external
field is determined by the Hamiltonian
\begin{equation}\label{Eq:Hamiltonian_TLS}
    H=H_0+ V_0f(t)= -\frac{1}{2}\varepsilon\sigma_z-vf(t)\sigma_x\;,
\end{equation}
where $H_0=-\varepsilon \sigma_z/2$  is the Hamiltonian of
the unperturbed system.
$v=d_{21}E_0$  where $d_{21}$ is the transition dipole.  $E_0$ is the pulse  amplitude.
 $\sigma_x$ and $\sigma_z$
are the  Pauli matrices in the standard notation and
$\varepsilon=\hslash\omega_{21}$ is the two level energy spacing.
In this case
Eq.~\eqref{Eq:U_result} takes on the form:
\begin{equation}\label{Eq:U_result_TLS}
    \mathcal{U}(t,t_{1},0)=\exp\left[i a_v s_1\sigma_x+i a_vb_\varepsilon s_2\sigma_y-i a_v b_\varepsilon^2\frac{s_3}{2}\sigma_x+i a_v^2 b_\varepsilon\frac{s'_3}{2}\sigma_z\right],
\end{equation}
where
\begin{equation}\label{Eq:TLS_ab}
    a_v=\frac{v\tau_{\rm
    d}}{\hslash}\;, \ \ \ b_\varepsilon=\frac{\varepsilon\tau_{\rm d}}{\hslash}\;.
\end{equation}
The fourth term in the exponent leads
to a time shift with respect to
the unperturbed evolution
%
\begin{equation}\label{Eq:TLS_time_shift}
    \Delta T=\tau_{\rm d}a_v^2\frac{s'_3}{2}\;.
\end{equation}
The validity of the approximation scheme leading to
Eq.~\eqref{Eq:U_result_TLS} requires $b_\varepsilon\ll 1$,
i.e., Eq.~\eqref{Eq:condition_IA_validity} should be satisfied. Apart from this, we must demand  $|\Delta T|
\ll \tau_{\rm d}$, meaning  $|a_v|
\ll 1$ that restricts the
 interaction strength $v$. The latter condition can be also deduced analyzing directly
the Magnus expansion \eqref{Eq:Omega}. Its $k$-th order  contains
a term proportional to $|a_v|^kb_\varepsilon$, entailing  that
fast convergence  requires $|a_v|\ll 1$.
 In some cases the expansion delivers
 reasonable results for
$b_\varepsilon\approx 1$ and $|a_v|\approx 1$ and even slightly
higher. This can be seen from Fig.~\ref{Fig:dip_TLS_weak},
where we compare several possible approximations with the exact
numerical solution in a case with $b_\varepsilon=1$ and $a_v=2$.

In  Fig.~\ref{Fig:dip_TLS_weak} the time dynamics of the induced
dipole moment initiated by a sine-square HCP \eqref{Eq:sine_pulse}
with a duration of $\tau_{\rm d}=\hslash/\varepsilon$ (meaning
$b_\varepsilon=1$) is shown. The pulse is centered at $t=t_1=5
\hslash/\varepsilon$. Its temporal profile in relation to the dynamics
of the induced dipole moment is shown in the inset to
Fig.~\ref{Fig:dip_TLS_weak}. The TLS is initially in the ground
state. Therefore, for its wave function, expressed as
$\psi(t)\equiv\big(c_1(t),c_2(t)\big)^{T}$ we have $c_1(0)=1$ and
$c_2(0)=0$. The induced dipole moment $\mu(t)$ is proportional to
$\mathrm{Re}\left[c_1(t)c_2^*(t)\right]$. We
see that the best agreement with the numerics is provided by the
approximation corresponding to Eq.~\eqref{Eq:U_result_TLS} with
all three orders in $\tau_{\rm d}$ included in the exponent of
this equation. The IA with an appropriate selection of $t_1$, such
that the second order term in $\tau_{\rm d}$ vanishes in the
exponent of Eq.~\eqref{Eq:U_result_TLS}, also provides a very good
agreement with the numerical solution. Other selections of $t_1$,
e.g. just at the end of the pulse (the sine-square pulse is
limited to a finite time range), as it is shown in
Fig.~\ref{Fig:dip_TLS_weak}, lead to considerably larger
deviations from the numerics. In the case of a TLS it is possible
to get an exact analytical expression for the first term
\eqref{Eq:Omega1} of the Magnus expansion (FME)
\cite{Pechukas1966}. It is interesting that the corresponding
evolution operator does not depend on the selection of $t_1$. In
such an approximation all terms in the Backer-Hausdorff expansion
\eqref{ident} are taken into account but all higher order terms in
the Magnus expansion \eqref{Eq:Omega} are neglected, i.e. it
corresponds to omitting  the time-ordering operator in the exact
solution given by Eq.~\eqref{Eq:U_time_ordered}. This
approximation improves slightly the result of the IA but misses
the time shift \eqref{Eq:TLS_time_shift} needed for a better
agreement with the numerics achieved in the case corresponding to
Eq.~\eqref{Eq:U_result_TLS}. We have to mention here that the
dynamics during the pulse is not resolved for the approximate
solutions in Fig.~\ref{Fig:dip_TLS_weak}. This means that we use,
e.g., Eqs.~\eqref{Eq:s_n_infinity} and \eqref{Eq:s_33_infinity} in
connection with Eq.~\eqref{Eq:U_result_TLS} and not
Eqs.~\eqref{Eq:s_n} and \eqref{Eq:s_33}. It is also possible to
resolve the dynamics during the pulse based on the same
approximations. However, it is out of the scope of the present
work.


In the case of a SVS interaction, when $|a_v|\gtrsim 1$,
Eqs.~\eqref{Eq:U_1} and \eqref{Eq:U_2_solution} lead to
\begin{equation}\label{Eq:U_1_TLS}
   U_1(t)=\exp\left[ia_v s_1(t)\sigma_x\right]=
    \begin{pmatrix}  &  \\[-1.8cm]
  \cos[a_vs_1(t)] \ & \ i \sin[a_vs_1(t)] \\
  i \sin[a_vs_1(t)] \ & \ \cos[a_vs_1(t)]
\end{pmatrix}
\end{equation}
and
\begin{equation}\label{Eq:U_2_TLS}
    U_2(t)=\exp
   \begin{pmatrix}  &  \\[-1.8cm]
  -iA(t) \ & \ B(t) \\[-0.5cm]
  -B(t) \ & \ iA(t)
   \end{pmatrix},
\end{equation}
where
\begin{eqnarray}
  A(t) &=& a_vb_\varepsilon\, \frac{1}{\tau_{\rm d}^2}\int_0^t (t'-t_1) f(t') \sin\left[2 a_v s_1(t')\right] {\rm
  d}t'\;, \label{Eq:A_TLS}
  \\
  B(t) &=& a_vb_\varepsilon\, \frac{1}{\tau_{\rm d}^2}\int_0^t (t'-t_1) f(t') \cos\left[2 a_v s_1(t')\right] {\rm
  d}t'\;. \label{Eq:B_TLS}
\end{eqnarray}
If we are not interested in the dynamics during the pulse, the
lower and upper integration limits in Eqs.~\eqref{Eq:A_TLS} and
\eqref{Eq:B_TLS} can be extended to the minus and plus infinity,
respectively.

\begin{figure*}
  \includegraphics[width=10cm]{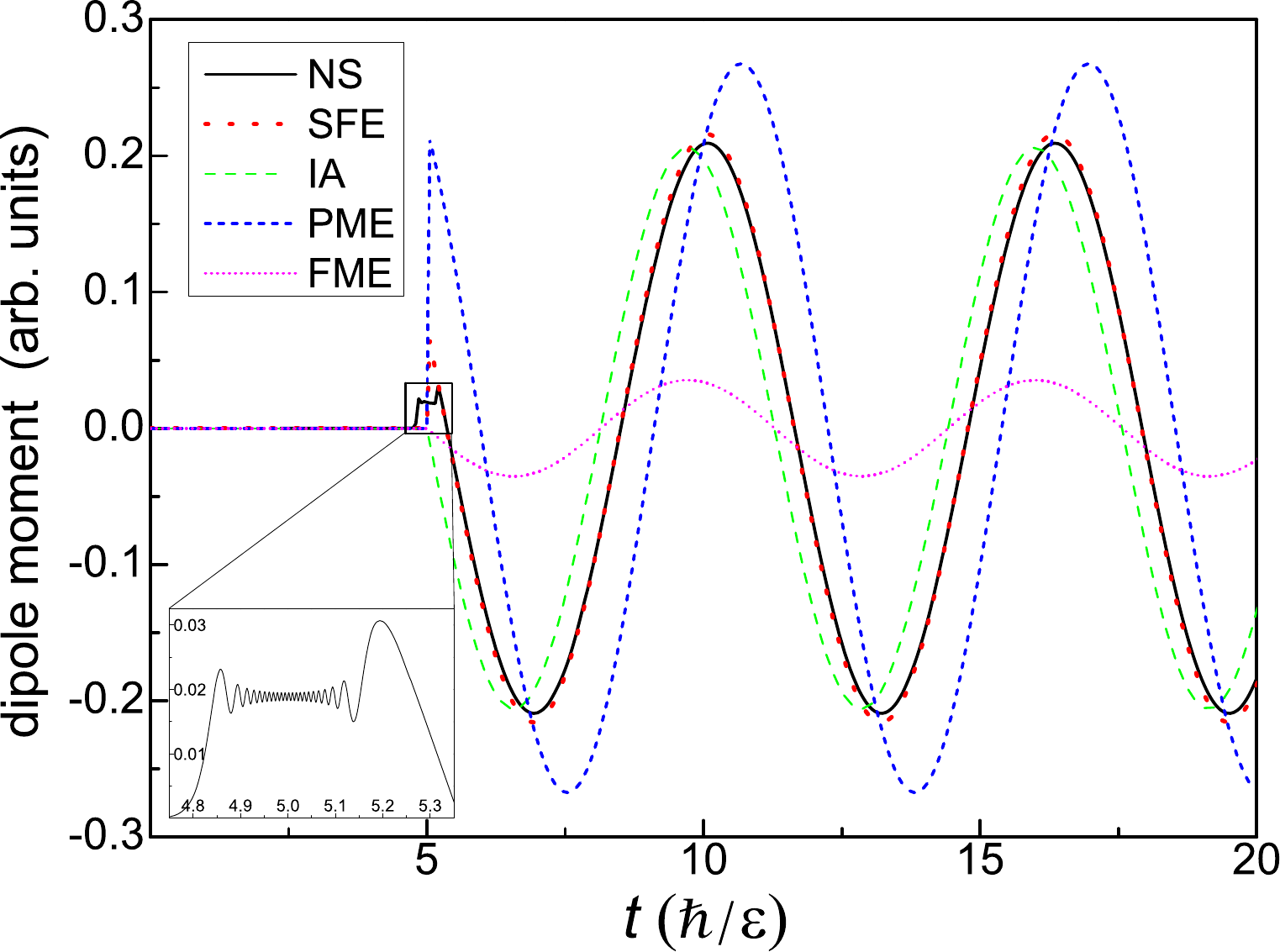}
  \caption{Dynamics of the dipole moment induced by a \,SVS\, Gaussian-shaped HCP \eqref{Eq:Gaussian_pulse}
  centered at $t=5\hslash/\varepsilon$. Parameters of the pulse are
  $\tau_{\rm d}=0.1 \hslash/\varepsilon$ and $v=400\varepsilon$
  corresponding to $b_\varepsilon=0.1$ and $a_v=40$.
  Red dotted line shows the result of the SVS field expansion (SFE) leading
  to Eqs.~\eqref{Eq:U_1_TLS}-\eqref{Eq:B_TLS} in the case of a TLS.
  Approximations denoted by abbreviations IA, PME, and FME (green dashed, blue short-dashed, and magenta short-dotted lines,
  respectively) are explained in the caption of Fig.~\ref{Fig:dip_TLS_weak}.
  Inset shows
  the dynamics of the dipole moment during the pulse (numerical solution with an increased time resolution), which is not modelled
  here for the approximate solutions.\label{Fig:dip_TLS_strong}}
\end{figure*}

The SVS case is illustrated in
Fig.~\ref{Fig:dip_TLS_strong}, where the dynamics of the dipole
moment induced by a Gaussian-shaped HCP is shown as it results
from the full numerical solution and different approximations. We
see that the IA ($t_1$ at the pulse center) demonstrates a
reasonable agreement with the numerics. Its improvement is
possible on the basis of the expansion proposed here for SVS fields. It is obvious from
Fig.~\ref{Fig:dip_TLS_strong} that the approximations based either
on Eq.~\eqref{Eq:U_result_TLS}, with more than just the first term
included in its exponent, or on the FME are inappropriate for the
modelling of the pulse-induced dynamics in this situation. They
should be also generally avoided as intermediate steps when
justifying the IA in the SVS case.

\subsection{Driving by periodic pulse sequences}
In the foregoing sections we have considered the electron dynamics
induced by the application of a single ultrashort broadband light pulse. The
characteristic time scales of the unperturbed system were much
larger than  the relevant pulse durations. In this consideration
we have also neglected possible relaxation processes assuming that
they take place on  larger time scales. In the case when an
electronic system is driven by a train of such pulses,
there are two possibilities: (i) the duration of the whole train
is still smaller than the relaxation time scales and (ii) the
relaxation influences the dynamics of the system.

For a single-electron or
a many-electron system in equilibrium with the environment,  the
density matrix $\rho(t)$ before the pulse application is
stationary and given by
\begin{equation}\label{Eq:rho_0}
    \rho_0=\sum_n f_n^0 |\psi_n^0\rangle\langle\psi_n^0|\;,
\end{equation}
where  $|\psi_n^0\rangle$ are  the
stationary single-particle eigenstates of the system with the
energies $E_n$ and $f_n^0$ is the distribution function reflecting
the initial statistics of the system. In the case (i) the pulse-induced
electron dynamics is coherent and therefore is determined by the evolution
operator $U(t,t')$:
\begin{equation}\label{Eq:rho_t}
    \rho(t)=U(t,t')\rho_0 U^\dagger(t,t')\;.
\end{equation}
Inserting Eq.~\eqref{Eq:rho_0} into Eq.~\eqref{Eq:rho_t} we get
\begin{equation}\label{Eq:rho_t_result}
    \rho(t)=\sum_n f_n^0 |\Psi_n(t)\rangle\langle\Psi_n(t)|\;,
\end{equation}
where
\begin{equation}\label{Eq:psi_n_t}
    |\Psi_n(t)\rangle=U(t,t')|\psi_n^0\rangle
    e^{-\frac{i}{\hslash}E_nt'}\;
\end{equation}
and $t'$ is any time moment before the excitation is applied. Thus
in the coherent case it is sufficient to study the time evolution
of each electron state and then to sum up over the initial
distribution according to Eq.~\eqref{Eq:rho_t_result} to obtain
 the density matrix which is sufficient to  calculate any
single-particle physical quantity. For a periodic
driving of a coherent state, the Floquet theorem can be used to
find the Floquet states and the corresponding quasienergies of the
driven system \cite[and references therein]{Grifoni1998,Chu2004}.

In the case (ii) the wave function formalism is not sufficient. The
dynamics can be described, e.g., using the density matrix
formalism taking the relaxation processes into account. Here a
generalization of the Floquet approach to the density matrix
dynamics is required that is provided, e.g., by the
Floquet-Liouville super-matrix formalism \cite{Chu2004}.

\subsection{Coherent quantum dynamics:  { Floquet approach, geometric phases, and}  {nonequilibrium} {sustainability}  }\label{Sec:Floquet}

\subsubsection{ {Periodic driving and periodic evolution}}
 Within the Floquet theory, any time-dependent wave function of a
periodically driven system can be developed in the Floquet basis:
\begin{equation}\label{Eq:Floquet_series}
  |\Psi(t)\rangle=\sum\limits_{\lambda}C_{\lambda}|\Psi_{\lambda}(t)\rangle\;,
\end{equation}
where the Floquet states $|\Psi_{\lambda}(t)\rangle$ can be
written as
\begin{equation}\label{Eq:Floquet_states}
  |\Psi_{\lambda}(t)\rangle=e^{-i\varepsilon_{\lambda}t/\hslash}|\Phi_{\lambda}(t)\rangle\;.
\end{equation}
The Floquet modes
$|\Phi_{\lambda}(t)\rangle=|\Phi_{\lambda}(t+T)\rangle$ are
 periodic in time with the same period $T$ as the
driving field. The quasienergies $\varepsilon_{\lambda}$ can be selected
inside the first Brillouin zone $\varepsilon_\lambda\in
[-\hslash\omega_T/2,\hslash\omega_T/2)$, where
\begin{equation}\label{Eq:Omega_T}
  \omega_T=\frac{2\pi}{T}\;.
\end{equation}
From Eq.~\eqref{Eq:Floquet_states} and the time-dependent
Schr\"{o}dinger equation (TDSE) follows that the Floquet modes
$|\Phi_{\lambda}(t)\rangle$ and the quasienergies
$\varepsilon_\lambda$ can be found from the eigenvalue equation:
\begin{equation}\label{Eq:Floquet_eingenvalue_problem}
  \left[ H-i\hslash\frac{\partial}{\partial t}\right]|\Phi_{\lambda}(t)\rangle
   =\varepsilon_{\lambda}|\Phi_{\lambda}(t)\rangle\;,
\end{equation}
where $H=H_0+V(t)$ with $V(t+T)=V(t)$.

The quasienergies and the Floquet modes at any time moment $t_0$ can
be also found from the evolution operator $U(t_{0}+T,t_{0})$
translating the system over one period in time. Indeed, from
$|\Psi_{\lambda}(t_0+T_0)\rangle=U(t_{0}+T,t_{0})|\Psi_{\lambda}(t_0)\rangle$,
Eq.~\eqref{Eq:Floquet_states} and the periodicity of the Floquet
modes we infer an alternative eigenvalue problem:
\begin{equation}\label{Eq:Floquet_eingenvalue_problem_U}
   U(t_{0}+T,t_{0})|\Phi_{\lambda}(t_{0})\rangle=e^{-i\varepsilon_{\lambda}T/\hslash}|\Phi_{\lambda}(t_{0})\rangle\;.
\end{equation}
The quasienergies $\varepsilon_{\lambda}$ are found then from the
eigenvalues $e^{-i\varepsilon_{\lambda}T/\hslash}$ of this equation.
 In comparison to a general
case of a continuous wave (CW) driving, for a periodic
short-pulsed driving Eq.~\eqref{Eq:Floquet_eingenvalue_problem_U}
offers an advantage that the evolution operator can be written
analytically in an explicit form with the help of the
approximations described in Section~\ref{Sec:Theory}. If $t=t_1^-$ is
the time moment just before the application of a light pulse then
$t=t_1^- +T$ is the time moment just before the application of the
next pulse and the propagation between these time moments is
determined by
\begin{equation}\label{Eq:Floquet_U_period}
   U(t_1^- +T,t_1^-)=U_0(T)\mathcal{U}(t_1) .
\end{equation}
 $\mathcal{U}(t_1)$ encapsulates  the action of the pulse
at $t=t_1$ and $U_0(T)$ is the contribution of the free
propagation between  consecutive light pulses. Using
appropriate expressions for $\mathcal{U}(t_1)$ from
Section~\ref{Sec:Theory} we can find analytical forms of $U(t_1^-
+T,t_1^-)$. With this, $\varepsilon_{\lambda}$ and
$|\Phi_{\lambda}(t_1^-)\rangle$ can be obtained from
Eq.~\eqref{Eq:Floquet_eingenvalue_problem_U}. In the range
$t_1<t<t_1+T$ any Floquet mode $|\Phi_{\lambda}(t)\rangle$ follows from $|\Phi_{\lambda}(t_1^-)\rangle$ as
$|\Phi_{\lambda}(t)\rangle=U_0(t-t_1)\mathcal{U}(t_1)|\Phi_{\lambda}(t_1^-)\rangle$.
Finally, if we know
the state of the system $|\Psi(t_1^-)\rangle$ at $t=t_1^-$, we can
calculate
the coefficients $C_\lambda$ in Eq.~\eqref{Eq:Floquet_series} as
\begin{equation}\label{Eq:Floquet_C}
   C_\lambda=\langle\Phi_{\lambda}(t_1^-)|\Psi(t_1^-)\rangle e^{i\varepsilon_{\lambda}t_1/\hslash}\;.
\end{equation}
Thus, using the solutions of
Eq.~\eqref{Eq:Floquet_eingenvalue_problem_U} and the initial state
of the system, the expansion of the corresponding solution of the
TDSE equation in the Floquet states can be determined.

In Ref.~\cite{Alex_Europhysics2005a} conditions for a
periodic cyclic evolution of a periodically driven quantum system
were formulated. Such an evolution takes place when the wave
function of the system fulfills the equation
\begin{equation}\label{Eq:periodicity_condition}
   |\Psi(t_0+k\mathcal{T})\rangle=e^{i k \phi}|\Psi(t_0)\rangle\;,
\end{equation}
where $\mathcal{T}$ is the duration of each cycle. The index
$k=1,2,3,\cdots$ numerates the cycles, and the real number $\phi$ is the
phase which is acquired by the wave function during a single
evolution cycle. With Eq.~\eqref{Eq:periodicity_condition} the
expectation value of any physical observable behaves
periodically~\cite{Alex_Europhysics2005a}. Comparing
Eq.~\eqref{Eq:periodicity_condition} with
Eqs.~\eqref{Eq:Floquet_series} and \eqref{Eq:Floquet_states} we
see that they are compatible if and only if one of the following conditions is
fulfilled:
\begin{itemize}
    \item[(a)] all but one coefficients $C_\lambda$ in
    Eq.~\eqref{Eq:Floquet_series} vanish, meaning that the system is in
    a pure Floquet state;
    \item[(b)]
    $\varepsilon_{\lambda}-\varepsilon_{\lambda'}=0$ for all (at least one) populated Floquet state pairs indexed by $\lambda$ and $\lambda'$
    (i.e. such that $C_\lambda \neq 0$ and $C_{\lambda'} \neq 0$),
    that means crossing of all populated quasienergy levels at just one point in the first Brillouin zone
    for the chosen parameters of the driving;
    \item[(c)] there are some populated Floquet state pairs with
    $|\varepsilon_{\lambda}-\varepsilon_{\lambda'}|=\frac{l_{\lambda,\lambda'}\hslash\omega_{_T}}{n_{\lambda,\lambda'}}$,
    where $l_{\lambda,\lambda'}$ and $n_{\lambda,\lambda'}>l_{\lambda,\lambda'}$ are coprime positive integer numbers,
    whereas $\varepsilon_{\lambda}-\varepsilon_{\lambda'}=0$ is valid for the rest of
    the populated Floquet
    state pairs.
\end{itemize}
The duration of the evolution cycle is $\mathcal{T}=T$ in the
cases (a) and (b), whereas the case (c) leads to $\mathcal{T}={\rm
lcm}(\{n_{\lambda,\lambda'}\})T$. Here ${\rm lcm}$ denotes the
least common multiple. In order to realize the condition (a), the
initial state should be prepared correspondingly, before applying
the periodic pulse sequence. If only two quasienergy levels are
populated, the conditions (b) and (c) both result in a single
algebraic equation, whose form depends on a particular realization. For
periodic driving provided by repeating
a single ultrashort HCP the
condition (b) can not be fulfilled, as a consequence of the lack
of well-defined generalized parity of the driven system. Both
quasienergy levels in the first Brillouin zone belong to the same
symmetry-related class, that leads to avoided crossings of the
quasienergy levels \cite{Neumann1929,Alex_QW2004,Grifoni1998}. In
the framework of the IA, the action of a single HCP is
characterized by only one parameter reflecting the effective strength of the
light pulse. Thus, for the case of two occupied Floquet states, we have two parameters,
the kick strength and the period of the driving $T$, to fulfill a
single algebraic equation for each different variant of the
condition (c), i.e. for each different pair of the numbers $n$ and
$l$ in this condition.

If the number of populated quasienergy levels is larger than two,
the conditions (b) and (c) both
require that several algebraic equations must be fulfilled at the
same time. In such a case realizations of a periodic evolution may be
possible if the period of the evolution
is much longer than the driving period
\cite{Alex_Europhysics2005a,Alex_PRA2006}, i.e. $\mathcal{T} \gg
T$,  that is only relevant for
(c). An alternative is to create more sophisticated sequences of HCPs containing
several pulses in a single period of driving, with different
strengths or/and different delays between the single HCPs in the
period of the driving pulse sequence.
\subsubsection{ {Measures of sustainability and Aharonov-Anandan phase}}
Periodic driving fields can be used to sustain a particular
desired state of the driven system, implicating that the expectation
values of all physical observables are sustained close to desired
target values \cite{Alex_Europhysics2005a}. It is clear that such
an ability is extremely important for various applications
\cite{Grossmann1991,Grifoni1998,Bavli1992,Dunlap1986,Alex_Molecules2003}.
Following to
Refs.~\cite{Alex_Europhysics2005a,Alex_PRA2006}, we consider
the case when the sustainability of the system state is
complemented by its periodic evolution and denote such states as
periodic quasistationary states. The sustainability
(quasistationarity) of a quantum state is closely related to the
fidelity \cite{Jozsa1994,Haug2005} of this state in a given time
range.  The fidelity function of a time-dependent pure state
$|\Psi(t)\rangle$ is given by
\begin{equation}\label{Eq:fidelity}
    \mathcal{F}\left[\Psi(t),\Psi(t_0)\right]=\left|\langle \Psi(t)|\Psi(t_0) \rangle\right|^2\;,
\end{equation}
where $|\Psi(t_0)\rangle$ is the reference state at $t=t_0$
\footnote{Sometimes a square root of the rhs of
Eq.~\eqref{Eq:fidelity} is used as an alternative definition of
the fidelity~\cite{Uhlmann2011,Uhlmann2009}}. $\mathcal{F}$ is connected with
several possible measures of the distance between the states
$|\Psi(t)\rangle$ and $|\Psi(t_0)\rangle$, e.g. with the
Fubini-Study distance
\cite{Wilczek_book,Pati1995,Pati1995_PRA,Uhlmann2009,Uhlmann2011}\footnote{Notice
that there is a different definition in
Refs.~\cite{Pati1992,Alex_PRA2006,Fubini1903,Study1905}
coinciding with the definition of the \textit{sine distance} as
discussed in Refs.~\cite{Rastegin2010,Mendonca2008}.}:
\begin{equation}\label{Eq:Fubini_Study_distance}
    d_{\rm FS}(t)=\sqrt{2}\sqrt{1- \sqrt{\mathcal{F}\left[\Psi(t),\Psi(t_0)\right]}}\;.
\end{equation}
Notice that $0 \le \mathcal{F}\left[\Psi(t),\Psi(t_0)\right]\le 1$ is valid,
leading to $0 \le d_{\rm FS}(t) \le \sqrt{2}$ (it is, of course,
possible to use different normalizations of this distance). An
additional geometrical interpretation of this quantity is provided
by the so-called  Bargmann
angle~\cite{Boya1989,Pati1992,Alex_PRA2006} defined as
\begin{equation}\label{Eq:Bargmann_angle}
    \beta(t)=\arccos \sqrt{\mathcal{F}\left[\Psi(t),\Psi(t_0)\right]}\;.
\end{equation}
It is confined to the range $0 \le \beta(t) \le \pi/2$ \footnote{In the
original works of Fubini and Study a double of this angle is used
as a distance measure \cite{Fubini1903,Study1905,Uhlmann2009}.}. A
geometrical analysis \cite{Alex_PRA2006} of the quantum evolution
allows to determine necessary and sufficient conditions for a
quantum system to be in a periodic quasistationary state. For that one of the
periodicity conditions (a)-(c) mentioned above should be fulfilled
and $\beta(t)$ should remain small during the period of the
evolution of the driven quantum system. In order to quantify the
condition on $\beta(t)$ via a single number, one can introduce
\footnote{Notice that there is a slightly different definition in
Ref.~\cite{Alex_PRA2006}, as a consequence of a different
definition of the distance.}, e.g.,
\begin{equation}\label{Eq:epsilon_measure}
    \varepsilon=\max_{t\in[t_0,t_0+\mathcal{T})}d_{\rm FS}^2(t)
    =2\sqrt{2}\max_{t\in[t_0,t_0+\mathcal{T})}\sin\left(\frac{\beta(t)}{2}\right).
\end{equation}
We see that $0 \le \varepsilon \le 2$ is valid. Thus a
necessary and sufficient condition of the quasistationarity of the
periodically evolving state can be formulated as $\varepsilon \ll
1$.

The concept of fidelity can be generalized to the case when the
states of the quantum system are described by a density matrix.
Following to Refs.~\cite{Jozsa1994,Uhlmann2009}, for a state
determined by a density matrix $\rho(t)$ and the reference state
at $t=t_0$ determined by $\rho(t_0)$, both normalized as ${\rm
Tr}\rho=1$, we can use
\begin{equation}\label{Eq:fidelity_density_matrix}
    \mathcal{F}\left[\rho(t),\rho(t_0)\right]=\left\{{\rm Tr}\left[\!\!\sqrt{\rho(t)}\:\rho(t_0) \sqrt{\rho(t)}
    \:\right]^{1/2}\right\}^2\;.
\end{equation}
Equations \eqref{Eq:Fubini_Study_distance} and
\eqref{Eq:Bargmann_angle} with this definition of the fidelity
function give then the so-called Bures distance $d_{\rm B}(t)$ and
Bures angle $\beta_{\rm B}(t)$
\cite{Bures1969,Uhlmann2009,Bengtsson_book}, the generalizations
of the Fubini-Study distance and the Bargmann angle, respectively.
Consequently, the measure of sustainability $\varepsilon$ can be
also calculated as it is determined by
Eq.~\eqref{Eq:epsilon_measure}. In the case of a coherent
dynamics, Eq.~\eqref{Eq:rho_t_result} can be used here to
find $\rho(t)$. Specifically for the case of a coherent
dynamics, it is significantly easier to use an alternative
function
\begin{equation}\label{Eq:fidelity_density_matrix_A}
    \mathcal{F}_{\rm A}\left[\rho(t),\rho(t_0)\right]={\rm Tr}\left[\frac{\rho(t)}{||\rho(t)||}\frac{\rho(t_0)}{||\rho(t_0)||}\right]
\end{equation}
in place of the fidelity function. Here  $||\rho||=\sqrt{{\rm
Tr}[\rho^2]}$ is the Frobenius norm \cite{Uhlmann2009} of  $\rho$. Such a function also satisfies all the axioms
\cite{Jozsa1994} for a fidelity function
\footnote{From the Cauchy-Schwarz inequality follows $0\le \mathcal{F}_{\rm
A}\left[\rho(t),\rho(t_0)\right]\le 1$ and $\mathcal{F}_{\rm
A}\left[\rho(t),\rho(t_0)\right]=1$ is achieved if and only if
$\rho(t)=\rho(t_0)$.}
except the one concerning the interpretation of the fidelity function
as a transition amplitude in a general case
\cite{Jozsa1994,Uhlmann2011}. However, we can use also the
function $\mathcal{F}_{\rm A}\left[\rho(t),\rho(t_0)\right]$ as an
alternative measure of the closeness of the quantum states
determined by $\rho(t)$ and $\rho(t_0)$. Consequently, we can also
calculate the corresponding distance and angle between these two
states, as they follow from Eqs.~\eqref{Eq:Fubini_Study_distance}
and Eq.~\eqref{Eq:Bargmann_angle}, based on the definition given
by Eq.~\eqref{Eq:fidelity_density_matrix_A}. Taking into account
Eq.~\eqref{Eq:rho_t_result}, we can rewrite
Eq.~\eqref{Eq:fidelity_density_matrix_A} as
\begin{equation}\label{Eq:fidelity_density_matrix_A_result}
    \mathcal{F}_{\rm A}\left[\rho(t),\rho(t_0)\right]=
    \frac{\sum_{nn'}f_n f_{n'}\left|\langle\Psi_{n'}(t)|\Psi_{n}(t_0)\rangle\right|^2}{\sum_n f_n^2}
\end{equation}
that can be calculated knowing the evolution of the wave function
of each initially populated state.


The quasistationarity of a periodically driven quantum system can
be also connected with properties of the phase $\phi$ in
Eq.~\eqref{Eq:periodicity_condition}, acquired by the driven
system during a cycle of the periodic evolution
\cite{Alex_Europhysics2005a,Alex_PRA2006}. It can be calculated as
\begin{equation}\label{Eq:total_phase}
    \phi=-{\rm arg}\left[\langle \Psi(t_0+\mathcal{T})| \Psi(t_0) \rangle\right].
\end{equation}
Looking at the conditions (a)-(c) for the periodic evolution of
the driven system, we notice that $\phi$ can be also written just
as
\begin{equation}\label{Eq:total_phase_quasienergy}
    \phi=-\varepsilon_\lambda\mathcal{T}/\hslash\;,
\end{equation}
where $\varepsilon_\lambda$ is the quasienergy of any of the
populated quantum states and unimportant phase shifts by integer
multiples of $2\pi$ are neglected.  On the other hand, the total
phase $\phi$ can be represented as a sum,
\begin{equation}\label{Eq:total_phase_sum}
    \phi=\phi_{\rm D} +\phi_{\rm AA}\;,
\end{equation}
of a dynamical phase
\begin{equation}\label{Eq:phi_D}
    \phi_{\rm D}=-\frac{1}{\hslash}\int_{t_0}^{t_0+\mathcal{T}}\langle\Psi(t)|H|\Psi(t) \rangle {\rm d}t=
           -i\int_{t_0}^{t_0+\mathcal{T}}\left\langle\Psi(t)\left|\frac{\partial}{\partial t}\right|\Psi(t) \right\rangle
           {\rm d}t
\end{equation}
and a geometric Aharonov-Anandan phase (non-adiabatic generalized
Berry phase) $\phi_{\rm AA}$
\cite{Aharonov1987,Wilczek_book,Grifoni1998,Alex_Europhysics2005a,Alex_PRA2006}.
Inserting Eq.~\eqref{Eq:Floquet_series} with
Eq.~\eqref{Eq:Floquet_states} into Eq.~\eqref{Eq:phi_D},
subtracting the result from Eq.~\eqref{Eq:total_phase_quasienergy}
and neglecting phase shifts by integer multiples of $2\pi$, we get
under conditions of a periodic evolution of the system
\begin{equation}\label{Eq:phi_AA}
    \phi_{\rm AA}=\sum_\lambda |C_\lambda|^2
    \frac{(\varepsilon_\lambda-\varepsilon_{\lambda_0})\mathcal{T}}{\hslash}+i
    \sum_{\lambda,\lambda'} C_\lambda^* C_{\lambda'}\int_{t_0}^{t_0+\mathcal{T}}
    e^{i(\varepsilon_\lambda-\varepsilon_{\lambda'})t/\hslash}
    \left\langle\Phi_\lambda(t)\left|\frac{\partial}{\partial t}\right|\Phi_{\lambda'}(t)\right\rangle {\rm d}t\;,
\end{equation}
where
$\lambda_0$ corresponds to an arbitrary occupied state. One can
see that $\phi_{\rm AA}$ is a real number and does not depend on
shifts of the energy scale, i.e. is gauge invariant. From Eq.~\eqref{Eq:phi_AA} follows that the Aharonov-Anandan phase also does not depend
on the selection of the starting point of the time integration
over the period $\mathcal{T}$. Thus it is a geometrical quantity
characterizing the closed path of the system in the Hilbert space
during a single period of the evolution. Usually the condition $|\phi_{\rm AA}|\ll 1$ is a strong indication of
the quasistationarity of the system evolution \cite{Alex_Europhysics2005a,Alex_PRA2006}.
\subsubsection{ {Implications for the periodic pulsed driving}}
{ {
If the evolution satisfies the periodicity condition (a) meaning that
 the system is in a pure Floquet state $\Psi_\lambda(t)$,
Eq.~\eqref{Eq:phi_AA} simplifies to
\begin{equation}\label{Eq:phi_AA_pure_Floquet}
    \phi_{\rm AA}\equiv \phi_{\rm AA}^{(\lambda)}= i
    \int_{t_0}^{t_0+T}
    \left\langle\Phi_\lambda(t)\left|\frac{\partial}{\partial t}\right|\Phi_{\lambda}(t)\right\rangle {\rm
    d}t\;.
\end{equation}
This expression simplifies using the
properties of Floquet states \cite{Grifoni1998} which   are eigenstates of the
operator
\begin{equation}\label{Eq:Schroedinger_operator}
    \mathcal{S}=H-i\hslash\frac{\partial}{\partial t}
\end{equation}
acting in the Sambe space $\mathbb{S}=\mathbb{H}\otimes\mathbb{T}$
\cite{Sambe1973} consisting of  a
direct product of the Hilbert space $\mathbb{H}$ of square
integrable complex functions} in the coordinate space and the
Hilbert space $\mathbb{T}$ of $T$-periodic functions in time. Here
the Hamiltionian of the periodically driven system is given by
\begin{equation}\label{Eq:H_general_pulse_sequence}
   H=H_0+V_0\sum_k f(t-kT)\;,
\end{equation}
where we have made use of Eq.~\eqref{Eq:V} expressing the
interaction corresponding to a single pulse. The Hamiltonian of
the unperturbed system $H_0$ and the interaction part $V_0$ do not
depend on time. $k$ enumerates the applied pulses. Using the
Schr\"{o}dinger operator,
Eq.~\eqref{Eq:Floquet_eingenvalue_problem} can be formulated as
$\mathcal{S}|\Phi_\lambda(t)\rangle=\varepsilon_\lambda
|\Phi_\lambda(t)\rangle$. With the help of the scalar product in
the Sambe space
$\;\langle\Big\langle\cdot\:|\:\cdot\Big\rangle\rangle\equiv\frac{1}{T}\int_0^T
\langle\cdot\:|\:\cdot\rangle{\rm d}t\;$
\cite{Sambe1973,Grifoni1998} Eq.~\eqref{Eq:phi_AA_pure_Floquet}
can be rewritten as
\begin{equation}\label{Eq:phi_AA_pure_Floquet_Sambe}
    \phi_{\rm AA}^{(\lambda)}= iT
    \Big\langle\!\left\langle\Phi_\lambda(t)\left|\frac{\partial}{\partial t}\right|\Phi_{\lambda}(t)\right\rangle\!\Big\rangle\;.
\end{equation}
Following the ideas of
Refs.~\cite{Fainshtein1978,Seleznyova1993,Grifoni1998} we
introduce a normalized time variable $\tilde{t}=\omega_{T} t$ and
write the Schr\"{o}dinger operator of the periodically driven
system then as
\begin{equation}\label{Eq:Schroedinger_operator_periodic}
    \mathcal{S}=H_0+V_0\sum_k f\left(\frac{\tilde{t}-2\pi k}{\omega_{T}}\right)-i\hslash \omega_{T}\frac{\partial}{\partial
    \tilde{t}}\;.
\end{equation}
Assume firstly that  the parameters of a single light pulse are
fixed. The second term in
Eq.~\eqref{Eq:Schroedinger_operator_periodic} depends on
$\omega_{T}$. A useful mathematical trick here, allowing to eliminate
this dependence, is to parameterize the pulse shape by
$\omega_{T}$ in such a way that the second term on the right
hand side of Eq.~\eqref{Eq:Schroedinger_operator_periodic} does
not change as $\omega_{T}$ is varied \footnote{It is the
same as to keep the coefficients of the Fourier expansion constant
as in Ref.~\cite{Seleznyova1993}. However, our suggestion is
more transparent for the case of the periodic pulsed driving.}.
This is achieved if we parameterize the pulse duration $\tau_{\rm
d}$ as
\begin{equation}\label{Eq:tau_d_parametrization}
    \tau_{\rm d}(\omega_T)=\tau^{(0)}_{\rm
    d}\frac{\omega^{(0)}_T}{\omega_T}\;,
\end{equation}
where $\omega_T$ and $\tau_{\rm d}$ are now changing variables
whereas $\tau^{(0)}_{\rm d}$ and $\omega^{(0)}_T$ denote their
values at which we calculate $\phi_{\rm AA}^{(\lambda)}$.
Introducing a new function $\tilde{f}(s)=f(s\tau_{\rm d})$ such
that  $\tilde{f}(s)$ has no dependence on $\tau_{\rm d}$ [cf.
Eqs.~\eqref{Eq:Gaussian_pulse}-\eqref{Eq:few_cycle_pulse_polynomial}],
we rewrite Eq.~\eqref{Eq:Schroedinger_operator_periodic} as
\begin{equation}\label{Eq:Schroedinger_operator_parametrized}
    \tilde{\mathcal{S}}(\tilde{t};\omega_T)=H_0
    +V_0\sum_k \tilde{f}\left(\frac{\tilde{t}-2\pi k}{\omega_{T}\tau_{\rm d}(\omega_T)}\right)
    -i\hslash \omega_{T}\frac{\partial}{\partial
    \tilde{t}}\;.
\end{equation}
Taking into account Eq.~\eqref{Eq:tau_d_parametrization}, we can
see that now the second term on the rhs of
Eq.~\eqref{Eq:Schroedinger_operator_parametrized} does not change
upon variation of $\omega_T$ that leads to
\begin{equation}\label{Eq:Schroedinger_operator_parametrized_derivative}
    \frac{\partial \tilde{\mathcal{S}}(\tilde{t};\omega_T)}{\partial \omega_T}=-i\hslash \frac{\partial}{\partial \tilde{t}}\;.
\end{equation}
Using this equation and $\tilde{t}=\omega_{T} t$ in
Eq.~\eqref{Eq:phi_AA_pure_Floquet_Sambe}, we get
\begin{equation}\label{Eq:phi_AA_pure_Floquet_via_S}
    \phi_{\rm AA}^{(\lambda)}= -\frac{2\pi}{\hslash}
    \Big\langle\!\left\langle\tilde{\Phi}_\lambda(\tilde{t})
    \left|\frac{\partial \tilde{\mathcal{S}}(\tilde{t};\omega_T)}{\partial \omega_T}
    \right|\tilde{\Phi}_{\lambda}(\tilde{t})\right\rangle\!\Big\rangle\;,
\end{equation}
where $\tilde{\Phi}_{\lambda}(\tilde{t})=\Phi_{\lambda}(t)$ are
eigenstates of $\tilde{\mathcal{S}}(\tilde{t};\omega_T)$ with the
same quasienergies $\varepsilon_\lambda$. Application
\cite{Seleznyova1993,Grifoni1998,Fernandez2004} of the
Hellmann-Feynman theorem \cite{Hellmann_book,Feynman1939} leads to
\begin{equation}\label{Eq:phi_AA_Hellmann_Feynman}
    \phi_{\rm AA}^{(\lambda)}=-\frac{2\pi}{\hslash}\left[\frac{\partial
    \varepsilon_\lambda(\omega_T,\tau_{\rm d})}{\partial \omega_T}
    -\frac{\partial
    \varepsilon_\lambda(\omega_T,\tau_{\rm d})}{\partial \tau_{\rm d}}\frac{\tau_{\rm
    d}}{\omega_T}\right],
\end{equation}
where it is important to express the quasienergy
$\varepsilon_\lambda$ as a function of $\omega_T$ and $\tau_{\rm
d}$ (here they are again considered as two independent parameters
of the periodic driving field) and to remember that it should be
taken from the first Brillouin zone.



In the case of a realization of the periodicity condition (b),
Eq.~\eqref{Eq:phi_AA} reduces to
\begin{equation}\label{Eq:phi_AA_condition_b}
    \phi_{\rm AA}=\sum_\lambda |C_\lambda|^2
    \left[\frac{(\varepsilon_\lambda-\varepsilon_{\lambda_0})T}{\hslash}
    +\phi_{\rm AA}^{(\lambda)}\right]+
    \sum_{\lambda,\lambda' (\lambda'\neq\lambda)} C_\lambda^* C_{\lambda'}i_{\lambda,\lambda'}\;,
\end{equation}
where $i_{\lambda,\lambda'}$ is given by
\begin{equation}\label{Eq:phi_AA_i}
    i_{\lambda,\lambda'}=i\int_{t_0}^{t_0+T}
    \left\langle\Phi_\lambda(t)\left|\frac{\partial}{\partial t}\right|\Phi_{\lambda'}(t)\right\rangle  {\rm d}t\;,
\end{equation}
Here we can proceed as for the derivation of
Eq.~\eqref{Eq:phi_AA_Hellmann_Feynman} and then apply a more
general formulation of the Hellmann-Feyman theorem including the
case of the non-diagonal matrix elements and the case of
degenerate levels \cite{Fernandez2004}. This implies then that we
get $i_{\lambda,\lambda'}=0$ for an appropriate choice of the
states $|\Phi_\lambda(t)\rangle$ corresponding to the degenerate
energy level. This is automatically the case when the energy level
crossing takes place, e.g., by a change of the pulse duration
$\tau_{\rm d}$ or of the driving period $T=2\pi/\omega_{T}$ and
the selected states correspond to the crossing
levels~\cite{Fernandez2004}. Then
Eq.~\eqref{Eq:phi_AA_condition_b} simplifies to
\begin{equation}\label{Eq:phi_AA_condition_b_simplified}
    \phi_{\rm AA}=\sum_\lambda |C_\lambda|^2
    \left[\frac{(\varepsilon_\lambda-\varepsilon_{\lambda_0})T}{\hslash}
    +\phi_{\rm AA}^{(\lambda)}\right].
\end{equation}


Considering the periodicity condition (c)  we
obtain
\begin{equation}\label{Eq:phi_AA_condition_c}
    \phi_{\rm AA}=\frac{\mathcal{T}}{T}\sum_\lambda |C_\lambda|^2
    \left[\frac{(\varepsilon_\lambda-\varepsilon_{\lambda_0})T}{\hslash}
    +\phi_{\rm AA}^{(\lambda)}\right]+\;\frac{\mathcal{T}}{T}\!\!\sum_{\lambda,\lambda'
(\lambda'\neq\lambda,\varepsilon_{\lambda'}=\varepsilon_{\lambda})}\!\!
    C_\lambda^* C_{\lambda'}i_{\lambda,\lambda'}\;,
\end{equation}
Here we have taken into account that the sum over
$\lambda,\lambda'$ with $\varepsilon_{\lambda'}\neq
\varepsilon_{\lambda}$ in Eq.~\eqref{Eq:phi_AA} vanishes for
$\mathcal{T}=nT$ ($n=2,3,4,\ldots$). The last term in
Eq.~\eqref{Eq:phi_AA_condition_c} can be treated in the same way
as for the case of the periodicity condition (b). Selecting the
appropriate eigenstates for each degenerate energy level, this term
can be eliminated that results in
\begin{equation}\label{Eq:phi_AA_condition_c_simplified}
    \phi_{\rm AA}=\frac{\mathcal{T}}{T}\sum_\lambda |C_\lambda|^2
    \left[\frac{(\varepsilon_\lambda-\varepsilon_{\lambda_0})T}{\hslash}
    +\phi_{\rm AA}^{(\lambda)}\right].
\end{equation}
This general equation is  useful for calculating
 $\phi_{\rm AA}$ in all cases when the system undergoes  a periodic evolution,
 i.e. for any of the conditions (a)-(c).
For the particular case of a harmonically driven two-level system
and $\mathcal{T}=T\ $,  Eq.~\eqref{Eq:phi_AA_condition_c_simplified}
leads to the same expression as obtained from an analytical
solution for the time-dependence of the Floquet modes
\cite[p.~993]{Seleznyova1993}.

We see that the Aharonov-Anandan phase can be determined without
really calculating the wave function dynamics during the whole
period of the evolution, only from the quasienergy spectrum and
initial populations of the Floquet states, and thus possesses a
certain predictive power. As it is mentioned above, in contrast to
a general CW driving, it is a property of a periodic driving by
ultrashort broadband light pulses that the quasienergy spectrum can be found
from Eq.~\eqref{Eq:Floquet_eingenvalue_problem_U} where the
evolution operator can be approximately expressed in an explicit
analytical form. In this sense, for realization of the
quasistationarity, in practice it can be more advantageous to
calculate the Aharonov-Anandan phase than to estimate the quantity
given by Eq.~\eqref{Eq:epsilon_measure}. An example of the
utilization of the Aharonov-Anandan phase for the characterization
of the sustainability of a periodic quantum evolution is discussed
in Section~\ref{Sec:Appl_DQW}.





\subsection{ {Quantum dynamics with dissipation: Floquet-Liouville
approach}}\label{Sec:FL_approach}
Next, let us consider periodically driven quantum systems with
dissipation. We assume that the system dynamics is
governed by a Markovian (convolutionless) master equation
\cite{Breuer_book,Graham1994,Grifoni1998,Sarandy2007,Sauvan2009}:
%
\begin{equation}\label{Eq:evolution_superoperator}
    i\hslash\frac{\partial}{\partial t} \rho(t) =\mathcal{L}(t) \rho(t)\;,
\end{equation}
where $\mathcal{L}(t)$ is a linear Liouville super-operator
(Liouvillian) which is periodic in time with a period $T$. Any
solution of Eq.~\eqref{Eq:evolution_superoperator} can be written
in the form \cite{Adrianova_book,Graham1994,Sauvan2009}
\begin{equation}\label{Eq:density_operator_Floquet}
    \rho(t)=\sum_{\mu\nu}
    \rho^{(\mu\nu)}(t)
    e^{-i\Omega_{\mu\nu}t/\hslash}\;,
\end{equation}
where the periodic operators
\begin{equation}\label{Eq:density_operator_Floquet_modes}
    \rho^{(\mu\nu)}(t+T)=\rho^{(\mu\nu)}(t)\;
\end{equation}
are the Floquet-Liouville modes and super-eigenvalues
$\Omega_{\mu\nu}$ are complex numbers. Introducing the
Floquet-Liouville super-operator
\begin{equation}\label{Eq:Floquet_Liouville_operator}
    \mathcal{L}_{\rm
    F}(t)=\mathcal{L}(t)-i\hslash\frac{\partial}{\partial t} \; ,
\end{equation}
the Floquet-Liouville modes and  the corresponding
super-eigenvalues can be found from the operator eigenvalue
equation:
\begin{equation}\label{Eq:superoperator_eigenvalues_equations}
    \mathcal{L}_{\rm F}(t)\rho^{(\mu\nu)}(t) = \Omega_{\mu\nu}
    \rho^{(\mu\nu)}(t)\;,
\end{equation}
i.e. $\rho^{(\mu\nu)}$ are right eigenstates of $\mathcal{L}_{\rm
F}$, which is in general non-Hermitian.  In the tetradic basis
$|\alpha \beta\rangle\!\rangle\equiv |\alpha \rangle\langle\beta|$
of the Liouville space one gets an eigenvalue problem for a system
of ordinary differential equations:
\begin{equation}\label{Eq:superoperator_eigenvalues_equations_ODE}
    \sum_{\delta \gamma}\langle\!\langle \alpha \beta |\mathcal{L}_{\rm F}(t)|\delta
    \gamma\rangle\!\rangle
    \langle\!\langle \delta \gamma
    |\rho^{(\mu\nu)}(t)\rangle\!\rangle
    = \Omega_{\mu\nu} \langle\!\langle \alpha \beta |
    \rho^{(\mu\nu)}(t)
    \rangle\!\rangle\;,
\end{equation}
where  $\langle\!\langle \alpha \beta | \rho^{(\mu\nu)}(t)
\rangle\!\rangle\equiv {\rm Tr}\left[\rho^{(\mu\nu)}(t)\ |\alpha
\rangle\langle\beta| \right]= \langle \alpha | \rho^{(\mu\nu)}(t)|
\beta \rangle$ is the projection of the state determined by
$\rho^{(\mu\nu)}(t)$ onto the basis state $|\alpha
\beta\rangle\!\rangle$ and $\langle\!\langle \alpha \beta
|\mathcal{L}_{\rm F}(t)|\delta \gamma\rangle\!\rangle$ is the
corresponding matrix element of the super-operator
$\mathcal{L}_{\rm F}$ \footnote{Notice that we chose slightly
different notations for the trace scalar product here and for the
scalar product in the Sambe space [cf., e.g.,
Eq.~\eqref{Eq:phi_AA_pure_Floquet_Sambe}]. In the literature both
scalar products are frequently denoted using double brackets. We
prefer to avoid a possible ambiguity here. Trace scalar products
in the tetradic Sambe space, requiring a mixture of both
notations, are not explicitly used in this work.}. Because both
$\mathcal{L}_{\rm F}(t)$ and $\rho^{(\mu\nu)}(t)$ are periodic in
time, Eq.~\eqref{Eq:superoperator_eigenvalues_equations_ODE} can
be expanded in Fourier series leading to a matrix eigenvalue
problem that is the essence of the Floquet-Liouville super-matrix
formalism \cite{Ho1986,Chu2004}. Note that in the limit of
vanishing dissipation the super-eigenvalues $\Omega_{\mu\nu}$ are
real and can be expressed via the quasienergies  of the
corresponding Schr\"odinger equation \cite{Chu2004}:
\begin{equation}\label{Eq:supereigenvalues_eigenvalues}
    \Omega_{\mu\nu}=\varepsilon_\mu-\varepsilon_\nu +\hslash\omega_{T}n\;,
\end{equation}
where $n$ is an arbitrary integer number and $\omega_{T}=2\pi/T$.
Selecting $n=-1,0,$ or $1$, possible values of $\Omega_{\mu\nu}$
can be also restricted to the first Brillouin zone. This is also
valid for $\mathrm{Re}[\Omega_{\mu\nu}]$ in the case of
nonvanishing dissipation. From
Eq.~\eqref{Eq:supereigenvalues_eigenvalues} it is clear why it is
practical to number the super-eigenvalues by two indices. Such a
convention reflects the fact that the related eigenstates belong
to the tetradic (Liouville) space whereas the quasienergy states
of the corresponding dissipationless system belong to the dyadic
(Hilbert) space.

For a quantum dissipative system the imaginary parts of the
super-eigenvalues can be only negative or zero. Following the
arguments of Ref.~\cite{Graham1994}, all the eigenstates
$\rho^{(\mu\nu)}$ with a negative imaginary part of the
super-eigenvalue $\mathrm{Im}[\Omega_{\mu\nu}]<0$ must have a
vanishing trace ${\rm Tr}[\rho^{(\mu\nu)}]=0$  to ensure
the particle number conservation (i.e. the conservation of the
trace of the density matrix). Thus they do not correspond to any
real physical state of the system. Let us consider then the states
with $\mathrm{Im}[\Omega_{\mu_{\rm s}\nu_{\rm s}}]=0$. These
eigenstates do not decay. The system prepared in such a state
remains in this state forever. They have to be real
physical states of the system and therefore possess a nonvanishing
trace: ${\rm Tr}[\rho^{(\mu_{\rm s}\nu_{\rm s})}]\neq 0$. From the
trace conservation follows that also the real part of their
super-eigenvalues must vanish, resulting in $\Omega_{\mu_{\rm
s}\nu_{\rm s}}=0$. In general, there can be more than one
such state \cite{Lidar1998}. From these considerations follows that in
the long time limit $t\rightarrow \infty$ we have
$\rho(t)\rightarrow  \rho^{\rm st}(t)$, with the property
$\rho^{\rm st}(t+T)=\rho^{\rm st}(t)$. Thus any measurable
physical quantity should have a periodic dynamics after a long
enough time since the initiation of the periodic driving, i.e. the
time that is considerably larger than the relaxation time scales.
The same conclusion is reached by the authors of
Refs.~\cite{Ho1986,Chu2004} for a particular case when there
is an additional inhomogeneous term on the rhs of
Eq.~\eqref{Eq:evolution_superoperator}. An example of the dynamics
induced in a periodically driven quantum many-body dissipative
system is discussed in Section~\ref{Sec:periodic_magnetic_pulses}.
Notice that the rotating wave approximation (RWA) leading in the
long time limit to a density matrix being diagonal in the basis of
the Floquet states \cite{Grifoni1998}, and therefore also periodic
in time, is clearly inappropriate for the case of the periodic
driving by ultrashort pulses \cite{Graham1994}.

In Section~\ref{Sec:Floquet} we have discussed how to characterize
the degree of quasistationarity of a quantum system describe
by a density matrix on a basis of the appropriate
fidelity function. Recently, also the concept of the geometric
phase was generalized for the density matrix case
\cite{Sarandy2007}. Utilization of this quantity for controlling
the dynamics of quantum systems with dissipation, which are driven
by periodic sequences of ultrashort pulses, in analogy to the case
of pure states discussed in the previous section, is a promising
topic for future studies.

\section{Broadband pulse induced charge polarization and currents in nanostructures}\label{Sec:Appl}
In this section we discuss applications of broadband ultrashort light pulses
for control of the charge polarization and localization properties
as well as currents in nanostructures. In particular examples we
consider semiconductor superlattices, semiconductor and graphene
quantum rings, and semiconductor-based one-dimensional double
quantum wells (double quantum dots). The action of the pulses
transferring a momentum (a kick) to the excited system is
described based on the IA, keeping in mind its limitations
discussed in the previous section. If not explicitly stated
otherwise, here and below we denote the applied light pulses as
HCPs for the aims of simplicity, meaning all suitable temporal
profiles discussed above.

\subsection{Indirect transitions and direct current generation in unbiased semiconductor superlattices}
Semiconductor superlattices (SLs) play an important role in the
engineering of materials with desired band-structure properties
\cite{Esaki1970,Ivchenko_Pikus}. A famous technological
application was the proposal and the following experimental
realization of quantum cascade lasers
\cite{Kazarinov_Suris1971,Capasso_review2001}. An interesting
fundamental phenomenon which can be observed in the
one-dimensional semiconductor SLs is the quantum ratchet effect
\cite{reimann,linke1,linke2,Ivchenko2011,Nalitov2012}. This effect
can be used for all-optical injection of direct currents in
unbiased SLs. Direct ballistic currents can be also generated as a
result of a nonlinear harmonic mixing of electromagnetic field
components with multiple frequencies, e.g. $\omega$ and $2\omega$,
in SLs as well as in bulk materials
\cite{Goychuk1998,Alekseev1999,Atanasov1996,Hache1997,Costa2007}.

 Ref.~\cite{Moskalenko_PLA2006}  investigated a
possibility of generation of direct ballistic currents in
one-dimensional SLs by application of HCPs. As a simple illustrative model for a
one-dimensional SL potential one uses the Dirac comb potential
$U(x)=\Omega\sum_{l=-\infty}^{\infty}  \delta(x-la)$ was used, where $a$ is
the period of the SL, $l$ is an integer and $\Omega$ reflects the strength of a
single peak. The wave function of an electron with an effective
mass $m^*$ moving in this potential obeys the Bloch theorem:
\begin{equation}\label{Eq:PLA_eigenfunctions_Bloch}
    \psi_q(x)=\frac{1}{\sqrt{L}}u_q(x)e^{iqx},
\end{equation}
where $u_q(x)=u_q(x+a)$ is the Bloch amplitude. The wave number
$q$, corresponding to the quasi-momentum $\hslash q$, belongs to
the first Brillouin zone $q\in[-\pi/a,\pi/a]$. The wave number is
discrete because of the normalization in the box of the length
$L$. It is convenient to express the energy $E$ of such a state
via an auxiliary quantity $k$ as $E=\hslash^2 k^2/(2m^*)$
to obtain the well-known equation determining the energy spectrum
\cite{Fluegge_textbook,Kronig_Penney}:
\begin{equation}\label{Eq:PLA_deltas_superlattice_spectrum}
    \cos(ka)+\frac{m^*\Omega}{k\hslash^2}\sin(ka)=\cos(qa)\;.
\end{equation}
In this model the parameter $\Omega$ can be used to regulate the
widths of the resulting energy bands $E_n(q)$ ($n=1,2,3,\ldots$)
and band gaps. In Fig.~\ref{Fig:PLA2006_energy_bands} the energy
band structure of a SL with a period $a=15$~nm,
$\Omega=23\hslash^2/(ma)$ and $m^*=0.067m_0$, where $m_0$ is the
free electron mass, is presented. Considering the three lowest
subbands ($n=1,2,3$), this selection allows to achieve a good
agreement with the result of a realistic modelling of the band
structure of the Al$_{0.32}$Ga$_{0.68}$As/GaAs SL based on the
Kronig-Penney model \cite{Vyborny_2002}.
\begin{figure*}
  \includegraphics[width=7.5cm]{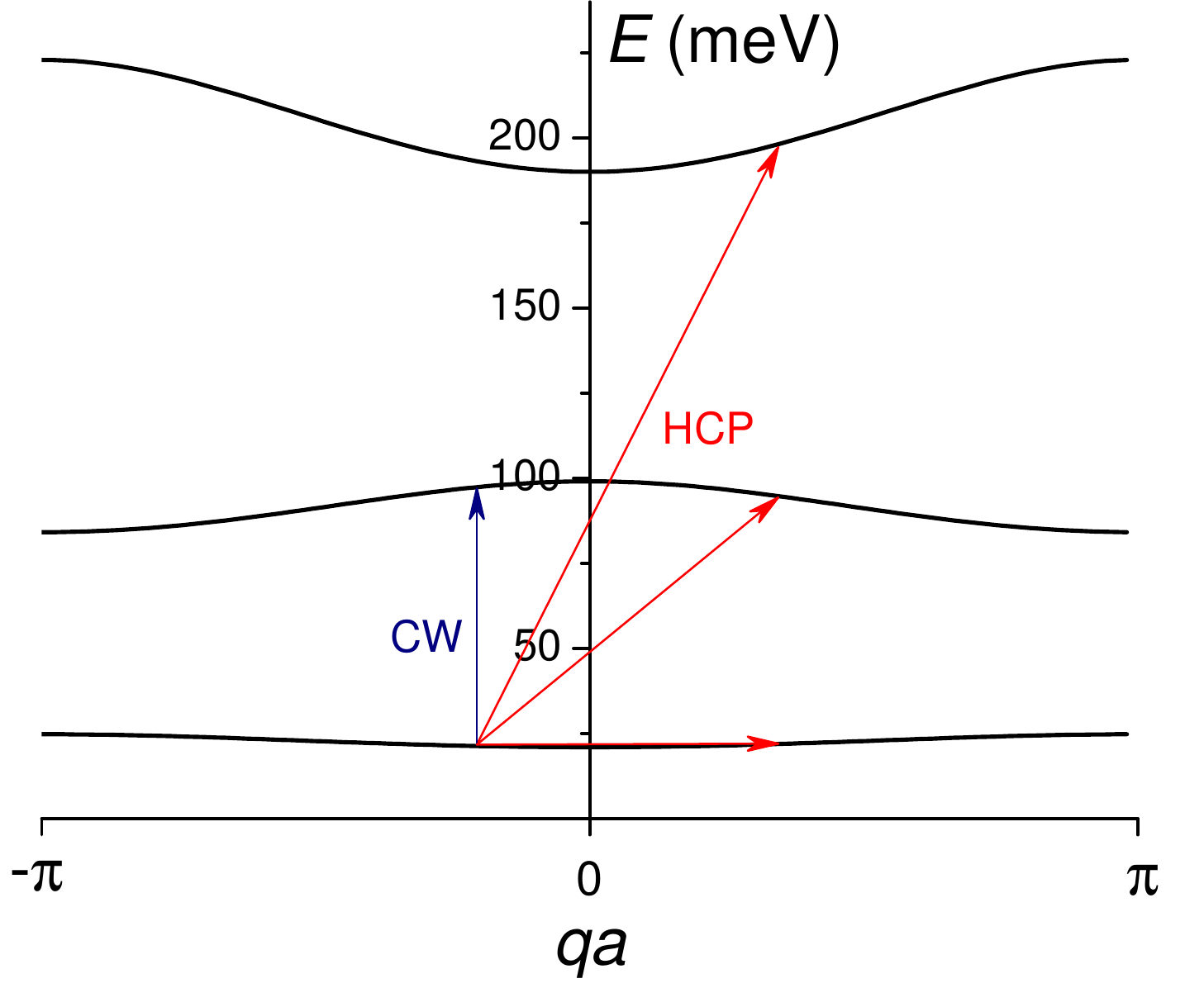}
  \caption{Band structure for the Dirac comb potential with $a=15$~nm and
  $\Omega=23\hslash^2/(m^*a)$. Vertical blue arrow shows
  the direct transition of an electron with a fixed wave number in the lowest subband
  due to a resonant CW excitation. Red lines illustrate
  indirect transitions of the same electron due to the application of HCPs (adapted from Ref.~\cite{Moskalenko_PLA2006}).
  \label{Fig:PLA2006_energy_bands}}
\end{figure*}

To ensure that the IA validity condition
\eqref{Eq:condition_IA_validity} is fulfilled for this system we
need to demand $\tau_{\rm d}
\ll\hslash/\max_{n_1,q_1;n_2,q_2}\{|E_{n_2}(q_2)-E_{n_1}(q_1)|\}$,
where the quantum numbers $n_1,q_1$ and $n_2,q_2$ numerate all
possible states involved by the excitation. For example, if the
excitation creates charge carriers in no more than three lowest
subbands, then in the case of the parameters of
Fig.~\ref{Fig:PLA2006_energy_bands} this condition means
$\tau_{\rm d} \ll 3~{\rm fs}$. HCPs with such durations recently
have became available experimentally \cite{Wu2012}. In the
framework of the IA the action of a HCP on an electron in the SL
results into a matching condition [cf.
Eq.~\eqref{Eq:IA_mapping_psi}] for the time-dependent wave
function $\Psi(x,t)$:
\begin{equation}\label{Eq:PLA_matching_condition}
  \Psi(x,t=t_1^+)=\Psi(x,t=t_1^-)\;e^{ipx},
\end{equation}
where the optimal selection of $t_1$ is discussed in
Sections \ref{Sec:HCPs} and  \ref{Sec:few_cycle}. $\hslash p$ is
the momentum transferred by the pulse, which is given by
Eq.~\eqref{Eq:Delta_p_free_space}.

Let us consider an electron being initially in the state
characterized by quantum numbers $n$ and $q$, i.e. having the
time-dependent wave function
$\Psi_{n,q}(x,t)=e^{-\frac{i}{\hslash}E_{n}(q)t}\psi_{n,q}(x)$
before the pulse application. Taking the action of the HCP into
account as it follows from Eq.~\eqref{Eq:PLA_matching_condition},
the wave function can be written as
\begin{equation}\label{Eq:PLA_psi_expansion}
  \Psi_{n,q}(x,t)=\sum_{n',q'}C_{n',q'}^{n,q}(t)
  e^{-\frac{i}{\hslash}E_{n'}(q')t}\psi_{n',q'}(x)\;,
\end{equation}
where the transition coefficients $C_{n',q'}^{n,q}(t)$ are given
by
\begin{equation}\label{Eq:PLA_psi_expansion_coeff}
  C_{n',q'}^{n,q}(t)=\Theta(t_1-t)\delta_{n,n'}\delta_{q,q'}+\Theta(t-t_1)T^{n,q}_{n',q'}\delta_{q+p,q'}\; .
\end{equation}
Here $\delta_{q,q'}$ denotes the Kronecker delta, $\Theta(t)$
denotes the Heaviside step function, and the coefficients
$T^{n,q}_{n',q'}$ are defined by
\begin{equation}\label{Eq:PLA_transition_ME_Bloch_definition}
    T^{n,q}_{n',q'}=\frac{1}{a}\int_0^a\!\! \mbox{d} x\
    u_{n',q'}^*(x)u_{n,q}(x)\;
\end{equation}
if $q'$ is inside the first Brillouin zone. Otherwise, for the
calculation one should use
$u_{n',q'}(x)=u_{n',q'+G_j}(x)e^{iG_jx}$,
%
where $G_j=j\frac{2\pi}{a}$ ($j\in\mathbb{Z}$) can be any
reciprocal lattice vector, and select $q'+G_j$ inside the first
Brillouin zone. Similarly, for the electron energy $E_{n'}(q')$ in
Eq.~\eqref{Eq:PLA_psi_expansion} we should use
$E_{n'}(q')=E_{n'}(q'+G_j)$ and select $q'+G_j$ such that it
belongs to the first Brillouin zone. From
Eqs.~\eqref{Eq:PLA_psi_expansion} and
\eqref{Eq:PLA_psi_expansion_coeff} we see a clear difference
between resonant transitions induced by a resonant
CW light and transitions induced by HCPs. It is illustrated in
Fig.~\ref{Fig:PLA2006_energy_bands}. Transitions due to the
resonant CW light are vertical. There is a single final state for
every initial eigenstate. Transitions induced by HCPs take place
with a momentum transfer, i.e. they are non-vertical. In this case
states in different energy subbands can be excited, with different
probabilities.

\begin{figure*}
  \includegraphics[width=9cm]{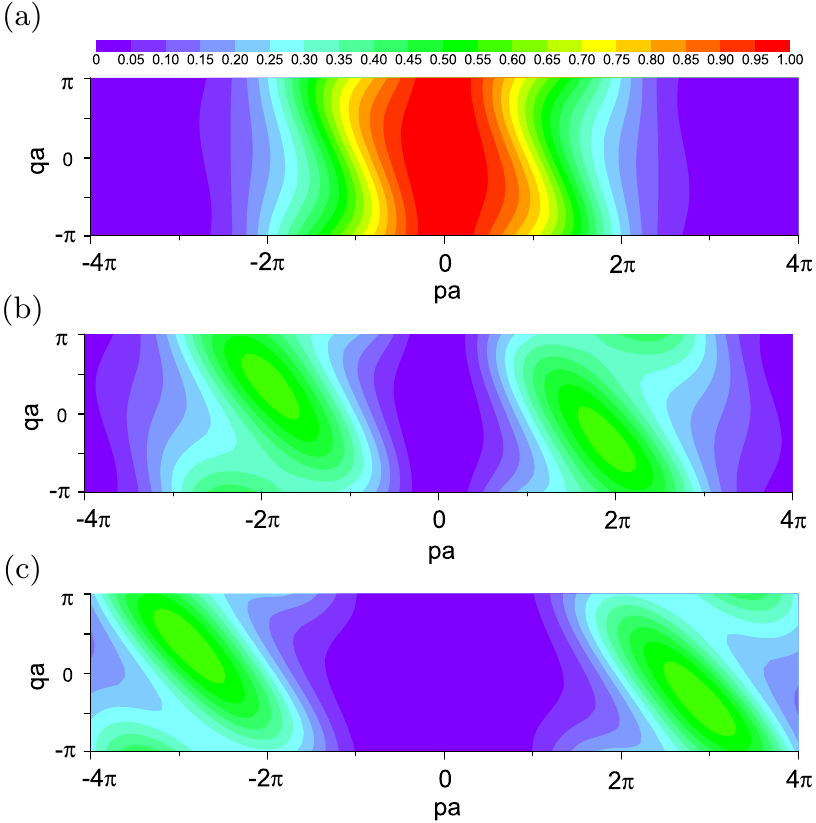}
  \caption{Occupation probabilities of  (a) first
  ($n=1$), (b) second ($n=2$), and (c) third ($n=3$)
  lowest subbands when a state in the first subband is excited
  by a HCP which transfers the momentum
  $\hslash p$. The probabilities are shown as functions of the
  initial wave number of the electron $q$ and the transferred wave
  number $p$ (momentum $\hslash p$), both normalized by the lattice
  constant $a$. Parameters of the SL are selected as in Fig.~\ref{Fig:PLA2006_energy_bands}
  (adapted from Ref.~\cite{Moskalenko_PLA2006}).\label{Fig:PLA2006_figure_Cnm}}
\end{figure*}

In Fig.~\ref{Fig:PLA2006_figure_Cnm} the occupation probabilities
of the three lowest subbands are shown in dependence on the
initial wave number $q$ of an electron in the lowest subband and
the transferred wave number $p$. We see that for small $p$ only
states in the same subband are excited. With increase of $p$, when
$pa$ becomes comparable with $\pi$, also states in upper subbands
begin to be susceptible to the excitation by a HCP. For $a=15$~nm
and $\tau_{\rm d}=1$~fs the value $pa=\pi$ is achieved for
amplitudes of the applied electric field on the order of 1~MV/cm
that is well within the range of accessibility in  experiment
\cite{Sell2008,Junginger2010,Wu2012}. Notice that the plots in
Fig.~\ref{Fig:PLA2006_figure_Cnm} are invariant upon the
transformation $p\rightarrow -p, q\rightarrow -q$, as it follows
from the symmetry of the system and the excitation. The invariance
to the transformation $q\rightarrow -q$, which is equivalent to
$x\rightarrow -x$, is however broken. This has consequences for
the charge transport in the system.

Let us consider an equilibrium electron gas in the SL at
temperature $T$. It is characterized by the Fermi-Dirac
distribution function $f_{n,q}(\mu,T)$, where $\mu$ is the
chemical potential determined by the electron density $n_{\rm
1D}=\frac{2}{L}\sum_{n,q}f_{n,q}(\mu,T)$ and $T$ is the
temperature. Application of a HCP induces a charge flow in the SL.
The corresponding ballistic current has an alternating and a
direct current (DC) component: $I_{\rm DC}+I_{\rm AC}(t)$. In
Ref.~\cite{Moskalenko_PLA2006} it was found that the induced DC
is given by
\begin{equation}\label{Eq:PLA_Idc}
    I_{\rm
    DC}=\frac{2e}{L}\sum_{n,n',q}f_{n,q}(\mu,T)|T_{n',q+p}^{n,q}|^{2}v_{n,q+p}\;,
\end{equation}
where $e$ is the electron charge,
\begin{equation}\label{Eq:PLA_v}
    v_{n,q}=\frac{1}{\hslash}\frac{\partial E_{n}(q)}{\partial q}\;
\end{equation}
is the electron velocity in the corresponding eigenstate. The
coefficients $T_{n',q+p}^{n,q}$ are defined by
Eq.~\eqref{Eq:PLA_transition_ME_Bloch_definition}. The factor 2
appears in Eq.~\eqref{Eq:PLA_Idc} and in the expression for the
electron density because the spin degeneracy was taken into
account. The alternating component of the current is determined exclusively by the induced intersubband coherences and (for $t>t_1$) is given by
\begin{equation}\label{iac}
I_{\rm AC}(t)=\!\!\sum_{n,n',n'',q\;(n'\neq
    n'')}\!\!\!\!f_{n,q}(\mu,T)T_{n',q+p}^{n,q}T_{n'',q+p}^{n,q}\;j_{n',n''}(q+p)\cos\!
    \left[\frac{(E_{n',q+p}\!-\!E_{n'',q+p})(t-t_1)}{\hslash}\right].
\end{equation}
The explicit analytic form of the coefficients (current matrix elements)  $j_{n',n''}(q+p)$ is somewhat lengthy so that we do not give it here. It can be  found in Ref.~\cite{Moskalenko_PLA2006}. Just after the pulse application, at $t=t_1^+$, both current components add up to
 a value $(n_{\rm 1D}aI_0) (pa)$, where $I_0=e\hslash/(m^*a^2)$.
 Notice that similar-looking expressions for DC and alternating contributions are found for the charge currents generated by few-cycle pulses in dielectrics \cite{Schiffrin2013,Yakovlev2016}. However, in the case considered here the resulting equations are field-free as far as the evolution during the interaction process is not resolved. More essentially, as discussed in the end of this section, we focus on the strong-field-driven electron dynamics in a different excitation regime which can not be captured by the physics of the Landau-Zener diabatic transitions, Bragg-like reflections at the edges of the Brillouin zone and inter(sub)band tunneling (cf. Ref.~\cite{Yakovlev2016} and the references therein).

\begin{figure*}
  \includegraphics[width=7.5cm]{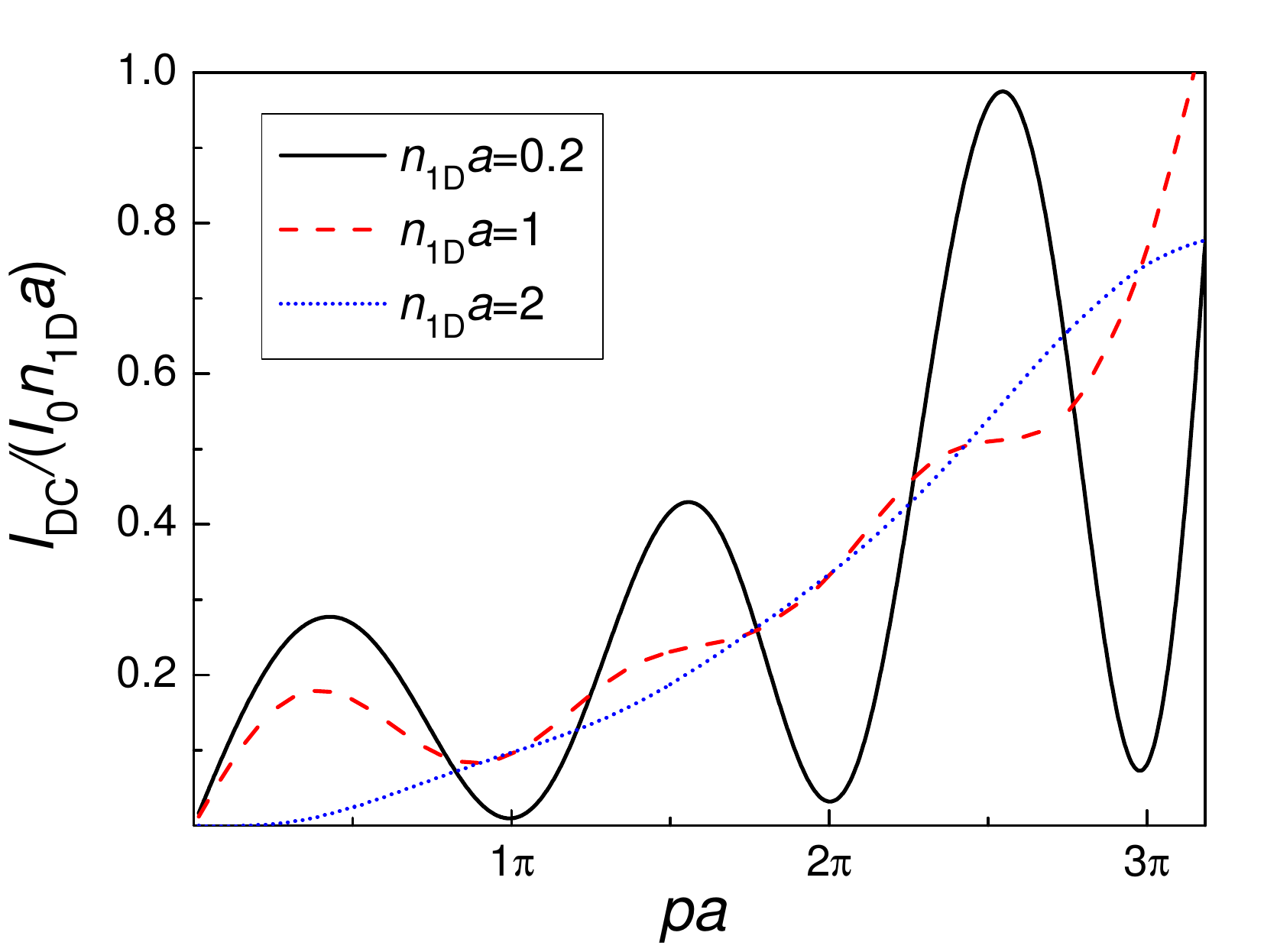}
  \caption{Ballistic direct current generated in the SL
  in dependence of the HCP strength characterized by the normalized transferred momentum
  $pa$. The case of  low  filling ($n_{\rm 1D}a$), half-filling, and full-filling of the lowest subband are shown.
  The current is normalized to $n_{\rm 1D}aI_0$, where
  $I_0=e\hslash/(m^*a^2)$.
  The parameters of the SL are as in Fig.~\ref{Fig:PLA2006_energy_bands} meaning that $|I_0|\approx 1.2~\mu$A
 (adapted from Ref.~\cite{Moskalenko_PLA2006}).\label{Fig:PLA2006_DC_current}}
\end{figure*}

Figure \ref{Fig:PLA2006_DC_current} shows the generated DC, which
is calculated from Eq.~\eqref{Eq:PLA_Idc}, in dependence on the
strength of the applied HCP characterized by the transferred
momentum. The three depicted curves correspond to different
populations in the lowest subband at zero temperature. The
oscillatory dependence of the generated DC on the transferred
momentum with minima around $pa=l\pi$ ($l\in \mathbb{Z}$), that is
strongly pronounced for a low filling of the subband and is also
present at the half-filling, is a consequence of the vanishing
electron velocity $v_{n,q}$ at the centrum of the Brillouin zone
and its boundaries. For low fillings, at equilibrium electrons
populate a part of the subband close to its centrum. In case of
$pa=l\pi$ they are transferred either to a Brillouin zone boundary
or again close to its centrum, where the electron velocity
vanishes. For the fully occupied first subband the DC generation
is possible only due to intersubband transitions to the
higher-laying subbands. For small pulse strengths intersubband
transitions practically do not occur (see
Fig.~\ref{Fig:PLA2006_energy_bands}). Only when the intersubband
transition probabilities become substantial, the current starts to
grow effectively with the increase of the HCP strength.

The time-dynamics of induced alternating current $I_{\rm AC}$  resolved in dependence
on the normalized transferred momentum $pa$ is illustrated in Fig.~\ref{Fig:AC_current}.   For this demonstration, no relaxation processes were taken into account. With the observed durations of oscillations being as short as some tens of fs, such an assumption is well justified \cite{Dekorsy1995,Leo1998,Dmitriev2002}. For low values of $pa$ only intrasubband coherences with close frequencies contribute to $I_{\rm AC}(t)$.
 With increase of $pa$ the oscillation amplitude grows and the dynamics becomes more anharmonic and irregular, since more higher subbands become involved, contributing with various larger frequencies.

\begin{figure*}
 \includegraphics[width=15cm]{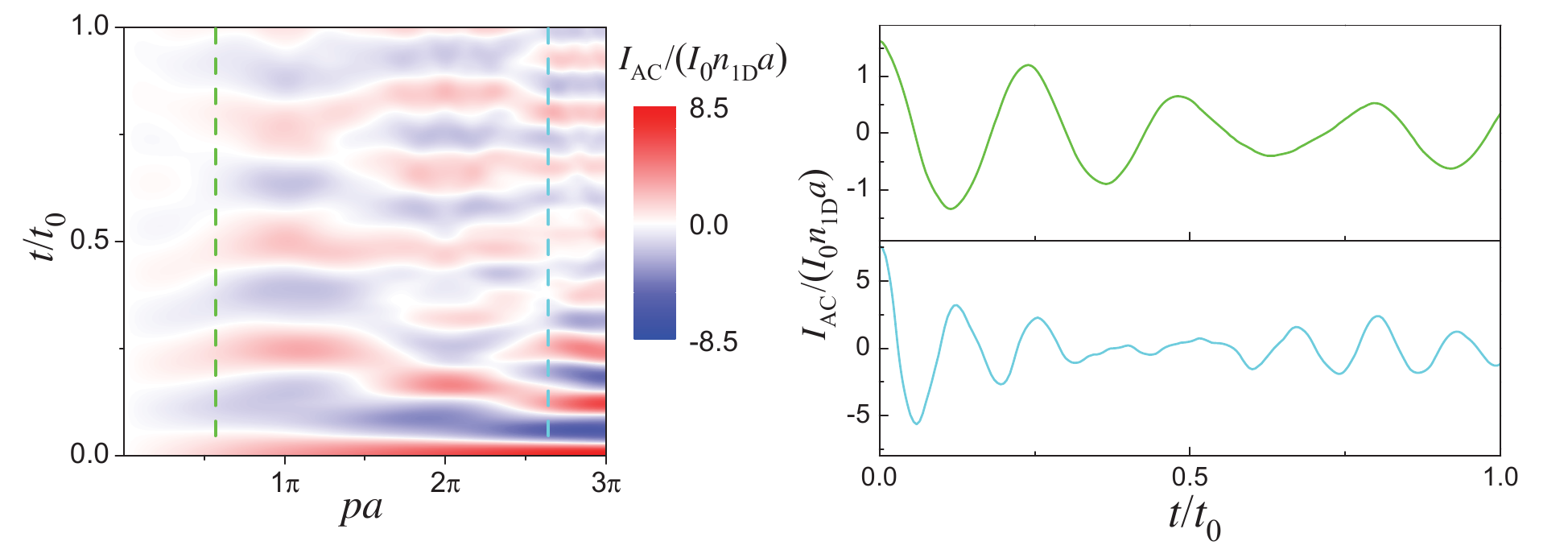}
  \caption{Left panel: Generated alternating current $I_{\rm AC}$ in dependence on the normalized time $t/t_0$  and transferred momentum
  $pa$. The parameters of the SL are as in Fig.~\ref{Fig:PLA2006_energy_bands},  $t_0=m^*a^2/\hslash\approx130$~fs. Right panel:
  Time-dependence $I_{\rm AC}(t)$ at fixed values of $pa$ corresponding to the vertical line cuts in the left panel [$pa=1.8$ ($pa=8.3$) for the top (bottom) plot] .
  \label{Fig:AC_current}}
\end{figure*}


The relaxation processes lead to decay of the generated ballistic current, typically on
the picosecond time scales at low temperatures. In order to
compensate for the effect of the relaxation of the DC, a periodic sequence
of HCPs can be used.  Finally, the obtained results suggest a new
mechanism for the generation of ratchet currents in SLs. This
should be possible to realize using the strongly asymmetric HCPs
(see Section~\ref{Sec:Strongly_as_HCPs}). For that their strong
subcycle has to be short enough, as discussed above. At the same time their
weak tail should be considerably longer than the characteristic
times of intersubband transitions determined by the corresponding
transition energies. For the SLs with parameters of
Fig.~\ref{Fig:PLA2006_energy_bands} and a HCP given by
Eq.~\eqref{Eq:pulse_shape} with the shape parameters described
below this equation and $\tau_{\rm d}=1$~fs, these requirements
are fulfilled. Then, on one hand, the integral over the temporal profile of the electric
field of such a pulse vanishes, as it is demanded for
the ratchet effect. On the other hand, the long negative subcycle
of such a HCP can not exactly compensate the DC generated by its
short positive subcycle because the former does not lead to the
intersubband transitions, in contrast to the latter. Notice that
for such a mechanism the asymmetry of the ratchet would be not
spatial but temporal.

\subsection{From short ultrabroadband to strong-field excitations}
It is interesting to contrast   the excitations following
 ultrashort, ultrabroadband pulses that we discussed above with
those caused by  ultrafast strong-field light-matter interactions
involving electrons in periodic crystalline lattices (bulk semiconductors or dielectrics but may be a semiconductor SL, as well \cite{vonPlessen1992,Feldmann1992}). This topic
is attracting currently much research \cite{Schultze2013,Schultze2014,Schiffrin2013,Yakovlev2016,Hohenleutner2015,Maag2015,Mayer2015,Schubert2014}.
A common key point for studies of the induced coherent phenomena and control is that the interaction should be short enough not to destroy the driven material even if the peak applied electric fields are extremely large (on the order of V/\AA). However, typically utilized few-cycle laser pulses are still much longer than (at least some of) the relevant fundamental time scales of the strongly-driven electrons in the lattice: $\tau_n=\hslash/\Delta E_{n}$ and $\tau_{\mathrm{g},n}=\hslash/E_{\mathrm{g},n}$. Here $\Delta E_{n}$ is the width of the $n$-th energy band and $E_{\mathrm{g},n}$ denotes the energy gap between this and the next band participating in the interaction process. Assume that such pulses have oscillation frequency $\omega_\mathrm{c}$ and electric field amplitude $\mathcal{E }_0$. Leaving the regime of Rabi floppings aside, for any values of the Keldysh's adiabaticity parameters $\gamma_n=\omega_\mathrm{c} \sqrt{m^*E_{\mathrm{g},n}}/(|e|\mathcal{E}_0)$ the state of electrons in the lattice is massively perturbed when the Bloch frequency $\omega_{_\mathrm{B}}=|e|\mathcal{E}_0a/\hslash$ is on the same order or exceeds $\omega_\mathrm{c}$ \cite{Yakovlev2016}.

For the reviewed case of extreme broadband ultrashort pulses, in this argument we should replace $\omega_\mathrm{c}$ by $\pi/\tau_\mathrm{d}$. Indeed, $pa\sim \omega_{_\mathrm{B}} \tau_\mathrm{d}$  and for $pa \gtrsim \pi$ we see that the electrons could be swept through the whole Brillouin zone and transferred to the next subbands (see Fig.~\ref{Fig:PLA2006_figure_Cnm}) and substantial intersubband coherences arise (see Fig.~\ref{Fig:AC_current}). Note that the  quantum states of electrons do not have time to ``adapt'' \cite{Yakovlev2016} themselves  to the external field and are ``frozen'' on its time scale, that is just an opposite limit case. The electron dynamics during the pulse is governed solely by the interaction part of the Hamiltonian.
As a consequence, no reflections at the zone boundaries and Bloch oscillations \cite{Feldmann1992,Schubert2014,Yakovlev2016} occur. The electron momentum increases continuously during the excitation process. Whereas approximations based on the direct and non-adiabatic (with energy absorption) tunneling, perturbative multiphoton transitions \cite{Keldysh1964_multi,Popov2004,Popruzhenko2014} and Landau-Zener approach \cite{Landau1932,Zener1932,Stueckelberg1932,Majorana1932} are not applicable in this regime, the unitary perturbation theory presented in Section \ref{Sec:Unitary_expansion} and used  above is an appropriate analytical tool for the description of the intersubband transitions and induced charge currents.

\subsection{Control of electronic motion in 1D semiconductor double
quantum wells}\label{Sec:Appl_DQW}

This section is devoted to the electron dynamics in
one-dimensional semiconductor double quantum wells (DQWs) induced
by ultrashort light pulses that was studied in
Refs.~\cite{Alex_APL_QW2004,Alex_QW2004,Alex_Europhysics2005a,Matos_Indian_paper}.
Coherent control of the electronic motion in DQWs by CW fields
was intensively investigated in the past
\cite{Grossmann1991,Grossmann1992,Bavli1992,Bavli1993,Holthaus1992,Grifoni1998}.
In the focus of interest are (a) coherent suppression of
tunneling between the wells \cite{Grossmann1991}, (b) controlled
persistent localization of the electron in one of the quantum
wells \cite{Bavli1992,Bavli1993} and (c) localization transfer
from one of the wells to another \cite{Holthaus1992}.
Consider an electron in a symmetric DQW (see Fig.~\ref{Fig:APL2004_DQW_structure}).
$E_1$ and $E_2$ are its two lowest energy levels, originating from the ground state
in each separate quantum well. $E_3$ is the next level in the DQW. A specially
simple and interesting case represents the situation when $E_2-E_1\ll E_3-E_1$ and
the electron populates initially only the two
lowest states.
Then, if the frequencies of the CW driving fields
are comparable with $(E_2-E_1)/\hslash$ or exceed it but are still much
smaller than $(E_3-E_2)/\hslash$, the two-level system
approximation (TLSA) can be applied
\cite{Bavli1992,Bavli1993,Grossmann1992,Grifoni1998}. The case of
a high-frequency field with an adiabatically changing envelope
can be also well described within this approximation
\cite{Holthaus1992,Grifoni1998}. The dynamics of the charge
carriers in the DQW is then mapped to the dynamics of the
corresponding two-level system
\cite{Holthaus1992,Gomez1992,Grifoni1998}. It was natural to
extend this consideration to the case when the driving is provided
by HCPs.

 {Such type of  driving} has several advantages. The time required
to localize an electron can be comparable with the time of its
tunneling between the wells in the absence of the driving field,
in contrast to the proposed CW driving schemes where this time is
much longer \cite{Bavli1992,Bavli1993}. Apart from this,
localization maintenance of the priorly  trapped electron can be
achieved in a wide range of the pulse parameters whereas for  CW
driving fields the coherent suppression of tunneling is a
resonance process \cite{Grossmann1991,Grifoni1998}. Finally, the
proposed scheme for the localization transfer between the wells by
means of CW driving fields \cite{Holthaus1992,Grifoni1998}
requires very high frequencies, which are many orders of
magnitudes higher than the reciprocal tunneling time of an
unperturbed state localized in one of the wells. However, for
sufficiently high frequencies the validity of the dipole approximation (which
is still assumed in those studies) is questionable.
 For example,
for a DQW with a tunneling time $\hslash\pi/(E_2-E_1)\approx
333$~fs, as considered in
Refs.~\cite{Alex_APL_QW2004,Alex_QW2004}, and a period of
the driving CW field of $5\times10^{4}$ times shorter, as in
Ref.~\cite{Holthaus1992}, the corresponding wavelength would
constitute $2$~nm that is much smaller than the size of the DQW.
Hence, along this direction, the dipole approximation is not applicable.
In contrast, the spectrum of the  short, broadband  pulses is
limited to much smaller frequencies for the same pulse duration
(1/100 of the tunneling time and 500 of the period of the CW
driving field in Ref.~\cite{Holthaus1992}) and the dipole
approximation  holds.

%

\begin{figure*}
  \includegraphics[width=7.5cm]{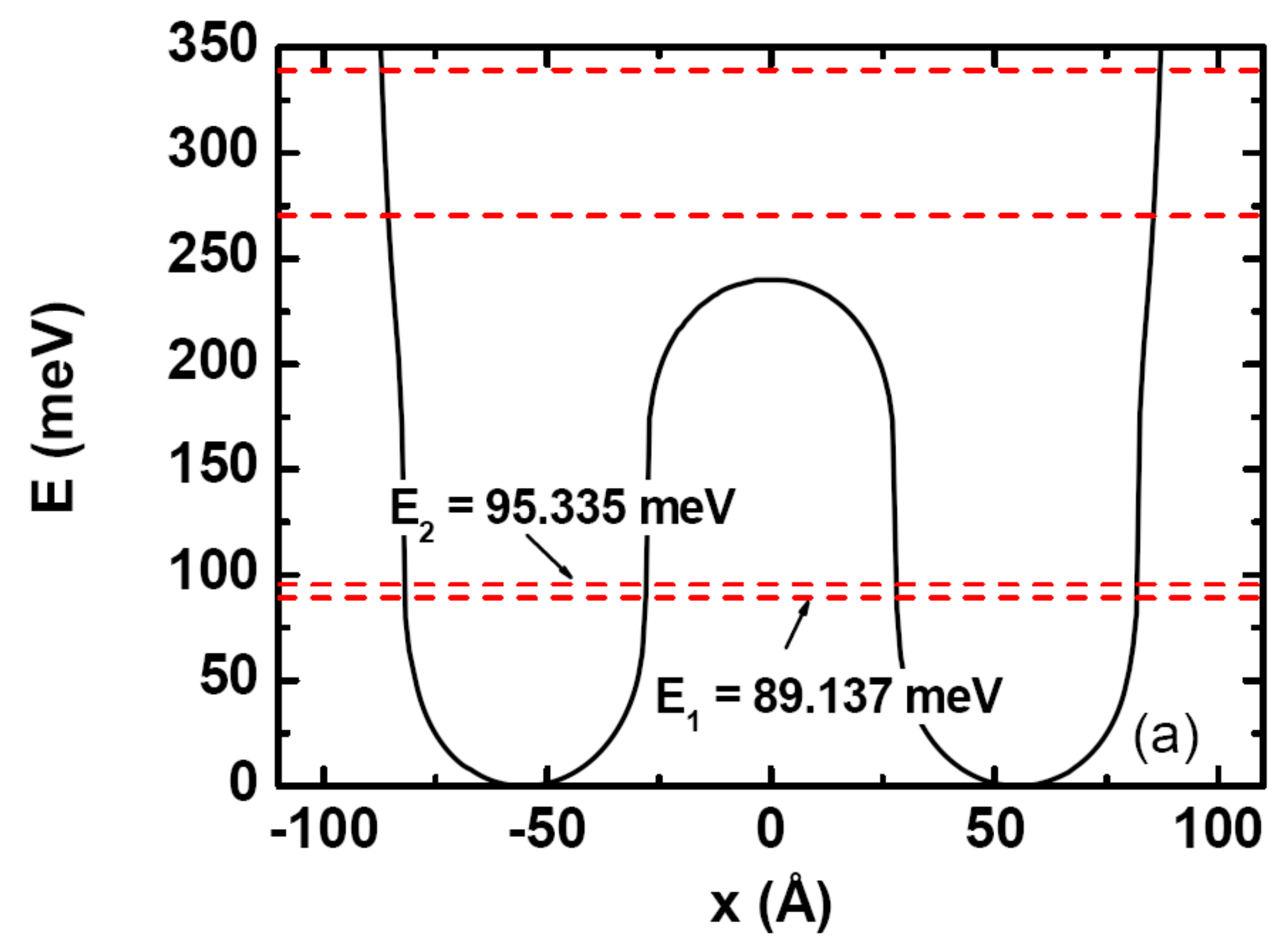}
  \caption{Model confinement potential $U(x)$ of a
  one-dimensional symmetrical semiconductor DQW based on a GaAs/Al$_x$Ga$_{1-x}$As heterostructure
 (adapted from Ref.~\cite{Alex_APL_QW2004}).
 \label{Fig:APL2004_DQW_structure}}
\end{figure*}

Let us consider an example of a model DQW system as employed in
Refs.~\cite{Alex_APL_QW2004,Alex_QW2004}. It is similar to
that used in Refs.~\cite{Bavli1992,Bavli1993}, but not
exactly the same. The corresponding confining potential $U(x)$ is
shown in Fig.~\ref{Fig:APL2004_DQW_structure}. The Hamiltonian of
the system is given by
\begin{equation}\label{EQ:APL2004_H}
H=H_{0}+V(x,t)\:, \mbox{\ } \\
\end{equation}
where $H_{0}=\frac{p^2}{2m^*}+U(x)$ represents the Hamiltonian of
a free electron with an effective mass $m^*$ in the DQW. The model
for the DQW resembles a GaAs/Al$_x$Ga$_{1-x}$As heterostructure.
The value of the effective mass is selected as $m^*=0.067m_0$,
where $m_0$ is the free electron mass. The width of each single
well is approximately $50$~\AA, and the barrier width is around
$60$~\AA\ whereas its height constitutes approximately $240$~meV.
As it can be seen from Fig.~\ref{Fig:APL2004_DQW_structure}, the
distance between the two lowest energy levels amounts to
$E_2-E_1\approx6$~meV whereas
$E_3-E_2\approx 200$~meV. Therefore the characteristic time scale,
which we define here as
\begin{equation}\label{Eq:T_c}
  T_{\rm c}=\hslash/(E_2-E_1)=1/\omega_{21}\;
\end{equation}
(by this definition $T_{\rm c}$ is a factor $2\pi$ smaller than in
Refs.~\cite{Alex_APL_QW2004,Alex_QW2004,Matos_Indian_paper},
and a factor $\pi$ smaller than the \textit{tunneling time} of
Refs.~\cite{Holthaus1992,Grifoni1998} that is here just a
matter of conventions), is calculated to $T_{\rm c}\approx
106$~fs.

The light-matter interaction part of the Hamiltonian is given by
\begin{equation}\label{Eq:V_DQW}
V(x,t)=-exE_0f(t)\;,
\end{equation}
where the electric field with the amplitude $E_0$ and the temporal
profile $f(t)$ is polarized along the $x$-axis. This form presumes
the dipole approximation \cite{CohenTannoudji_book1}. We deal here
with a special case belonging to the general class of light-driven
one-dimensional quantum systems (cf.
Section~\ref{Sec:motion_1D}). The relaxation processes are not
included in the treatment as far as they should take place on
much longer time scales than the considered coherent control of
the electronic motion \cite{Alex_QW2004,Bastard_book}.

The  quantity that characterizes the
localization properties of the electron  which is described by the wave function
$\Psi(x,t)$, is the time-dependent probability to find the
electron in the left (or right) quantum well
\begin{equation}\label{Eq:P_L_def}
P_{\rm L}(t)=\int_{-\infty}^0\Psi^*(x,t)\Psi(x,t) {\rm d}x\ .
\end{equation}
Its  average over a reasonably long time $\tau$ is
\begin{equation}\label{Eq:P_L_av}
\langle P_{\rm L}\rangle_{\tau}=\frac{1}{\tau}\int_{0}^\tau P_{\rm
L}(t) {\rm d}t\ .
\end{equation}
 The probability to
find the electron in the right well is just $P_{\rm R}(t)=1-P_{\rm
L}(t)$.

For a high degree of localization, the
electronic population should be limited to the two lowest energy
levels. Occupations of  higher levels would  lead to  electronic distribution in both wells. To fulfill this requirement the duration of the driving
light pulse $\tau_{\rm d}$ has to be selected in such a way that
it is much larger than $\hslash/(E_3-E_2)$. It is the condition
justifying the TLSA. For the parameters of the DQW in question   this
means $\tau_{\rm d}\gg 4$~fs. Furthermore, it is advantageous
to reduce  the dynamics to that of the IA. For
this $\tau_{\rm d}\ll T_{\rm c}$ should apply. In calculations
of Ref.~\cite{Alex_APL_QW2004} a Gaussian shape
\eqref{Eq:Gaussian_pulse} of the temporal profile of the electric field was
used, with a duration $\tau_{\rm d}=14.14$~fs. For
this choice both conditions, the first required for the TLSA,
and the second required for the IA, are satisfied.

In Ref.~\cite{Alex_QW2004} an alternative shape of the pulse
with a short and a strong positive subcycle and a weak but also
short negative subcycle was used:
\begin{equation}\label{Eq:f_Alex_PRB_QW_2004}
   f(t)=\exp(-t^2/\tau_{\rm d}^2)\cos(\Omega t)\,\Theta\!\left(t+\frac{\pi}{2\Omega}\right),
\end{equation}
where $\Omega=\sqrt{2}\pi/(3\tau_{\rm d}\sqrt{\ln 2})$ and
$\Theta(t)$ denotes the Heaviside step function. The pulse
duration parameter $\tau_{\rm d}$ was selected as $\tau_{\rm
d}=28.28$~fs. For a periodic train of the pulses in
Ref.~\cite{Alex_QW2004}, $f(t)$ was set to zero at the time
moment when the next pulse sets in. However, if the time distance
between the pulses $T$ is large enough, as in
Ref.~\cite{Alex_QW2004} ($T=100$~fs), there is no
significant difference between calculations based on this
assumption and calculations just using
Eq.~\eqref{Eq:f_Alex_PRB_QW_2004} for all $t>0$. Because the
negative subcycle has here the same duration as the positive
one such a pulse is not treated in a way suitable for the
strongly asymmetric HCPs described in
Section~\ref{Sec:Strongly_as_HCPs}. The negative tail is not
neglected but is taken into account when calculating the parameter
$s_1$ [see Eqs.~\eqref{Eq:s_n_infinity} and
\eqref{Eq:Delta_p_free_space}] determining the magnitude of the
transferred momentum. In
Ref.~\cite{Alex_QW2004} the value of $t_1$ was selected at
the center of the positive subsycle, i.e. at $t_1=0$. Following
our arguments in Section~\ref{Sec:Unitary_expansion}, an even better
choice in this case would be the center of gravity of the pulse at
$t_1\approx 0.138\tau_{\rm d}$ but the corresponding corrections
to the induced dynamics of the system are vanishingly small
because of the smallness of this shift. This is a consequence of
the $1:8$ relation between the amplitudes of the negative and
positive subcycles having the same durations. The mentioned above
conditions for the IA and TLSA are also fulfilled for this pulse
type and the described choice of parameters.

Within the TLSA the electronic wave function can be expressed
as
\begin{equation}\label{Eq:TLSA_wave_function}
\Psi(x,t)=\sum\limits_{i=1}^{2}C_{i}(t)\psi_i(x)\;,
\end{equation}
where $\psi_i(x)$ ($i=1,2$) are the stationary wave functions of the
unperturbed DQW corresponding to the two lowest energy levels. One
can select their phases in such a way that both of them are real
and have positive values in the right well. The state in the left
well is then given by
$\psi_{L}(x)=\frac{1}{\sqrt{2}}\left[\psi_1(x)-\psi_2(x)\right]$,
whereas the wave function of the state in the right well reads as
$\psi_{R}(x)=\frac{1}{\sqrt{2}}\left[\psi_1(x)+\psi_2(x)\right]$
\cite{Bavli1992}. Therefore, the time-dependent probability for finding
the electron in the left well can be written as
\begin{equation}
P_{\rm L}(t)=\frac{1}{2}
-\mathrm{Re}\left[C_{1}^{*}(t)C_{2}(t)\right]\:.
\end{equation}

\subsubsection{Suppression of tunneling: the short broadband driving case}\label{Sec:suppress_tun}
Assume that  initially (at $t=0$) the electron is located in one of the
quantum wells, say  the left quantum well, i.e.
  $\Psi(x,t=0)=\psi_{L}(x)$. A
sequence of HCPs is applied to suppress the process of tunneling
of the electronic density to the other quantum well. It is clear
that the time period $T$ of the pulse sequence should be much
smaller than the tunneling time, i.e. than $\pi T_{\rm c}$ in our
notations, but not too small as  HCPs should not overlap
significantly. In accordance with these requirements, for the HCP
shapes and DQW parameters discussed above, $T=100~$ps was selected
in all simulations of
Ref.~\cite{Alex_APL_QW2004,Alex_QW2004}. Then the remaining
free parameter which can be tuned as to  suppress the
tunneling process, i.e. to keep the electron in the same quantum
well, is the strength of a single HCP. An illustration
is provided by Fig.~\ref{Fig:Fig1_AlexPRB_QW_2004} which shows
 the time-averaged probability $\langle
P_{\rm L}\rangle_{\rm 2 ps}$ for finding the electron in the left well
during the first two picoseconds as  the peak
electric field of the HCPs changes. The  HCP shape is given by
Eq.~\eqref{Eq:f_Alex_PRB_QW_2004}. The  averaging time $\tau=2$~ps is
much larger than the characteristic time scales of the system. Both
the result of a full numerical simulation of the TDSE and the
approximate results following from the TLSA with the IA are shown.
We notice  a very good agreement between  exact and
approximate results up to significantly large values of the
peak electric field.

\begin{figure*}
  \includegraphics[width=7.5cm]{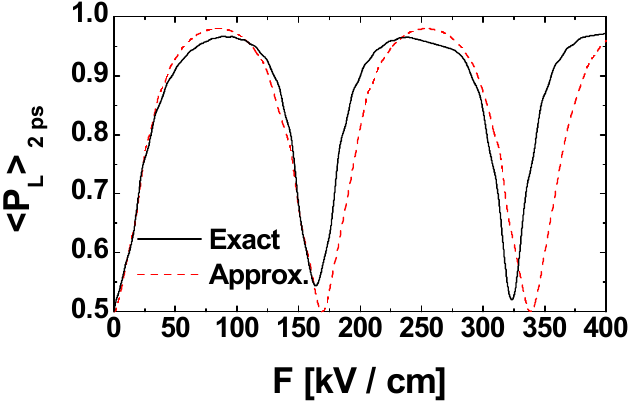}
  \caption{Time-averaged probability $\langle P_{\rm
L}\rangle_{\rm 2 ps}$  is shown in dependence of the peak electric
field $F$ of a single HCP of the applied periodic pulse train. At
the time moment $t=0$ the electron is situated in the left well.
The averaging is performed in the time interval from 0 to 2~ps
 (adapted from Ref.~\cite{Alex_QW2004}). 
 \label{Fig:Fig1_AlexPRB_QW_2004}}
\end{figure*}

Figure \ref{Fig:Fig3_AlexPRB_QW_2004} shows the dynamics of the
localization probability $P_{\rm L}$ for particular values of the
peak electric field $F$ corresponding to a maximum or a minimum of
the time-averaged probability $\langle P_{\rm L}\rangle_{\rm 2
ps}$ in Fig.~\ref{Fig:Fig1_AlexPRB_QW_2004}, respectively. It is
evident that in the first case shown in
Fig.~\ref{Fig:Fig3_AlexPRB_QW_2004}a the electron is well
localized in the left well during the whole time interval, whereas
in the second case shown in Fig.~\ref{Fig:Fig3_AlexPRB_QW_2004}b
it oscillates between the two wells. Again, both
full numerical and  approximate results
within the TLSA and IA are presented.

\begin{figure*}
  \includegraphics[width=7.5cm]{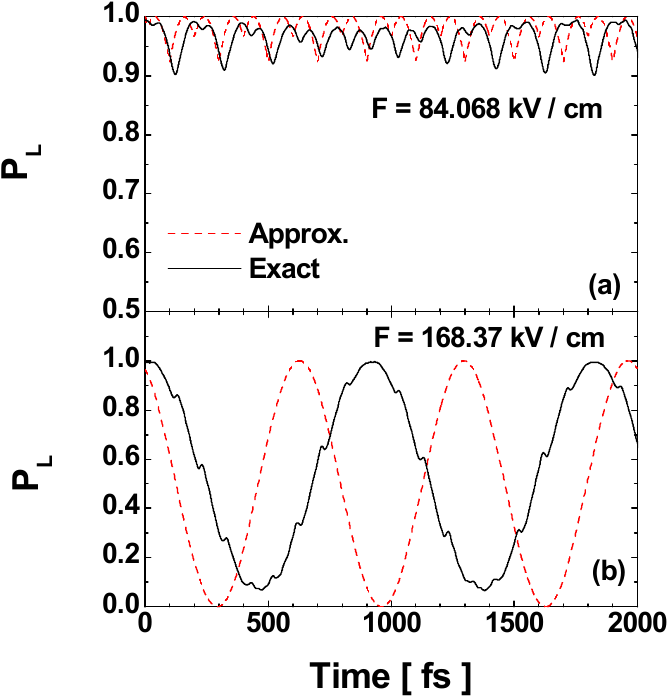}
  \caption{Dynamics of the localization probability $P_\mathrm{L}(t)$ for two particular values of the peak electric field
  of the HCP, corresponding to (a) the first maximum and (b) the following
  minimum in
  Fig.~\ref{Fig:Fig1_AlexPRB_QW_2004}
 (adapted from Ref.~\cite{Alex_QW2004}). 
 \label{Fig:Fig3_AlexPRB_QW_2004}}
\end{figure*}

The approximate solution reproduces the localization
and the delocalization behavior.
Discrepancies with respect to the exact numerical solution are visible in
Fig.~\ref{Fig:Fig3_AlexPRB_QW_2004}, however.
Their origin lies in the IA. In
the case of localization, the agreement
is better due to a smaller value of $F$.
In the second case, when the electron delocalizes across the DQW in
the course of time, the oscillation periods are
somewhat different and  the numerical solution is not given by
just a single harmonic, as in the case of the approximate
solution. One actually already approaches the regime of SVS
excitations by single HCPs, as  described in
Section~\ref{Sec:ultrastrong_theory} (cf. also
Fig.~\ref{Fig:dip_TLS_strong}).
 The IA is still valid but not accurate and the difference to the exact solution accumulates
with each applied HCP. Thus, a small phase difference after the
application of a single HCP results into a noticeable change of the
oscillation period for a periodic HCP train. The deviation from a
single harmonic oscillation of the exact solution can be traced
back to the fact that the value of $F$ in
Fig.~\ref{Fig:Fig3_AlexPRB_QW_2004}b (chosen as to ensure  the corresponding
minimum in Fig.~\ref{Fig:Fig1_AlexPRB_QW_2004})
 was selected in
Ref.~\cite{Alex_QW2004}, actually, according to the
approximate solution. Within the TLSA and IA, the excitation by a
HCP is described by Eq.~\eqref{Eq:U_result_TLS} where only the
first term in the exponent is retained. In the case when
$a_vs_1=n\pi$ ($n$ is an integer)
is selected, in the framework of these approximations the electron
dynamics is unperturbed by any of the applied HCPs and the
electron oscillates freely between the two wells. This leads to
the minimum of the time-averaged probability to find it in the
left well in Fig.~\ref{Fig:Fig1_AlexPRB_QW_2004}. However, for the
exact solution the first minimum of the time-averaged probability
$\langle P_{\rm L}\rangle_{\rm 2 ps}$ occurs at a slightly
different value of the peak electric field of the pulse. Moreover,
as it can be seen from
Eqs.~\eqref{Eq:Ansatz_V_strong},\eqref{Eq:U_1_TLS} and
\eqref{Eq:U_2_TLS}, with the inclusion of the first correction to
the IA in this regime, there is no possibility to achieve a
situation that the electron dynamics is left unperturbed when the
HCP is gone by just increasing the strength of the HCP. In
general, there are always changes in the dynamics after
application of each HCP that can be observed as small jerks in
Fig.~\ref{Fig:Fig3_AlexPRB_QW_2004}b at each time moment when a
HCP is applied. This is also the reason why in the case of the
exact solution $\langle P_{\rm L}\rangle_{\rm 2 ps}$ does not
reach the value $0.5$ at the minima in
Fig.~\ref{Fig:Fig1_AlexPRB_QW_2004}.

For a deeper insight into the physics of the observed
suppression of tunneling induced by a periodic train of HCPs, the
knowledge gained from the Floquet approach described in
Section~\ref{Sec:Floquet} can be used. To obtain analytical
results we make use of the TLSA and apply the IA that leads to the
interaction part of the evolution operator in the form of
Eq.~\eqref{Eq:U_result_TLS} where only the first term in the
exponent is taken into account [leading also to
Eq.~\eqref{Eq:U_1_TLS} where $s_1(t)$ should be replaced by
$s_1(t=\infty)$]. This can be then used in
Eqs.~\eqref{Eq:Floquet_eingenvalue_problem_U} and
\eqref{Eq:Floquet_U_period} to obtain the quasienergies
$\varepsilon_\lambda$ ($\lambda=1,2$):
\begin{equation}\label{Eq:quasienergies_TLS}
  \varepsilon_{\lambda}=(-1)^{\lambda}\frac{\hslash\omega_{T}}{2\pi}\arccos\left( \cos\alpha\cos\theta\right),
\end{equation}
where $\omega_T$ is given by Eq.~\eqref{Eq:Omega_T}. We use
 notations \footnote{Notice a factor of 2 difference in the definition of $\alpha$ with respect to
        Refs.~\cite{Alex_QW2004,Schueler2010}.}
\begin{equation}\label{Eq:quasienergies_TLS_alpha}
  \alpha=a_v s_1\;,
\end{equation}
\begin{equation}\label{Eq:quasienergies_TLS_theta}
  \theta=\pi
 \frac{1}{\omega_T T_{\rm c}}\;.
\end{equation}
$a_v$ is given by Eq.~\eqref{Eq:TLS_ab}, $s_1$ is calculated
after Eq.~\eqref{Eq:s_n_infinity} taking a particular pulse shape
into account, e.g., Eq.~\eqref{Eq:f_Alex_PRB_QW_2004} in
Ref.~\cite{Alex_QW2004}, and $T_{\rm c}$ is defined by
Eq.~\eqref{Eq:T_c}. The calculated quasienergies are shown in
Fig.~\ref{Fig:quasienergies} as a function of  the pulse strength
parameter $\alpha$ for several values of $\theta$: $\theta=0.472$,
which corresponds to the selected values $T_{\rm c}=106~$fs and
$T=100~$fs, $\theta=\pi/2$ (meaning, e.g., that $T$ is increased
by approximately 3.33 times), and $\theta=2.83$ (e.g., $T$ is
increased by the factor 6).  For the following consideration we assume $\theta<\pi$.

\begin{figure*}
  \includegraphics[width=7.5cm]{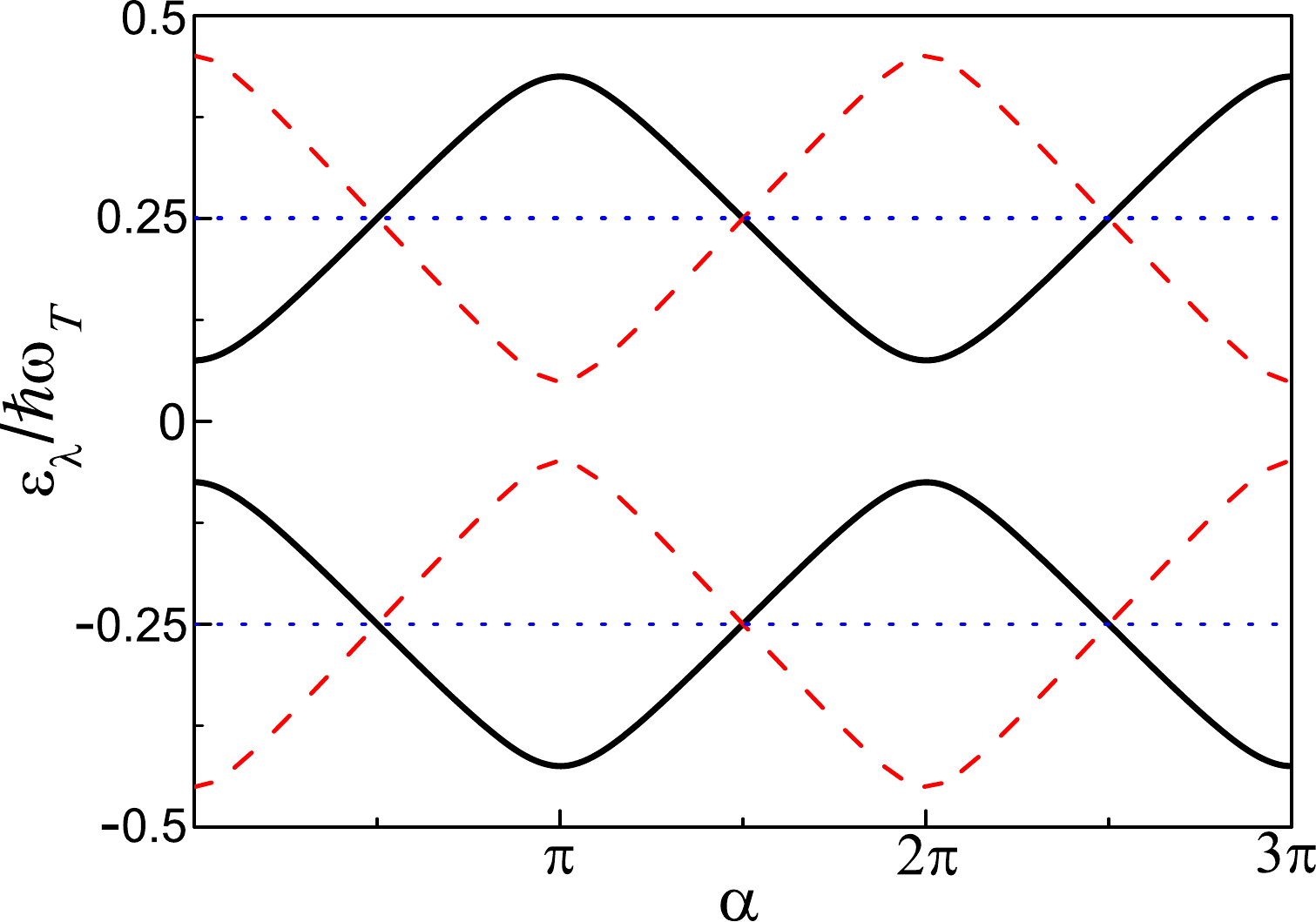}
  \caption{Dependence of the quasienergy levels
  of the periodically driven DQW $\varepsilon_{\lambda}$ ($\lambda=1,2$) in the first Brillouin zone
   on the pulse strength parameter $\alpha$  that is proportional to the peak electric field of the HCPs in the applied HCP train.
   The TLSA and IA are used. Values of $\varepsilon_{1}$
   ($\varepsilon_{2}$) are negative (positive). Full black lines
   correspond to the pulse train parameters discussed in the text
   and used in Ref.~\cite{Alex_QW2004}, with $\theta=0.472$. Accordingly,
   $\alpha=\pi$ yields the peak HCP electric field
   $F=168.37$~kV/cm. Blue  dotted lines represent the case with
   $\theta=\pi/2$ whereas $\theta=2.83$ is chosen for red dashed lines.
 \label{Fig:quasienergies}}
\end{figure*}

\subsubsection{ {Aharonov-Anandan phase as an indicator for}  {nonequilibrium}  {charge localization}}
We can use Eqs.~\eqref{Eq:Floquet_eingenvalue_problem_U} and
\eqref{Eq:Floquet_U_period} to calculate the Floquet modes at the
time moment $t_1^-$ just before the application of the first HCP
of the periodic pulse sequence, resulting in
\begin{equation}\label{Eq:Floquet_modes_TLS}
    |\Phi_\lambda(t_1^-)\rangle=\frac{1}{N_\lambda}\left(%
\begin{array}{c}
  \\[-2cm]
  e^{i\theta/2} \\[-0.5cm]
  -e^{-i\theta/2}\left[\gamma-(-1)^\lambda{\rm sgn}(\sin\alpha)\sqrt{1+\gamma^2}\right] \\
\end{array}%
\right),
\end{equation}
where we have introduced a notation
\begin{equation}\label{Eq:Floquet_modes_TLS_gamma}
   \gamma=\cot\alpha\:\sin\theta\;,
\end{equation}
the normalization constants are given by
\begin{equation}\label{Eq:Floquet_modes_TLS_normalization}
   N_\lambda=\sqrt{2}\sqrt{1+\gamma^2-(-1)^\lambda{\rm sgn}(\sin\alpha)\gamma\sqrt{1+\gamma^2}}\;.
\end{equation}
and $\alpha\neq n\pi$ ($n\in \mathbb{Z}$) has been assumed. The
case
\begin{equation}\label{Eq:alpha_delocalization}
    \alpha=n\pi \hspace{0.5cm}
    (n\in\mathbb{Z})
\end{equation}
is trivial because the HCPs do not influence the system dynamics. Under the action of such a HCP the wave function either remains the same (even $n$) or just changes its sign (odd $n$). In the former case the difference in
quasienergies coincides with the energy spacing of the
unperturbed TLS $E_2-E_1=\hslash\omega_{21}$. In the latter case it equals to $\hslash\omega_T-\hslash\omega_{21}$ in the first Brillouin but remains the same if the Brillouin zone is shifted by $\hslash\omega_T/2$ (cf.
Fig. ~\ref{Fig:quasienergies}).
 Actually, with the mentioned
approximations, Eq.~\eqref{Eq:alpha_delocalization} represents a
delocalization condition for an electron which at some time moment
is situated in one of the quantum wells \cite{Alex_QW2004}.

Knowing the wave function $|\Psi_\lambda(t_1^-)\rangle$ at the
time moment $t_1^-$, we can project it onto the Floquet modes and
find the expansion coefficients $C_\lambda$ [cf.
Eq.~\eqref{Eq:Floquet_C}]. Then in the case of a periodic
evolution of the system we can calculate the corresponding
Aharonov-Anandan phase $\phi_{\rm AA}$ after
Eq.~\eqref{Eq:phi_AA_condition_c_simplified}, where
$\phi^{(\lambda)}_{\rm AA}$ can be obtained from
Eq.~\eqref{Eq:phi_AA_Hellmann_Feynman} and
Eq.~\eqref{Eq:quasienergies_TLS}. The value of $\phi_{\rm AA}$ serves as an indicator of
the sustainability of the system state. Notice that as far as the
effective HCP strength $\alpha$ is proportional to its duration
$\tau_{\rm d}$ we can write the quasienergies
$\varepsilon_\lambda$ as a function of $\omega_T$ and $\alpha$ and
use Eq.~\eqref{Eq:phi_AA_Hellmann_Feynman}, where $\tau_{\rm d}$
should be  replaced by $\alpha$. In this way we calculate
\begin{equation}\label{Eq:phi_AA_TLS}
    \phi^{(\lambda)}_{\rm AA}=-(-1)^\lambda\left[\arccos(\cos\alpha\cos\theta)
                                 -\frac{\theta\cos\alpha\sin\theta+\alpha\sin\alpha\cos\theta}{\sqrt{1-\cos^2\alpha\cos^2\theta}}\right].
\end{equation}
If the electron is initially located in one of the wells and the
periodic quantum evolution is sustainable then the tunneling from
this well is obviously suppressed.

Let us consider a particular case with
\begin{equation}\label{Eq:alpha_localization}
    \alpha=\left(\frac{1}{2}+n\right)\pi \hspace{0.5cm}
    (n\in\mathbb{Z})\;.
\end{equation}
For $n=0$ this includes the case of
Fig.~\ref{Fig:Fig3_AlexPRB_QW_2004}a, where we have observed the
suppression of tunneling.
Given Eq.~\eqref{Eq:alpha_localization},
Eq.~\eqref{Eq:quasienergies_TLS} simplifies to
$\varepsilon_{\lambda}=(-1)^{\lambda}\hslash\omega_{T}/4$ (cf.
Fig.~\ref{Fig:quasienergies}) and Eq.~\eqref{Eq:Floquet_modes_TLS}
reduces to
$|\Phi_\lambda(t_1^-)\rangle=2^{-1/2}\left(e^{i\theta/2},(-1)^{\lambda+n}
e^{-i\theta/2}\right)^T$. We can also calculate:
\begin{equation}\label{Eq:phi_AA_localization}
    \phi^{(\lambda)}_{\rm AA}=-(-1)^{\lambda+n}\frac{(2n+1)\pi}{2}\left[1-\cos\theta\right],
\end{equation}
where shifts by a multiple of $2\pi$ have been taken into
account. For example, for $n=0$ and $\lambda=1$ we have

$\phi^{(\lambda)}_{\rm AA}=(\pi/2)(1-\cos\theta)$. Then we see
that in the case of $\theta\ll \pi/2$, meaning $T_{\rm c}\ll T$,
we can expect sustainability of the electron state if it exactly
coincides with a pure Floquet state at the time moment just before
the first HCP of the periodic pulse sequence is applied. If the
initial state in the left well at $t=0$ is assumed, i.e.
$|\Psi(t=0)\rangle=2^{-1/2}(1,-1)^T$, we infer
$|\Psi(t=T/2)\rangle=2^{-1/2}\left(e^{i\theta/2},
-e^{-i\theta/2}\right)^T=|\Phi_1(t_1^-)\rangle$ after $T/2$
duration of a free propagation. Thus, the sustainability of the
electronic state is achieved when $t_1^-=T/2$ is selected. This
result, as well as Eq.~\eqref{Eq:phi_AA_localization} for $n=1$
and $\lambda=1$, was obtained in Ref.~\cite{Alex_PhD} from a
geometrical reasoning, by mapping the electron dynamics to a
trajectory onto the
 Bloch sphere and calculating the enclosed solid
angle (see Ref.~\cite{Alex_PRA2006} for a general
description of such an approach) \footnote{In
Ref.~\cite{Alex_Europhysics2005a} the validity of the
corresponding expression is restricted to $n=0$.}. In
this case $\phi_{\rm AA}=\phi^{(1)}_{\rm AA}$ can be expressed as
unambiguous function of the driving period: $\phi_{\rm
AA}=\phi_{\rm AA}(T)$. The dependence of the system dynamics
on $T$ can so be mapped onto its dependence on $\phi_{\rm AA}$. Figure
\ref{Fig:Fig_Alex_EPL2005_AA} shows the time-dependent probability
to find the electron in the left well $P_{\rm L}(t)$ depending on
the Aharonov-Anandan phase $\phi_{\rm AA}$. We see that the
electron spends more and more time in the left well as $\phi_{\rm
AA}$ is decreased. For the effective localization of the electron
in the left well at all times $\phi_{\rm AA}\approx 0$ is
required. For $\phi_{\rm AA}\approx \pi$ the electron oscillates
between the two quantum wells with the oscillation period of the
unperturbed system $T_0=2\pi T_{\rm c}$.

\begin{figure*}
  \includegraphics[width=0.6\textwidth]{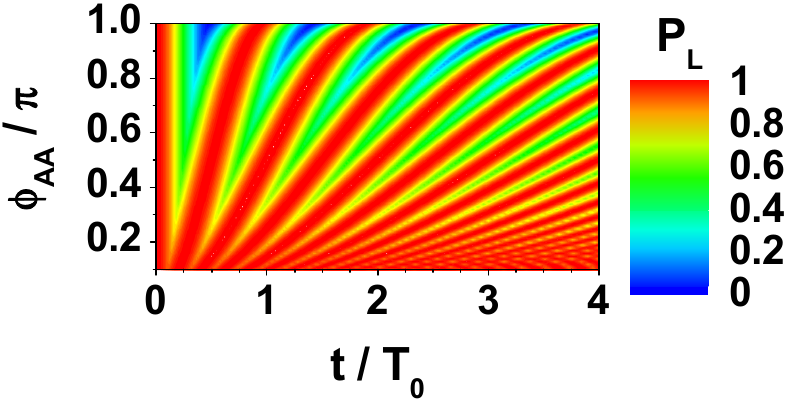}
  \caption{Localization probability $P_\mathrm{L}$ is shown as  a function of  the time and Aharonov-Anandan phase $\phi_{\rm
  AA}$. The electron resides in a pure Floquet state with the lower energy in the first Brillouin  zone
  (cf. Fig.~\ref{Fig:quasienergies}).
  The calculation is produced in the framework of the TLSA and IA for  the  effective
  HCP strength $\alpha=\pi/2$. Here $T_0=2\pi T_c$ with $T_c$
  given by Eq.~\eqref{Eq:T_c}
 (adapted from Ref.~\cite{Alex_Europhysics2005a}).
 \label{Fig:Fig_Alex_EPL2005_AA}}
\end{figure*}

The dynamics of the excited TLS in the case determined by
Eq.~\eqref{Eq:alpha_localization} is always periodic. Depending on
the initial condition, either the periodicity condition (a) or the
periodicity condition (c) with $l_{1,2}=1$ and $n_{1,2}=2$, listed
in Section~\ref{Sec:Floquet}, is fulfilled
\cite{Alex_Europhysics2005a}. In the former case the period of the
system dynamics coincides with the excitation period:
$\mathcal{T}=T$. In the latter case the system is in a
superposition of two Floquet states and the period of the system
dynamics is two times larger: $\mathcal{T}=2T$. For the
superposition state the Aharonov-Anandan phase should be
calculated after Eq.~\eqref{Eq:phi_AA_condition_c_simplified}. We
can see that if the occupation probability of the second Floquet
state $|C_2|^2$ is small compared with the occupation probability
of the first Floquet state $|C_1|^2$ and $\phi^{(1)}_{\rm AA}$ is
also small, as discussed above, then the value of the resulting
Aharonov-Anandan phase $\phi_{\rm AA}$ is close to zero. Thus for
a certain degree of admixture of the second state the system is
still sustained in the neighborhood of the initial state. This can
be observed in Fig.~\ref{Fig:Fig3_AlexPRB_QW_2004}a, where the TLS
is not exactly in a pure Floquet state. Therefore, the observed
period of the probability dynamics is twice the driving period.
However, $P_{\rm L}(t)$ remains close to one at all time moments.
For the exact solution, the system is not only in a superposition
of the Floquet states but also the periodicity condition (c) is
not  met accurately. Therefore, the periodicity of the solution is
lost. However, the tunneling to the right quantum well is still
efficiently suppressed, because both the admixture of the second
Floquet state and the mismatch in the periodicity condition are
relatively small.

It should be mentioned that, similarly to the discussed case of
the DQW in the TLSA, the Aharonov-Anandan phase was utilized for
the characterization of the sustainability of quantum states in
the case of a kicked rotor in the four-level approximation
\cite{Alex_Europhysics2005a} and in the general case of a kicked
quantum rotor where many rotational levels can be involved
\cite{Alex_PRA2006}. It looks promising to generalize the results
of the present work, allowing to predict pulse train parameters
required for the sustainability of target states based on the
properties of the quasienergy spectrum, also to these quantum
systems.

\subsubsection{Persistent localization}\label{Sec:persistent_localization}
The next important process which can be realized in DQWs with the
help of HCPs is the persistent localization of an initially
delocalized electron. Before the application of the driving light
pulses the electron is situated in the ground state with the wave
function $|1\rangle$ that implicates equal probabilities to find
the electron in the left and right quantum wells at any time moment
before the pulses are applied. This case corresponds to the
so-called optical initial condition \cite{Alex_QW2004}. The
persistent localization can be considered as a two-step process.
At first, the localization of the electron in one of the quantum
wells is achieved that is immediately followed by the suppression
of tunneling from this quantum well.

In the framework of the IA it is impossible to achieve the
localization of the electron by a HCP at the time moment just
after the pulse is gone. In this approximation, the HCP delivers a
certain amount of momentum to the electron. An additional time is
required for the electron to move into one of the wells. This time
is comparable with the characteristic time of the system $T_{\rm
c}$. In the TLSA, the first step can be realized by application of
a HCP with the effective strength $\alpha=-\pi/4$ [see
Eqs.~\eqref{Eq:U_1_TLS} and \eqref{Eq:quasienergies_TLS_alpha}]
that transforms the electronic state $|1\rangle$ to the state
$2^{-1/2}(|1\rangle-i|2\rangle)$. The latter evolves then to
$|\psi_{\rm
L}\rangle=2^{-1/2}\left(|1\rangle-|2\rangle\right)$ in a
quarter of the free evolution period, up to an unimportant phase
factor. Thus at the corresponding time moment the electron is
localized in the left well. Then,  to suppress the
tunneling from this state, the scheme using a periodic train of
HCPs with $\alpha=\pi/2$, as discussed in the previous section,
can be applied. Figure \ref{Fig:Fig8_AlexPRB_QW_2004} shows the
result of the corresponding full numerical simulation of the
time-dependent probability to find the electron in the left well,
where the parameters of the first auxiliary HCP and the following
HCP train were selected according to the described scheme
\cite{Alex_QW2004}. The time needed to achieve the persistent
localization of the electron is approximately equal to the
quarter-period of the free evolution. Although it is considerably
larger than the duration of a single pulse, as mentioned above, it
is still much shorter than the duration required to realize this
process with the help of CW driving fields
\cite{Holthaus1992,Grifoni1998}.

\begin{figure*}
  \includegraphics[width=7.5cm]{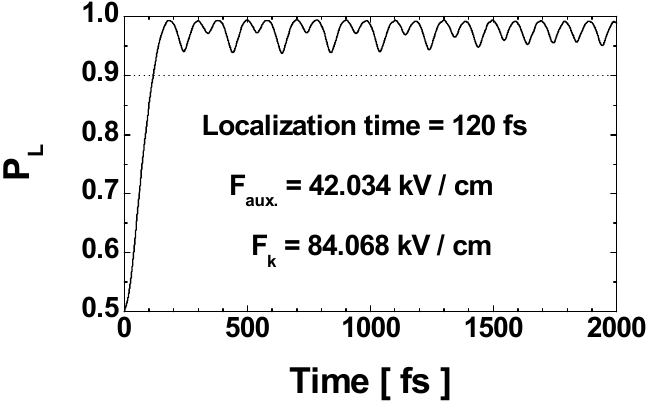}
  \caption{Persistent localization of an initially delocalized electron to the left quantum well, as it follows from a full
  numerical simulation of the TDSE. The indicated
  peak electric field value $F_{\rm aux}$ ($F_{\rm k}$) of the auxiliary HCP
  (HCPs in the pulse train) corresponds to the effective pulse strength $|\alpha|=\pi/4$ ($|\alpha|=\pi/2$)
 (adapted from Ref.~\cite{Alex_QW2004}). 
 \label{Fig:Fig8_AlexPRB_QW_2004}}
\end{figure*}

\subsubsection{Population transfer}
In the framework of the IA it is impossible to transfer an
electron from one quantum well to another just by a single light
pulse, so that the electron changes the well at the end of the
pulse, for the same reason  as for the localization process: the
electronic density needs time to redistribute  after it attains
a momentum kick from the pulse. The time of such a redistribution
is always on the order of the tunneling time in the field-free system
$\pi T_{\rm c}$. However, in this time an electron initially
completely localized in one of the quantum wells changes the
quantum well even if no light pulse is applied. The real challenge
is to reduce the transfer time significantly below $\pi T_{\rm
c}$, as it was theoretically investigated for CW driving fields
\cite{Holthaus1992,Grifoni1998}.

At first glance, there is an easy solution for this problem beyond
the IA if a short single-cycle pulse is utilized for it. As we
already could see for a general one-dimensional potential [cf.
Eqs.~\eqref{Eq:U_result_1d_potential},
\eqref{Eq:Delta_x_free_space} and \eqref{Eq:single_cycle_pulse}],
the action of a single-cycle pulse leads to a coordinate shift of
the electron. For simplicity, we consider the problem in the
framework of the TLSA. Then the first term in the exponent of
Eq.~\eqref{Eq:U_result_TLS} vanishes and the leading contribution
is given by the second term so that the action of the single-cycle
pulse is determined by
\begin{equation}\label{Eq:U_TLS_single_cycle}
    \mathcal{U}=\exp\left[i \beta\sigma_y\right]=\left(
\begin{array}{cc}
  \\[-2.0cm]
  \cos\beta\ \  & \ \ \sin\beta \\[-0.5cm]
  -\sin\beta\ \ & \ \ \cos\beta \\[-0.2cm]
\end{array}%
\right)\;,
\end{equation}
where $\beta$ is given by
\begin{equation}\label{Eq:U_T}
    \beta=a_vb_\varepsilon s_2\;,
\end{equation}
$a_v$ and $b_\varepsilon$ are defined by Eq.~\eqref{Eq:TLS_ab},
and $s_2$ is given by Eq.~\eqref{Eq:s_n_infinity} [see
Section~\ref{Sec:single_cycle} for a particular example]. If
$\beta=\pi/2$ is selected, such a propagator transforms the wave
function of the electron localized in the left well $|\psi_{\rm
L}\rangle$ to the wave function of the electron localized in the
right well $|\psi_{\rm R}\rangle$ and vice versa. However, if the
pulse duration is much smaller than the characteristic system time
then we have  $b_\varepsilon \ll 1$ and therefore $a_v \gg 1$ is
required in order to fulfill the population transfer condition.
With that, the perturbation expansion of
Eq.~\eqref{Eq:U_result_TLS} is not valid beyond the first term in
the exponent anymore because we enter the regime of SVS
excitations. Consequently,
Eqs.~\eqref{Eq:Ansatz_V_strong},\eqref{Eq:U_1_TLS} and
\eqref{Eq:U_2_TLS} should be used. The population transfer
condition in this regime is given by
\begin{equation}\label{Eq:condition_transfer}
    \sqrt{A^2+B^2}=\pi/2+n\pi\; \ \ \ (n=0,1,2,\ldots),
\end{equation}
where we denote $A\equiv A(t=\infty)$ and $B\equiv A(t=\infty)$.
$A(t)$ and $B(t)$ are given by Eqs.~\eqref{Eq:A_TLS} and
\eqref{Eq:B_TLS}, respectively. For SVS pulses
$A(t)$ and $B(t)$ are determined by integrals containing strongly
oscillating functions. In result, the magnitudes of $A$ and $B$
are limited from above so that the condition
\eqref{Eq:condition_transfer} can not be reached even for $n=0$.
For example, selecting a particular single-cycle pulse shape
\eqref{Eq:single_cycle_pulse}, it is not possible to fulfill Eq.~\eqref{Eq:condition_transfer} for pulse durations $\tau_{\rm d}<0.5 T_{\rm c}$ and
realistic pulse strengths. A way out of this problem might be a
special design of the pulse shape aiming at increasing the value of
the quantity $\sqrt{A^2+B^2}$ for a given pulse duration and
strength.

%
%
%
\subsubsection{Persistent localization in presence of relaxation}\label{Sec:persistent_localization_relax}
%
%
 %
A relevant question here is whether the achieved sustainability of the nonequilibrium state, e.g. in a DQW, is resistant to
residual perturbations such as phonons, impurities, etc. For a first insight let us consider the case where the
electronic motion is coupled to a vibrational mode located within the well \cite{Schueler2010} and revisit the pulse-induced
charge localization. Mathematically,  this scenario can be captured by introducing
an additional degree of freedom ($y$) associated with intrinsic  vibrations.  Their Hamiltonian reads $H_R = p_y^2/(2m_\mathrm{ph}) + m_\mathrm{ph}\omega_\mathrm{ph}^2 y^2/2$ with the attributed  mass $m_\mathrm{ph}$ and frequency $\omega_\mathrm{ph}$. The vibrational mode  couples linearly via the potential $H_\mathrm{C} = (y/y_0)\,v_\mathrm{c}(x)$
to the bare electronic motion along the $x$-direction in the DQW. The function $v_\mathrm{c}(x)$ quantifies
the interaction strength along the structure, $y_0$ is an appropriate length scale $\sqrt{\hslash/(m_\mathrm{ph}\omega_\mathrm{ph})}$
of the  vibrating impurity.
\begin{figure*}[t]
	\centering
		\includegraphics[width=12cm]{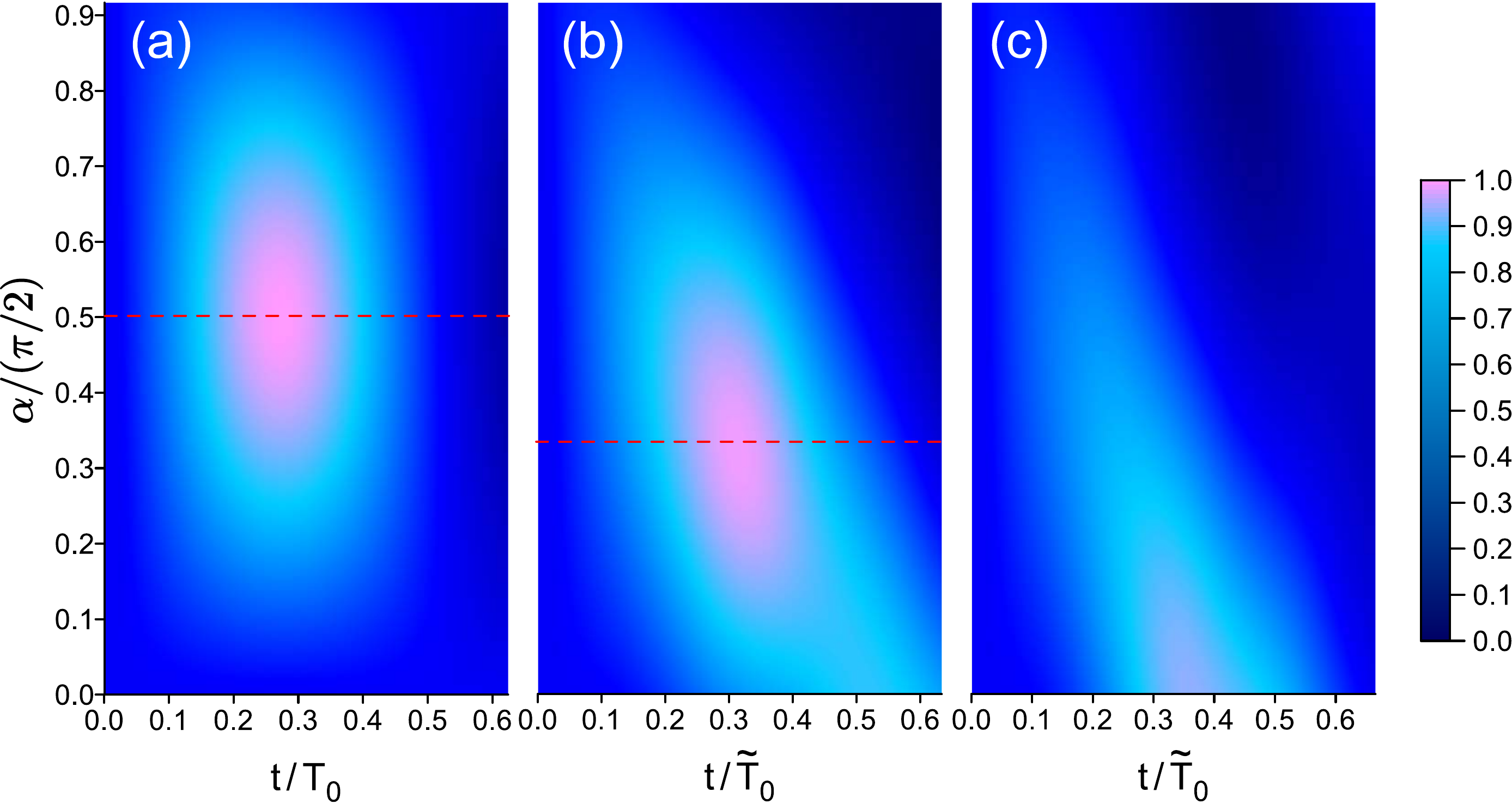}\vspace{-0.2cm}
		\caption{The probability $P_\mathrm{L}$  for finding the electron in the left well as a function of time $t$
       and the effective HCP strength $\alpha$, defined in Eq.~\eqref{Eq:quasienergies_TLS_alpha}.
       Time  is measured in units of $\tilde{T}_0=2\pi\hslash/(\Delta E)$ where $\Delta E$ is the
       energy splitting between the two lowest energy levels (in the presence of the vibrational modes). Panels correspond to different
		values of the scaled coupling parameter $\gamma= 2 V_0 / \hslash \omega_\mathrm{ph}$: (a) $\gamma= 0$, (b) $\gamma = 0.6$, and (c) $\gamma = 1$ (adapted from Ref.~\cite{Schueler2010}).}\vspace{-0.3cm}
		\label{apl_michael_dqw:fig}
\end{figure*}
Let us discuss the case $v_\mathrm{c}(x) =V_0 \exp[-(x-x_\mathrm{c})^2/2 \lambda^2]$ which is suitable for the situation when
the tunneling electron scatters inelastically from a pinned impurity triggering on its vibrational excitations.
 $V_0$ is the characteristic interaction energy, $x_\mathrm{c}$ is the impurity position and $\lambda$ characterizes the localization length of the coupling (taken as 2.5 nm in the calculations shown here).
If the  impurity is
located around the potential barrier, meaning $x_\mathrm{c} = 0$,  the two degrees of freedom become effectively
decoupled, for the electronic density there is marginal.
 Hence the strongest destruction of the electron localization due to the impurity coupling is expected when the impurity is in the middle of one of the wells
  where the electronic  wave packet is localized. In that case the localization scheme (Section~\ref{Sec:persistent_localization}) needs to be adjusted \cite{Schueler2010}, but nevertheless is
  still possible; the point here is then whether it is still sustainable. As detailed in Ref.~\cite{Schueler2010} the stability  of the localized wave packet
  is achievable with an adjusted pulse sequence or by applying a constant electric field  oriented in the
 $x$-direction.
A typical result is shown in Fig.~\ref{apl_michael_dqw:fig}a,b evidencing that the localization induced by a single HCP is satisfactorily robust, at least for the
type of scattering events taken into account. Considering another type of scattering yielded also similar findings  \cite{Schueler2010}.
Clearly if the coupling to the vibrating impurity is very strong the localization becomes ineffective  (cf. Fig.~\ref{apl_michael_dqw:fig}c).
Maintenance of the initially prepared localized state, $P_\mathrm{L}(t=0)=1$, by  applying a sequence of appropriate HCPs is demonstrated in Fig.~\ref{apl_michael_dqw:fig2}.
\begin{figure*}[t]
    \centering
		\vspace{-0.2cm}\includegraphics[width=8cm]{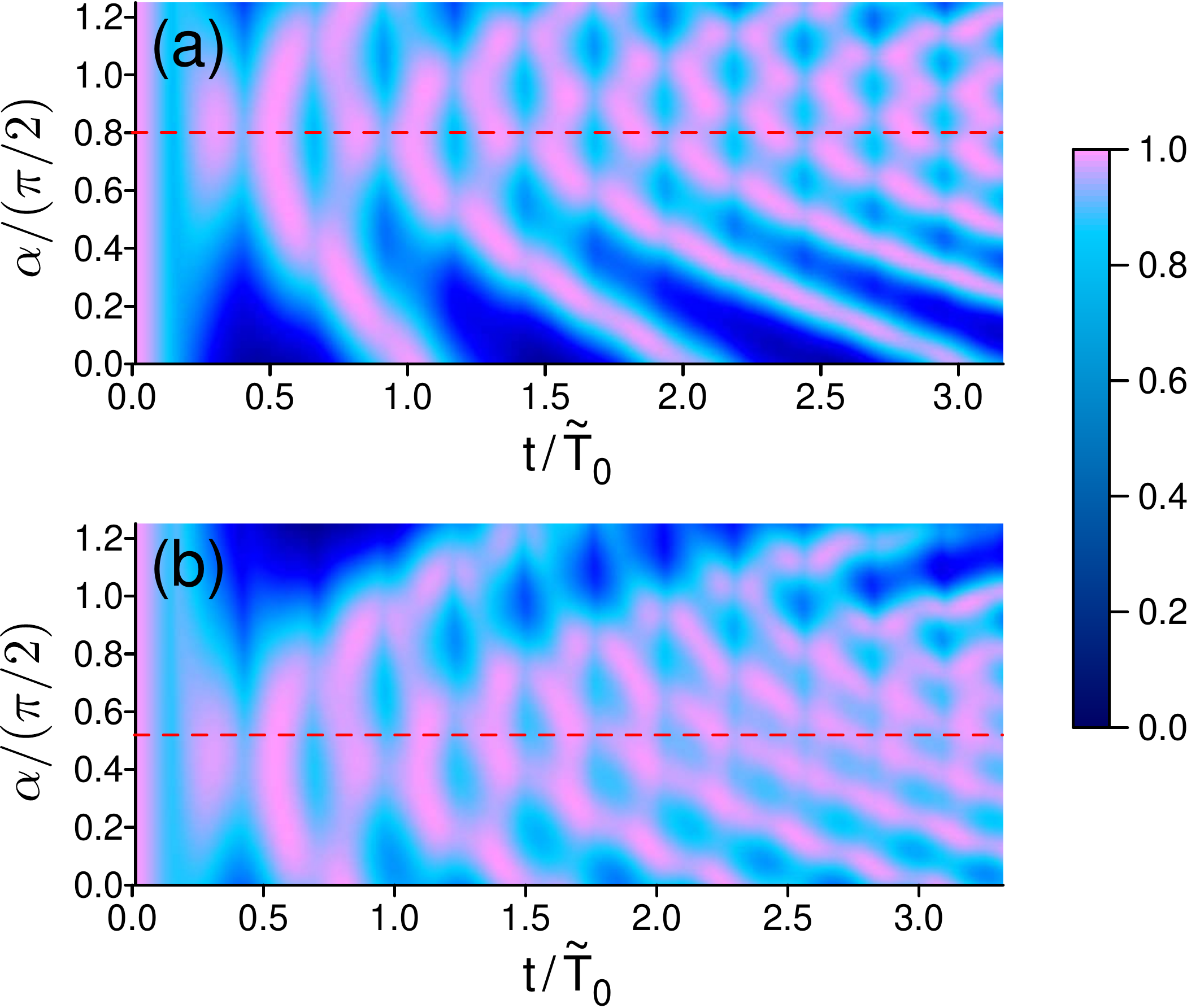}\vspace{-0.2cm}
		\caption{The probability $P_\mathrm{L}$ as function of time $t$ and kick parameter $\alpha$. The electron  is initially (at $t=0$) is assumed to be
perfectly localized to the left well and  then it is subjected to
        a train of HCPs with the effective strength of each pulse $\alpha$ and a repetition period $T = 80$~fs.
        In (a) $\gamma = 0.6$, in (b) $\gamma = 1$.
  Other parameters are as in Fig.~\ref{apl_michael_dqw:fig}.}
		\label{apl_michael_dqw:fig2}
\end{figure*}
%
%
%
\subsection{Pulse-driven charge polarization, currents and magnetic moments in semiconductor
quantum rings}\label{Sec:Appl_QR}

Quantum rings (QRs) play a special role in  mesoscopic physics
bridging classical and quantum-mechanical phenomena \cite{imry}.
Many intriguing physical effects like the Aharonov-Bohm effect,
persistent charge and spin currents, etc. were discovered in these
structures \cite{imry,washburn,Russo2008}. In many
cases their physical
properties can be captured using single-particle states
corresponding to a free angular motion of the electron having an
appropriate effective mass and confined to a single radial channel
\cite{Fuhrer2001,Lorke2000,mailly,Moskalets2002,imry,tan,Alex_PRB2004,Alex_PRL2005}
(see Section~\ref{Sec:1d_ring}). Generalization to a multichannel
model can be constructed straightforwardly if the radius of the
ring is considerably larger than its width and height
\cite{Alex_Europhysics2005,Alex_PRL2005}.

\begin{figure*}[t]
\includegraphics[width=5.0cm]{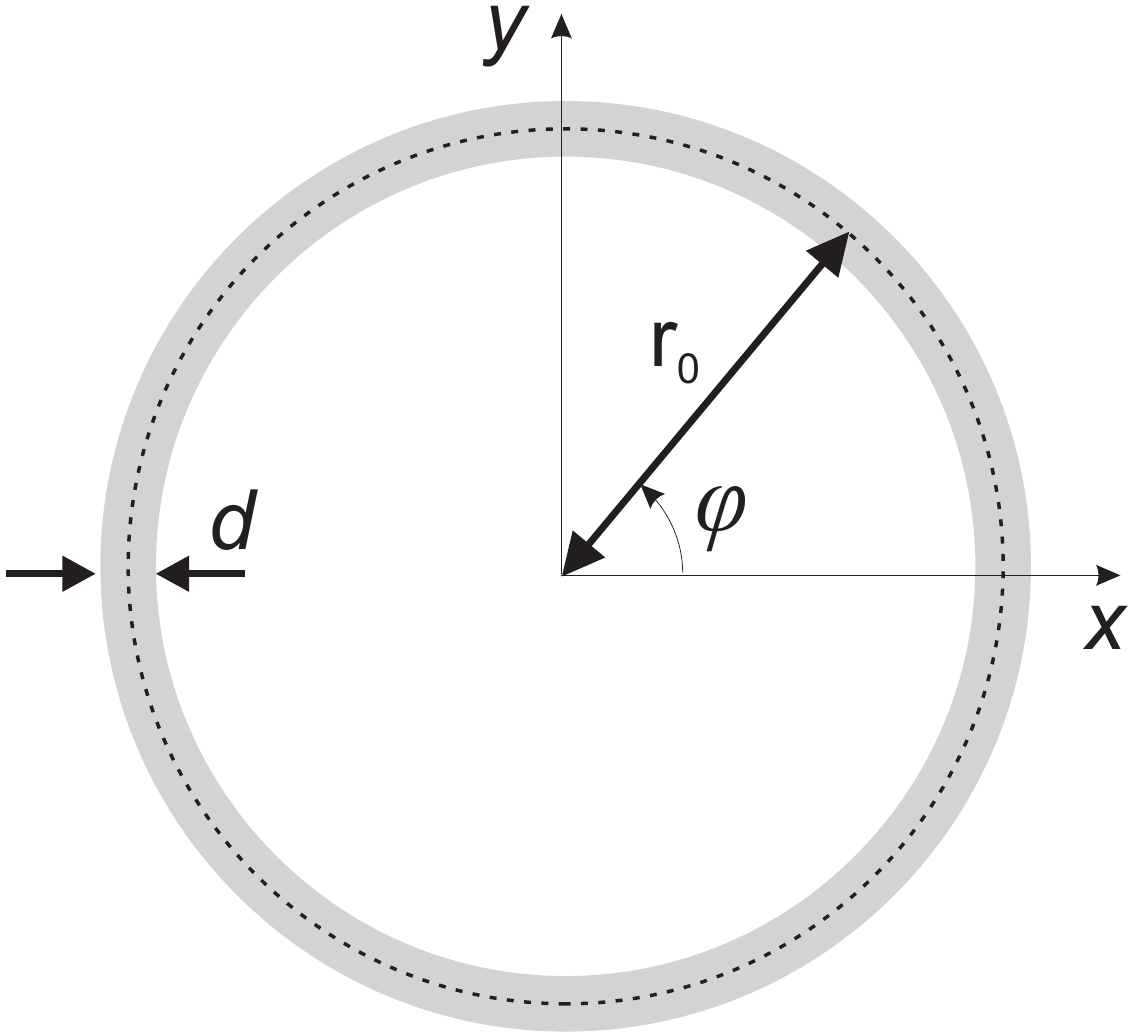}
\caption{Sketch of a QR and the coordinate system. The $z$-axis
is directed perpendicular to the plane of the ring.}
\label{Fig:ring}
\end{figure*}

Let us consider electrons in a thin isolated (without contacts)
semiconductor QR with a mean radius $r_0$ and width $d$ ($d\ll
r_0$) at low temperatures (see Fig.~\ref{Fig:ring}). It can be
made, for example, from GaAs surrounded by AlGaAs having a wider
band gap \cite{mailly} so that up to a certain energy the charge
carriers are confined inside the ring. Other material choices are
also possible \cite{Offermans2005}. Its single electron wave
functions can be written using cylindrical coordinates
$(z,\rho,\varphi)$ as
\begin{equation}\label{Eq:ring_wave_function}
  \psi_{l,m}(z,\rho,\varphi)=Z_0(z)R_l(\rho)\Phi_m(\varphi)\;,
\end{equation}
where $\Phi_m(\phi)=\frac{1}{\sqrt{2\pi}}\mbox{e}^{im\varphi}$ is
the angular wave function, $R_l(\rho)$ denotes the radial wave
function; $l=1,2,3\ldots$ and $m=0,\pm 1,\pm 2,\ldots$ are the
radial and orbital quantum numbers, respectively. In the $z$
direction the width of the ring is much smaller than $d$.
Therefore, we consider only electrons residing in the lowest
$z$-channel with wave function $Z_0(z)$. If we also assume that
$d$ is sufficiently small and the particle number in the ring $N$
is not too large, only single-electron states in  the lowest
radial subband with $l=1$ are relevant. Such assumptions mean that
the thickness and the width of the ring are much smaller than the
Fermi wavelength of the confined electrons. This situation is
experimentally feasible for semiconductor QRs. Then the
single-particle basis is given by
$|m\rangle \equiv|\psi_{1,m}\rangle$ so that the free-electron
Hamiltonian in the second quantization reads
\begin{equation}\label{Eq:H0_carr}
  \hat{H}^{\rm el}_0=\sum_m
        \varepsilon_m
        \hat{a}_m^{\dagger} \hat{a}_m\;,
\end{equation}
where $\hat{a}_m^\dagger$ and $\hat{a}_m$ denote the carrier creation
and annihilation operators, respectively. The single-particle
energies $\varepsilon_m$ are given by
Eq.~\eqref{Eq:energies_ring}. The Hamiltonian of the interaction
with a time-dependent external light field, which is polarized in
the plane of the QR, is given by
\begin{equation}\label{Eq:V_external}
  \begin{split}
   \hat{V}=& -\frac{1}{2}eE_x(t) r_0 \sum_{m}
   (\hat{a}_m^\dagger \hat{a}_{m-1}+\hat{a}_m^\dagger \hat{a}_{m-1})\\
   & -\frac{1}{2i}eE_y(t) r_0 \sum_{m}
   (\hat{a}_m^\dagger \hat{a}_{m-1}-\hat{a}_m^\dagger \hat{a}_{m+1}),
  \end{split}
\end{equation}
where $E_x(t)=E_{0x}f_x(t)$ and  $E_y(t)=E_{0y}f_y(t)$ are the
electric field components. Here $E_{0x}$, $E_{0y}$ and
$f_x(t),f_y(t)$ denote the  corresponding amplitudes and temporal
profiles of the ultrashort  pulses driving the
electron dynamics in the QR, respectively. In the case of
semiconductor QRs the relaxation time scales may become comparable
with the characteristic system time $\tau_{_\mathrm{F}}$ (see
Section~\ref{Sec:1d_ring}). To account for  relaxation one should resort to the density matrix formalism
\cite{Rossi_Kuhn,Haug_Koch,Bonitz_book}. The density operator
components are expressed as $\hat{\rho}_{mm'}(t)=
\hat{a}_{m}^{\dag}(t)\hat{a}_{m'}(t)$. Taking the expectation
value of it with respect to the initial state of the system gives
the density matrix components $\rho_{mm'}(t)=\langle
\hat{\rho}_{mm'}(t)\rangle$. The density matrix provides full
information on single-particle evolution.

\subsubsection{Relaxation and dephasing in driven quantum rings }\label{Sec:relaxation_QRs}
Before we start with the description of the system dynamics
induced by short broadband pulses, it is important to
discuss here the relevant relaxation processes to
determine the corresponding relaxation time scales. We consider
the semiconductor QR to be free from impurities. Redistribution of
the excitation in the system due to the electron-electron
collisions is limited by the Pauli blocking and the energy
conservation \cite{chakraborty}. For typical semiconductor QRs
optical phonon energies are much larger than the interlevel energy
distance in the neighborhood of the Fermi level. Therefore the
main relaxation channels of the excited electronic system of the QR
should be governed by the interactions with photon and acoustic
phonon fields. Specifically, emission of coherent radiation,
spontaneous emission of photons, emission of coherent phonon
waves, and scattering by incoherent phonons may play an essential
role for the relaxation. In case of semiconductor QRs, these
processes were considered in
Refs.~\cite{Moskalenko_PRB2006,Moskalenko_PRA2008} and are
described in Appendices \ref{Sec:radiative_damping} and
\ref{Sec:relaxation_phonons}. One can see that the scattering by
incoherent phonons determines the largest contribution. At low
temperatures the corresponding decay rates may vary from
microseconds to picoseconds depending on the temperature and QR
parameters (cf. Fig.~\ref{Fig:eff_rate} in Appendix
\ref{Sec:relaxation_phonons}).

\subsubsection{Charge polarization dynamics}\label{Sec:polarization_dynamics}
The most evident way to induce a field-free charge polarization in a
semiconductor QR is to apply appropriate ultrashort HCPs. As
far as such  pulses deliver momentum  \textit{kicks} to the
electronic system of the ring, it is clear that after some time
delay the charge distribution of the QR becomes polarized. This,
in essence classical, effect was firstly theoretically
demonstrated in Ref.~\cite{Alex_PRB2004} for a
single-channel semiconductor QR. The model was generalized to the
case of a narrow multichannel ring in
Ref.~\cite{Alex_Europhysics2005}.

Let us consider the field-free polarization dynamics generated by
an applied HCP in more details. The charge polarization of the QR
is characterized by its dipole moment that is given by
\begin{equation}\label{Eq:dip_mom_ring}
    \boldsymbol{\mu}=\textrm{Tr}[e \mathbf{r}\hat{\rho}]\;,
\end{equation}
where  $e$ is the electron charge. The components of $\boldsymbol{\mu}$ can be expressed
via near-diagonal density matrix elements as
\begin{eqnarray}
  \mu_x &=& e r_0 \sum_m \mbox{Re}[\rho_{m,m-1}], \label{Eq:dip_mom_ring_x}\\
  \mu_y &=& e r_0 \sum_m \mbox{Im}[\rho_{m,m-1}]\;
  \label{Eq:dip_mom_ring_y}.
\end{eqnarray}
Thus, to access the charge polarization of the QR, one
can  solve numerically the equation of motion for the
single-particle density matrix. Its dynamics is governed by the Hamiltonians
\eqref{Eq:H0_carr} and \eqref{Eq:V_external} with an appropriate
relaxation contribution, as discussed above.   The result has
to be substituted into Eqs.~\eqref{Eq:dip_mom_ring_x} and
\eqref{Eq:dip_mom_ring_y}.

As it was discussed in Section~\ref{Sec:1d_ring}, the IA provides a
good description of the excitation process if the parameters of
the applied HCP and of the QR are selected properly. In this
approximation, selecting the $x$-axis being parallel to the
polarization direction of the applied HCP and using
Eq.~\eqref{Eq:IA_mapping_rho}, where the evolution operator is
given by Eq.~\eqref{Eq:U_1_ring_result} with
Eq.~\eqref{Eq:alpha_def}, leads to
\cite{Moskalenko_PRB2006,Moskalenko_EPL2007}
\begin{equation}\label{Eq:rho_basis_free_result}
   \begin{split}
    \rho_{m_1,m_2}\!(t)=&\Theta(t)
    {\rm e}^{\frac{i}{\hslash}(\varepsilon_{m_1}-\varepsilon_{m_2})t}
    \sum_{mm'}C_{m_2,m}^* C_{m_1,m'}\rho_{m'\!,m}^0(0)\;\\
    &+\Theta(-t)\rho_{m'\!,m}^0(t)\;.
   \end{split}
\end{equation}
Here the origin of the time axis was selected to coincide with the
HCP parameter $t_1$ and $\rho_{m'\!,m}^0(t)$ describes the
evolution of the density matrix of the unperturbed QR. $\Theta(t)$
denotes the Heaviside step function. The coefficients $C_{m_1,m}$
are determined by $C_{m_1,m}\equiv\langle m|{\rm
e}^{i\alpha\cos\phi}|m_1 \rangle=i^{m_1-m}J_{m_1-m}(\alpha)$,
where $J_l(x)$ is the Bessel function and the parameter
$\alpha$ characterizes the strength of the HCP, as it is described
in Section~\ref{Sec:1d_ring}. Notice that
Eq.~\eqref{Eq:rho_basis_free_result} does not take account of the
relaxation processes and can be therefore used only for sufficiently short
 times after the excitation. When calculating the induced
charge polarization dynamics, we can introduce the corresponding
relaxation time via the inverse of the effective decay rate
$\tau_{\rm rel}=1/\gamma_{\rm eff}$ of the charge dipole moment of
the QR, as it is described in details in Appendix
\ref{Sec:relaxation_phonons}. Then, assuming that the unperturbed
QR has been in an equilibrium state with the density matrix
$\rho_{m'\!,m}^0=f_m^0\delta_{m'\!,m}$, where $f^0_m\equiv
f^0(\varepsilon_m)$ is the Fermi-Dirac distribution function
determined by the temperature and number of the electrons in the
QR, and inserting Eq.~\eqref{Eq:rho_basis_free_result} into
Eqs.~\eqref{Eq:dip_mom_ring_x} and \eqref{Eq:dip_mom_ring_y} gives \cite{Moskalenko_EPL2007}
\begin{equation}\label{Eq:mu_general}
    \mu_x(t)=\Theta(t)er_0\alpha h_1(t)\sin\!\left(\frac{2\pi
    t}{t_{p}}\right)
    e^{-t/\tau_{\rm rel}}
            \sum_m f_m^0\cos\!\left(\frac{4\pi t}{t_{p}}m\right)
\end{equation}
and $\mu_y(t)=0$. Here the function $h_1(t)$ is defined as
\begin{equation}\label{Eq:h_function}
    h_1(t)=J_0\big(\Omega_1(t)\big)+J_2\big(\Omega_1(t)\big)\;,
\end{equation}
where the function $\Omega_1(t)$ reads
\begin{equation}\label{Eq:Omega_function}
    \Omega_1(t)=2\alpha\sin(2\pi t/t_{p})
\end{equation}
with
\begin{equation}\label{Eq:t_p}
    t_{p}=4\pi m^{\ast}r_{0}^{2}/\hslash\;.
\end{equation}
 $m^{\ast}$ is the electron effective mass and $r_0$ is the QR radius
(see Fig.~\ref{Fig:ring}).

\begin{figure*}[t]
\includegraphics[width=10.0cm]{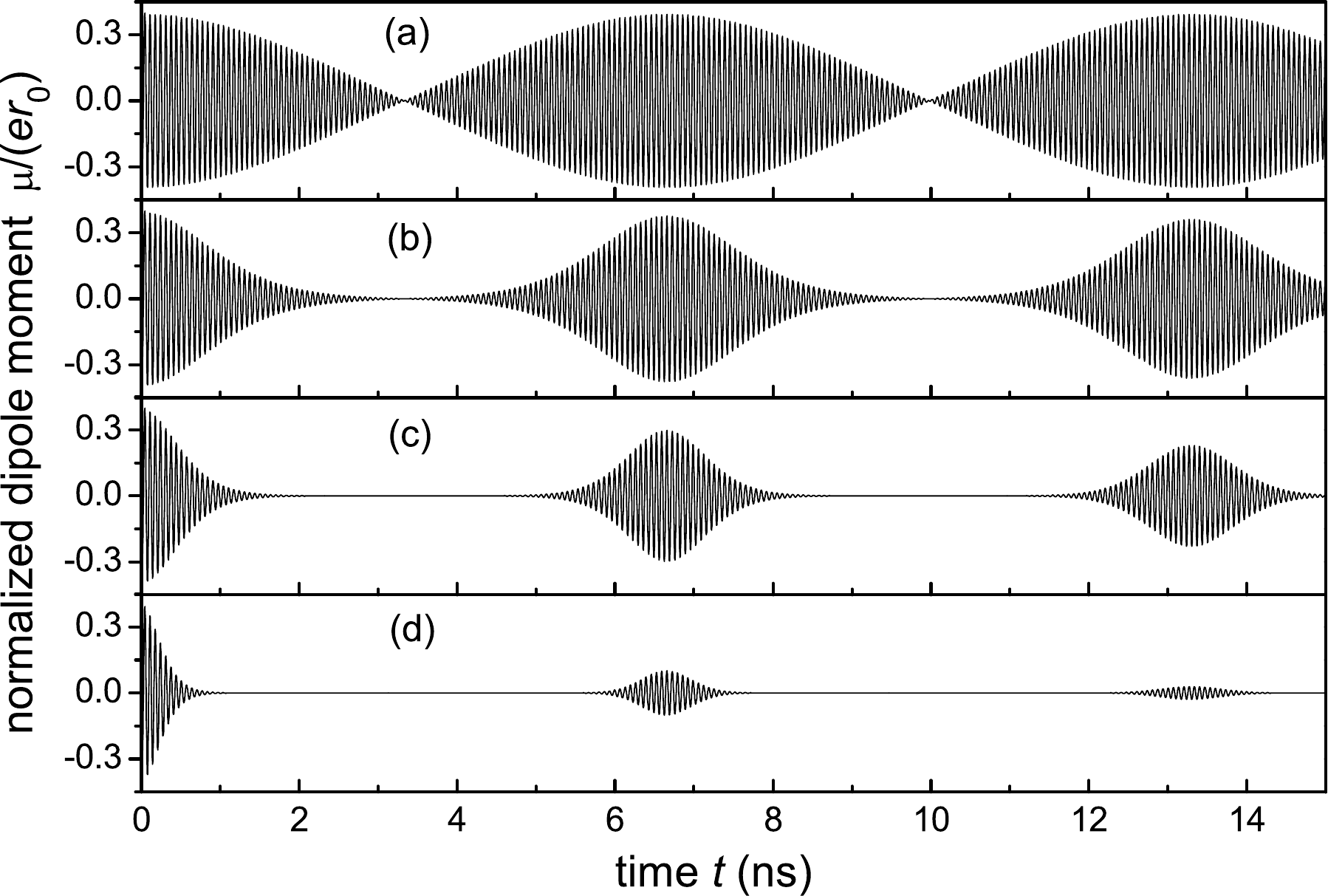}
\caption{Numerically calculated field-free dynamics of the QR
dipole moment $\mu$ induced by a single HCP is shown for different
temperatures: (a) $T=0.1$~K, (b) $T=0.5$~K, (c) $T=1$~K and (d)
$T=2$~K. The HCP shape is determined by Eq.~\eqref{Eq:pulse_shape}
with $\tau_{\rm d}=3$~ps and the kick strength $\alpha=0.2$.
Parameters of the QR: $r_0=1.35~\mu{\rm m}$, $d=50~{\rm nm}$ and
$N=400$. The relaxation due to the scattering by phonons is taken into
account in the way described in Appendix \ref{Sec:relaxation_phonons} for
a n-GaAs QR with $c_{_{\rm LA}}=4.79\times10^5~\mbox{cm/s},\
\rho_{\rm s}=5.32~\mbox{g/cm}^{-3},\ |D|=8.6~\mbox{eV}, \
\hslash\omega_{_{\rm D}}=30~\mbox{meV}$, and $m^*=0.067m_0$
(adapted from Ref.~\cite{Moskalenko_PRB2006}).}
\label{Fig:revivals_PRB}
\end{figure*}

The field-free dynamics of the QR dipole moment induced by a
single HCP is illustrated in Fig.~\ref{Fig:revivals_PRB}, where
the result of a numerical solution, including the relaxation as
described in Appendix \ref{Sec:relaxation_phonons}, is shown for
different temperatures. For the chosen parameters of the
simulation, the approximate analytical solution
\eqref{Eq:mu_general} gives practically the same dependence. For
all temperatures we observe fast (classical) oscillations of the dipole
moment (better resolved in Fig.~\ref{Fig:IA_pulses}) with a period
around \cite{Robinett2004}
\begin{equation}\label{Eq:fast_oscillations_time}
    T_{\rm cl}=2\pi\hslash/\big|\partial \varepsilon_m/\partial
m\big|_{m=m_{_\mathrm{F}}}\approx \tau_{_\mathrm{F}}\;,
\end{equation}
where $m_{_\mathrm{F}}$ is the value of the orbital quantum number
close to the Fermi level and $\tau_{_\mathrm{F}}$ is the ballistic
time.

In the case of low temperatures and low excitation strengths only
the states nearest to the Fermi level contribute to the dynamics. Thus
if the highest occupied state of the QR is fully populated before
the HCP is applied, only one oscillation frequency is observed. If
this state is only partly populated, then oscillations with two
close frequencies corresponding to two possible allowed dipole
transitions between the populated (partly populated) states and the next higher
laying partly empty (empty) states contribute to the dynamics. Therefore, the
long-time behavior exhibits beating (see
Fig.~\ref{Fig:revivals_PRB}a).
With increasing $T$ more levels are only partly populated
initially. Therefore, more transitions to the neighboring states
become possible. As a consequence, the dynamics of the dipole
moment shows alternating collapse  and quantum revivals
(see Fig.~\ref{Fig:revivals_PRB}b-d). The revival time is given by
\cite{Robinett2004}
\begin{equation}\label{Eq:revival_time}
    T_{\rm rev}=4\pi\hslash/\big|\partial^2 \varepsilon_m/\partial
m^2\big|_{m=m_{_\mathrm{F}}}\;
\end{equation}
that agrees with the calculation results shown in
Fig.~\ref{Fig:revivals_PRB}. The observed decay of the
peak values of the revivals is governed by  relaxation processes.

\subsubsection{Switching on and off the charge currents}\label{Sec:current_generation}
The generation mechanism of charge currents in semiconductor QRs
by means of HCPs was firstly  described in
Ref.~\cite{Alex_PRL2005}. It is a two-step process (see
Fig.~\ref{Fig:current_generation_mechanism}). In the first step, the
application of a HCP to the ring results in transient charge polarization,
as described above. In the second step, another HCP applied with
an appropriate time delay with respect to the first HCP leads
to the emergence of a circular charge current in the QR. In the optimal situation, the
second HCP should be polarized in the plane of the QR
perpendicular to the first HCP.

\begin{figure*}[t]
  \includegraphics[width=14cm]{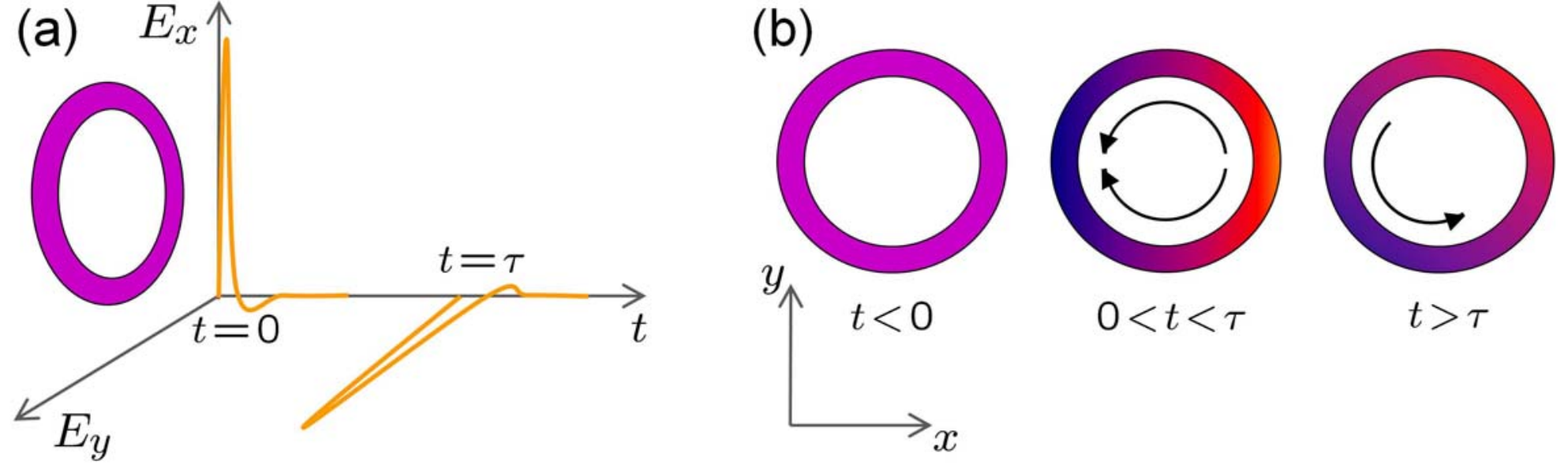}
  \caption{Illustration of the mechanism of the charge current generation in the QR by applying  two HCPs.
  The second HCP is delayed with respect to the first one by the
  time $\tau$. Both pulses are linearly polarized. The polarization vectors are  orthogonal to each other  and both lie in the plane of the
   QR (a).
  Action of the HCPs leads to a dynamical charge redistribution in the
  QR (b). The charge density is indicated by the color code, where the red (blue) color corresponds to the higher (lower) density,  and
  its motion direction is shown by  the arrows.}\label{Fig:current_generation_mechanism}
\end{figure*}

In the single-channel model the charge current flowing in the QR
$I(t)$ can be expressed via the diagonal density matrix components
$\rho_{m\!,m}(t)$ as \cite{Moskalenko_EPL2007}
\begin{equation}\label{Eq:current_general}
I(t)=I_{0}\sum_{m} m\rho_{m\!,m}(t)\;,
\end{equation}
where
\begin{equation}\label{Eq:I0}
I_{0}=e\hslash/(m^{\ast}r_{0}^{2})\;.
\end{equation}
Here $e$ denotes the electron charge, i.e. $I_0$ is negative. Then
with the help of these equations and
Eqs.~\eqref{Eq:rho_basis_free_result}, \eqref{Eq:dip_mom_ring_x}
and \eqref{Eq:dip_mom_ring_y} one can find that the charge current
generated by means of the above-described mechanism is given by
\begin{equation}\label{Eq:generated_current_HCPs}
    I(t)=\Theta(t-\tau)I_{0}\alpha_{\bot}\frac{\mu(\tau)}{er_{0}}e^{-t/\tau_{\rm
    rel}^{\rm cur}}\;,
\end{equation}
where $\tau$ is the time moment of the application of the second
HCP and $\alpha_{\bot}$ is the corresponding kick strength.
$\mu(\tau)$ is the charge dipole moment of the QR at $t=\tau$. Further,
$\tau_{\rm rel}^{\rm cur}$ denotes the current relaxation time, an
effective relaxation time connected with the relaxation of the
diagonal density matrix components \cite{Moskalenko_EPL2007}.
A similar expression to \eqref{Eq:generated_current_HCPs} can be
obtained also for a multichannel model
\cite{Alex_PRL2005,Alex_PhD,Matos_Indian_paper}, where a sum over
the different radial channels should be performed additionally. In
the limit $d\ll r_0$ (see Fig.~\ref{Fig:ring}), considered here,
the charge current $I$ flowing in the QR is associated with an orbital
magnetic dipole moment $M$ via the relation
\begin{equation}\label{Eq:magnetic_moment}
    M=\pi r_0^2 I\;.
\end{equation}
Therefore we may interchangeably discuss the generated current and the
generated magnetic moment of the QR.

\begin{figure*}[t]
\includegraphics[width=9.0cm]{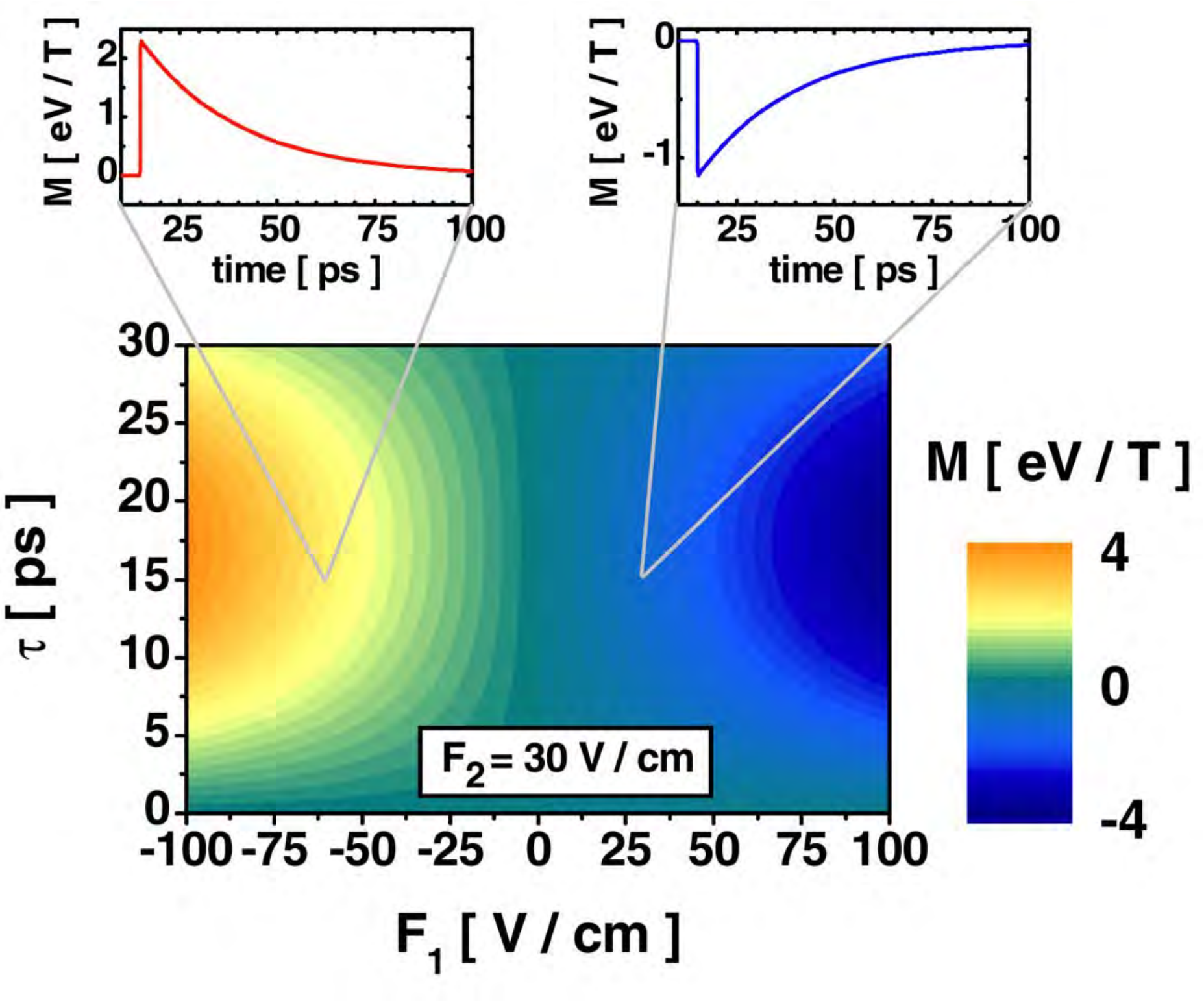}
\caption{Induced magnetic dipole moment $M$ is shown as a function
of the peak electric field of the first HCP $F_1$ and the time
delay $\tau$ between the two applied HCPs. The  peak electric
field of the second HCP $F_2$ is fixed to 30~V/cm. For both HCPs
the pulse shape is given by Eq.~\eqref{Eq:f_Alex_PRB_QW_2004} with
$\tau_{\rm d}=1$~ps. The insets show the dynamics of the
generated magnetic moment for $\tau=30$~ps and two values of
$F_1$, $F_1=-60$~V/cm and $F_1=30$~V/cm. Parameters of the
simulation: QR radius $r_0=1.35~\mu{\rm m}$, width $d=10~{\rm
nm}$, electron number $N=1400$, and temperature $T=0$~K. For these
parameters a magnetic moment of 1~eV/T corresponds to a QR current
of 8.9~nA. Current relaxation time
$\tau_{\rm rel}^{\rm cur}=30$~ps (adapted from
Ref.~\cite{Alex_PRL2005}).} \label{Fig:Fig3_Alex_PRL2005}
\end{figure*}

Dependence of $M$ on the time
delay $\tau$ between the two applied HCPs and its decay for two
particular choices of $\tau$ are illustrated in
Fig.~\ref{Fig:Fig3_Alex_PRL2005}. We see that there is an optimal
time delay between the pulses around $17$~ps leading to the
maximal peak value of the emerging current and magnetic moment.
This value of the time delay corresponds to $T_{\rm cl}/4$ such
that the charge dipole moment generated by the first HCP achieves
its maximum value at the time moment of the application of the
second HCP. The time dynamics of the magnetic moment
shown in the insets demonstrates the relaxation process, which in
Ref.~\cite{Alex_PRL2005} was modeled with the help of a single
relaxation time $\tau_{\rm rel}^{\rm cur}$ [cf. Eq.~\eqref{Eq:generated_current_HCPs}]. One can show that an
increase of the QR temperature leads to a decrease of the peak
value of the generated magnetic moment \cite{Alex_PRL2005},
whereas a larger particle number leads to a smaller $T_{\rm cl}$
\cite{Alex_PRB2004} and thus enables shorter optimal delay times.

An additional amount of the charge current produced in a QR,
where some charge current is already flowing, is independent of
the existing current. It can be again described by
Eq.~\eqref{Eq:generated_current_HCPs} and is proportional to the
effective strength of the applied HCP and the charge dipole moment of the QR
at the time moment when this pulse arrives. This fact together with
the property of the dipole moment dynamics to have collapsed
phases (see Fig.~\ref{Fig:revivals_PRB}) can be used to switch off completely
the current flowing in the ring  at a selected time
moment. A possibility to create the charge current in the QR and
then to eliminate it after a time delay allows to produce short
pulses of magnetic field in the neighborhood of the QR. Then one
HCP pair is used to switch on the magnetic moment and the second
one, coming with a time delay $\tau_{\rm p-p}$ with respect to the
first one, is used to switch it off. For a complete switch off we
need to know the current value at the time moment when the second
pulse pair is applied during the collapsed phase of the charge
dipole moment. Thus an appropriate modeling of the relaxation is required  in
the time range between the HCP sequences for both the charge
dipole moment and the current (see
Section~\ref{Sec:relaxation_QRs}).

\begin{figure*}[t]
\includegraphics[width=6.0cm]{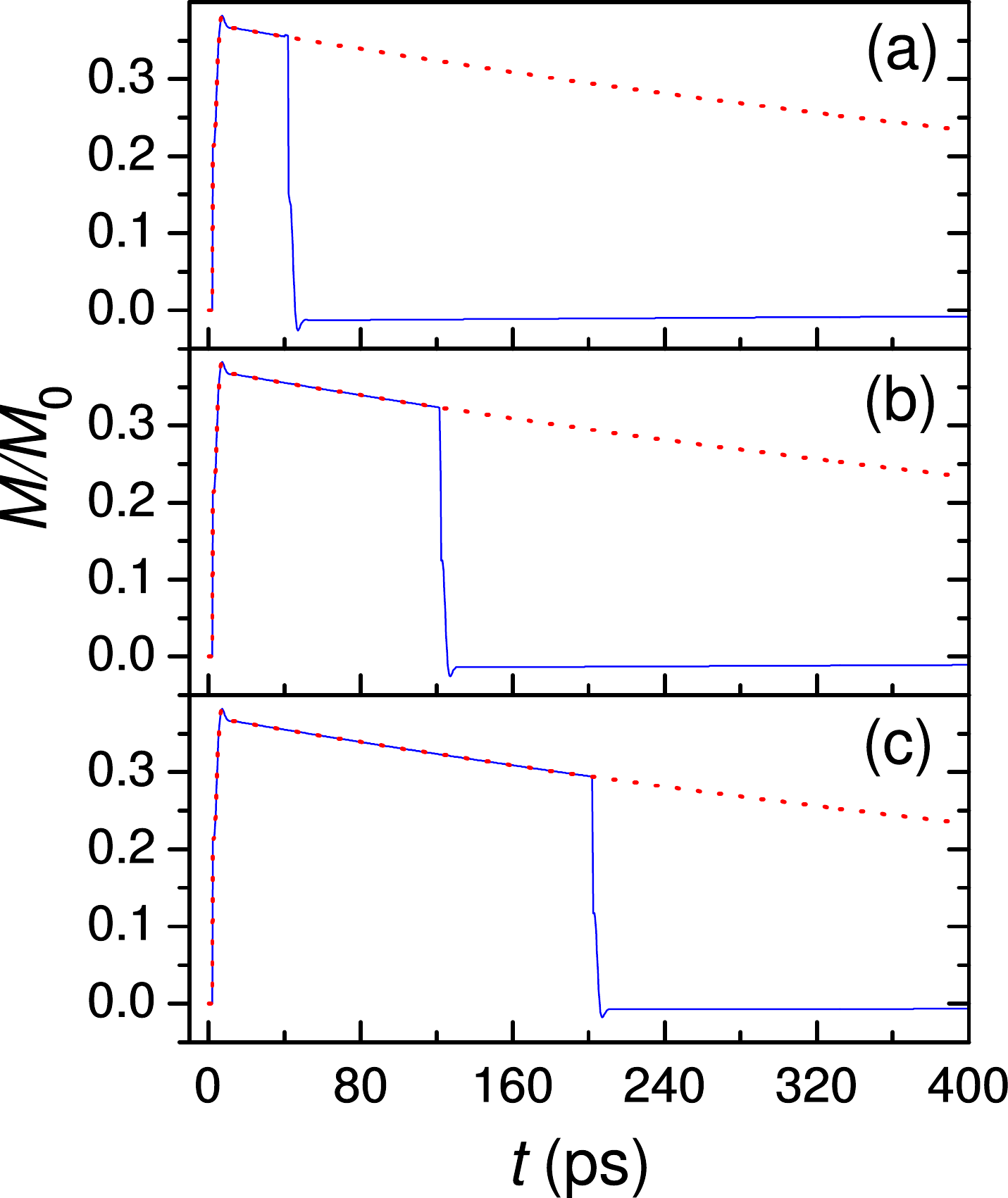}
\caption{Magnetization pulses generated in the n-GaAs QR by
application of two HCP sequences delayed in respect to each other
by $\tau_{\rm p-p}=40$~ps in (a), $\tau_{\rm p-p}=120$~ps in (b),
and $\tau_{\rm p-p}=200$~ps in (c). Kick strengths of the first
three HCPs are the same and are equal to $0.4$, the last HCP has
the kick strength: $\alpha_\bot'=-0.39$ in (a),
$\alpha_\bot'=-0.35$ in (b), and $\alpha_\bot'=-0.32$ in (c).
Dotted lines show the result in the absence of the eliminating
pulse sequence. Parameters of the QR: $r_0=0.3~\mu{\rm m}$,
$d=20~{\rm nm}$, $N=160$, and $T=20$~K. Parameters of the
phonon-induced relaxation are as in Fig.~\ref{Fig:revivals_PRB}
(adapted from Ref.~\cite{Moskalenko_PRB2006}).}
\label{Fig:current_control2}
\end{figure*}

The generation of magnetic pulses and its tunability was
demonstrated in Ref.~\cite{Moskalenko_PRB2006} and is
illustrated in Fig.~\ref{Fig:current_control2}. The magnetic
dipole moment shown in this figure is normalized by  $M_0=\pi r_0^2 I_0$,
where $I_0$ is given by Eq.~\eqref{Eq:I0}. The shape of the HCPs
used in the simulations corresponds to Eq.~\eqref{Eq:pulse_shape}
with $\tau_{\rm d}=0.5$~ps. Here the first HCP of the generating
pulse pair and the first HCP of the eliminating pulse pair are
polarized in the same direction and have kick strengths
$\alpha=0.4$, i.e., initiate an identical dynamics of the charge dipole moment
in the QR. The second HCP of the generating pulse pair also has
the same kick strength  $\alpha_\bot=\alpha=0.4$ and is polarized
in the perpendicular direction in the plane of the QR,
antiparallel to the last HCP (second HCP of the eliminating pulse
pair). The kick strength of the last HCP $\alpha'_\bot$ is
attenuated with respect to $\alpha_\bot$, in order to compensate for
the current relaxation taking place during the time $\tau_{\rm
p-p}$ elapsed between the pulse pairs. Possible durations of the generated
magnetic pulses are restricted by the relaxation time scale and
extension of the collapsed phase of the charge dipole moment
dynamics.


\subsubsection{Generation of periodic magnetic pulses}\label{Sec:periodic_magnetic_pulses}
The mechanism of the generation and elimination of the charge
current and magnetization in the QR can be adapted for the
generation of periodic magnetic pulses \cite{Moskalenko_PRB2006}.
One possibility is to repeat both the generating and eliminating
pulse sequences periodically in time (see
Fig.~\ref{Fig:magnetic_sequences2}c). As it was discussed in
Section~\ref{Sec:FL_approach}, as any other observable, the
magnetization should demonstrate a periodic dynamics after some
equilibration stage following the onset of the periodic driving,
due to the dissipation in the QR. The resulting periodic
magnetization dynamics is shown in
Fig.~\ref{Fig:magnetic_sequences2}a.

\begin{figure*}[t]
\includegraphics[width=12.0cm]{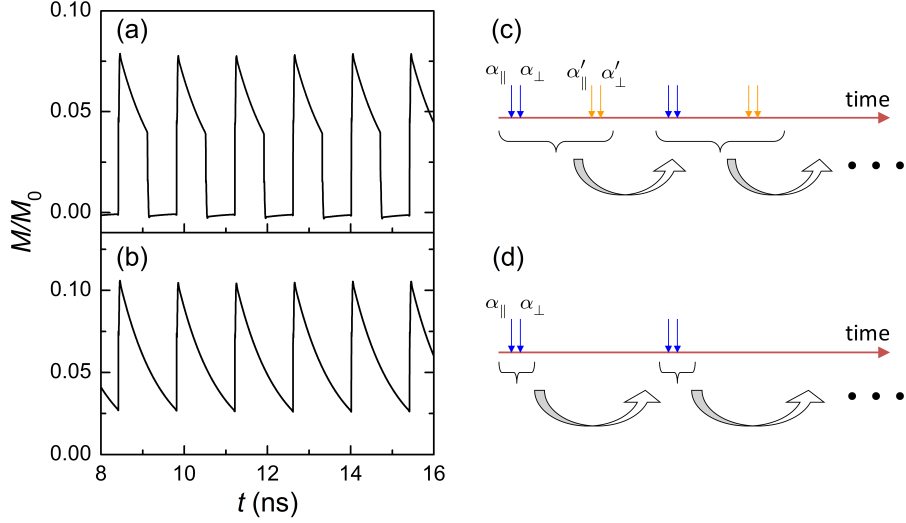}
\caption{Magnetization dynamics generated in the QR by
  applying a of HCPs sequences, delayed with respect to each other by $\tau_{\rm p-p}=1.4$~ns.
  In (a) each sequence contains four HCPs having
  kick strengths $\alpha_\|=0.2,\alpha_\bot=0.2,\alpha'_\|=0.2,$ and
  $\alpha'_\bot=-0.1$,
  respectively, as it is illustrated in (c).
  In (b) each sequence contains two HCPs having
  kick strengths $\alpha_\|=0.2$ and $\alpha_\bot=0.2$,
  respectively, as it is illustrated in (d).
  Parameters of the QR, relaxation and temporal profile of the HCPs are as in
  Fig.~\ref{Fig:revivals_PRB}, except for
  $T=4$~K here.}
  \label{Fig:magnetic_sequences2}
\end{figure*}

Another possibility is to use only the generating pulse sequences
(see Fig.~\ref{Fig:magnetic_sequences2}d). Each pulse sequence
compensates for the ongoing current relaxation during the time
$\tau_{\rm p-p}$. Again, an exact compensation, leading to a
periodic magnetization dynamics, is achieved after the preceding
equilibration stage. The result is shown in
Fig.~\ref{Fig:magnetic_sequences2}b. If $\tau_{\rm p-p}$ is small
enough in comparison with the charge current relaxation time
scale, a characteristic sawtooth periodic time dependence of the
QR magnetization is observed.


\subsubsection{Influence of the magnetic flux on the generated charge polarization and
currents}\label{Sec:QR_flux}
The effect of an external static magnetic flux $\Phi$ on the
charge polarization and currents generated by HCPs in QRs was
studied in Ref.~\cite{Moskalenko_EPL2007}. In such a case
the expression for the single-particle energies $\varepsilon_m$ in
the free-electron Hamiltonian of a one-dimensional QR
\eqref{Eq:H0_carr} changes to
\begin{equation}\label{Eq:energies_ring_flux}
    \varepsilon_m=\frac{\hslash^{2}}{2m^{\ast}r_{0}^{2}}\left(m+\tilde{\Phi}\right)^2\;,
\end{equation}
where
\begin{equation}\label{Eq:normalized_flux}
    \tilde{\Phi}=\Phi/\Phi_0
\end{equation}
is the normalized external static magnetic flux with
$\Phi_0=-hc/e$ being the flux quantum. Then, assuming that the QR
is at the thermal equilibrium at $t<0$, a single HCP polarized
along the $x$-axis and applied to the QR at the time moment $t=0$
induces the charge dipole moment
\begin{equation}\label{Eq:mu_general_flux}
    \mu_x(t)=\Theta(t)er_0\alpha h_1(t)\sin\!\left(\frac{2\pi
    t}{t_{p}}\right)
    e^{-t/\tau_{\rm rel}}
            \sum_m f_m^0\cos\!\left[\frac{4\pi t}{t_{p}}(m+\tilde{\Phi})\right]
\end{equation}
instead  of Eq.~\eqref{Eq:mu_general}, which is valid for
$\Phi=0$. From Eqs.~\eqref{Eq:energies_ring_flux} and
\eqref{Eq:mu_general_flux} it is clear that $\mu_x(t)$ is periodic
in $\tilde{\Phi}$ with the period being equal to 1. In the case of
zero temperature $T=0$ and $N$ electrons in the QR
Eq.~\eqref{Eq:mu_general_flux} leads to \cite{Moskalenko_EPL2007}
\begin{equation}\label{Eq:mu_general_flux_T0}
    \mu_x(t)=\Theta(t)er_0\alpha h_1(t)s(t)e^{-t/\tau_{\rm rel}},
\end{equation}
where
  \begin{align}
    s(t)\!=&2\cos\!\left[\frac{4\pi t}{t_p}\left(\left|\tilde{\Phi}\right|\!-\!\frac{1}{2}\right)\right]
    \sin\!\left[\frac{N\pi t}{t_p}\right],\hspace{0.1cm} &N\!=0\ ({\rm
    mod}\; 4);\label{Eq:s_t_0}\\
    s(t)\!=&\cos\!\left[\frac{4\pi t}{t_p}\tilde{\Phi}\right]
    \!\sin\!\left[\frac{(N+1)\pi t}{t_p}\right] & \nonumber\\
         &+\cos\!\left[\frac{4\pi t}{t_p}\left(\left|\tilde{\Phi}\right|\!-\!\frac{1}{2}\right)\right]
    \!\sin\!\left[\frac{(N-1)\pi t}{t_p}\right],\hspace{0.1cm} &N\!=1\ ({\rm
    mod}\; 4); \label{Eq:s_t_1}\\
    s(t)\!=&2\cos\!\left[\frac{4\pi t}{t_p}\tilde{\Phi}\right]
    \sin\!\left[\frac{N\pi t}{t_p}\right],\hspace{0.1cm} &N\!=2\ ({\rm
    mod}\; 4); \label{Eq:s_t_2}\\
    s(t)\!=&\cos\!\left[\frac{4\pi t}{t_p}\left(\left|\tilde{\Phi}\right|\!-\!\frac{1}{2}\right)\right]
    \!\sin\!\left[\frac{(N+1)\pi t}{t_p}\right] & \nonumber \\
         &+\cos\!\left[\frac{4\pi t}{t_p}\tilde{\Phi}\right]
    \!\sin\!\left[\frac{(N-1)\pi t}{t_p}\right],\hspace{0.1cm} &N\!=3\ ({\rm
    mod}\; 4); \label{Eq:s_t_3}
  \end{align}
for  $\tilde{\Phi}\in[-1/2,1/2]$. If $\tilde{\Phi}$ is outside of
this range when $s(t;\tilde{\Phi})$ should be determined from the
periodicity condition $s(t;\tilde{\Phi}+1)$=$s(t;\tilde{\Phi})$.
In the absence of the magnetic flux
Eqs.~\eqref{Eq:mu_general_flux_T0}--\eqref{Eq:s_t_3} simplify to
the result obtained in Ref.~\cite{Alex_PRB2004}.

Variations of the dipole moment dynamics when sweeping the
magnetic flux are illustrated in Fig.~\ref{Fig:figure1_EPL2007}
for different values of the HCP kick strength $\alpha$. In the case of
small  $\alpha$ we observe a beating behavior.
The beating period  can be
tuned by varying the value of the magnetic flux or the electron
number from $t_p/2$ to infinity. Higher values of $\alpha$ lead
to more complicated time patterns of the dipole moment dynamics,
determined by the higher anharmonicity of the function $h_1(t)$ in
this case. Very high kick strengths give rise to  collapses and
revivals, due to the large number of states affected by the
excitation.

\begin{figure*}[t]
\includegraphics[width=12.0cm]{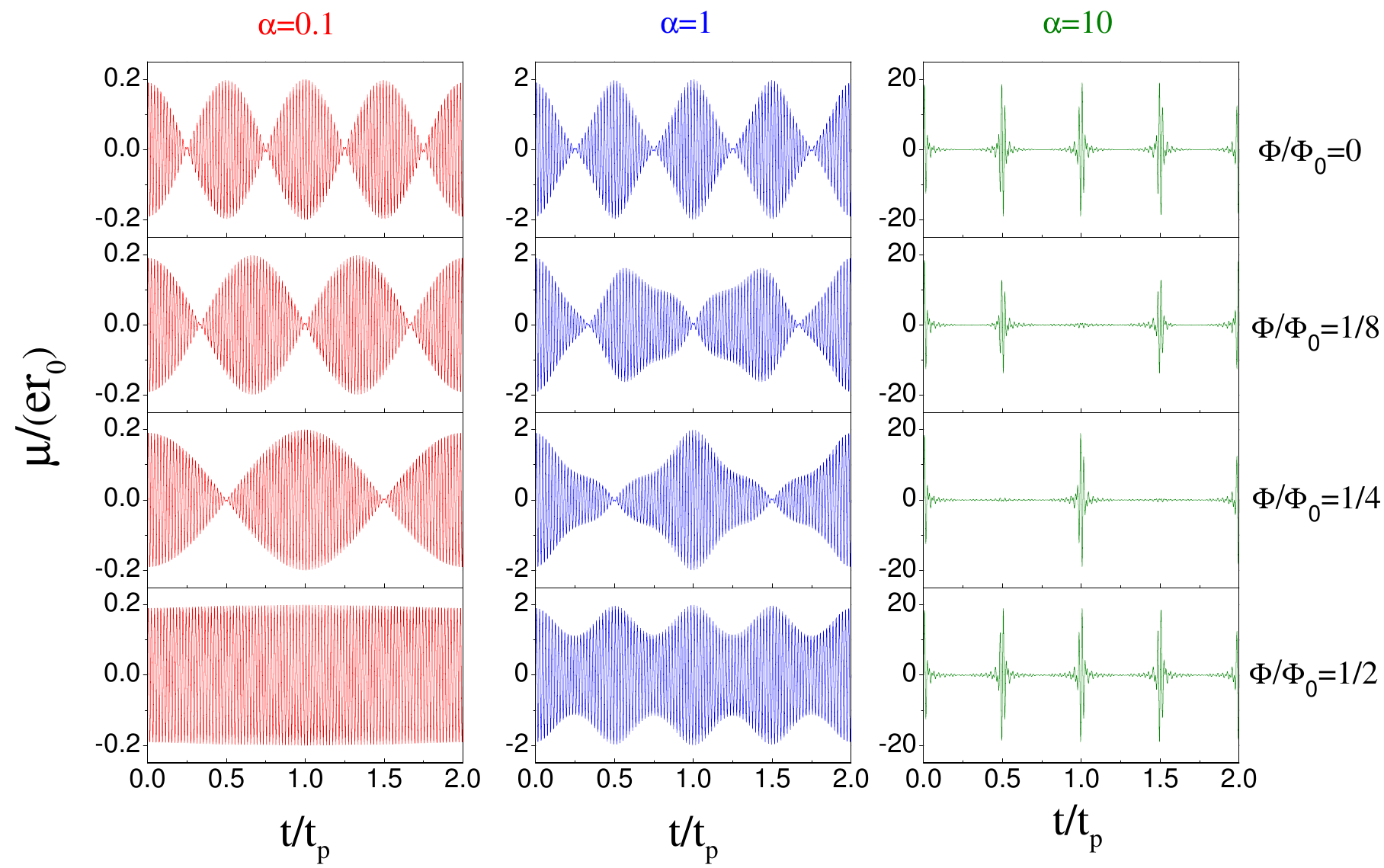}
\caption{Dynamics of the charge dipole moment $\mu(t)$ generated
in the QR by a single HCP, calculated from
Eq.~\eqref{Eq:mu_general_flux_T0} for various values of
the magnetic flux $\Phi$ and HCP kick strength $\alpha$. $t_{p}$
is given
   by Eq.~\eqref{Eq:t_p}. We assume $N=100$, $T=0$~K, and $\tau_{\rm rel}\gg t_{p}$
   (adapted from Ref.~\cite{Moskalenko_EPL2007}).  } \label{Fig:figure1_EPL2007}
\end{figure*}

In the presence of the static magnetic flux the charge current
flowing in the QR is given by
\begin{equation}\label{Eq:current_general_flux}
I(t)=I_{0}\sum_{m} (m+\tilde{\Phi})\rho_{m\!,m}(t)\;
\end{equation}
as compared to  Eq.~\eqref{Eq:current_general}. Then the current
generation mechanism described in
Section~\ref{Sec:current_generation} leads to the following
expression for the total charge current
\begin{equation}\label{Eq:itot}
    I(t)=I_{\rm pers}+I_{\rm dyn}(t),
\end{equation}
where
\begin{equation}\label{Eq:ipers}
    I_{\rm pers}=I_{0}\sum_{m}\left(m+\tilde{\Phi}\right)f_m^0
\end{equation}
is a static (persistent current) component induced by the magnetic
flux \cite{buttiker,Cheung1988,mailly,Rabaud2001}, which was
explicitly calculated in Ref.~\cite{Loss1991}, and $I_{\rm
dyn}(t)$ is a dynamical component, which is given by
Eq.~\eqref{Eq:generated_current_HCPs}. The static component is
determined by the equilibrium distribution function $f_m^0$ and
the magnetic flux $\Phi$. It remains unaffected by the application
of the HCPs. The dynamical component, induced by the HCPs, is a
function of the magnetic flux through the corresponding dependence
of the dipole moment $\mu(\tau)$, dictated by
Eqs.~\eqref{Eq:mu_general_flux}-\eqref{Eq:s_t_3}.

\begin{figure*}[t]
   \centering
  \includegraphics[width=14cm]{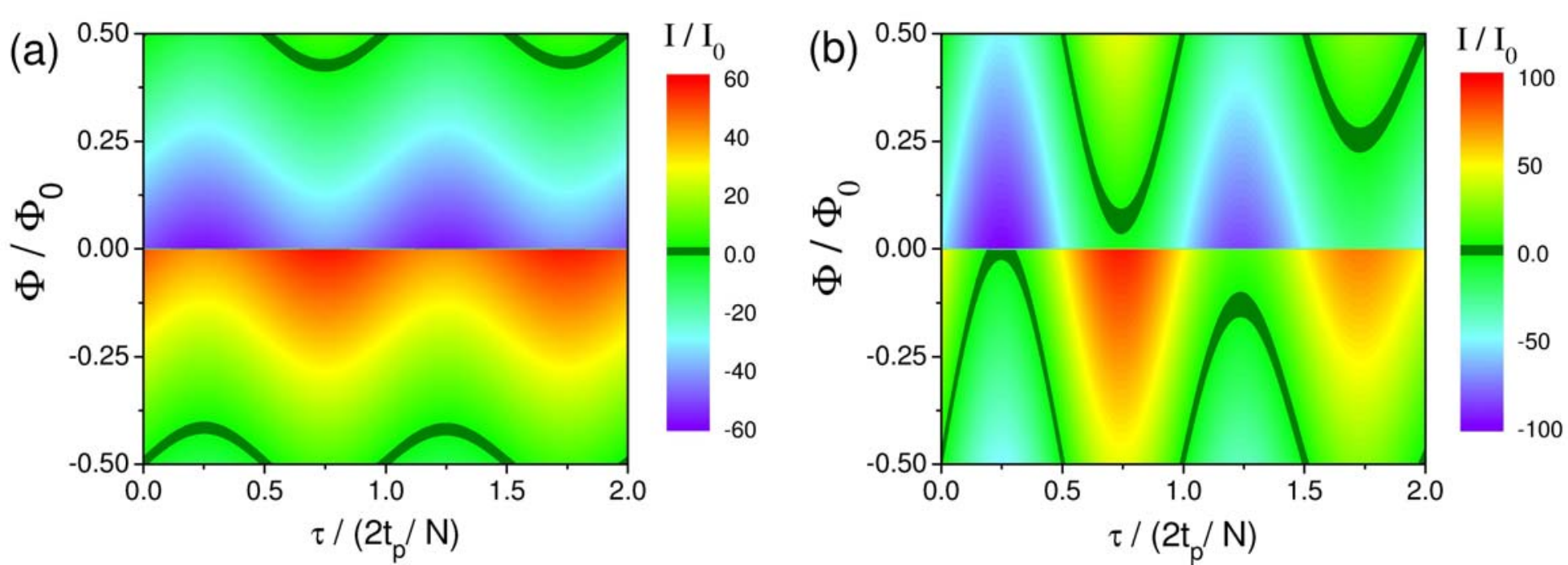}
   \caption{
   Total charge current $I$, flowing in the QR just after the application of the generating HCP pair,
   is shown as a function of   the magnetic flux $\Phi$ and the time
   delay $\tau$ between the HCPs for different values of their kick strengths:
   (a) $\alpha=\alpha_\bot=2$ and (b) $\alpha=\alpha_\bot=5$ (adapted from Ref.~\cite{Moskalenko_EPL2007}).
   \label{Fig:figure2_EPL2007}}
\end{figure*}

Figure \ref{Fig:figure2_EPL2007} shows the dependence of the
charge current flowing in the QR shortly after the application of
the generating HCP pair on the time delay $\tau$ between the HCPs
and the magnetic flux $\Phi$. We observe an oscillatory behavior with
respect to both $\tau$ and $\Phi$. An interesting feature observed
in Fig.~\ref{Fig:figure2_EPL2007} is that the current, which flows
in the QR at the thermal equilibrium with nonzero magnetic flux,
can be temporarily stopped by an appropriate tuning of the HCP
strengths and their mutual delay $\tau$. The reoccurrence of the
persistent current is then determined by the following relaxation
of the dynamical component of the current. The total current can
be eliminated again if a further HCP pair is applied. A train of
HCP, as described in Section~\ref{Sec:periodic_magnetic_pulses}, can
be used to induce a desired periodic temporal behavior of the current.

\subsection{Dynamics of the charge and valley polarization and currents in graphene rings}
Graphene QRs represent a particularly interesting type of QRs. One key
property of the charge carriers in bulk graphene is their
massless quasi-relativistic energy dispersion leading to a number
of fascinating phenomena
\cite{Castro_Neto2009,Novoselov_Nature2005,Geim2007,Beenakker_RMP2008,Morozov2008,Mikhailov2008}.
Another one is that they offer an additional degree of freedom,
provided by the valley quantum number $\varkappa=\pm 1$, which is
related to the two nonequivalent positions of the $K$ points in
the Brillouin zone. Due to the suppressed intervalley scattering
the control of the valley degree of freedom allows for
valley-based optoelectronic applications (\textit{valleytronics})
\cite{Rycerz2007,Xiao2007,Wang2008}, wherein the generation of
pure valley currents is a fundamental goal \cite{Jiang2013}.
Graphene nanostructures driven by ultrashort light pulses may be
utilized for the development of novel ultrafast valleytronic
devices. The fabrication of graphene QRs was already demonstrated
and the Aharonov-Bohm effect was studied in them \cite{Russo2008}.

Graphene QRs driven by HCPs were studied in
Ref.~\cite{Moskalenko_PRB2009}. An approach for describing the
single-particle states in graphene QRs was
proposed in Ref.~\cite{Recher2007} (see also
Ref.~\cite{Zarenia2010}). Let us  consider a graphene QR of
radius $r_0$ and width $d$ (cf. Fig.~\ref{Fig:ring}). The ring is
threaded by a static magnetic flux $\Phi$. Properties of the free
charge carriers inside the QR are governed by the effective
Dirac-Weyl Hamiltonian $H_0=v_{_\mathrm{F}}\boldsymbol{\sigma}\cdot\mathbf{P}$
\cite{Novoselov_Nature2005,Geim2007,Beenakker_RMP2008}, where
$\mathbf{P}=\mathbf{p}-\frac{e}{c}\mathbf{A}_{\rm stat}$ is the
momentum operator, modified by the vector potential
$\mathbf{A}_{\rm stat}$ of the static magnetic field, and
$v_{_\mathrm{F}}\approx10^6$~m/s is the Fermi velocity in bulk
graphene. $\boldsymbol{\sigma}=(\sigma_x,\sigma_y)$, where
$\sigma_x$ and $\sigma_y$ are the corresponding Pauli matrices.
They are built on the basis of the pseudospin eigenstates corresponding to the different sublattices of the
graphene lattice. The boundary of the ring, causing the quantum
confinement of the carriers, is modeled by a potential $\varkappa
U(\mathbf{r})\sigma_z$ \cite{Berry1987,Recher2007,Akhmerov2008},
where $U(\mathbf{r})=0$ inside the QR and $U(\mathbf{r})=U_0$
outside of it. Thus the Dirac fermions acquire a mass outside of
the QR \cite{Giovannetti2007,Hunt2013}. Such a valley-dependent
potential  can originate from the interaction of the graphene
layer with a nanostructured substrate
\cite{Zhou2008,Rotenberg_2008,Giovannetti2007,Hunt2013}.

The single-particle energies of the charge carriers in the
graphene QR are degenerate in spin ($\pm 1/2$) and  in the limit
$d\ll r_0$ (see Fig.~\ref{Fig:ring}) can be approximately found as
\cite{Moskalenko_PRB2009}
\begin{equation}\label{Eq:energies_graphene}
    E_{nm}^{s\varkappa}=s\varepsilon_n+s\lambda_n\left[m+\tilde{\Phi}^{s\varkappa n}_{\rm eff}\right]^2
                        -\frac{s\lambda_n}{4\pi^2(n+1/2)^2}\;,
\end{equation}
with
\begin{eqnarray}
 \varepsilon_n&=&(1-\gamma_0)\frac{\hslash v_{_\mathrm{F}}}{d}(n+1/2)\;,\label{Eq:graphene_eps_n}\\
 \lambda_n&=&(1+\gamma_0)\frac{\hslash
 v_{_\mathrm{F}}}{d}\left(\frac{d}{r_0}\right)^2\frac{1}{\pi(2n+1)}\;.\label{Eq:lambda_n}
\end{eqnarray}
Here we have $s=1$ ($s=-1$) for electrons (holes); $n=0,1,2,...$
is the radial quantum number;
$m=\pm\frac{1}{2},\pm\frac{3}{2},...$ numerates the angular
states. The quantity $\tilde{\Phi}^{s\varkappa n}_{\rm eff}$ in
Eq.~\eqref{Eq:energies_graphene} is given by
\begin{equation}\label{Eq:Phi_tilde}
    \tilde{\Phi}^{s\varkappa n}_{\rm
eff}=\tilde{\Phi}-\frac{1}{2}\frac{s\varkappa}{(n+1/2)\pi}\;.
\end{equation}
For fixed $s,\varkappa$, and $n$ it plays a role of an effective
normalized magnetic flux. Notice that with respect to
$\tilde{\Phi}$ [as given by Eq.~\eqref{Eq:normalized_flux}] it contains a
valley-dependent shift. The dimensionless factor $\gamma_0$ in
Eqs.~\eqref{Eq:graphene_eps_n} and \eqref{Eq:lambda_n} is determined by
\begin{equation}\label{Eq:gamma_graphene}
    \gamma_0= \frac{\hslash v_{_\mathrm{F}}}{d U_0}\;
\end{equation}
and we assumed $\gamma_0\ll 1$. In the case of a graphene QR with
a typical width $d=0.1~\mu$m this implies $U_0\gg7$~meV that is
reasonable, e.g., for a hexagonal boron nitride substrate
\cite{Giovannetti2007,Hunt2013}.

For brevity, we limit our consideration to the
lowest radial channel of the positive energy branch, i.e. $n=0$
and $s=1$ \footnote{Experimentally, the filling of quantized
energy states in unexcited graphene rings can be controlled by the
gate voltage.}. Let us find the magnetic dipole moment $\mu(t)$
induced in the graphene QR by the application of a single HCP,
polarized along the $x$-axis, at $t=0$. Performing analogous steps
as in Sections \ref{Sec:polarization_dynamics} and \ref{Sec:QR_flux},
we get
\begin{eqnarray}
    \mu(t)&=&\sum_\varkappa \mu_{\varkappa}(t)\;,\label{Eq:mu_sum}\\
    \mu_\varkappa(t)&=&er_0\alpha
    h_1(t)\sin\left(\frac{2\pi t}{t_p}\right)
    \sum_m f^\varkappa_{m}\cos\left[\frac{4\pi t}{t_p}(m+\Phi^{\varkappa}_{\rm
    eff})\right],
    \label{Eq:mu_tau}
\end{eqnarray}
where $\Phi^{\varkappa}_{\rm eff}\equiv
\tilde{\Phi}^{s=1,\varkappa, n=0}_{\rm eff}$,
\begin{equation}\label{Eq:t_p_graphene}
    t_p=2\pi\frac{\hslash}{\lambda_0}\;,
\end{equation}
$h_1(t)$ is given by Eq.~\eqref{Eq:h_function}, however with $t_p$
from Eq.~\eqref{Eq:t_p_graphene}, and $\alpha$ is the HCP kick
strength. $f^\varkappa_{m}$ is the equilibrium distribution
function for the quantum states numbered by $\varkappa$ and $m$.
Here we have neglected the relaxation processes. The dipole moment
\eqref{Eq:mu_sum} as well as its valley-resolved parts
\eqref{Eq:mu_tau} are periodic in $\tilde{\Phi}$. The period
equals to 1.

Figure \ref{Fig:graphene_mu} shows the dynamics of the generated
dipole moment as a function of the magnetic flux for $N=8$
electrons and two different HCP kick strengths. Prior to the
excitation the electrons in the QR are in the equilibrium at zero
temperature $T=0$. The complex oscillation structured depending
on $t$ and $\tilde{\Phi}$ (Fig.~\ref{Fig:graphene_mu}a) is a consequence of the interplay
between the constituents of
Eq.~\eqref{Eq:mu_tau}. Firstly, the sine and cosine factors oscillate with $t$ and $\tilde{\Phi}$. Secondly, the discontinuous jumps in the equilibrium
distribution  $f^\varkappa_{m}$ as a function of
$\tilde{\Phi}$ occur each time when one of the energy
levels [cf.
Eq.~\eqref{Eq:energies_graphene}] crosses the Fermi level
\cite{Moskalenko_PRB2009}.
 When the HCP kick strength is increased
we observe a higher number of oscillation nodes as a function of
$t$ (see Fig.~\ref{Fig:graphene_mu}b), owed to the
pronounced anharmonicity of the function $h_1(t)$ entering
Eq.~\eqref{Eq:mu_tau}.

\begin{figure*}[t]
  \includegraphics[width=15cm]{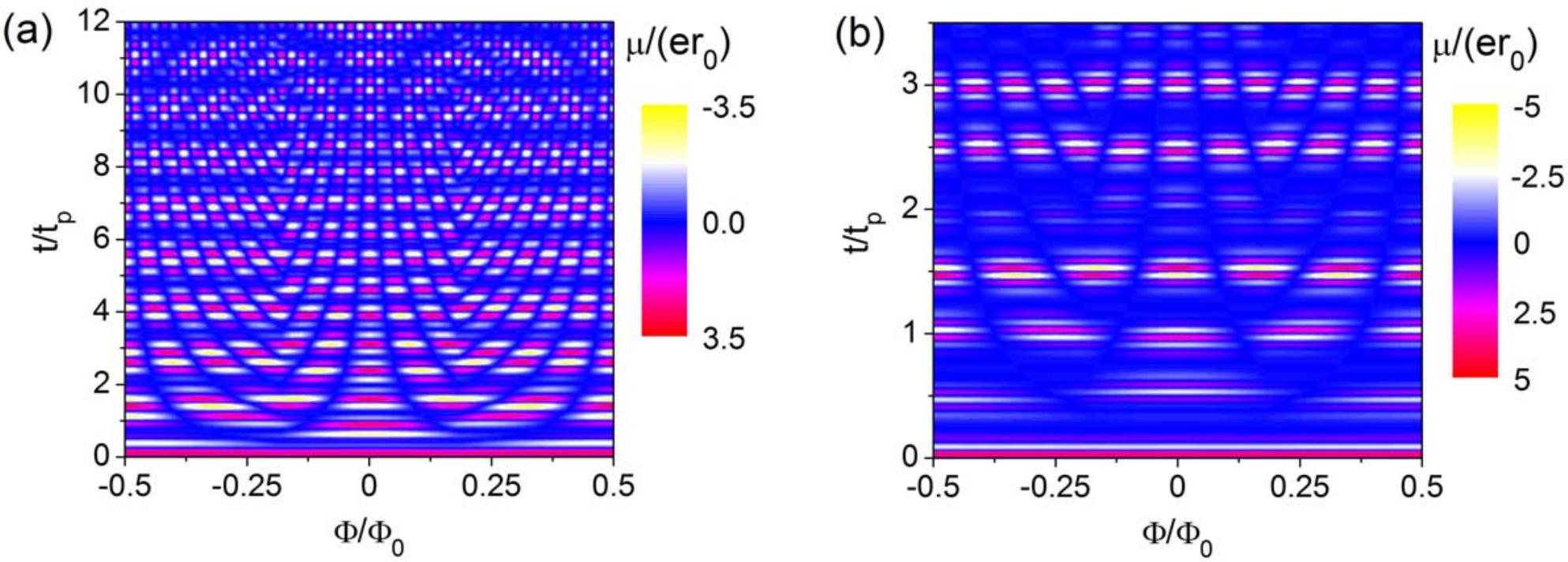}
  \caption{Post-pulse polarization dynamics induced in a single-channel graphene QR with $N=8$ electrons at $T=0$
  by a HCP having a kick strength: (a) $\alpha=1$ and (b) $\alpha=5$.}
  \label{Fig:graphene_mu}
\end{figure*}

Although the effective time-reversal symmetry is broken by the
boundary conditions \cite{Recher2007} and therefore the energy
levels are not invariant under the transformation
$\varkappa\rightarrow -\varkappa$ \cite{Recher2007}, we  deduce
$\mu_{\varkappa=+1}(t)=\mu_{\varkappa=-1}(t)$ for $\tilde{\Phi}=0$
because of their invariance under the transformation
$\varkappa\rightarrow -\varkappa$, $m \rightarrow -m$. Only the
application of a stationary magnetic flux $\tilde{\Phi}\neq 0$
lifts also this degeneracy between the quantum states, leading to
$\mu_{\varkappa=+1}(t)\neq\mu_{\varkappa=-1}(t)$. This property
can be then utilized for the generation of valley-polarized
currents in graphene QRs.

The total charge current flowing in the light-driven ring consists
of the persistent/static (equilibrium) $I_{\rm pers}$ and
dynamical (nonequilibrium) $I_{\rm dyn}(t)$ components [see
Eq.~\eqref{Eq:itot}], where the persistent charge current is given
by \cite{Recher2007,Moskalenko_PRB2009}
\begin{eqnarray}
    I_{\rm pers}&=&\sum_\varkappa I_{\rm pers,\varkappa}\;,\label{Eq:I_sum}\\
    I_{\rm pers,\varkappa}&=&I_0\sum_m f^\varkappa_{m} \left(m+\tilde{\Phi}^\varkappa_{\rm eff}\right)
    \label{Eq:I_tau}
\end{eqnarray}
[cf. Eq.~\eqref{Eq:ipers}]. Here $I_{\rm pers,\varkappa}$ denotes
the contributions from the two different valleys and $I_0$ is
given by
\begin{equation}\label{Eq:I0_graphene}
   I_0=\frac{1}{\pi^2}\frac{ev_{_\mathrm{F}}d}{r_0^2}.
\end{equation}
In contrast to the case of single-channel semiconductor QRs [see
Eq.~\eqref{Eq:I0}], here $I_0$ depends not only on the ring radius
$r_0$ but also on its width $d$. For a ring with
$r_0=1~\mu\mbox{m}$ and $d=100~\mbox{nm}$ ($r_0=425~\mbox{nm}$ and
$d=150~\mbox{nm}$) we have $I_0=-0.16~\mbox{nA}$
($I_0=-1~\mbox{nA}$). Utilizing the same mechanism for the current
generation as in the case of the semiconductor QRs (see
Fig.~\ref{Fig:current_generation_mechanism}) we find that the
dynamical charge current component $I_{\rm dyn}(t)$ is given by
Eq.~\eqref{Eq:generated_current_HCPs}, where in the considered
case of the graphene QRs $I_0$ should be calculated from
Eq.~\eqref{Eq:I0_graphene} and the dipole moment $\mu(\tau)$ at
the time moment $t=\tau$ should be determined from
Eqs.~\eqref{Eq:mu_sum} and \eqref{Eq:mu_tau}. As mentioned above,
for a nonzero magnetic flux the polarization of the pulse-driven
graphene QR becomes valley-dependent. We introduce then the valley
current in the ring as $I^{\rm
v}=(I_{\varkappa=+1}-I_{\varkappa=-1})/e$ and calculate that its
dynamical component induced by the same generating HCP pair as $I_{\rm dyn}(t)$ is
given by
\begin{equation}\label{Eq:I_valley_dyn}
   I^{\rm v}_{\rm
   dyn}(t)=\Theta(t-\tau)\frac{I_0}{e}\alpha_\bot\frac{\mu_{\varkappa=+1}(\tau)-\mu_{\varkappa=-1}(\tau)}{er_0}\;.
\end{equation}

\begin{figure*}[t]
  \includegraphics[width=8cm]{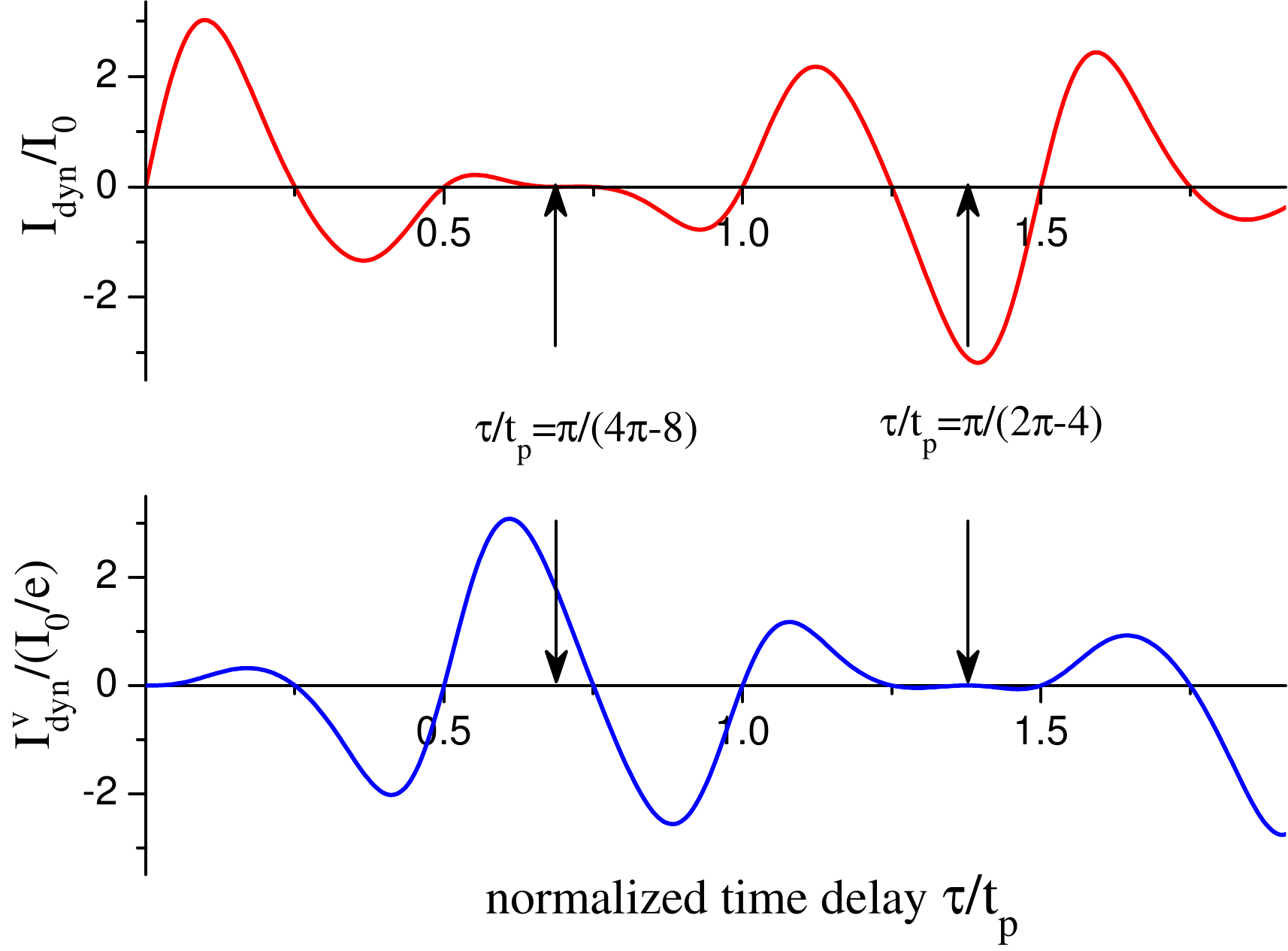}
  \caption{The upper panel shows the dependence of the nonequilibrium charge current
  generated in the graphene QR by application of two mutually
  perpendicular HCPs, delayed with respect to each other by
  $\tau$ and having kick strengths $\alpha=\alpha_\bot=1$. The lower panel shows the nonequilibrium valley current
  generated by the same pulse sequence. The presented
  case corresponds to  $N=8$ and $\tilde{\Phi}=1/\pi$. The arrows
  indicate time moments when either the generated charge current or
  the generated valley current vanishes (adapted from
Ref.~\cite{Moskalenko_PRB2009}).}
   \label{Fig:graphene_currents}
\end{figure*}

An example of the dependencies of the induced nonequilibrium charge and valley
currents on the time delay $\tau$ between the HCPs of
the generating pulse sequence is presented in
Fig.~\ref{Fig:graphene_currents}. Comparing the upper and the lower panels of
this figure, one can see that there are delay times $\tau$ such
that either the charge current is present and the valley current
vanishes or vice versa. Especially, the second case is of interest
because it provides a possibility for a swift generation of pure
(nonequilibrium) valley currents in graphene QRs, paving the way
towards ultrafast valleytronics.

\section{Control of the spin dynamics in semiconductor
nanostructures}\label{Chapter:Zhengang}

Light-matter interaction effects  in
conventional semiconducting nanostructures have been of great interest in view of the  diverse applications for
photo-electronic devices such as radiation-controlled field-effect
transistors, light-emitting diodes, photodiodes, and single-mode
optical fiber
\cite{Duan2001,Wang2001,Freitag2003,Ohno2004,Marcus2006,Davanco2011}.
The meanwhile well-established field spintronics utilizes in addition the spin degree of freedom
exploring  its potential  for new
spin-based devices \cite[and references therein]{Zutic2004}.
A variety of materials were under study, including  dilute magnetic semiconductors,
topological insulators, heterojunctions with ferromagnetic
materials, semiconductors, graphene, oxide-based heterostructures and others.
 Also in many
conventional semiconductors, such as GaAs and InSb, the spin
degree of freedom is exploitable by means of the  strong spin-orbit
interaction (SOI). The SOI in bulk semiconductors described by the
Luttinger Hamiltonian was predicted to be the key for generating
dissipationless quantum spin current at room temperature when
applying an electric field in hole-doped semiconductors such as
Si, Ge, and GaAs \cite{Murakami2003}. A similar effect was
predicted in a two-dimensional electron gas with Rashba SOI due to
a broken inversion symmetry at the interface \cite{Sinova2004}.
The SOI is also central to the spin field-effect transistor
(SFET) \cite{Datta1990} and spin-interference device
\cite{Nitta1999}. A large conductance modulation effect was
predicted in both devices giving impetus to a vast number of studies
on the implications and possible applications.
Here we discuss  semiconductor QRs with SOI and SFET
driven by ultrashort  electromagnetic pulses.
In addition we describe the predicted pulse-induced spatial separation of  spin states and coherent control
of pure spin currents in double quantum dots. Possible applications
are in the areas of the ultrafast spintronics and spin-based
quantum computing \cite{Allwood2002,Loss1998,Kane1998,Zutic2004}.

\subsection{Spin dynamics in semiconductor quantum rings triggered
by HCPs}

\subsubsection{Rashba spin-orbit interaction}

The Rashba spin-orbit interaction, SOI, appears in material structures as a consequence of
the inversion symmetry breaking. At a heterojunction of two
semiconductors this symmetry breaking can be effectively viewed as being induced by a normally oriented electric
field due to band bending at the interface. Selecting the $z$-axis
perpendicular to the interface, the Rashba SOI contribution to the
Hamiltonian has the form
\cite{Rashba1960,Bychkov1984,Winkler2003}:
\begin{equation}
H_{\mathrm{R}}=\alpha_{\rm
R}[\boldsymbol{\sigma}\times\mathbf{k}]\cdot\mathbf{e}_{z}\;,
\label{soi1}
\end{equation}
where $\alpha_{\rm R}$ is a parameter reflecting the strength of
the SOI, $\mathbf{e}_{z}$ is a unit vector perpendicular to the
surface, $\mathbf{k}$ is the wave vector of the charge carrier,
and $\boldsymbol{\sigma}$ is the vector of the Pauli matrices.
Writing the interface-induced effective electric field as
$\mathbf{E}=E_{z}\mathbf{e}_{z}$, we have $H_{\mathrm{R}}
=\frac{\alpha_{\rm R}}{\hslash
E_{z}}\boldsymbol{\sigma}\cdot\left[\mathbf{p}\times\mathbf{E}\right],$
where $\mathbf{p}=\hslash\mathbf{k}$ is the carrier momentum
\footnote{Please, note that there is a minus sign difference in the
definition of the Rashba coefficient in respect to Eq.~(1) of
Ref.~\cite{Meijer2002} and Eq.~(4) of
Ref.~\cite{Datta1990}.}.

\subsubsection{Hamiltonian of a light-driven 1D quantum ring with Rashba
effect}\label{Eq:Hamiltonian_QR_SOI_driven} Nitta \textit{et al.}
\cite{Nitta1999} proposed a spin-interference device, namely an
Aharonov-Bohm ring with a uniform  Rashba SOI, which causes a
phase difference between the electron waves with opposite spins traveling in the
clockwise and anticlockwise direction. The SOI strength can be
controlled by a gate covering the whole ring. A large conductance
modulation effect can be expected due to the spin interference.
The tunable spin interference in a ring geometry with the SOI thus
motivated many studies in the recent past
\cite{Meijer2002,Splettstoesser2003,Molnar2004,Nikolic2005,Liu2006,Foldi2005,Foldi2009,Foldi2010,Sheng2006,Sheng2008,Frustaglia2004,Frustaglia2004_2,zgzhu2008,zgzhu2009,zgzhu2010}.
In presence of an external electromagnetic field and a static
magnetic field, directed perpendicular to the plane of the QR
($\mathbf{B}_{\rm stat}=B_{\rm stat}\mathbf{e}_z$), the
single-particle Hamiltonian in the minimal coupling scheme for an
electron confined to the QR reads
\begin{equation}
H=\frac{\mathbf{\Pi}^{2}}{2m^{*}}+U(\mathbf{r})+e\phi+\frac{\alpha_{\rm
R}}{\hslash} (\boldsymbol{\sigma}\times\mathbf{\Pi})_{z}+
\frac{1}{2}g\mathbf{\mu}_{B}\boldsymbol{\sigma}\cdot(\mathbf{B}_{\rm
stat}+\mathbf{B})\;. \label{h1}
\end{equation}
Here the momentum operator $\mathbf{\Pi}$ can be represented as
$\mathbf{\Pi}=\mathbf{P}-\frac{e}{c}\mathbf{A}$, where
$\mathbf{A}$ is the vector potential of the electromagnetic field
and
$\mathbf{P}=-i\hslash\boldsymbol{\nabla}-\frac{e}{c}\mathbf{A}_{\rm
stat}$; $-i\hslash \boldsymbol{\nabla}$ is the canonical momentum operator and
$\mathbf{A}_{\rm stat}$  is the vector potential of the static
magnetic field. $e$ is the electron charge \footnote{In the case
of holes the charge $q=|e|$ should be used in place of the
negative electron charge $e$.}, $c$ is the speed of light, $m^*$
is the electron effective mass and $g$ is its g-factor in the QR.
$\phi$ denotes the scalar potential of the electromagnetic field
and $\mathbf{B}$ is its magnetic component. $U(\mathbf{r})$ is the
confining potential of the QR. The last term in Eq.~\eqref{h1} is
the Zeeman term describing the coupling between the electron
magnetic moment $-\frac{1}{2}g\mu_B{\boldsymbol{\sigma}}$ and the
total external magnetic field $\mathbf{B}_{\rm stat}+\mathbf{B}$.

Upon a  gauge transformation, we can recast the Hamiltonian
in the form \cite{zgzhu2008}
\begin{equation}\label{Eq:Hamiltonian_SOI_transformed}
    H' = H_0+V(t)\;,
\end{equation}
where
\begin{eqnarray}
 H_0&=&\frac{\mathbf{P}^{2}}{2m^{*}}+U(\mathbf{r})
+\frac{\alpha_{\rm R}}{\hslash}(\boldsymbol{\sigma}\times\mathbf{P})_{z}
+\frac{1}{2}g\mathbf{\mu}_{B}\boldsymbol{\sigma}\cdot\mathbf{B}_{\rm stat}\;, \label{Eq:SOI}\\
V(t) &=&
-e\mathbf{r}\cdot\mathbf{E}(t)+\frac{1}{2}g\mathbf{\mu}_{B}\boldsymbol{\sigma}\cdot\mathbf{B}(t)\;.
\label{h1t}
\end{eqnarray}
 $\mathbf{E}(t)$ and $\mathbf{B}(t)$ are the electric and
magnetic fields  of the light pulse, respectively. Using polar
coordinates and integrating out the radial dependence in the limit
$d\ll r_0$ (see Fig.~\ref{Fig:ring}), we find for  $H_0$
\cite{Meijer2002,Splettstoesser2003,Molnar2004,Foldi2005,Sheng2006,Frustaglia2004,zgzhu2008,zgzhu2009,zgzhu2010}
\begin{equation}
H_0=\frac{\hslash\omega_{0}}{2}\left[\left(-i\frac{\partial}{\partial\varphi}+\tilde{\Phi}+\frac{\omega_{R}}{2\omega_{0}}\sigma_{r}\right)^{2}-
\left(\frac{\omega_{\mathrm
R}}{2\omega_{0}}\right)^{2}+\frac{\omega_{B}}{\omega_{0}}\sigma_{z}\right].
\label{hsoi1}
\end{equation}
 $\sigma_{r}=\sigma_{x}\cos\varphi+\sigma_{y}\sin\varphi$,
$\tilde{\Phi}$ is the normalized external static magnetic flux
threading the ring \eqref{Eq:normalized_flux},
$\hslash\omega_{0}=\hslash^{2}/(m^{*}r_0^{2})$,
$\hslash\omega_{\mathrm R}=2\alpha_{\mathrm R}/r_0$, and
$\hslash\omega_{B}=g\mu_{B}B$.

The stationary eigenstates of $H_0$ are
\begin{equation}\label{Eq:stat_states_QR_spin}
  \psi_{m}^{S}(\varphi)=\frac{1}{\sqrt{2\pi}}e^{i(m+1/2)\varphi}\nu_{S}(\gamma,\varphi)\;,
\end{equation}
where $\nu_{S}(\gamma,\varphi)=( a_{S}e^{-i\varphi/2},
b_{S}e^{i\varphi/2})^{T}$ are spinors in the angle-dependent local
frame and \textit{T} means transposed. The angular quantum number
$m\in\mathbb{N}$ determines the eigenvalue $m+1/2$ of the
$z$-component of the total angular momentum. $S=+1\, (S=-1)$
stands for spin up (spin down) in the local frame. The spinor
components can be written as $a_{+1}=\cos(\gamma/2),\;
b_{+1}=\sin(\gamma/2),\; a_{-1}=-\sin(\gamma/2),$ and
$b_{-1}=\cos(\gamma/2),$ where the angle
$\gamma=\gamma(\omega_{\mathrm R}/\omega_{0},\omega_B/\omega_{0})$
gives the direction of the spin quantization axis, as illustrated
in Fig.~\ref{rfig1}. The corresponding energy spectrum of the QR
with the SOI reads
\begin{equation}\label{eigenenergy}
E_{m}^{S}=\frac{\hslash\omega_{0}}{2}\left(m+\tilde{\Phi}^{S}_{\rm
eff}\right)^{2}-\hslash\omega_{0}\frac{Q_{\mathrm R}^{2}}{8}\;,
\end{equation}
where $Q_{\mathrm R}=\omega_{\mathrm R}/\omega_{0}$, $Q_
B=\omega_B/\omega_{0}$. We have introduced an effective
spin-dependent normalized magnetic flux
\begin{equation}\label{Eq:Phi_eff_spin}
\tilde{\Phi}^{S}_{\rm eff}=\tilde{\Phi}+\frac{1-Sw}{2}
\end{equation}
with
\begin{equation}\label{eigenenergy_w}
w = \sqrt{1+Q_{\mathrm R}^{2}+4\left(Q_B^2-Q_B\right)}\;,
\end{equation}
in analogy to the energy spectrum of the graphene QRs [cf.
Eq.~\eqref{Eq:energies_graphene}]. If the static magnetic field
threading the QR is absent in the QR itself, i.e. there is no
Zeeman splitting implicating $Q_B=0$, we can write $Q_{\mathrm
R}=-\tan\gamma$ and $w =1/\cos\gamma$.

\begin{figure*}[t]
\includegraphics[width=4cm]{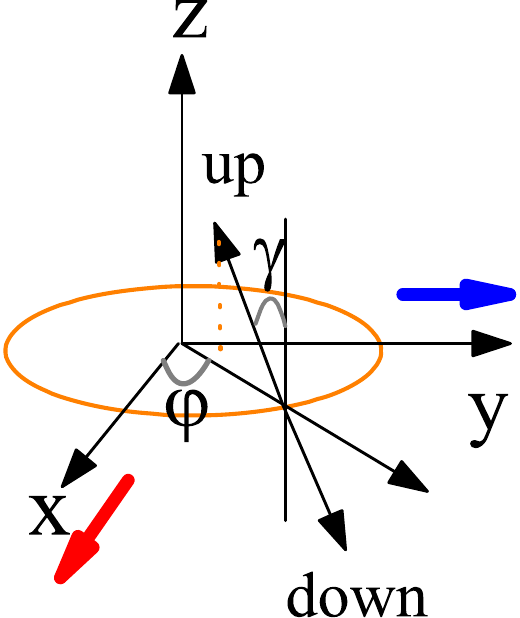}
\caption{Schematic graph of the geometry, spin configuration and
considered electric field polarization directions (indicated by
arrows) of the applied HCPs.
\label{rfig1}}
\end{figure*}

\subsubsection{Pulse-driven spin-dependent dynamics and THz emission as  indicator for spin precession}\label{Sec:QR_SOI_one_HCP}
For clarity let us focus on the time scale on which
 relaxation processes are not dominant, in which case one may equally use both
 the density-matrix approach discussed
  or  a wave function formalism to describe
the spin and charge  dynamic induced by a  HCP \footnote{In many cases spin relaxation proceeds on a much
longer time scale than charge relaxation.}.
Within the IA the action of a
single HCP polarized in the $x$-direction on a single electron in
the QR is determined by Eq.~\eqref{Eq:IA_mapping_psi}, where the
evolution operator is given by Eq.~\eqref{Eq:U_1_ring_result} and
the kick strength $\alpha$ is determined by
Eq.~\eqref{Eq:alpha_def}. The part of the interaction determined
by the magnetic field component of the HCP in Eq.~\eqref{h1t} is marginal.
In fact, the consideration of the second term in
Eq.~\eqref{h1t} within the IA leads to an additional term
$-i\beta\sigma_y$ in the exponent of the evolution operator, where
$\beta=s_1\frac{g\mathbf{\mu}_{B}B_0\tau_{\rm d}}{2\hslash}$.
$B_0$ denotes the amplitude of the magnetic field and the dimensionless factor $s_1$ is defined by Eq.~\eqref{Eq:s_n_infinity}. For a typical HCP with
electric field amplitude 1~kV/cm and duration $\tau_{\rm
d}=1$~ps, the numerical factor $\beta$ characterizing the
effective strength of the interaction does not exceed $10^{-5}$ \footnote{In general this component can be indeed very
interesting for exploring the magnetic dynamics at high fields, as demonstrated by recent experiments \cite{Kampfrath2015,Kampfrath2011}.}.
%
%
%
%
Assume that the  HCP is applied at $t=0$ \footnote{This means that the origin
of the time axis is selected so that we have $t_1=0$ (cf.
Section~\ref{Sec:HCPs})}. Let us consider an electron which occupies the
eigenstate characterized by the quantum numbers $m_{0}$ and
$S_{0}$ before the HCP application. Then its time-dependent wave
function is expressible in terms of the QR stationary
eigenstates \eqref{Eq:stat_states_QR_spin} as
\begin{equation}
\Psi_{m_{0}}^{S_{0}}(\varphi,t)=
\sum_{mS}C_{m}^{S}(m_{0},S_{0},t)e^{-iE_{m}^{S}t/\hslash}\psi_{m}^{S}(\varphi)\;,
\label{wavef1}
\end{equation}
where the expansion coefficients are given by
\begin{equation}\label{coeff}
C_{m}^{S}(m_{0},S_{0},t)=\delta_{SS_{0}}\left[\Theta(-t)\delta_{mm_{0}}
       +\Theta(t)i^{m_{0}-m}J_{m_{0}-m}(\alpha)\right].
\end{equation}
Here $\Theta(t)$ is the Heaviside step function and $J_n(x)$
denotes the Bessel function of the order $n$. The time-dependent
energy
$E_{m}^{S}(t)=\langle\Psi_{m}^{S}(t)|H|\Psi_{m}^{S}(t)\rangle$ of
the electron being in the state \eqref{wavef1}  is readily inferred as
\cite{Alex_PRB2004,zgzhu2008}
\begin{equation}\label{energy2}
E_{m}^{S}(t)=E_{m}^{S}+\Theta(t)\frac{\hslash\omega_{0}}{2}\frac{\alpha^{2}}{2}\;,
\end{equation}
where $E_{m}^{S}$ is given by Eq.~\eqref{eigenenergy}.
Thus, effectively each electron attains an additional amount of energy
$\frac{\hslash\omega_{0}}{2}\frac{\alpha^{2}}{2}$ from the HCP. This quantity serves also as the upper limit of energy
gain, which is valid irrespective of the
situation that a very short  HCP might contain very high frequency components.


The  charge dipole moment $\boldsymbol{\mu}(t)=\mathbf{e}_x\mu(t)$
 associated with the intraband coherence induced by the HCP
is deduced  from
Eq.~\eqref{Eq:dip_mom_ring} which can be represented as [cf.
Eqs.~\eqref{Eq:mu_sum} and \eqref{Eq:mu_tau}]:
\begin{eqnarray}
    \mu(t)&=&\sum_S \mu^S(t)\;,\label{Eq:mu_sum_S}\\
    \mu^S(t)&=& \sum_m f^S_{m} \mu_{m}^{S}(t)\;,
    \label{Eq:mu_S}
\end{eqnarray}
where $\mu^S(t)$ is the spin-resolved dipole moment and $f^S_{m}$
stands for the equilibrium distribution function. The contribution
to the dipole moment
$\mu_m^S(t)=er_0\langle\Psi_{m}^{S}(t)|\cos\varphi|\Psi_{m}^{S}(t)\rangle$
from an electron with the initial state $\psi_{m}^{S}(\varphi)$
can be found as
\begin{equation}\label{Eq:mu_Sm_res}
 \mu_m^S(t)=\Theta(t)er_0\alpha
    h_1(t)\sin\left(\frac{2\pi t}{t_p}\right)
    \cos\left[\frac{4\pi t}{t_p}(m+\tilde{\Phi}^{S}_{\rm eff})\right],
\end{equation}
where $h_1(t)$ is determined by Eq.~\eqref{Eq:h_function} and
$t_p$ is given by Eq.~\eqref{Eq:t_p}, i.e. $t_p=4\pi/\omega_0$.

In Fig.~\ref{rfig2}, the time dependence of the dipole moment  is
shown for different spin-orbit angles $\gamma$ and  HCP kick
strengths $\alpha$ in absence of the static magnetic field
($Q_B=0,\Phi=0$), when we have $\mu^{S=+1}(t)=\mu^{S=-1}(t)$. The
QR contains  $N=100$  {electrons (i.e. $N=0$ mod 4, whereas the value of $N$ modulo 4 characterizes distinct configurations for the occupation of the states in the neighborhood of the Fermi level)} being initially in
equilibrium at zero temperature. The quick oscillations have
period $T_{\rm cl}$ and frequency $\omega_{_\mathrm{F}}=2\pi/T_{\rm
cl}$, with $T_{\rm cl}$ given by
Eq.~\eqref{Eq:fast_oscillations_time}. In the case $N\gg1$, which
is valid here, we have $T_{\rm cl}= 2t_p/N$ and
$\omega_\mathrm{F}= N\omega_0/4$.

\begin{figure*}[t]
\includegraphics[width=11cm]{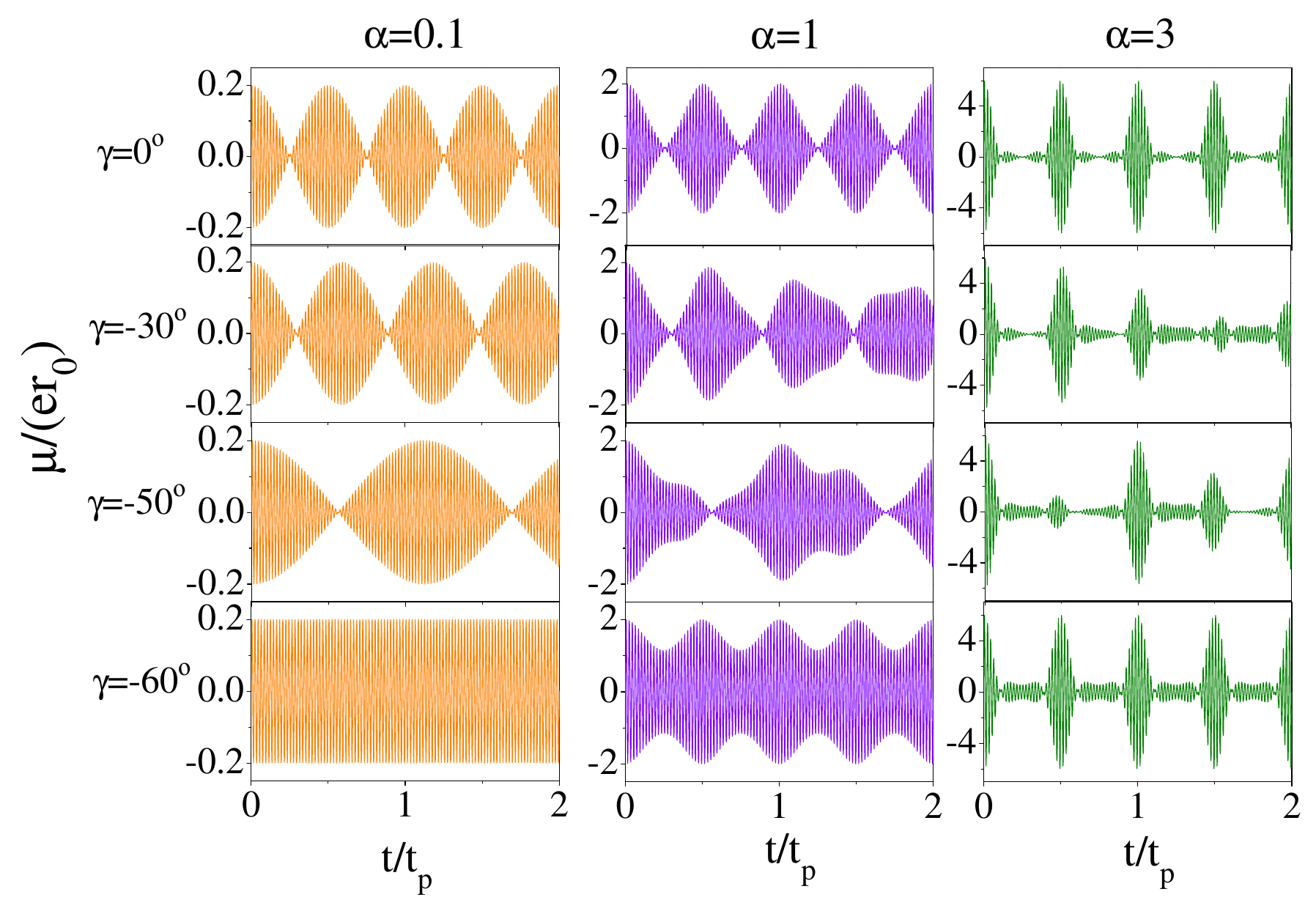}
\caption{Time dependence  of the HCP-induced dipole moment
$\mu(t)$ at different values of $\gamma$ and $\alpha$. There are
$N=100$ electrons in the QR at $T=0$, $Q_B=0$ and $\Phi=0$
(adapted from Ref.~\cite{zgzhu2008}). \label{rfig2}}
\end{figure*}

At a low excitation strength, the beating  is
determined by the difference between the frequencies corresponding
to the transitions between the involved states near the Fermi
level. Therefore, from Eq.~\eqref{eigenenergy} it is clear that
the SOI strength (quantified by $\gamma$) has a dramatic influence
on the low frequency modulation of the envelope of the dipole moment and
can lead to a variation of the corresponding low frequency in the
range from 0 to $\omega_0$. This means that the beating period can be
varied from 0 to $t_p/2$. This behavior in dependence on $\gamma$
is observed in Fig.~\ref{rfig2}. We see that $\gamma$ can be tuned
in such a way that the slow envelope oscillations vanish
completely. This happens at $\gamma=-60^\circ$. When $|\gamma|$ is
increased further the beating appears again. This is insofar
important as $\gamma$ can be modified by an external gate voltage
offering thus the possibility of engineering the emission spectrum
via an applied static electric field and opening a way for
testing experimentally these theoretical predictions.
On the other hand the THz emission (power) spectrum deduced from Fig.~\ref{rfig2}  contains evidently
components related to $\gamma$ (cf. Fig.~\ref{Fig:spectra_QR_rfig3}c). Recalling  that
 the angle
$\gamma$ sets  the direction of the spin quantization axis (cf.  Fig.~\ref{rfig1}), theoretically,
 one may pick up the spin precession by monitoring the relevant emission (which can even be changed, e.g., by
 a gate voltage that effectively changes the Rashba spin-orbit coupling constant)
 \footnote{Note, also in this spin-dependent case the emission is solely electric dipolar in character. Contributions from the magnetic dipole emission are marginal.}.

Inspecting Fig.~\ref{rfig2} we note that
with increasing $\alpha$ more states around the Fermi level are
excited giving rise to higher contributing harmonics (cf.
$\alpha=1$ and $\alpha=3$ in Fig.~\ref{rfig2}). This affects
the dipole moment dynamics in two ways. On one hand, the
corresponding anharmonicity makes the time structure more
sophisticated. On the other hand, at high enough HCP strengths, we
observe a periodic appearance of revivals. The period $T_{\rm
rev}$ is determined by Eq.~\eqref{Eq:revival_time}, so that we infer that
$T_{\rm rev}=t_p/2$. The corresponding frequency is equal to
$\omega_0$ and thus coincides with the maximum beating frequency.
From Fig.~\ref{rfig2} we see that the amplitude of the revivals is
constant if the beating frequency vanishes ($\gamma=-60^\circ$) or if
it attains its maximum value ($\gamma=0^\circ$). In other cases
(e.g. for $\gamma=-30^\circ$ and $\gamma=-50^\circ$ shown in
Fig.~\ref{rfig2}), a low frequency modulation of the amplitude of
the revivals is observed.

\subsection{Spin dynamics in 1D semiconductor quantum wires
triggered by HCPs and single-cycle pulses}

The SFET proposed in 1990
\cite{Datta1990} triggered  extensive research to validate the performance
of this device. Two serious difficulties were overcome in the
past. The first is the control of the Rashba SOI in a conductive channel
which was demonstrated by Nitta \textit{et al.} in 1997 using an
inverted InGaAs/InAlAs quantum well with a top gate
\cite{Nitta1997}. The second more difficult challenge is the
spin-polarized injection into the channel with solutions suggested by
various groups
\cite{Schmidt2005,Fert2007,Schmidt2000,Rashba2000,zgzhu2008pla,Jiang2003,Dijken2003}.
 Based on these advances Koo \textit{et al.}
constructed an operational SFET \cite{Koo2009}.
The analysis was based on the model proposed in
Ref.~\cite{Datta1990}. However this was extended to take
into account the angular spectrum in the confinement direction
\cite{Zainuddin2011}. On the theory side, the SOI effect in a 1D
or quasi-1D channel was investigated extensively
\cite{Moroz1999,Hausler2001,Governale2002,Malshukov2003,Silva2003,Bellucci2003,Debald2005,Wu2006,Gelabert2010}.
Here we review how the spin dynamics and transport in the SFET
with a 1D channel can be controlled by the application of
broadband ultrashort light pulses \cite{zgzhu2010qw}.

\begin{figure*}[t]
\includegraphics[width=10cm]{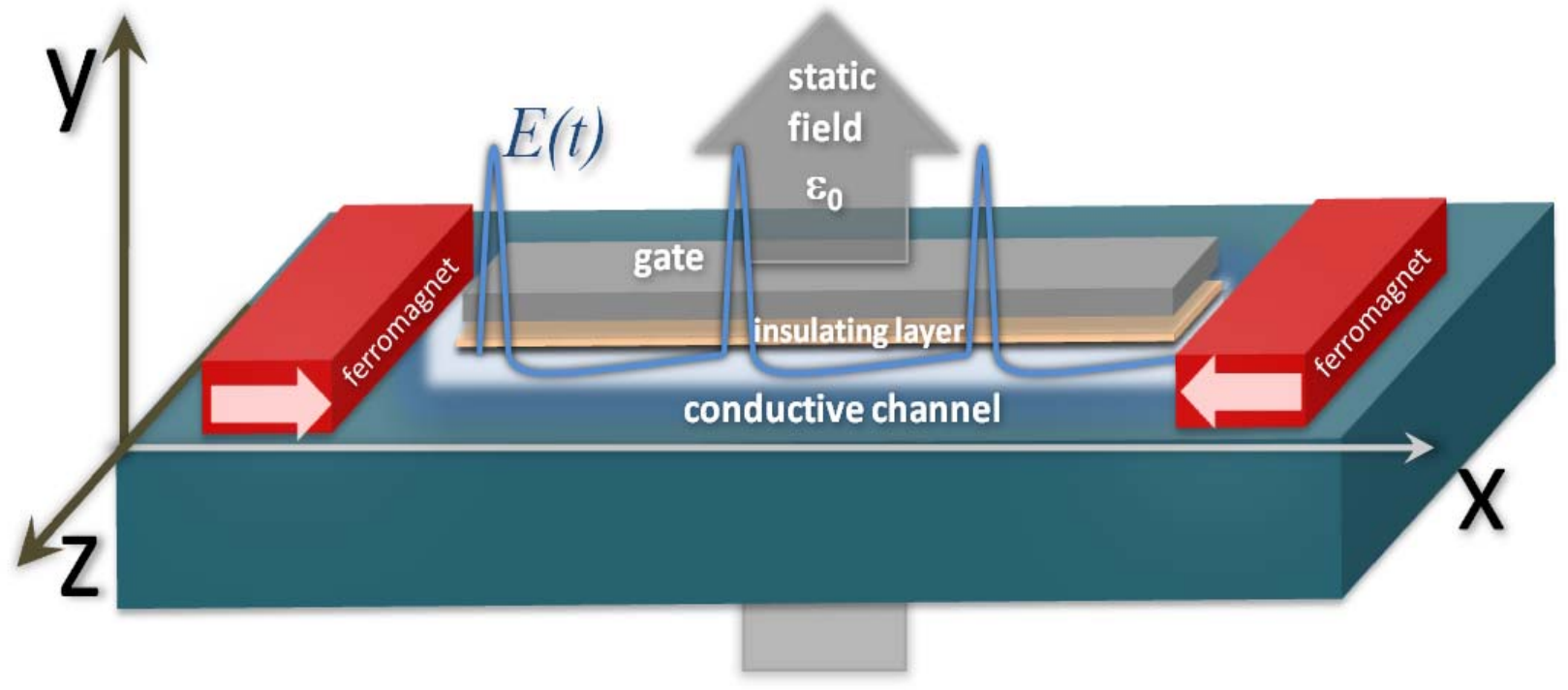}
\caption{Schematics of the optically driven SFET.
 Ferromagnetic leads are separated from the conductive channel
by a tunneling barrier to enhance the spin injection efficiency.
A
 metallic gate is used to tune the Rashba SOI via a static field $\mathcal{E}_0$. $E(t)$ is the time-dependent
 electric field (adapted from Ref.~\cite{zgzhu2010qw}).  \label{rfig8}}
\end{figure*}

We focus on a 1D quantum wire channel and ignore the spin-flip
transitions between the first and second subbands (the energy
spacing between these subbands is considered to be too large for
the activation of the intersubband transitions). The inversion
asymmetry of the confining potential results in the Rashba SOI
$H_{\mathrm{R}}$. For a two-dimensional electron gas in the
$xy$-plane we have
$H_{\mathrm{R}}=\alpha^{0}_{\mathrm{R}}(\sigma_{x}k_{z}-\sigma_{z}k_{x})$,
where $\alpha^{0}_{\mathrm{R}}=r_{\mathrm{R}}\mathcal{E}_0$ is the
static Rashba SOI coefficient, which is proportional to the
perpendicular electric field $\mathcal{E}_{0}$ resulting from the
band bending,  and $r_{\mathrm{R}}$ is a material-specific prefactor
\cite{Andrada1994,Andrada1997,Zhang2006,Gvozdic2006,Ekenberg2008,Winkler2003}.
Under equilibrium conditions, spin transport in such a device was
investigated extensively \cite{Datta1990,Winkler2003}; in brief,
one chooses the $z$-axis as the spin quantization axis
(Fig.~\ref{rfig8}) and $x$ for the quantum wire direction, so that
$H_{\mathrm{R}}=-\alpha^{0}_{\mathrm{R}}\sigma_{z}k=-\mu_{\mathrm{B}}\sigma_{z}\tilde{B}_{z}$,
where $\mu_{\mathrm{B}}$ is the Bohr magneton,
$\tilde{B}_{z}=\alpha^{0}_{\mathrm{R}}k/\mu_{\mathrm{B}}$ is an
effective magnetic field along $z$ and $k\equiv k_x$. Usually, one
should distinguish two situations: spins are injected from the
ferromagnetic leads are aligned (a) parallel to the $z$-axis and (b)
aligned parallel to the $x$- or $y$-axis. In the case (a), spins
do not precess under the effective magnetic field but feel a
spin-dependent potential shift. This results in the splitting
$2\alpha^{0}_{\mathrm{R}}k$ between the energies of the injected spins oriented in
the $-z$ (down) and $z$ (up) directions. When electrons with spins
in the down and up states and the same energy pass through the
length $L$, they acquire an accumulated phase difference
\begin{equation}
\Delta\theta_{0}=(k_{\downarrow}-k_{\uparrow})L=k_{\mathrm{so}}L\;,
\label{Eq:theta0}
\end{equation}
where $k_{\mathrm{so}}$ is given by
\begin{equation}
k_{\mathrm{so}}=\frac{2m^{*}\alpha^{0}_{\mathrm{R}}}{\hslash^2}\;.
\label{Eq:p_so}
\end{equation}
The eigenenergies and the stationary eigenstates are approximately
given by
\begin{equation}
E_{kS}^0=\frac{\hslash^{2}k^{2}}{2m^*}-S\alpha_{\mathrm{R}}^{0}k
\label{1Denergy}
\end{equation}
and
\begin{equation}
 \psi_{k}^{S}(x)=\frac{1}{\sqrt{L}}e^{ikx} \nu_{S}\;, \label{1Dwf}
\end{equation}
respectively, where $\nu_{S}$ ($S=\pm 1$) denotes the spin states
and $m^*$ is the effective mass.

Considering the spins to  be injected  aligned along the  $x$- or
$y$-directions, they precess around $\tilde{\mathbf{B}}_{z}$.
In
the Heisenberg picture, the spin operators vary with time as
$$\dot{\sigma}_{x(y)}(t)=\pm\omega_{k}\sigma_{y(x)}(t)\;, $$
where
\begin{equation}\label{Eq:omega_k}
 \omega_{k}=\frac{2\alpha^{0}_{\mathrm{R}}k}{\hslash}\;,
\end{equation}
is the precession frequency. It was used that close to the
Fermi level we have $k\gg k_{\mathrm{so}}\;$, i.e. the Fermi
energy $E_{\mathrm{F}}$ is much larger than the corresponding spin
splitting $2\alpha^{0}_{\mathrm{R}}k$. The solution of the
differential equation above reads
 $$\sigma_{\pm}(t)\equiv \sigma_{x}(t)\pm i\sigma_{y}(t)=\sigma_{\pm}(0)e^{\mp i\omega_{k}t}\;.$$
Let us specify the initial orientation, say $\sigma_{y}(0)=0$,
then
$$\sigma_{x}(t)=\sigma_{x}(0)\cos(\omega_{k}t) \mbox{\ \ and\ \ }
\sigma_{y}(t)=-\sigma_{x}(0)\sin(\omega_{k}t)\;.$$  
A spin aligned initially along the $x$-direction  rotates
clockwise with the angular frequency $\omega_{k}$. The accumulated
angle through the length  $L$ amounts to
$\Delta\theta_{0}=\frac{2m^{*}\alpha^{0}_{\mathrm{R}}L}{\hslash^{2}}$,
which is exactly equal to the phase shift for the injected spins
with the alignment along $z$.

\subsubsection{First dynamic case}

Having outlined the equilibrium case, we consider the quantum wire
driven by a linearly polarized electromagnetic pulse with the
vector potential $\mathbf{A}=\mathbf{e}_{x}A(t)$. The polarization
vector $\mathbf{e}_{x}$  is oriented along the $x$-direction.  We
may employ the Coulomb (radiation, velocity) gauge
\cite{Yamanouchi_book,Faisal_book} and make use of the dipole
approximation. Thus,
\begin{equation}
\mathbf{E}(t)=-\frac{1}{c}\frac{\partial A(t)}{\partial
t}\mathbf{e}_{x}=E_0 f(t)\mathbf{e}_{x}\;, \label{Eq:A_SFET}
\end{equation}
where $E_0$ is the peak amplitude of the electric field $E(t)$
and $f(t)$ describes its temporal profile. The single-particle
Hamiltonian describing the 1D motion of the electron and its spin
dynamics reads \cite{zgzhu2010qw}
\begin{equation}
H=\frac{\Pi^{2}}{2m^{*}}-
\frac{\alpha^{0}_{\mathrm{R}}}{\hslash}\sigma_z\Pi,
\label{Eq:H_SFET}
\end{equation}
where $\Pi=-i\hslash\frac{\mathrm{d}}{\mathrm{d}x}-\frac{e}{c}A$.
In the Heisenberg picture, it is straightforward to calculate the
time-dependent energy expectation value
$E_{k_0S_0}(t)=\langle\psi_{k_0}^{S_0}|U^\dagger(t,0) H
U(t,0)|\psi_{k_0}^{S_0}\rangle$ of an electron being initially in
the eigenstate $|\psi_{k_0}^{S_0}\rangle$. Here the stationary states
$|\psi_{k_0}^{S_0}\rangle$ are determined by Eq.~\eqref{1Dwf}. The
evolution operator $U(t,0)$ commutes with $H$ and we get
\begin{equation}
 E_{k_0S_0}(t)= \frac{1}{2m^{*}}\left[\left(\hslash
k_0-\frac{e}{c}A(t)-S_0\frac{\hslash
k_{\mathrm{so}}}{2}\right)^{2}-\frac{\hslash^2
k^{2}_{\mathrm{so}}}{4}\right], \label{hsecondquantization}
\end{equation}
where $k_{\mathrm{so}}$ is given by Eq.~\eqref{Eq:p_so} and $A(t)$
is selected such that it vanishes before the pulse application.
Using Eq.~\eqref{Eq:A_SFET} we can see from
Eq.~\eqref{hsecondquantization} that after the light pulse is gone
the electron energy ${E_{k_0S_0}(t=\infty)}$ equals to an
eigenenergy $E^0_{kS_0}$ of the unperturbed system, determined by
Eq.~\eqref{1Denergy}. This eigenenergy corresponds to the state
with the same spin $S_0$ but with a shifted wave vector
\begin{equation}
  k=k_0+\frac{\Delta p}{\hslash}\;,
\label{Eq:k_shift}
\end{equation}
where the transferred momentum $\Delta p=-\frac{e}{c}A(t=\infty)$
is given by Eq.~\eqref{Eq:Delta_p_free_space}.

This result can be also obtained if we transform the Hamiltonian
\eqref{Eq:H_SFET} into the length gauge
\cite{Yamanouchi_book,Faisal_book}, like in
Section~\ref{Eq:Hamiltonian_QR_SOI_driven} and
Ref.~\cite{zgzhu2008}, leading to $H'=H_0+V(t)$ with
\begin{equation}
H_0=\frac{p^2}{2m^{*}}-
\frac{\alpha^{0}_{\mathrm{R}}}{\hslash}\sigma_z p\;
\label{Eq:H0_SFET_lg}
\end{equation}
and
\begin{equation}
V(t)=-exE(t)\;, \label{Eq:V_SFET_lg}
\end{equation}
where $p=-i\hslash\frac{\mathrm{d}}{\mathrm{d}x}$. Then we use
Eq.~\eqref{Eq:IA_mapping_psi} and  proceed like in
Section~\ref{Sec:1d_free_driven} taking now into account also the
spin-dependent part of $H_0$ present in Eq.~\eqref{Eq:H0_SFET_lg}.
In this way we get that the operator entering the matching
condition \eqref{Eq:IA_mapping_psi} for the wave functions before
and after the pulse is determined by
Eq.~\eqref{Eq:U_result_free_space}, where $\Delta p$ is given by
Eq.~\eqref{Eq:Delta_p_free_space}, $\Delta x$ is given by
\eqref{Eq:Delta_x_free_space} and $\Delta \phi$ is given by
\begin{equation}\label{Eq:Delta_phi_SFET}
    \Delta \phi=\Delta \phi_0-\frac{1}{2}k_{\mathrm{so}} \Delta x \sigma_z\;,
\end{equation}
with $\Delta \phi_0=s'_3 e^2E_0^2\tau_{\rm
    d}^3/(4m^*\hslash)$, in place of Eq.~\eqref{Eq:Delta_phi_free_space}. Thus by
this method we see that each electron may get not only a momentum
boost $\Delta p$ but also a coordinate shift $\Delta x$ and a
spin-dependent phase shift $\propto \Delta x$ from the applied
pulse. The spin-independent phase shift, given by the first term
on the rhs of Eq.~\eqref{Eq:Delta_phi_SFET}, has no
physical effect and can be dropped from the present consideration.
As it is discussed in Section~\ref{Sec:1d_free_driven}, depending on
the pulse shape we can select the time moment for the matching of
the wave functions such that either $\Delta x$ or $\Delta p$
vanish; e.g., for a HCP pulse $\Delta p\neq 0$ and $\Delta x=0$,
whereas for a single-cycle pulse $\Delta p=0$ and $\Delta x\neq
0$. It is interesting that in any case the accumulated phase
difference between the spin-up and spin-down electrons,
Eq.~\eqref{Eq:theta0}, remains unchanged by the pulse action.
Indeed, the difference $k_{\downarrow}-k_{\uparrow}$ is unaffected
by the momentum shift $\Delta p$ experienced by all electrons, and
the change in the accumulated phase shift due to the coordinate
shift $-k_{\mathrm{so}}\Delta x$, induced by the pulse, is exactly
compensated by the phase difference gained directly from the pulse
$k_{\mathrm{so}}\Delta x$.

Let us consider a particular example of a SFET built on the basis
of the GaSb/InAs/GaSb heterostructure \cite{Luo1990} with the
electron effective mass $m^*=0.055m_0$, where $m_0$ is the free
electron mass. Assuming that the system is driven by single-cycle
pulses described in Section~\ref{Sec:single_cycle} with the pulse
shape parameter $s_2=\sqrt{\pi}/2$ and duration $\tau_{\rm
d}=50$~fs, such a pulse with a relatively small amplitude of
$E_0=2$~keV/cm induces a coordinate shift $\Delta x\approx
140$~nm, which is on the order of a typical device length. Thus
the operation time can be reduced down to the pulse duration of
50~fs.

Summarizing we conclude that whereas the phase difference $\Delta
\theta_0$ is maintained as for the static case, the ultrafast
momentum boost or coordinate shift experienced by all the
electrons can be used to speed up the device operation. This fact
is promising for the realization of an ultrafast SFET.




\subsubsection{Second dynamic case}
Let us consider the case when the electric field polarization
direction coincides with the $y$-axis (see Fig.~\ref{rfig8}). This
electric field $\mathbf{e}_yE(t)=\mathbf{e}_yE_0f(t)$ induces an
additional time-dependent SOI
\cite{Andrada1994,Andrada1997,Zhang2006,Gvozdic2006,Ekenberg2008}.
The single-particle Hamiltonian in the length gauge can be written
as $H=H_0+V(t)$ with $H_0$ given by Eq.~\eqref{Eq:H0_SFET_lg} and
\begin{equation}\label{Eq:V_SFET_lg_Ey}
 V(t)=-\frac{\alpha^{\mathrm{ind}}_{\mathrm{R}}(t)}{\hslash}\sigma_z
 p\;.
\end{equation}
The time-dependent pulse-induced Rashba SOI coefficient
$\alpha^{\mathrm{ind}}_{\mathrm{R}}(t)$ is given by
\begin{equation}\label{Eq:Raschba_coeff_ind}
 \alpha^{\mathrm{ind}}_{\mathrm{R}}(t)=r_{\mathrm{R}}E(t)\;.
\end{equation}
From Eq.~\eqref{Eq:U_result} we obtain for the operator
$\mathcal{U}(t_1)$ describing the action of the pulse on the
electrons in the wire
\begin{equation}\label{Eq:U_second_dynamic_case}
 \mathcal{U}(t_1)=\exp\left(\frac{i}{\hslash}\Delta
 x_{\mathrm{R}}p\sigma_z\right),
\end{equation}
where
\begin{equation}\label{Eq:Delta_x_R}
 \Delta x_{\mathrm{R}}=\frac{s_1r_{\mathrm{R}}E_0\tau_{\mathrm{d}}}{\hslash}\;.
\end{equation}
Here $\tau_{\mathrm{d}}$ is the pulse duration and the pulse shape
parameter $s_1$ is determined by Eq.~\eqref{Eq:s_n_infinity}.
Because $[V(t),H_0]$ vanishes, there are no other terms in the
exponent of Eq.~\eqref{Eq:U_second_dynamic_case}. Inserted into
Eq.~\eqref{Eq:IA_mapping_psi}, it provides an exact solution for
the electron wave function after the pulse.

In the case of injected electrons with spins aligned parallel to
the $z$-axis, we see from Eq.~\eqref{Eq:U_second_dynamic_case}
that the pulse induces a spin-dependent coordinate shift. The
spin-down electrons are shifted by $-\Delta x_{\mathrm{R}}$
whereas the spin-up electrons experience a coordinate shift
$\Delta x_{\mathrm{R}}$. In effect, the accumulated phase
difference between the spin-down and spin-up electrons with the
same energy while traversing  the conducting channel is
\begin{equation}
\Delta\theta=\Delta\theta_{0}+\theta_{\mathrm{p}}\;,
\label{Eq:theta_total}
\end{equation}
where $\Delta\theta_{0}$ is given by Eq.~\eqref{Eq:theta0} and the
additional phase difference $\theta_{\mathrm{p}}$ induced by the
pulse amounts to
\begin{equation}
\theta_{\mathrm{p}}= (k_{\downarrow}+k_{\uparrow})\Delta
x_{\mathrm{R}}\;. \label{thetap interm}
\end{equation}
Using Eqs.~\eqref{Eq:omega_k} and \eqref{Eq:Delta_x_R}, for the
electrons in the neighborhood of the Fermi level with $k\gg
k_{\mathrm{so}}$ we infer
\begin{equation}
\theta_{\mathrm{p}}= \omega_{k}\tau_{\rm
d}\frac{s_1E_0}{\mathcal{E}_0}\;, \label{thetap}
\end{equation}
where $\omega_{k}$ is given by Eq.~\eqref{Eq:omega_k}. For the estimation of the effectiveness of the optical control of
the SFET, we analyze the ratio between the pulse-induced and
stationary accumulated phase shifts. Using Eq.~\eqref{Eq:p_so} and
\eqref{Eq:omega_k} in Eqs.~\eqref{Eq:theta0} and \eqref{thetap} we
can write
\begin{equation}
\frac{\theta_{\mathrm{p}}}{\Delta\theta_{0}}=\frac{2k\Delta
x_{\mathrm{R}}}{k_{\mathrm{so}}L}=\frac{ v_{\mathrm{F}}\tau_{\rm
d}}{L}\frac{s_1E_0}{\mathcal{E}_0}\;, \label{lamda}
\end{equation}
where $v_{\mathrm{F}}=\hslash k/m^*$ is the electron velocity at the Fermi
level. In order to keep the electrons inside the conductive
channel upon the pulse application, the condition $\Delta
x_{\mathrm{R}}\ll L$ must be fulfilled that limits the pulse
strength. Nevertheless, $\theta_{\mathrm{p}}/\Delta\theta_{0}$ can
be large if the smallness of $\Delta x_{\mathrm{R}}/L$ is
overcompensated by the largeness of $k/k_{\mathrm{so}}$.

Considering the case of the electron injection with spins aligned
along the $x$ or $y$ directions, with the help of
Eq.~\eqref{Eq:U_second_dynamic_case} and the same assumptions as
above, we find for the spin operators ${\sigma}_{\pm}(t)$ at a
time moment $t$ after the pulse
\begin{equation}
\sigma_{\pm}(t)=\sigma_{\pm}(0)e^{\mp i\omega_k t}e^{\mp
i\theta_{\mathrm{p}}}. \label{sigmapmt}
\end{equation}
The pulse induces a clockwise rotation of the spin by the angle
$\theta_{\mathrm{p}}$, additionally to the rotation angle
$\omega_k L/v_{\mathrm{F}}$ accumulated over the length of the
conductive channel $L$. Consequently, the total precession angle in the clockwise direction
amounts also to $\Delta\theta$ given by
Eq.~\eqref{Eq:theta_total}.

Let us discuss again a SFET built on the basis of the
GaSb/InAs/GaSb heterostructure  with $\alpha^{0}_{R}= 0.9\times
10^{-9}$~eV cm, the charge density $n=10^{12}$~cm$^{-2}$ and
$2\alpha^{0}_{R}k=4$~meV at the Fermi level \cite{Luo1990}. These
values result then in $\omega_{k}\approx 2\pi$~ps$^{-1}$ and
$v_{\mathrm{F}}=0.4 ~\mu$m/ps. Thus in a $L=0.2~\mu$m long 1D
quantum wire there are 20 distributed electrons and the transport
time $t_{\mathrm{tr}}$ for the electrons with the Fermi velocity
is about 500~fs.

%

\begin{figure*}[t]
\includegraphics[width=8cm]{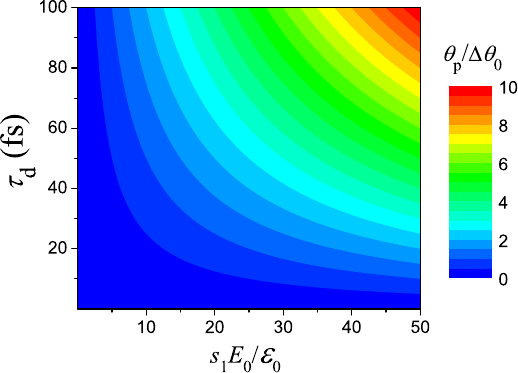}
\caption{Ratio between the pulse-induced and stationary
accumulated phase shifts (precession angles)
$\theta_{\mathrm{p}}/\Delta\theta_{0}$ in dependence on the HCP
duration $\tau_{\mathrm{d}}$ and the normalized HCP amplitude
$s_1E_0/\mathcal{E}_0$, calculated from Eq.~\eqref{lamda}.
Parameters of the heterostructure and driving HCP are given in the
text.
\label{rfig9}}
\end{figure*}

Considering the pulse durations $\tau_{\rm d}$ below 100~fs we
have $\tau_{\rm d}\ll t_{\mathrm{tr}}$ and $v_{\mathrm{F}}\tau_{\rm
d}<40$~nm, i.e. the factor $v_{\mathrm{F}}\tau_{\rm d}/L$ in
Eq.~\eqref{lamda} is less than 0.2. HCPs with such durations can
be generated with amplitudes $E_0$ up to several hundreds of
kV/cm.
The static fields $\mathcal{E}_{0}$ are typically of the order of
several kV/cm
\cite{Winkler2003}.  Therefore, taking into account that the HCP
pulse shape parameter $s_1$ is about one, the factor
$s_1E_0/\mathcal{E}_0$ in Eq.~\eqref{lamda} can be tuned as high
as 100 (without inducing intersubband transitions), so that the
ratio $\theta_{\mathrm{p}}/\Delta\theta_{0}$, calculated after
Eq.~\eqref{lamda}, can be significantly larger than 1.
 The spin-dependent coordinate shift $\Delta x_{\mathrm{R}}$
does not exceed a value on the order of $100$~nm so that the
condition $\Delta x_{\mathrm{R}}\ll L$ can be satisfied.
%
%

In Fig.~\ref{rfig9} we show a contour plot of the ratio
$\theta_{\mathrm{p}}/\Delta\theta_{0}$ as a function of the external
field parameters.  We clearly see that the pulse-induced phase
shift (or rotation angle) is comparable to the static one or may
even be significantly larger. We conclude that HCPs can control
the spin dynamic in a 1D SFET even playing a dominate role
\cite{zgzhu2010qw}. This opens the way  to
design an optical SFET, where spin can be engineered via appropriate shaping of
electromagnetic pulses.


\subsection{Ultrafast spin filtering and its maintenance in a double quantum dot}

With the help of additional gate contacts the 1D conductive
channel of a heterostructure system like the one shown in
Fig.~\ref{rfig8} can be transformed into a 1D double quantum dot
(DQD) system (see Fig.~\ref{Fig:fig1_APL}). The zero of the
$x$-axis is selected in such a way that the left (right) quantum
dot is located in the region $x<0$ ($x>0$). To control the carrier
motion we utilize HCPs polarized either along the $x$-axis
($x$-HCPs) or along the $y$-axis ($y$-HCPs), as illustrated in
Fig.~\ref{Fig:fig1_APL}. Like in Section~\ref{Sec:Appl_DQW}, the aim
of our consideration is the control of the electron dynamics and
its spatial localization, depending now on its spin
\cite{Waetzel2011}.

\begin{figure*}[t]
\centering
\includegraphics[width=10.0cm]{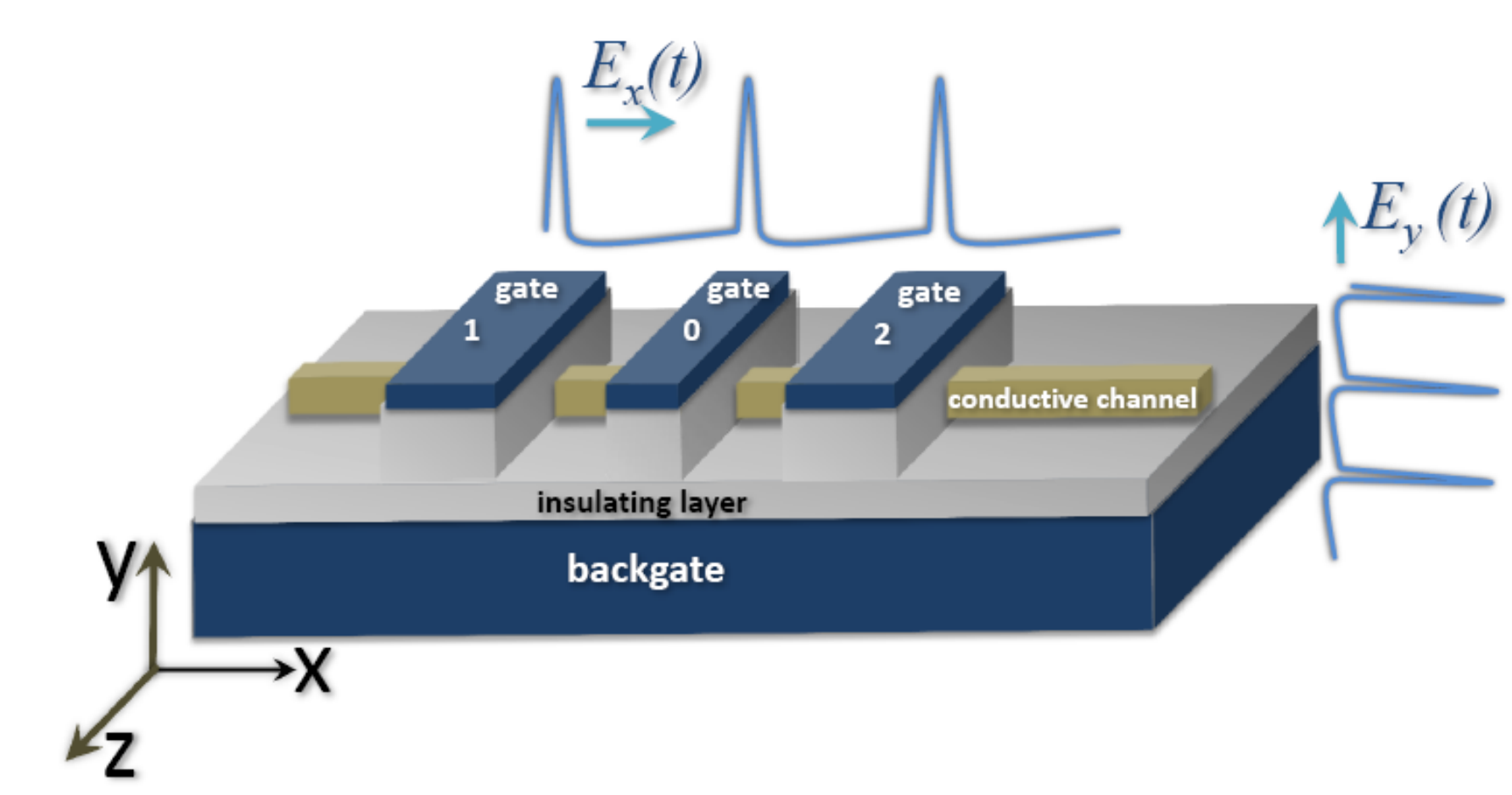}
\caption{Scheme of the pulse-driven DQD. The two quantum dots are
created and are controllable  by two local depletion gates: ''gate
1'' and ''gate 2''.  A further ''gate 0'' can tune the tunnel
coupling between the dots. The optical driving of the system is
provided by linearly polarized HCPs. Their polarization direction
can be along the $x$- or along the $y$-axis (adapted from
Ref.~\cite{Waetzel2011}). } \label{Fig:fig1_APL}
\end{figure*}

The system Hamiltonian is given by $H=H_{0}+V(t),$ where $H_{0}$
represents the Hamiltonian of a free electron in the DQD and
$V(t)$ is given by Eq.~\eqref{Eq:V_SFET_lg} in the case of
$x$-HCPs or by Eq.~\eqref{Eq:V_SFET_lg_Ey} in the case of
$y$-HCPs. As it was discussed in Section~\ref{Sec:Appl_DQW}, for a
certain range of the parameters of the DQD and HCPs we can model
the electron dynamics using the two-level system approximation
(TLSA), which is extended here to include the spin degree of
freedom. Within such an approximation the electron wave function
can be represented as
\begin{equation}\label{Eq:TLSA_wave_function_spin}
|\Psi(t)\rangle=\sum_{i,s}C_{i,s}(t)|i\rangle|\nu_s\rangle\;,
\end{equation}
there $|i\rangle$ with $i=1,2$ denotes the stationary spatial
states and $|\nu_s\rangle$ with $s=\uparrow,\downarrow$ denotes
the spin states. The Hamiltonian of the unperturbed system is
given by $H_0=-\frac{1}{2}\varepsilon \varsigma_z$ [cf.
Eq.~\eqref{Eq:Hamiltonian_TLS}], where
$\varepsilon=\hslash\omega_{21}$ is the distance between the two
lowest energy levels of the DQD. We reserve the notation
$\boldsymbol{\sigma}=(\sigma_x,\sigma_y,\sigma_z)^T$ for the
vector of the Pauli matrices in the basis of the spinor states
$|\nu_\uparrow\rangle,|\nu_\downarrow\rangle$, whereas for the
vector of the Pauli matrices in the basis of the pseudospinors
states $|i\rangle$ we are going to use here the notation
$\boldsymbol{\varsigma}=(\varsigma_x,\varsigma_y,\varsigma_z)^T$
\footnote{Notice the difference in the notation in respect to
Section~\ref{Sec:driven_TLS}. We need here different notations for
the Pauli matrices in the basis of the spinor states and the
Pauli matrices in the basis of the pseudospinor states.}. The
electron interaction with a $x$-HCP is  given by
$V_v(t)=-vf_v(t)\varsigma_z$ [cf. Eq.~\eqref{Eq:Hamiltonian_TLS}],
where $v= eE^v_0\langle 1|x|2\rangle$, $E^v_0$ is the amplitude of
the $x$-HCP and $f_v(t)$ describes its temporal profile. We select the
phases of the basis functions such that the coordinate matrix
element $\langle 1|x|2\rangle=x_{12}$ is real and positive,
whereas the momentum matrix element $\langle 1|p|2\rangle=i
p_{12}$ is imaginary. The interaction with a $y$-HCP reads
\begin{equation}\label{Eq:V_u}
 V_u(t)=-uf_u(t)\sigma_z\varsigma_y\;,
\end{equation}
where
\begin{equation}\label{Eq:u_in_V_u}
  u=\frac{r_{\mathrm{R}}E_0^u p_{12}}{\hslash}\;,
\end{equation}
$E_0^u$ and $f_u(t)$ are the $y$-HCP amplitude and its temporal
profile, respectively.

Like in Section~\ref{Sec:Appl_DQW}, we also utilize the impulsive
approximation (IA). In effect the action of the $x$-HCP on the
electron can be reduced to the matching condition
\eqref{Eq:IA_mapping_psi} with $\mathcal{U}(t_{1})=\exp(i
\alpha_v\varsigma_x)$ [cf. Eq.~\eqref{Eq:U_result_TLS}].
Here the dimensionless parameter $\alpha_v$ characterizing the
strength of the $x$-HCP is given by $\alpha_v=s_1^v a_v$, where
$a_v$ is determined by Eq.~\eqref{Eq:TLS_ab} and $s_1^v$ should be
calculated using Eq.~\eqref{Eq:s_n_infinity}. Alternatively,
$\alpha_v$ can be expressed as $\alpha_v=\Delta p x_{12}/\hslash$
with $\Delta p$ from Eq.~\eqref{Eq:Delta_p_free_space}. The action
of the $y$-HCP is determined by
\begin{equation}\label{Eq:U_DQD_yHCP}
    \mathcal{U}(t_{1})=\exp(i \alpha_u \sigma_z\varsigma_x)\;,
\end{equation}
where the dimensionless parameter $\alpha_u$ characterizing the effective
strength of the $y$-HCP can be found as
\begin{equation}\label{Eq:alpha_u}
    \alpha_u=\frac{\Delta x_R p_{12}}{\hslash}\;.
\end{equation}
Here $\Delta x_R$ should be calculated from
Eq.~\eqref{Eq:Delta_x_R} using the parameters of the $y$-HCP.

In order to characterize the spin-resolved electron dynamics, we
introduce the time-dependent spin-resolved probability to find the
electron in the left quantum dot
\begin{equation}\label{Eq:P_Ls}
    P_{\mathrm{L}s}(t)=\langle
    \Psi(t)|\nu_s\rangle\langle\nu_s|\Theta(-x)|\Psi(t)\rangle\;,
\end{equation}
whereas the total probability to find the electron in the left
quantum dot reads $P_{\mathrm{L}}(t)=P_{\mathrm{L}\uparrow}(t)+P_{\mathrm{L}\downarrow}(t)$.
Here $\Theta(x)$ denotes the Heaviside step function. The spin
polarization vector in the left quantum dot is given by
\begin{equation}\label{Eq:Lambda_L}
   \boldsymbol{\Lambda}_{\mathrm{L}}(t)
   =\frac{\langle
   \Psi(t)|\boldsymbol{\sigma}\Theta(-x)|\Psi(t)\rangle}{P_{\mathrm{L}}(t)}\;.
\end{equation}
In particular,
${\Lambda}_{\mathrm{L},z}(t)=\left[P_{\mathrm{L}\uparrow}(t)-P_{\mathrm{L}\downarrow}(t)\right]\big/P_{\mathrm{L}}(t)$.

The time evolution of the system strongly depends on the initial
condition. Here it is useful to consider two situations: (a) the
initial condition, which corresponds to an electron being
completely localized in the left well (tunneling initial
condition) and a vanishing average spin along the $z$-axis, i.e.
$\langle \sigma_z\rangle=0$, and (b) the initially delocalized
state (optical initial condition) with the same spin properties,
which belongs to the ground state of the system. In
Ref.~\cite{Alex_QW2004} it was shown  that the optical
initial condition can be created from the tunneling initial
condition with the help of an appropriate $x$-HCP applied at an
appropriate time moment. Of course, no spin-polarization can be
created only by $x$-HCPs because the excitation is not
spin-dependent. The spin polarization can emerge if we apply a
$y$-HCP at the time moment just after a state corresponding to the
optical initial condition has been produced. In order to maintain
the spin polarization, an appropriate periodic train of $x$-HCPs
can be used, in analogy to the maintenance of spin-unpolarized
states (see Sections \ref{Sec:suppress_tun} and
\ref{Sec:persistent_localization}).

\begin{figure*}[t!]
\centering
\includegraphics[width=10.0cm]{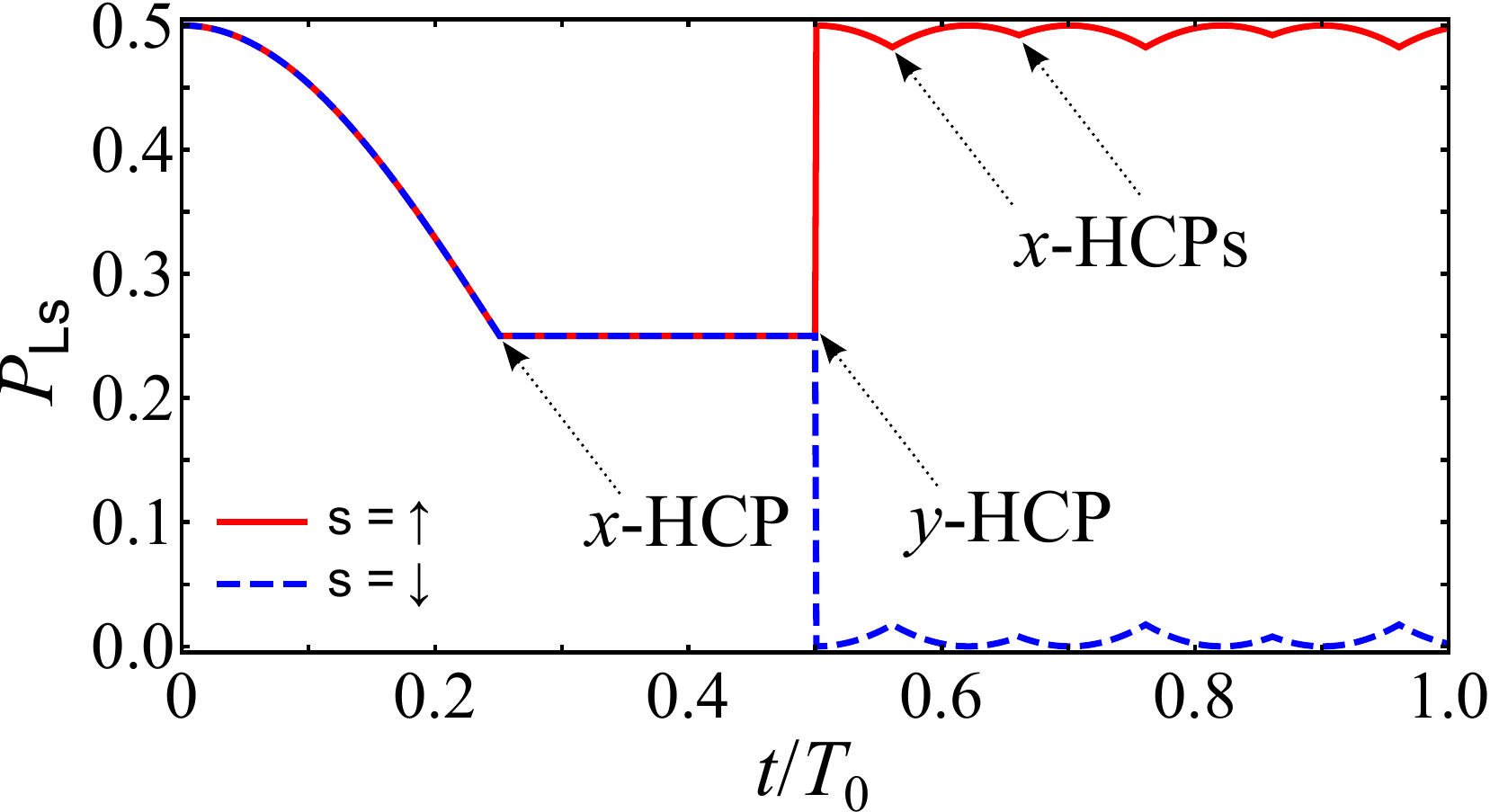}
\caption{Time-dependent probabilities to find a spin-up
$P_{\mathrm{L}\uparrow}(t)$ (red solid line) and to find a spin-down
$P_{\mathrm{L}\downarrow}(t)$ (blue dashed line) electron in the left dot.
Parameters of the excitation by HCPs are given in the text. For
the discussed GaAs-based DQD $T_0=0.67$~ps (adapted from
Ref.~\cite{Waetzel2011}). } \label{Fig:fig2_APL}
\end{figure*}

The results of a numerical simulation realizing this scenario are
illustrated in Fig.~\ref{Fig:fig2_APL} where the dynamics of the
spin-resolved localization probability is shown.
We start with the system in the state corresponding to the
tunneling initial condition (both spin states have the same
probability). After a fixed time $0.25T_0$ of the free
propagation, where $T_0=2\pi/\omega_{21}$, a $x$-HCP with
$\alpha_v=\pi/4$ is applied, which transfers the system into the
ground state (both spin states have the same probability
$P_{\mathrm{L}\uparrow}=P_{\mathrm{L}\downarrow}=0.25$). At $t=0.5T_0$ we apply a
$y$-HCP with $\alpha_u=\pi/4$ and obtain immediately a nearly
perfect spin separation, that means the two spin states are
localized in different wells: $P_{\mathrm{L}\uparrow}=0.5$ and
$P_{\mathrm{L}\downarrow}=0$. This separation is then maintained by
applying a periodic $x$-HCP train with the effective strength
$\alpha_v=\pi/2$, period $T=0.1T_0$ and the first HCP centered at
$t=0.56T_0$. In result, afterwards we find for the mean values of the
probabilities averaged over the time
period $2T_0$: $\langle P_{\mathrm{L}\uparrow}\rangle=0.495$ and $\langle
P_{\mathrm{L}\uparrow}\rangle=0.005$, correspondingly. Thus a very good
spin separation is stabilized in time. This is also illustrated in
Fig.~\ref{Fig:fig3_APL}a where the dynamics of the components of
the spin polarization $\boldsymbol{\Lambda}_{\rm{L}}(t)$ in the
left well is shown. Until the time moment of applying the $y$-HCP
at $t=0.5 T_0$ the spin polarization in the left well is oriented
along the $y$-axis. Just after this time moment the spin
polarization turns abruptly into the $z$-direction. Its
orientation oscillates then inside a small solid angle close to
the $z$-axis under the action of the maintaining $x$-HCP train.
The time average (over $2T_0$) of the $z$-component of the spin
polarization amounts to $0.982$. The corresponding trajectory of  the polarization vector
on the unit sphere is shown in
Fig.~\ref{Fig:fig3_APL}b.

\begin{figure*}[t!]
\centering
\includegraphics[width=12.0cm]{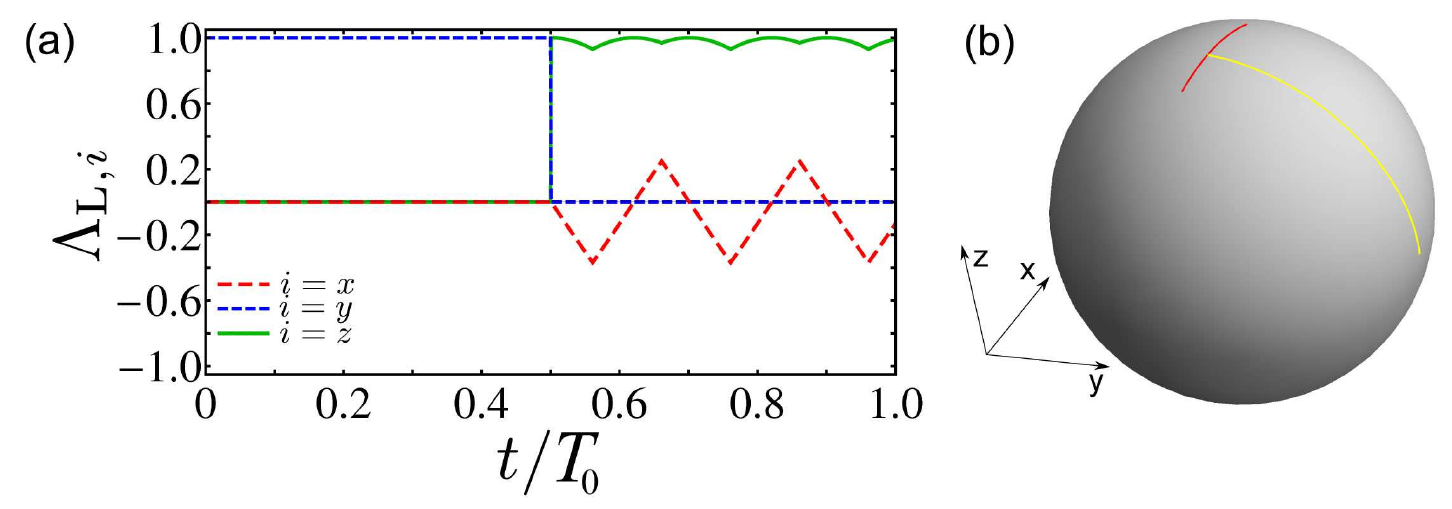}
\caption{(a) Components of the time-dependent spin polarization in the
left well $\boldsymbol{\Lambda}_{\rm{L}}(t)$. The same driving
fields as in Fig.~\ref{Fig:fig2_APL} are used. (b) Trajectory on
the unit sphere determined by the dynamics of
$\boldsymbol{\Lambda}_{\rm{L}}(t)$. The part in yellow (red) color
corresponds to $t\le 0.5 T_0$ ($t> 0.5 T_0)$ (adapted from
Ref.~\cite{Waetzel2011}).} \label{Fig:fig3_APL}
\end{figure*}

For the experimental realization one might consider a GaAs-based
DQD, e.g., with parameters as in Ref.~\cite{Alex_QW2004},
giving $T_0\approx 0.67$~ps, and $r_{\mathrm{R}}/|e|\approx 4
~\mbox{\AA}^2$ \cite{Andrada1997}. Assuming a HCP duration of
$40~\rm fs$, we estimate that the amplitude of the $y$-HCPs
required to observe the predicted dynamics should be on the order
of $10^7~{\rm V/cm}$. Such high and short fields are currently
available with the help of the modern short pulse generating techniques
\cite{Sell2008,Junginger2010} but may lead to undesirable effects
in the semiconductor heterostructure. Materials with higher values
of $r_{\mathrm{R}}$ are required. Narrow gap semiconductors, e.g.
InSb, would be more promising in this respect because of the much
higher values of $r_{\mathrm{R}}$ compared  to GaAs
\cite{Andrada1997}.

We have discussed how to separate the spin states
of electrons in a DQD by application of two delayed HCPs. In this
way a nearly perfect spin polarization can be achieved and then
maintained for a desired period of time by applying an additional
HCP train. Such light-induced spin filtering can be realized on a
sub-picosecond time scale that can be of relevance for designing
ultrafast spintronic and spin-qubit devices
\cite{Allwood2002,Loss1998,Kane1998,Zutic2004}.

%
\subsection{Generation and coherent control of pure
spin current via THz pulses}
\begin{figure*}[ht!]
  \centering
  \includegraphics[width=12cm]{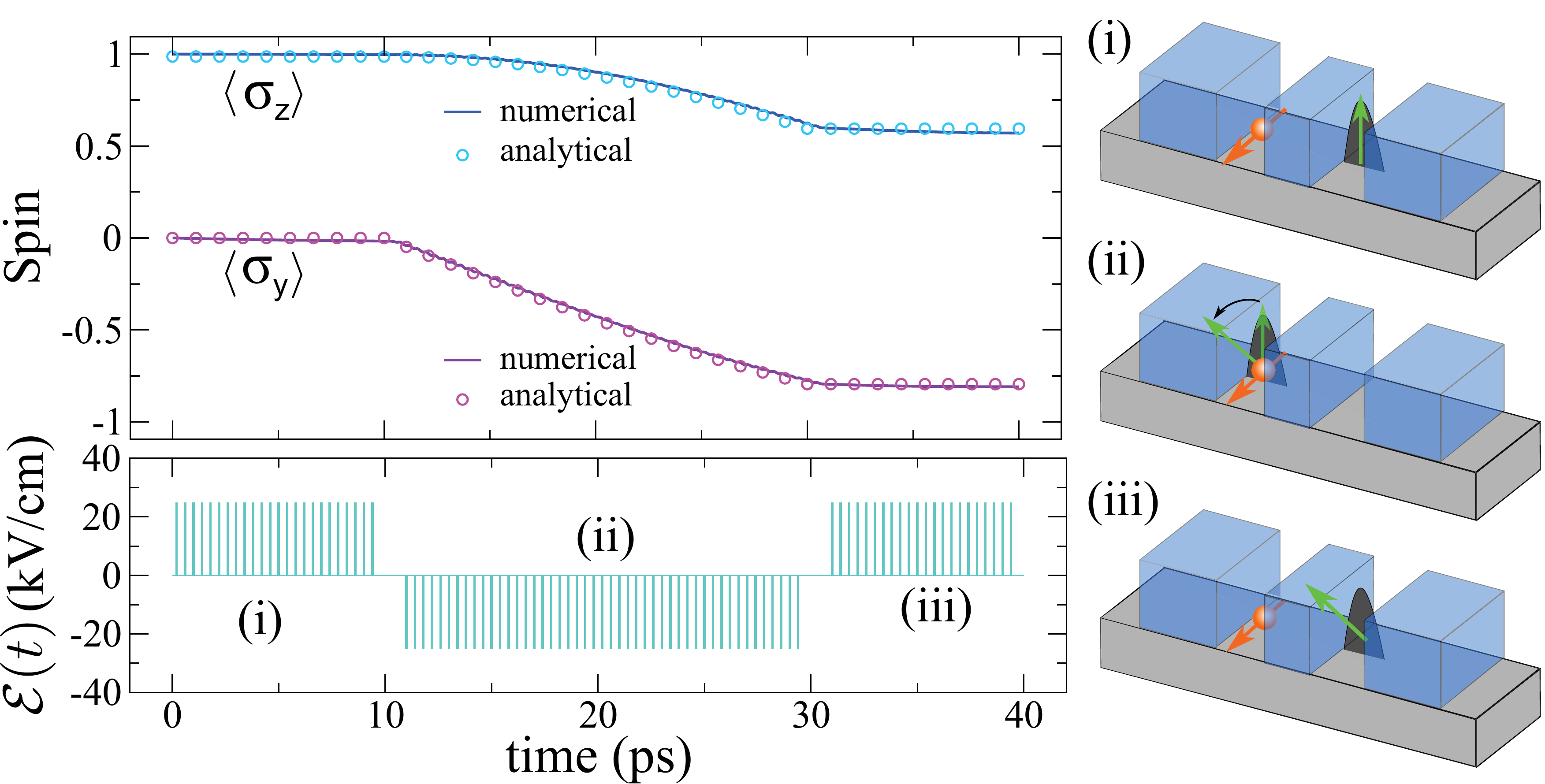}
  \caption{The double quantum well structure  is considered  with the
   additional feature that in the left well the charge carriers experience a local
   spin-orbital coupling (red arrow), i.e. a wave packet with a spin up (green arrow) in the right well changes
   spin orientation when traversing the left well.
   As discussed in the text, however, starting with a spin-polarized population in the right well [i.e., $\langle \sigma_z \rangle  (t=0)=1$], we can apply short pulses stabilizing this state in time, i.e.  blocking tunneling.
   The appropriate pulse sequence is marked (i). No spin dynamics takes place during this time.
  Allowing for tunneling (time interval between pulse sequences i and ii) and maintaining the charge localization in the left well (pulse sequence ii) initiates the spin dynamics leading to a steady decrease in $\langle \sigma_z \rangle$
  and hence an increase in $\langle \sigma_y \rangle$, which means in turn a net pure spin current. The results are averaged over the fast
  electronic transient oscillations.  It is even possible to use the pulse sequence (iii) and stop
  the  spin dynamics within picoseconds. The full lines in the left top panel
    are results of a full numerical time propagation,  whereas the circles
    are the outcome of an analytical adiabatic decoupling of the electronic and spin dynamics. (Figure is due to Ref.~\cite{Schueler2014}).
    \label{fig2apl14}}
\end{figure*}
We have seen that pulse trains are capable of triggering directed charge currents and spin-polarized currents.
A relatively recent development is the generation and utilization of pure spin currents in an open circuit geometry, i.e. a flow of angular momentum without a net charge current
\cite{Shikoh2013,Uchida2008,Jaworski2010,Jaworski2012,Qu2013,Schmid2013,Xiao2010,Adachi2013,Kikkawa2013,Chotorlishvili2013}.
The main focus has been on triggering this current by a temperature gradient in materials
that exhibit an exchange spin torque with a non-uniform temperature distribution,
 or by spin pumping (e.g., via ferromagnetic resonance) in a ferromagnetic/paramagnetic structure, and exploiting the
 spin-orbital coupling that has to be present in the paramagnet (e.g., platinum). The interest is motivated by potential applications
in spintronic devices with low-energy consumption. Another way to avoid the use of magnetic fields or high
current-density, spin-polarized
current  for magnetic  switching is to generate the pure spin current by optical  means using ultrashort THz pulses \cite{Schueler2014}.
The general idea exploits the  different time scales of the charge  (generally in the atto- to femtosecond time scale) and
spin dynamics (usually in the pico- to nanosecond time scale). Thus, driving the system with a residual spin-orbit coupling
 by a sequence of  ultrashort  HCPs of moderate intensity results in a fast pure electronic transient accompanied by a much slower transversal spin dynamics (longitudinal spin dynamics is much higher in energy and requires orders of magnitude  higher intensity than discussed here).
 Therefore in an open circuit geometry, a time average on the time scale of the spin dynamics results in pure spin phenomena.
 An explicit demonstration was given in Ref.~\cite{Schueler2014} for the generation of a pure spin current by a sequence of THz HCPs of moderate intensity.
 A typical result is shown in Fig.~\ref{fig2apl14}
 for the DQD  structure which we discussed above with the difference that a SOI is
 operational in one of the wells, say the left well. Thus, starting with a spin-polarized (green arrow, which also defines the $z$-direction in Fig.~\ref{fig2apl14})
  state localized in the right well and driving the charge dynamics by the pulse sequence plotted in the left bottom panel of Fig.~\ref{fig2apl14} results
  in a well-defined spin rotation once the wave packet is in the left well. The rotation proceeds on a much longer time scale than the tunneling. Hence, we may
  adiabatically decouple these two types of  motions, as shown in Ref.~\cite{Schueler2014}. This yields   semi-analytical expressions that agree
  very well with the full numerical solution, as evident from Fig.~\ref{fig2apl14} endorsing a posteriori the possibility
   of a dynamic decoupling of charge and spin. Indeed, as demonstrated in Fig.~\ref{fig2apl14}, upon averaging over the
    fast oscillations due to the electronic transient, the pulse sequence leads to a pure spin current.
    Interestingly this current can even be controlled coherently by applying an appropriate pulse sequence, which is a clear advantage
    over thermal schemes for generating pure spin currents by  utilizing a thermal gradient.
\section{Light emission from quantum systems driven by short broadband  pulses }\label{Sec:emission}
How to capture theoretically the radiation emitted from electrons moving along
confined trajectories is an old problem, probably first addressed  in 1907 by Schott
in connection with the question of as how to describe the discrete atomic emission spectra in the
framework of the classical electrodynamics \cite{Schott}. It is
clear that the emission spectra encode  certain information about
the electronic system and also about its dynamics. Thus they
provide a possibility for the experimental characterization and
control of the system dynamics. This is especially important for
ultrafast processes in small quantum systems where a direct
observation of this dynamics can be a challenge.

On one hand, in connection with the spectral measurements,
ultrashort broadband pulses provide an excellent tool for the
characterization of  electronic systems on ultrafast time
scales, allowing for the separation of the excitation process and
the following field-free system dynamics. The generated emission
can be detected in the absence of the influence of the driving
fields. On the other hand, the electron dynamics, controlled by
such electromagnetic pulses, can be monitored via the emission
spectra. It is possible to check if the system is in a particular target
state or if it follows the prescribed dynamics. Finally, the light emitted from the pulse-driven quantum
systems, e.g., nanostructures, might have interesting properties so
that it, in its turn, can be used for further application.

Here we review several examples of the stationary spectra of the
pulse-driven quantum systems. Then we discuss the time-dependent
spectrum and the control of the
circular polarization degree by means of broadband, ultrashort
pulses.

\subsection{Stationary spectra}
Firstly, we are going to study the stationary (time-integrated) emission
spectra $I_s(\omega)$ of the pulse-driven quantum systems. In a
non-relativistic case at large distances from the emitting system
we have (see Appendix \ref{Sec:intensity_spectrum_def})
\begin{equation}\label{Eq:Intensity_stat}
    I_{\mathrm{s}}(\omega)\propto \omega^4|\mu_{\omega}|^2\;,
\end{equation}
where $\boldsymbol{\mu}_\omega$ is the Fourier component of the
dipole moment. If the spectrum is limited to a some frequency range
such that its width $\Delta\omega$ is much smaller than the frequencies
$\omega$ inside this range, then it is sufficient to consider just
$|\mu_{\omega}|^2$. However, if this condition is not fulfilled,
it is important to remember about the additional factor $\omega^4$
in Eq.~\eqref{Eq:Intensity_stat} when comparing the calculations
with experimentally measured spectra.

\subsubsection{Spectra of 1D double quantum wells driven by periodic HCP trains}
The emission spectrum of 1D DQWs driven by periodic trains of HCPs
was analyzed in Refs.~\cite{Alex_QW2004,Alex_Scripta2005}.
The dipole moment $\mu(t)$ is connected with the time-dependent
probability to find the electron in the left well $P_{\mathrm{L}}(t)$ as
\begin{equation}\label{Eq:mu_DQW}
   \mu(t)= d_{12}\left[2P_{\mathrm{L}}(t)-1\right],
\end{equation}
where $d_{12}$ is the (real) transition dipole matrix element.
Thus the emission spectra can be obtained from the numerically
calculated dynamics of $P_{\mathrm{L}}(t)$, which was discussed in
Section~\ref{Sec:Appl_DQW}. In Fig.~\ref{Fig:Fig5_Alex_PRB_QW}
$|\mu_\omega|^2$, i.e. $I_{\mathrm{s}}(\omega)/\omega^4$, is
presented for a DQW driven by a periodic train of HCPs. Different
HCP strengths are selected in
Figs.~\ref{Fig:Fig5_Alex_PRB_QW}a-\ref{Fig:Fig5_Alex_PRB_QW}c,
whereas other simulation parameters are the same as for the full
numerical calculations of Fig.~\ref{Fig:Fig3_AlexPRB_QW_2004}. In
particular, in the considered case the pulse train frequency
$\omega_T$, defined by Eq.~\eqref{Eq:Omega_T}, is much larger than
characteristic system frequency $\omega_{21}$, determined by
Eq.~\eqref{Eq:T_c}: $\omega_{21}/\omega_T\approx 0.15$.

\begin{figure*}[t!]
\centering
\includegraphics[width=16.0cm]{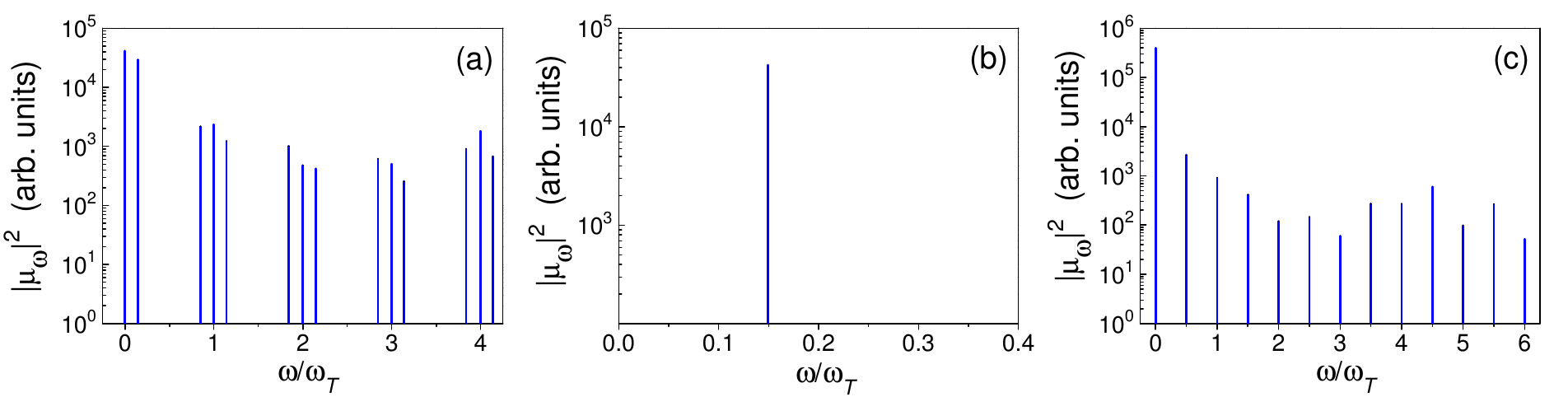}
\caption{Spectra $|\mu_\omega|^2$ from a DQW driven by periodic
trains of HCPs with peak electric fields: (a) 150~kV/cm, (b)
168.37~kV/cm, and (c) 84.068~kV/cm. The corresponding values of
the parameter $\alpha$, defined by
Eq.~\eqref{Eq:quasienergies_TLS_alpha}, are $\alpha=0.89\pi$ in
(a), $\alpha=\pi$ in (b), and $\alpha=\pi/2$ in (c). Other
parameters are as in Fig.~\eqref{Fig:Fig3_AlexPRB_QW_2004}. The
frequency is normalized by $\omega_T=2\pi/T$, where $T$ is the
period of the HCP train (adapted from
Ref.~\cite{Alex_QW2004}).} \label{Fig:Fig5_Alex_PRB_QW}
\end{figure*}

To comprehend the structure of these spectra, it is
worthwhile to consider the Floquet representation
\eqref{Eq:Floquet_states} of the wave function of the periodically
driven electron within the TLSA. Decomposing the Floquet modes
into the Fourier series,
\begin{equation}\label{Eq:Floquet_modes_Fourier}
    |\Phi_\lambda(t)\rangle=\sum_{n=-\infty}^\infty |b_{\lambda
    n}\rangle e^{i n\omega_T t}\;,
\end{equation}
leads to \cite{Bavli1993,Alex_QW2004}
\begin{equation}\label{Eq:mu_3i}
    |\mu_\omega|^2=4\pi^2|i_1+i_2+i_3|^2\;,
\end{equation}
where
\begin{eqnarray}
    i_1&=&\sum_{\lambda=1}^2 |C_\lambda|^2 \sum_{n,n'=-\infty}^\infty \langle b_{\lambda n}|x|b_{\lambda n'}\rangle
      \delta\big((n'-n)\omega_T-\omega\big)\;,\label{Eq:i1_emission}\\
    i_2&=&C_1C_2^*\sum_{n,n'=-\infty}^\infty \langle b_{2n}|x|b_{1n'}\rangle
      \delta\big((n'-n)\omega_T+(\epsilon_2-\epsilon_1)/\hslash-\omega\big)\;,\label{Eq:i2_emission}\\
    i_3&=&C_2C_1^*\sum_{n,n'=-\infty}^\infty  \langle b_{1n}|x|b_{2n'}\rangle
      \delta\big((n'-n)\omega_T+(\epsilon_1-\epsilon_2)/\hslash-\omega\big)\;.\label{Eq:i3_emission}
\end{eqnarray}
Here $\epsilon_1$ and $\epsilon_2$ are the quasienergies
determined by Eq.~\eqref{Eq:quasienergies_TLS}. One can see from
these equations that in the general case the spectrum ($\omega\ge
0$) consists of peaks located at (i) $\omega=N\omega_T$, (ii)
$\omega=(N+1)\omega_T\pm (\epsilon_2-\epsilon_1)/\hslash$, and
(iii) $\omega=(\epsilon_2-\epsilon_1)/\hslash$. Here
$N=0,1,2,\ldots$\ . Note that $(\epsilon_2-\epsilon_1)/\hslash$ is
positive and does not exceed $\omega_T$ (see
Fig.~\ref{Fig:quasienergies}). The peaks of all these three types
can be observed in Fig.~\ref{Fig:Fig5_Alex_PRB_QW}a.

Assuming $\omega_{21}<\omega_T/2$, the minimum of both $(\epsilon_2-\epsilon_1)/\hslash$ and $\omega_T-(\epsilon_2-\epsilon_1)/\hslash$ equals to $\omega_{21}$. It is achieved when the delocalization condition
\eqref{Eq:alpha_delocalization} is fulfilled. In this case HCPs do
not influence the dynamics of the electron that is oscillating between the
quantum wells.
%
Then only one peak in the
spectrum at the frequency $\omega_{21}$ survives.
This situation corresponds to Fig.~\ref{Fig:Fig5_Alex_PRB_QW}b.

Finally, in the case $(\epsilon_2-\epsilon_1)/\hslash=\omega_T/2$
the peaks of the types (ii) and (iii) are positioned at
half-integers of $\omega_T$. The peak of the type (ii) with the
lowest frequency coincides with the peak of the type (iii). Each
of the other peaks of the type (ii) coincides with one another
peak of the same type. Thus there are two times less peaks per
frequency interval than in a general case. This situation
corresponds to the condition \eqref{Eq:alpha_localization}, which
is necessary for the suppression of tunneling. If another
necessary condition $\omega_{21}/\omega_T\ll 1$ is also fulfilled,
that is the case here, then choosing an appropriate initial
condition we have an electron persistently localized in the left
well (cf. Fig.~\ref{Fig:Fig3_AlexPRB_QW_2004}a). The corresponding
spectrum is displayed in Fig.~\ref{Fig:Fig5_Alex_PRB_QW}c.

\subsubsection{Driven quantum rings as THz emitter}
Stationary spectra of semiconductor quantum rings excited by a
single HCP were discussed in Ref.~\cite{Moskalenko_PRB2006}.
In particular, their temperature dependence was analyzed in
connection to the beating, revivals and collapses observed in the
polarization dynamics (cf. Section~\ref{Sec:polarization_dynamics}).
An example, of the temperature-dependent spectra is shown in
Fig.~\ref{Fig:spectra_QR_rfig3}a. Figure
\ref{Fig:spectra_QR_rfig3}b from Ref.~\cite{zgzhu2008}
illustrates the dependence  of the spectra on the HCP kick
strength $\alpha$. At low temperature and low excitation strengths
we observe two peaks. This follows from the properties of the
distribution function in the neighborhood of the Fermi level. Here
we have the electron number $N=400$ for
Fig.~\ref{Fig:spectra_QR_rfig3}a and $N=100$ for
Figs.~\ref{Fig:spectra_QR_rfig3}b, i.e. there is an even number of
electrons in each spin channel. Therefore, at zero temperature all
levels with angular quantum numbers $|m|<N/4-1$ are occupied, the
occupation of the levels with $m=\pm N/4$ equals to 1/2, and the
levels with $|m|>N/4$ are empty. As a consequence, at low
excitation strengths only the transitions between the states with
$|m|=N/4-1$ and $|m|=N/4$ and between the states with $|m|=N/4$
and $|m|=N/4+1$ (whereas $m$ does not change its sign) contribute to the dipole moment dynamics, where we
observe the beating in this regime (see Fig.~\ref{Fig:revivals_PRB}),
leading to the emission spectrum with the two corresponding peaks.
They are located at $\omega=(N/2-1)\omega_0/2$ and
$\omega=(N/2+1)\omega_0/2$. Increasing the temperature, we increase
the number of levels which are only partly occupied. Therefore
more transitions are involved in the dynamics, resulting in the
interchanging revivals and collapses (cf.
Fig.~\ref{Fig:revivals_PRB}), and contribute to the emission
spectrum (see Fig.~\ref{Fig:spectra_QR_rfig3}a). The same happens
when increasing the effective HCP strength $\alpha$ (see
Fig.~\ref{Fig:spectra_QR_rfig3}b) because more states in the
neighborhood of the Fermi level are affected by the excitation.
Peaks at $(N/2\pm 3)\omega_0/2,(N/2\pm 5)\omega_0/2,...$ appear. The
corresponding energy range is determined by the energy
$\hslash\omega_0\alpha^2/4$ transferred by the HCP to an electron in the QR [cf.
Eq.~\eqref{energy2}]. In Fig.~\ref{Fig:spectra_QR_rfig3}a we see
that the temperature increases not only multiplies
the number of peaks but also makes them less pronounced due to the
smaller relaxation times (cf. Fig.~\ref{Fig:eff_rate}),
which determine the spectral broadening. Small relaxation times lead
to the smearing out of the peaks so that finally only one broad
peak can be distinguished.

\begin{figure*}[t!]
\centering
\includegraphics[width=16.0cm]{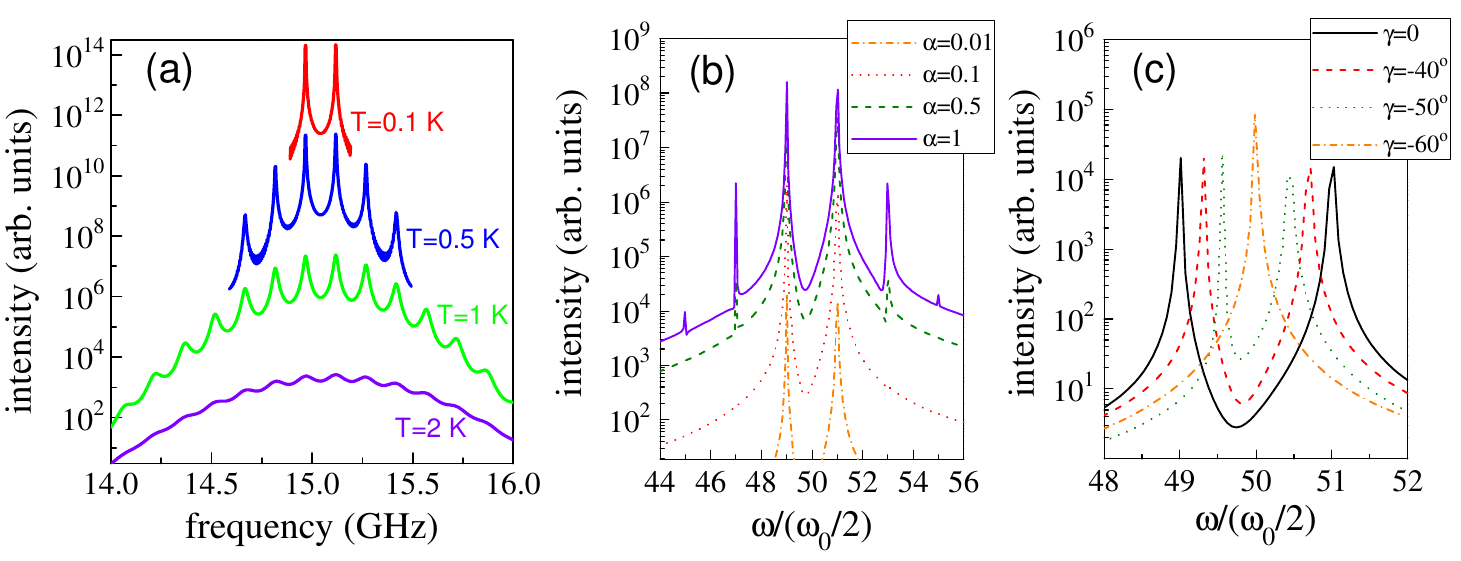}
\caption{(a) Emission spectra $I_{\mathrm{s}}(\omega)$ of a
semiconductor QR excited by a single HCP for different values of
temperature. The spectra are calculated for the charge
polarization dynamics shown in Fig.~\ref{Fig:revivals_PRB}. They
are offset vertically for clarity.
(b) Emission spectra for different values of the HCP kick
strengths $\alpha$. No SOI is present in this case. (c) Emission
spectra for different SOI strengths, quantified by the angle
$\gamma$ (see Fig.~\ref{rfig1}). In (b) $\gamma=0$ whereas in (c)
$\alpha=0.01$ is fixed; in both cases the frequency $\omega$ is
normalized by $\omega_0/2$, $\omega_0=\hslash/(m^*r_0^2)$ [cf.
Eq.~\eqref{hsoi1}], there are $N=100$ electrons in the QR at zero
temperature [(a) is adapted from
Ref.~\cite{Moskalenko_PRB2006}, (b) and (c) are adapted from
Ref.~\cite{zgzhu2008}].} \label{Fig:spectra_QR_rfig3}
\end{figure*}

Figure~\ref{Fig:spectra_QR_rfig3}c illustrates how the SOI
modifies the spectra at zero temperature in the low excitation
regime. Enahncing $|\gamma|$, which characterizes the SOI
strength, we observe a blue shift of the lower frequency peak and
a red shift of the higher frequency peak, until they merge at
$\gamma=-60^{\circ}$ (cf. also the left bottom panel of
Fig.~\ref{rfig2}). A further increase of the SOI strength (not
shown in Fig.~\ref{Fig:spectra_QR_rfig3}c) leads to the
reappearance of the peak splitting and their motion towards the
initial positions at $\gamma=0$.
%

In Refs.~\cite{Alex_PRB2004,Alex_PLA2004} stationary spectra of
semiconductor quantum rings excited by a HCP train were studied in
the limit case when the period of the train $T$ is much larger
than the relaxation time $\tau_{\rm rel}$. In such a case, one
easily finds:
\begin{equation}\label{Eq:mu_QR_kHCPs}
    |\mu_\omega|^2=\big|\mu^{\rm 1HCP}_\omega\big|^2\left[\frac{\sin(k\pi\omega/\omega_T)}{\sin(\pi\omega/\omega_T)}\right]^2\;,
\end{equation}
where $\mu^{\rm 1HCP}(t)$ is the time-dependent dipole moment
created in the QR by a single HCP.

\begin{figure*}[t!]
\centering
\includegraphics[width=8.0cm]{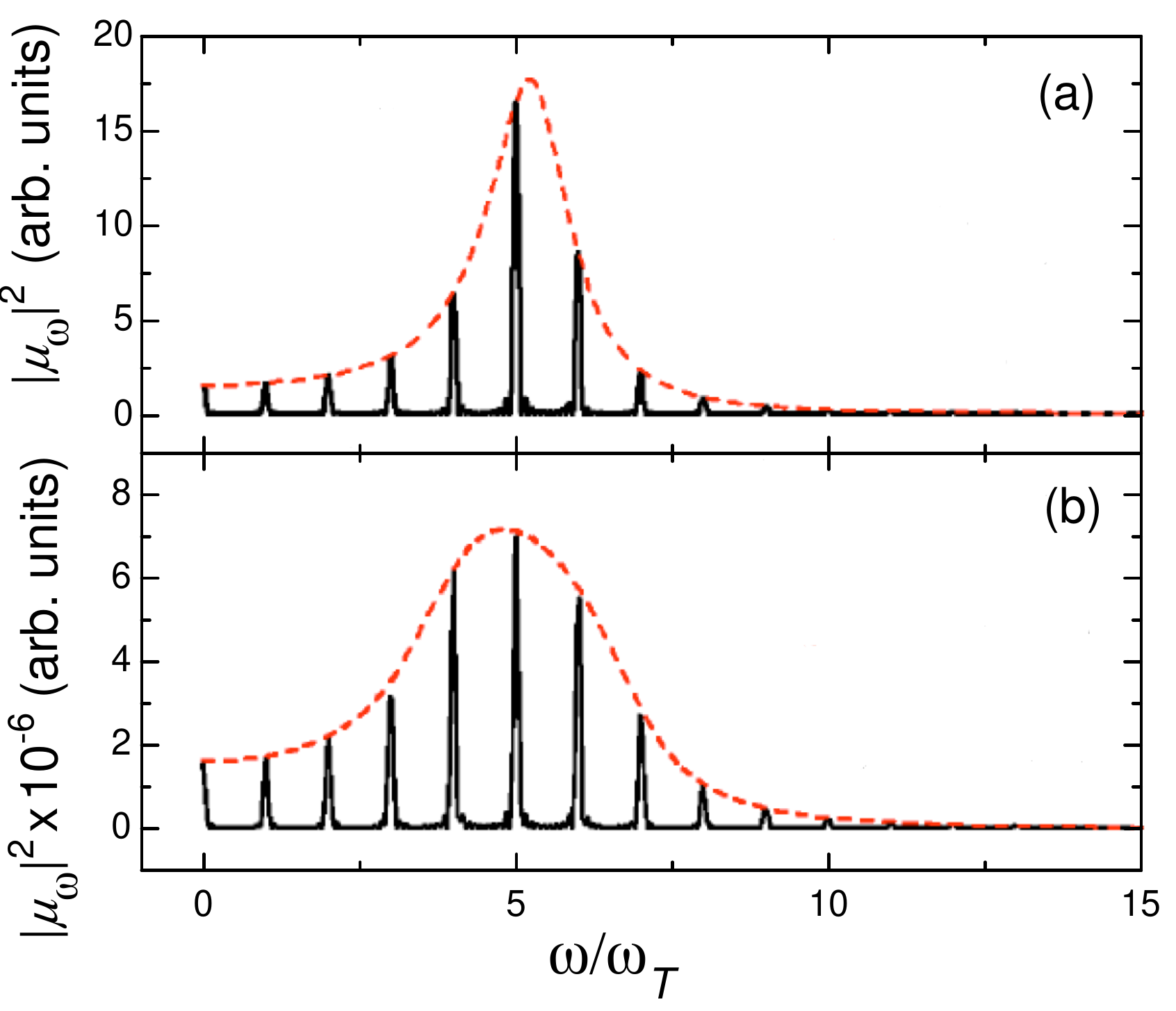}
\caption{Spectrum $|\mu_\omega|^2$ (full black line) of a
semiconductor QR driven by a train of HCPs with a period
$T=100$~ps. There are $k=10$ HCPs in the train. The dashed red
line shows the modulation function $k^2\big|\mu^{\rm
1HCP}_\omega\big|^2$. The underlaying dipole dynamics is
calculated after Eq.~\eqref{Eq:mu_general} with $\tau_{\rm
rel}=20$~ps and zero temperature. Other QR parameters:
$r_0=1.35~\mu$m, $m^*=0.067m_0$ and $N=1400$. The HCPs have the
sine-square temporal profile [see Eq.~\eqref{Eq:sine_pulse}], duration
$\tau_{\rm d}=1$~ps and peak electric field 1~V/cm in (a) and
1~kV/cm in (b) (adapted from Ref.~\cite{Alex_PRB2004}).}
\label{Fig:Fig7_Alex_PRB2004_QW}
\end{figure*}

Figure~\ref{Fig:Fig7_Alex_PRB2004_QW} shows $|\mu_\omega|^2$
calculated for a train with $k=10$ HCPs. One can observe peaks
positioned at multiples of $\omega_T=2\pi/T$. This reflects the
behavior of the second factor on the rhs of
Eq.~\eqref{Eq:mu_QR_kHCPs}. The heights of the peaks are modulated
by the function $k^2\big|\mu^{\rm 1HCP}_\omega\big|^2$,
representing the amplified spectrum of the QR excited by a single
HCP. Using a train of HCPs one can considerably amplify the
radiation intensity and, at the same time, via the modulation
function retrieve the spectrum generated by a single HCP.
Figure~\ref{Fig:Fig7_Alex_PRB2004_QW}a corresponds to the low
excitation regime ($\alpha\ll 1$) whereas
Fig.~\ref{Fig:Fig7_Alex_PRB2004_QW}b is calculated for the strong
excitation regime. In the latter case the broadening of the
modulation function is larger due to the reasons discussed above
for the single-pulse excitation.

\subsubsection{High-harmonic emission from  quantum rings driven by THz broadband pulses}\label{Sec:HHG}

As we could see above, the emission spectrum of HCP-driven QRs is
confined to a frequency range around the central frequency
$\omega_{_\mathrm{F}}=N\omega_0/4$, where
$\omega_0=\hslash/(m^*r_0^2)$. Interestingly, inspired by the rescattering model for high-harmonics generation \cite{Corkum1993}, one may expect a strong modulation of the emission spectra in a range  well beyond
$\omega_{_\mathrm{F}}$ due  to scattering impurities or engineering  a scattering/tunneling potential in the
QR. This idea was followed in Ref.~\cite{Hinsche2009}, where a QR with a localized impurity was considered.
Such an impurity can be realized, e.g., by cutting the QR at one
place \cite{Clark2008} or by fabricating a small part of the ring
from a material with a broader band gap. The single-particle
Hamiltonian for a free electron \eqref{Eq:Hamiltonian_ring} is
then modified by an additional angular potential
$V_{\mathrm{imp}}(\varphi)$, which for an impurity located at
$\varphi=0$ (see Fig.~\ref{Fig:ring}) can be modelled as
$V_{\mathrm{imp}}(\varphi)=\Omega \delta(\varphi)$,
where the barrier strength $\Omega$ is measured  in units of
energy times angle interval. For this model a semi-analytical treatment is possible \cite{Hinsche2009}.

In case of an excitation by a HCP with the electric field
component polarized along the $x$-axis, the emission spectrum is
practically the same as for a ring without the impurity. The reason lies in the symmetry of the
problem \cite{Hinsche2009}. The electron density driven by the HCP bounces off the
impurity that leads to the generation of the electromagnetic
field. However, in the far-field region the field generated by
electrons moving towards the impurity in the clockwise direction
and then scattered by it is compensated by the field generated by
electrons moving in the opposite direction.

\begin{figure*}[t!]
\centering
\includegraphics[width=10.0cm]{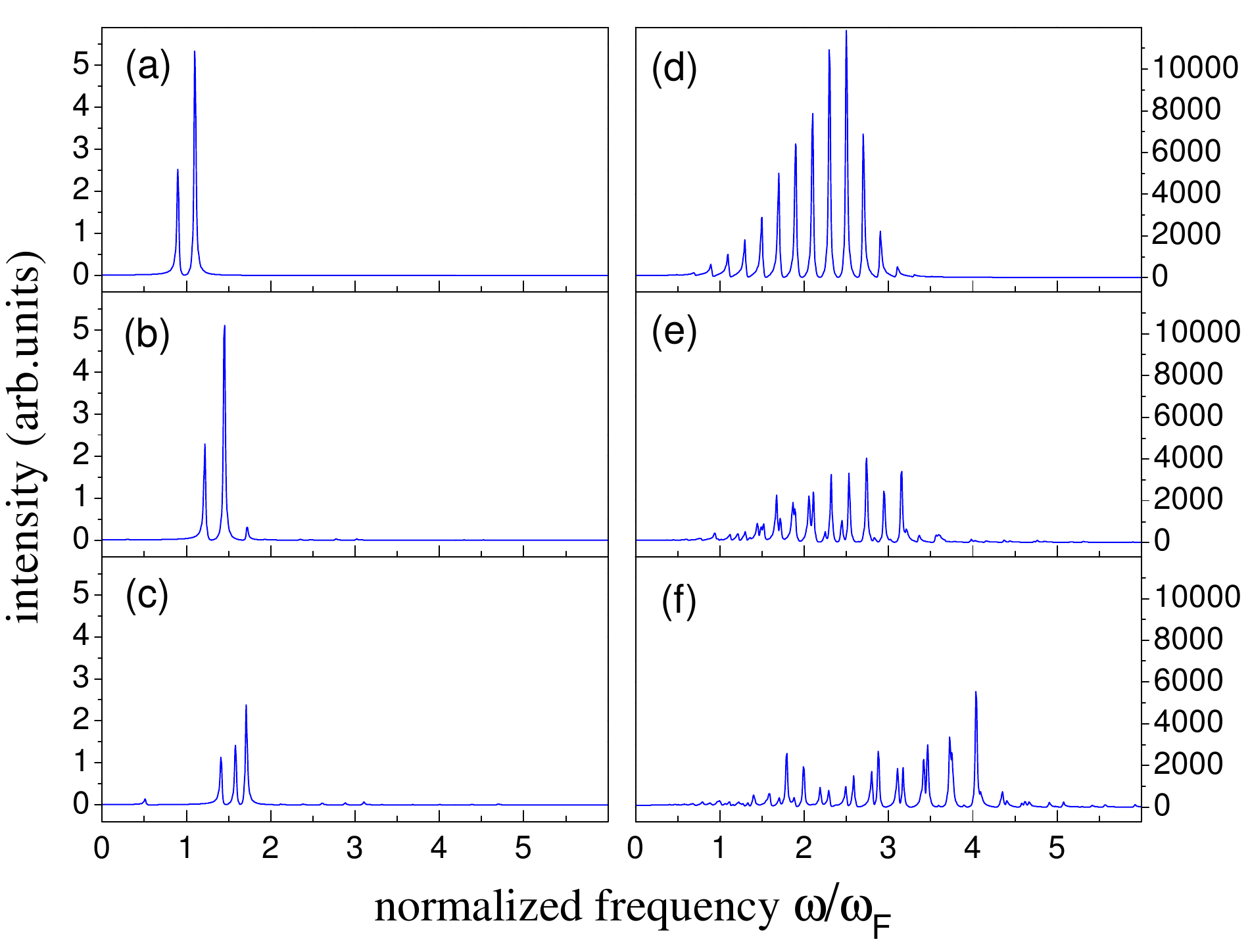}
\caption{Emission spectra of a HCP-driven QR with a localized
impurity for different values of the HCP kick strength $\alpha$
and impurity barrier strength $\Omega$: (a) $\alpha=0.1$,
$\Omega=0$; (b) $\alpha=0.1$, $\Omega=0.5$~meV\,rad; (c)
$\alpha=0.1$, $\Omega=10$~meV\,rad;  (d) $\alpha=10$, $\Omega=0$;
(e) $\alpha=10$, $\Omega=0.5$~meV\,rad; (f) $\alpha=10$,
$\Omega=10$~meV\,rad. The calculations are produced assuming the
dipole moment relaxation time $\tau_{\rm rel}=1$~ns and zero
temperature. Other QR parameters: $r_0=132~$nm, $m^*=0.067m_0$ and
$N=20$, resulting in $\omega_{_\mathrm{F}}=0.08~$THz (adapted from
Ref.~\cite{Hinsche2009}).} \label{Fig:Nicki_spectra1}
\end{figure*}

The situation changes if a HCP with the polarization along the
$y$-axis is used. Here the mirror symmetry of the nonequilibrium
electron density in respect to $\varphi=0$ is broken that leads to
the generation of coherent light bursts by the scattered
electrons. In Fig.~\ref{Fig:Nicki_spectra1} the emission spectrum
of the driven QR is shown for different HCP kick strengths
$\alpha$ and impurity barrier strengths $\Omega$. In
Figs.~\ref{Fig:Nicki_spectra1}a-c the
excitation strength is low ($\alpha=0.1$) and the barrier strength
is varied. In the absence of the impurity
(Fig.~\ref{Fig:Nicki_spectra1}a) we observe two peaks in the
spectrum, as already discussed above. The increase of the impurity
barrier strength leads to a blue shift of the peaks and the
appearance of additional peaks both below $\omega_{_\mathrm{F}}$ and
above it (see Figs.~\ref{Fig:Nicki_spectra1}b,c).

In the case of a strong excitation ($\alpha=10$), illustrated by
Figs.~\ref{Fig:Nicki_spectra1}d-f, we see
many peaks already without the impurity because many levels are
affected by the excitation [cf. Eq.~\eqref{energy2}]. This,
together with a relatively small number of electrons ($N=20$) in
the QR, is also the reason for the higher frequencies present in
the spectrum in respect to the low excitation case. The
introduction of the impurity enables transitions not only between
neighboring levels but also between distant levels affected by the
excitation. Therefore, apart from the blue shift of the spectrum
and appearance of additional frequencies in the frequency range of
Fig.~\ref{Fig:Nicki_spectra1}, also much higher frequencies can be
observed in the spectrum.

\begin{figure*}[t!]
\centering
\includegraphics[width=10.0cm]{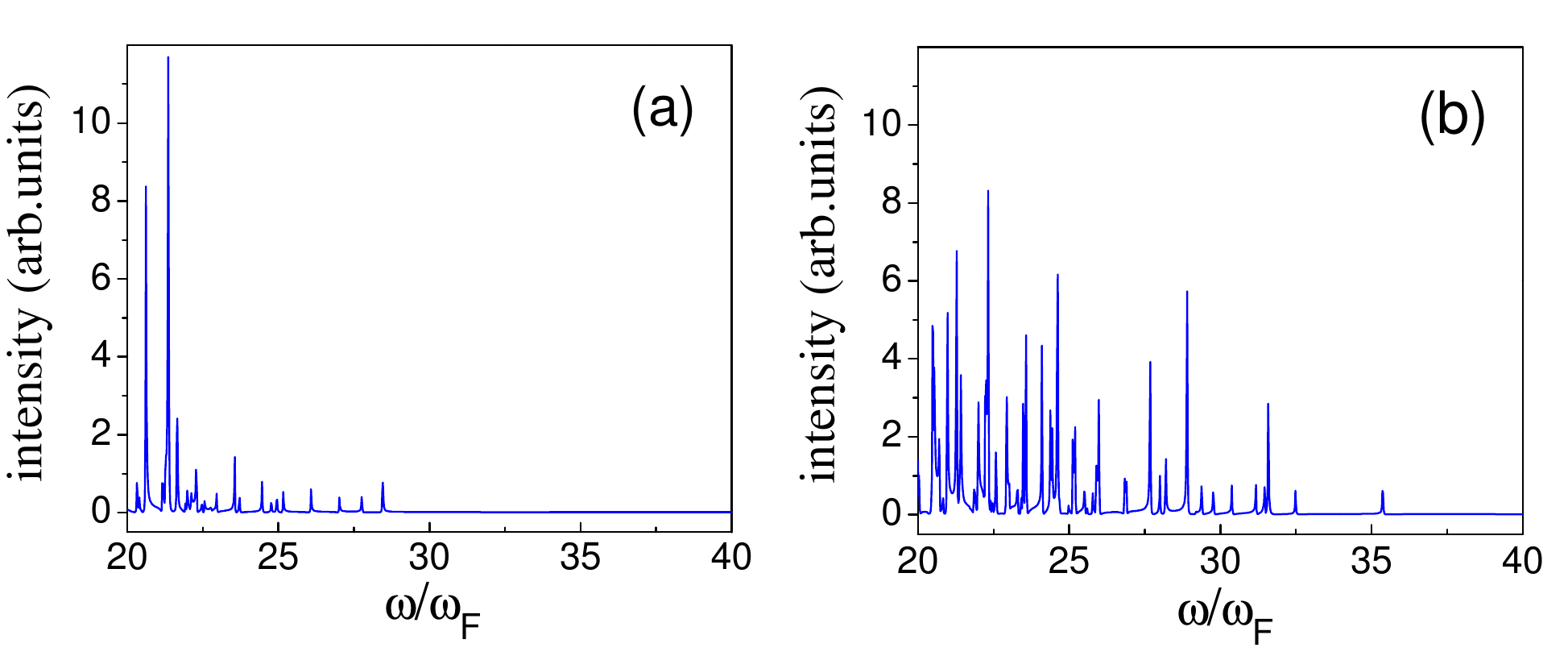}
\caption{High frequency range of the emission spectra
corresponding (a) to Fig.~\ref{Fig:Nicki_spectra1}e and (b) to
Fig.~\ref{Fig:Nicki_spectra1}f (adapted from
Ref.~\cite{Hinsche2009}).} \label{Fig:Nicki_spectra2}
\end{figure*}

In Fig.~\ref{Fig:Nicki_spectra2}a we can see that for
$\Omega=0.5$~meV\,rad frequencies up to around $28\omega_{_\mathrm{F}}$  and for $\Omega=10$~meV\,rad (and also larger impurity
barriers \cite{Hinsche2009}) frequencies up to around
$35\omega_{_\mathrm{F}}$ are present in the spectrum. For the
parameters of the simulations the latter value corresponds
approximately to 3~THz. Therefore, appropriately manufactured
arrays of the QR with impurities can be potentially used as
sources of coherent THz radiation \cite{Hinsche2009}. As far as
the high harmonics are absent for the $x$-polarized HCPs, the
effect could be also utilized for the detection of positions of
localized defects in QRs by measuring the emission spectrum while
rotating the HCP polarization.

\subsection{Time-dependent spectra}\label{Sec:TD_spectra}

\begin{figure*}[t!]
\centering
\includegraphics[width=12.0cm]{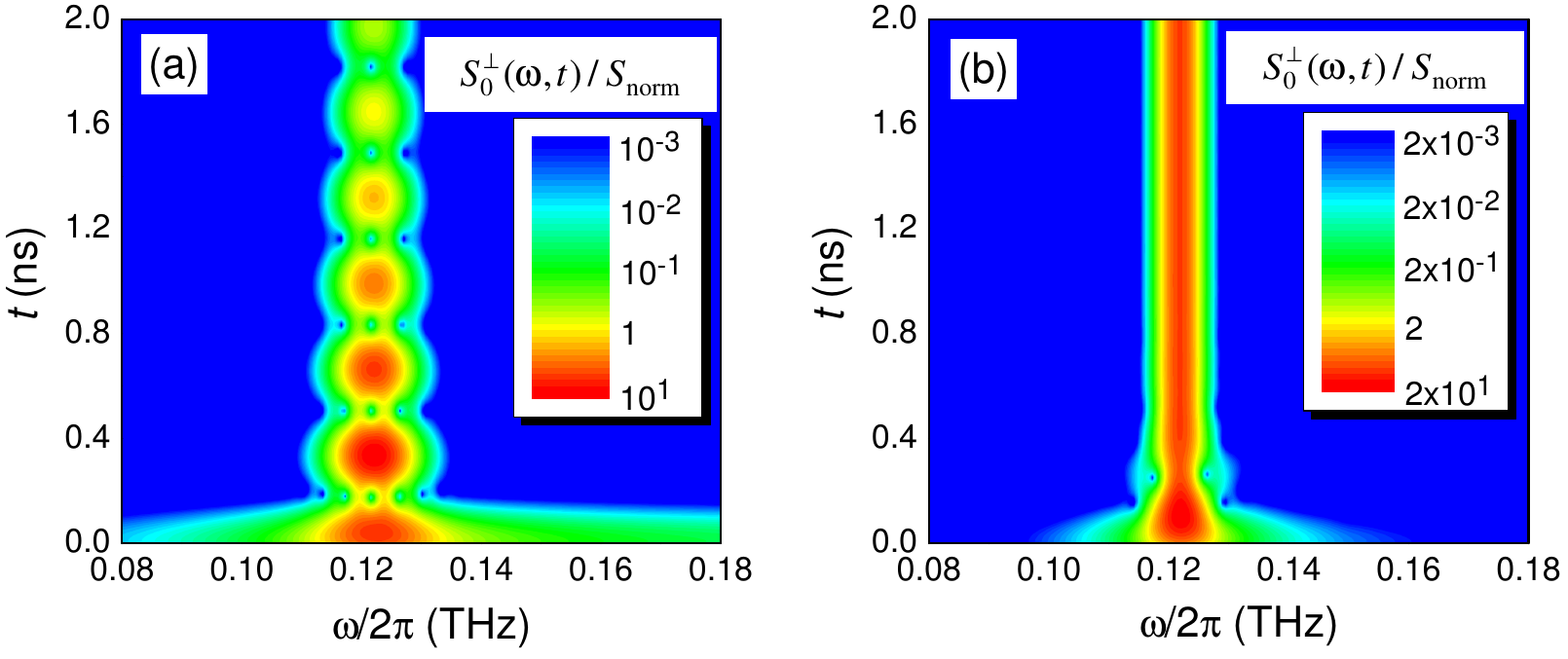}
\caption{Time-dependent spectra of a HCP-driven QR calculated
after Eq.~\eqref{Eq:S0_bot}. The spectra are
  normalized to $S_{\rm norm}=\frac{e^2r_0^2\omega_{\mathrm{F}}^3}{4\pi c^3}$. The QR
  is excited (a) by two mutually perpendicular HCPs with kick strength $\alpha=0.4$ and delay
  $T_{\rm cl}/4$ separating them in time and (b) by
  similar HCP pairs, but with kick strength $\alpha=0.1$, which
  are repeatedly applied with the period equal to $T_{\rm cl}$ (cf. Fig.~\ref{Fig:magnetic_sequences2}d).
  The temporal profile of the HCPs used in the
  simulations is determined by Eq.~\eqref{Eq:pulse_shape} with
  $\tau_{\rm d}=0.5$~ps and the peak electric field equal to 33.6~V/cm in (a) and 8.4~V/cm in (b). Detector
  time used in Eq.~\eqref{Eq:detector_function} amounts to $\Delta T=100$~ps. Parameters
  of the QR and phonon-induced relaxation are as in Fig.~\ref{Fig:current_control2},
  except for the temperature, which is $4$~K in (a) and $10$~K in (b) (adapted from Ref.~\cite{Moskalenko_PRA2008}).}
\label{Fig:TD_spectrum}
\end{figure*}

In Ref.~\cite{Moskalenko_PRA2008} the time-dependent
spectrum (see Appendix \ref{Sec:intensity_spectrum_def}) was studied for HCP-driven QRs.
Figure~\ref{Fig:TD_spectrum}a shows the time-dependent spectrum
for a QR after the excitation by a sequence of two HCPs, which are
polarized in the plane of the QR perpendicular to each other and
separated by the time interval $T_{\rm cl}/4$, where $T_{\rm
cl}/4$ is determined by Eq.~\eqref{Eq:fast_oscillations_time}. The
spectrum exhibits repeated  light bursts, which are centered in
frequency at approximately $\omega_{_\mathrm{F}}$ and have peak
values decaying with time due to the relaxation. The bursts
correspond to revivals of the charge polarization dynamics
\cite{Moskalenko_PRB2006}. The abrupt radiation switch-on, when
the charge polarization is generated by the HCPs, results in a
relatively broad spectrum at short times.
Figure~\ref{Fig:TD_spectrum}b illustrates that the emission can be
effectively stabilized (cf. Section~\ref{Sec:FL_approach}) if the QR
is driven by a periodic train of HCPs so that the time-dependent
spectrum reaches a practically stationary state.

\subsection{Ultrafast control of the circular polarization degree of the emitted radiation}
The driving by HCP pairs used in the numerical calculations of
Fig.~\ref{Fig:TD_spectrum} is such that it creates a rotating
dipole moment in the QR. The magnitude of the dipole moment
oscillates and decays in time for Fig.~\ref{Fig:TD_spectrum}a
whereas it is stabilized for Fig.~\ref{Fig:TD_spectrum}b. Such a
dynamics of the dipole moment leads to the circular polarization
of the radiation emitted in the direction perpendicular to the QR,
similar to the synchrotron radiation \cite{Ternov_review}. This
can be detected by measuring the corresponding time-dependent
degree of circular polarization $P_\mathrm{circ}^\bot(t)$ at any
time moment $t$ after the excitation (see
Appendix~\ref{Sec:Stokes} for an appropriate definition of this
quantity). For the frequency range of detection
$[0.5\omega_{_\mathrm{F}},1.5\omega_{_\mathrm{F}}]$ and detector
time $\Delta T=100$~ps, we find $P_\mathrm{circ}^\bot(t)> 0.99$ in
the case of the oscillating and decaying magnitude of the induced
rotating dipole moment in the QR (corresponding to
Fig.~\ref{Fig:TD_spectrum}a) and $P_\mathrm{circ}^\bot(t)> 0.999$
in the case of the stabilized magnitude (corresponding to
Fig.~\ref{Fig:TD_spectrum}b). In both cases an ultrafast
generation of the circularly polarized light from the QR is
realized.

\begin{figure*}[t!]
  \centering
  \includegraphics[width=7.0cm]{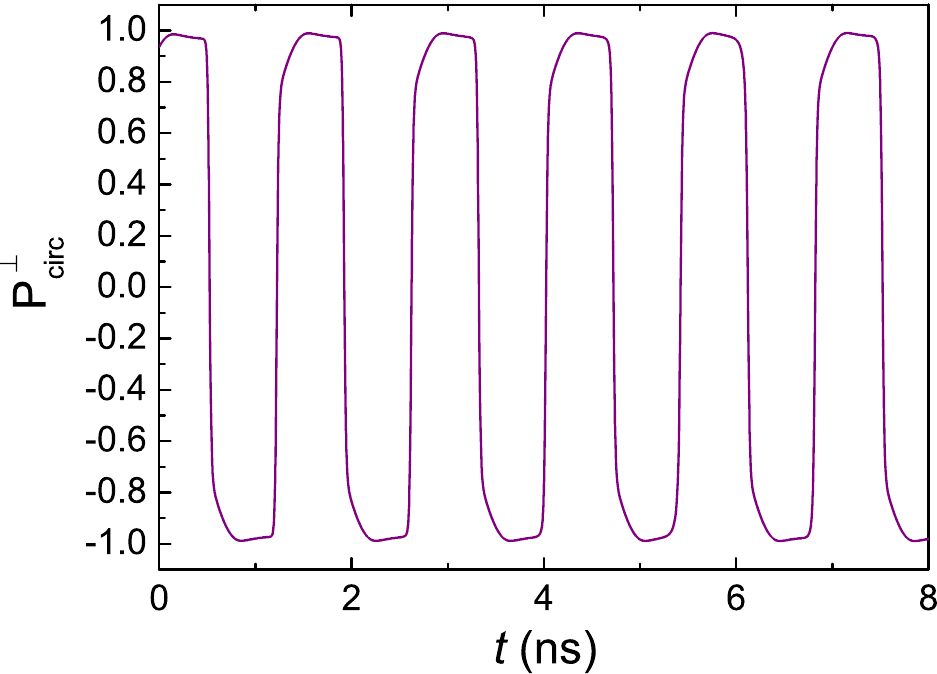}
  \caption{\label{Fig:circular_polarization} Time-dependent degree of circular polarization characterizing the emission
  of the HCP-driven QR.  The QR
  is driven by a periodic train of HCPs. There are four HCPs in each period of the train. The first (second) pulse pair in
  the period consists of two mutually perpendicular HCPs with kick strengths $\alpha_\|=0.2,\ \alpha_\bot=0.2$
  ($\alpha_\|=0.2,\ \alpha_\bot=-0.2$) and delay time
  $T_{\rm cl}/4$ between them (cf. Fig.~\ref{Fig:magnetic_sequences2}c).
  Time delay between the HCP pairs amounts to 0.7 ns (which is large in respect to $T_{\rm cl}/4$).
  Duration of each HCP is $\tau_{\rm d}=3$~ps.
  Detector time is $\Delta T=100$~ps. Temperature is $4$~K and the other QR and relaxation
  parameters are as in Fig.~\ref{Fig:revivals_PRB} (adapted from Ref.~\cite{Moskalenko_PRA2008}).}
\end{figure*}

As it was proposed in Ref.~\cite{Moskalenko_PRA2008}, by an
appropriate driving by HCPs the degree of circular polarization
can be manipulated on ultrafast time scales. For example, we can
design a train of HCPs so that each next pair of the HCPs in the
train changes the sense of rotation of the QR dipole moment. This
leads to the corresponding sign changes of
$P_\mathrm{circ}^\bot(t)$, as demonstrated in
Fig.~\ref{Fig:circular_polarization}. Thus the photons are emitted
in portions with an alternating helicity. Generally,  more
complicated pulse sequences allow to control the chirality of each
emitted photon portion.

\section{Correlated many-body systems driven by ultrashort pulses}\label{Sec:many_body}
As shown in detail in the preceding sections, an ultrashort HCP acting on an electronic system may result in a momentum shift of the states.
Choosing the pulse parameters appropriately, this may happen  without exciting any relevant optical phonon modes \cite{Kuebler2007,Kim2015}, leading
so to a net increase in the average kinetic energy with minor energy dissipation in form of the lattice heating.
Giving that many phenomena in electronic systems are determined by the competition between kinetic and correlation energies, these
pulses offer thus exciting opportunities to tune and disentangle the intrinsic competing interactions and trace
of these interactions unfold in time \cite{Kuebler2007,Kim2015,Tao2012,Liu2012,Schultze2013,Kampfrath2013,Aetukuri2013,Grupp2015,Mayer2015,Cueff2015}.
A paradigm example is the metal-insulator transition in correlated systems \cite{Mott1968,Imada1998,Georges1996,Kotliar2004}. In this context
 our scope here is very limited (for this particular case we refer to the  dedicated review article \cite{Aoki2014} and references therein)
 in that we shall  just elucidate how the ultrashort broadband pulses act on a correlated many-body state.
  After the pulse, the state evolves in a field-free manner and can be dealt with  by established methods appropriate to the nature of the system.

Let us consider the fully correlated ground state  of an $N$-electron system  $|\Psi_0(1,2,\ldots,N)\rangle$.
Selecting a suitable one-particle basis,  the semiclassical (i.e. treating the photon field classically but quantizing matter) \ light-matter
interaction in {the} second-quantized form and in the dipole approximation reads
$$H_\mathrm{int}(t)=
-\sum_{i,j}d_{ij}
E(t)\hat c_i^\dagger \hat c_j.$$
${\hat c_j^\dagger},\hat c_j$ are creation and annihilation operators of the orbital $j$. The applied electric field is linearly polarized and has otherwise the form  $E(t)=E_0f(t)$, where $E_0$ is its amplitude and $f(t)$ is its temporal profile.
 $d_{ij}$ are the transition dipole matrix elements between the respective orbitals in the
  polarization direction of the field.
Under the conditions when the IA is applicable for the description of the interaction of the broadband ultrashort light pulse with the considered correlated many-body system  (i.e. the system
evolution within the  pulse duration is marginal) and the pulse effectively transfers a non-vanishing momentum (e.g., a HCP),  the action of the pulse is determined by Eq.~\eqref{Eq:IA_mapping_psi} with
\begin{equation}
\label{eq:exp}
 \mathcal{U}_1=e^{i\sum_{i,j}S_{ij} \hat c_i^\dagger \hat c_j}\equiv e^{i\hat S}.
\end{equation}
Here $\mathcal{U}_1\equiv\mathcal{U}(t_{1})$, i.e. for the IA the result does not depend on the selection of the matching time moment $t_1$ within the pulse, and $S_{ij}=(1/\hslash) s_1d_{ij}E_0$, where $s_1$ is determined by Eq.~\eqref{Eq:s_n_infinity}.
In a time interval on the order of the pulse duration $\tau_\mathrm{d}$ after the pulse the state of the system is given by $|\Psi\rangle=\mathcal{U}_1|\Psi_0\rangle$.

One of the physically relevant quantities is the survival (recurrence) probability of the ground state that is the
same as the fidelity function $\mathcal{F}$ [see Eq.~\eqref{Eq:fidelity}], viewed here from a different perspective. E.g., imagine we start from
a magnetic system
whose magnetization is quantified for example by the Kerr spectroscopy. Applying a moderate intensity HCP and tracing a time lag after the HCP
the nonequilibrium magnetic properties by the time-resolved Kerr spectroscopy we may access to some extent
 the  survival, i.e. the  magnetization dynamics,  of the initial magnetic state.
  Another option is the time-resolved photoemission that maps the   spectral function
at the ground state energy while the system is evolving upon the excitation with  a HCP.
In any case,  the survival probability is given by the square of the overlap $\langle \Psi_0 |\Psi\rangle = \langle \Psi_0 |e^{i\hat S}|\Psi_0\rangle$. The latter is nothing else but   the \emph{quantum-mechanical} average of   $e^{i\hat S}$   which reads
\begin{equation}
\label{eq:cumulant}
\langle \Psi_0|\Psi\rangle=\exp\Bigl\{\frac{i}{1!}\mathfrak{S}_1
+\frac{i^2}{2!}\mathfrak{S}_2+\frac{i^3}{3!}\mathfrak{S}_3+\cdots\Bigr\}.
\end{equation}
Interestingly, the central moments which can also be
viewed as a measure for electronic correlations in the system  $\mathfrak{S}_1=\langle\hat S\rangle$, {$\mathfrak{S}_n=\big\langle(\hat S-\langle\hat S\rangle)^n\big\rangle$} (for $n>1$) with their established meanings are the essential ingredients
 of the HCP-induced survival probability. $\mathfrak{S}_n$ represent mean values of $n$-body operators and hence characterize $n$-particle correlations in the system.
 %
%

Equation \eqref{eq:cumulant}  evidences not only that we can tune the kinetic energy  by a HCP but also
by increasing $E_0$ we may allow step by step for higher order correlations, for $\mathfrak{S}_n$ scale with ${E_0}^n$, albeit
isolating these correlation is not possible. Some explicit calculations are provided in Ref.~\cite{Pavlyukh2014}.  Furthermore, higher order correlations may even dominate, even at low HCP intensity, e.g. in the vicinity of a continuous phase transition in which case
fluctuations dominate.

A further point is that when the HCP-system interaction is spin-active. As elucidated by Eq.~\eqref{Eq:Hamiltonian_SOI_transformed}
 and consequences thereof, in the presence of the SOI and/or if the magnetic component of the HCP  couples in a Zeeman-type to the  spin degrees of freedom of the system, the interaction with the HCP, given for example by Eq.~\eqref{h1t}, is still of a single-particle nature.
This means that in this case the operator $\hat S$ in Eq.~\eqref{eq:exp} becomes ($\sigma_j$ quantifies the spin of the single particle orbitals)
\begin{equation}
\label{eq:exps}
   \hat{S}=\sum_{i,\sigma_i\, j,\sigma_j}S_{i\sigma_i\, j\sigma_j} \hat c_{i\sigma_i}^\dagger \hat c_{j\sigma_j}\;,
\end{equation}
%
yielding so the post-pulse spin-dependent coherent state $|\Psi\rangle$ that evolved from the spin-active state $|\Psi_0\rangle$.
 Components $S_{i\sigma_i\, j\sigma_j}$ are given by the same expression as $S_{i, j}$ in the case of Eq.~\eqref{eq:cumulant},
 with the modification that the matrix elements  $d_{ij}$ are now spin-dependent
 and involve further terms, such as the Zeeman term in Eq.~\eqref{h1t}. In fact, one may exploit the conventionally different
 energy scales of  the charge and spin dynamics and design an  experiment in which the orbital motion
 remains frozen and zoom so into the magnetic dynamics. Obviously, Eq.~\eqref{eq:cumulant} applies in this case but the relevant dynamics in this situation is a pure magnetic one [i.e. Eq.~\eqref{eq:exps} should be used for $\hat S$]. Clearly, increasing the pulse intensity the orbital motion becomes more relevant and modifies the magnetic signal.

 Experimentally there has been an upsurge in activities in recent years  utilizing ultrashort, THz and/or broadband
 pulses to gain an insight into the  spin dynamics of spin-ordered systems \cite{Ulbricht2011,Kampfrath2011,Tanaka2011,Liu2012,Zaks2012,Vicario2013,Harold2015,Schubert2014}.

\section{Conclusion and outlook}\label{Sec:Conclusion}
The focus of this report has been on the theoretical aspects of the charge and spin dynamics in nanostructures driven by broadband pulses with a duration shorter than the typical
pulse-free time scales of relevant processes. We have seen that depending on their parameters
such pulses can shift the states in momentum or space.  In the presence of spin-orbit coupling the pulses provide also an effective mean for controlling the spin dynamics on
time scales much shorter than the field-free, precessional time. Further, they allow to access completely new phenomena, such as the pulse-induced generation of a pure spin current.
With the impressive advances in the generation and control of the pulses in shape, duration, and intensity, it is anticipated, and indeed evidenced by the numerous recent
exciting experiments on the driving and control of the  spin and charge  dynamics in condensed matter, that broadband, intense, ultrashort pulses are becoming  a major tool for the exploration of
material properties. In particular,  the behavior of complex materials of technological relevance is determined by a competition of spin and  orbital orders as well as of the kinetic energy vs. the electron-correlation-induced localization. Since  the pulses can be engineered as to address these degrees of freedom almost separately,  we expect a dynamic development of future research based on the corresponding ultrafast techniques.
 Further development of the reviewed approach towards description of the change in correlations upon excitation within the impulsive approximation as well as a general treatment of many-particle dynamics in nanostructures driven by ultrashort ultrabroadband pulses beyond the impulsive approximation (e.g. within the mean-field, Hartree-Fock, DFT, cluster expansion density matrix or Green's function approaches) is a fascinating and challenging goal for future research.

%

\section*{Acknowledgements}
The work was partly  supported by the DFG under SFB 762 and SFB 767 as well as SPP 1840  ``Quantum Dynamics in Tailored Intense Fields". Consultations and discussions with I. Barth, J.S. Briggs, H. D\"urr, E. Goulielmakis, E.K.U. Gross, R. Huber, C.L. Jia, A. Leitenstorfer, J. Manz, A. Matos-Abiague, Y. Pavlyukh, M. Sch\"uler, D.V. Seletskiy, A. Sukhov, J. W\"atzel are gratefully acknowledged.

\appendix

\section{Radiative damping in semiconductor quantum rings}\label{Sec:radiative_damping}

Let us consider the effect of the radiative damping that is
introduced by the back action of the emitted radiation on the
nonequilibrium carriers in a semiconductor QR. The emitting
electrons lose energy and this leads to the relaxation of the
density matrix. Here we are going to determine the corresponding
time scales. The Hamiltonian of the light-matter interaction in
the second-quantized form can be written as
\begin{equation}\label{Eq:H_D_start}
    \hat{H}_{\rm D}=-\int \!\!{\rm d}^3r\; \hat{\Psi}^\dagger(\mathbf{r},t)
    e\mathbf{r}\cdot \hat{\mathbf{E}}(\mathbf{r},t)
    \hat{\Psi}(\mathbf{r},t)
\end{equation}
where the quantized  {electric field} reads
\begin{equation}\label{Eq:E_quantized}
    \hat{\mathbf{E}}(\mathbf{r},t)=\sum_{\sigma
    \mathbf{q}}i\mathbf{e}_{\sigma\mathbf{q}}{\cal E}_q
    u_{\sigma\mathbf{q}}(\mathbf{r})\hat{b}_{\sigma\mathbf{q}}+\mbox{H.c.}
    \;
\end{equation}
with ${\cal E}_q=\sqrt{\frac{2\pi\hslash\omega_q}{\kappa V}}$. $\mbox{H.c.}$ denotes the Hermitian conjugate.
$\hat{b}_{\sigma\mathbf{q}}$ is the creation operator of the
electromagnetic field mode with the polarization vector
$\mathbf{e}_{\sigma\mathbf{q}}$ and mode function
$u_{\sigma\mathbf{q}}(\mathbf{r})$, $\omega_q=cq/\sqrt{\kappa}$ is
the frequency, $\mathbf{q}$ is the wave vector, $V$ is the
normalization volume, $c$ is the speed of light in vacuum, $e$ is
the electron charge and $\kappa$ is the dielectric constant of the
medium. The electronic field operator is given by
$\hat{\Psi}(\mathbf{r},t)=\sum_m
\hat{a}_m(t)\psi_{1,m}(\mathbf{r})$. The single-particle wave
functions $\psi_{1,m}(\mathbf{r})$ are determined by
Eq.~\eqref{Eq:ring_wave_function}. Substituting
Eq.~\eqref{Eq:E_quantized} into Eq.~\eqref{Eq:H_D_start} leads to
\begin{equation}\label{Eq:H_D}
    \hat{H}_{\rm D}=-\sum_{mm'}\sum_{\sigma \mathbf{q}}
    i{\cal
    E}_{q}d_{\sigma,mm'}(\mathbf{q})\hat{b}_{\sigma\mathbf{q}}\hat{a}_m^\dagger
    \hat{a}_{m'}^\dagger+\mbox{H.c.} \;,
\end{equation}
where  the
matrix element $d_{\sigma,mm'}(\mathbf{q})$ is given by
\begin{equation}\label{Eq:dipole_def}
    d_{\sigma,mm'}(\mathbf{q})=e\int\!\!{\rm d}^3 r\;
    \psi_{1,m}^*(\mathbf{r})u_{\sigma\mathbf{q}}(\mathbf{r})(\mathbf{e}_{\sigma\mathbf{q}}\cdot\mathbf{r})\psi_{1,m'}(\mathbf{r})\;.
\end{equation}

Apart from $\hat{H}_{\rm D}$,  to describe the radiative
damping, we have to take into account the free-photon Hamiltonian
\begin{equation}\label{Eq:H_free_photon}
    \hat{H}_0^{\rm phot}=\sum_{\sigma\mathbf{q}}\left(\hat{b}_{\sigma\mathbf{q}}^\dagger \hat{b}_{\sigma\mathbf{q}}+\frac{1}{2}\right)
\end{equation}
and the free-carrier Hamiltonian $\hat{H}_0^{\rm el}$ which is
given by Eq.~\eqref{Eq:H0_carr}. The total Hamiltonian governing
the dynamics of the electrons in the QR and photons is given by
$\hat{H}^{\rm tot}=\hat{H}_0^{\rm el}+\hat{H}_0^{\rm
phot}+\hat{H}_{\rm D}$. Writing the Heisenberg equations of motion
corresponding to $\hat{H}^{\rm tot}$ for operators
$\hat{b}_{\sigma \mathbf{q}}$ and $\hat{a}_m^\dagger \hat{a}_{m'}$
we arrive at the following equations
\begin{equation}\label{Eq:b}
    i\hslash
    \frac{\partial}{\partial t}\hat{b}_{\sigma\mathbf{q}}=\hslash\omega_{q}\hat{b}_{\sigma\mathbf{q}}+\sum_{mm'}i{\cal
    E}_{q}^*d_{\sigma,m'm}^*(\mathbf{q})\hat{a}_m^\dagger \hat{a}_{m'},
\end{equation}
\begin{equation}\label{Eq:aa}
  \begin{split}
    i\hslash
    \frac{\partial}{\partial t}\hat{a}_m^\dagger
    \hat{a}_{m'}\!=&(\varepsilon_{m'}-\varepsilon_m)\hat{a}_m^\dagger\hat{a}_{m'}\\
    &\hspace{-0.1cm}+\hspace{-0.1cm}\sum_{m''\sigma\mathbf{q}}\hspace{-0.1cm}\left[i{\cal
    E}_{q}d_{\sigma,m''m}(\mathbf{q})\hat{b}_{\sigma\mathbf{q}}\right.\\
                                  &\hspace{1.2cm}\left.-i{\cal E}_{q}^*d_{\sigma,mm''}^*(\mathbf{q})\hat{b}^\dagger_{\sigma\mathbf{q}}\right]
                                  \hat{a}^\dagger_{m''}\hat{a}_{m'}\\
    &\hspace{-0.1cm}-\hspace{-0.1cm}\sum_{m''\sigma\mathbf{q}}\hspace{-0.1cm}\left[i{\cal
    E}_{q}d_{\sigma,m'm''}(\mathbf{q})\hat{b}_{\sigma\mathbf{q}}\right.\\
                                  &\hspace{1.2cm}\left.-i{\cal E}_{q}^*d_{\sigma,m''m'}^*(\mathbf{q})\hat{b}^\dagger_{\sigma\mathbf{q}}\right]
                                  \hat{a}^\dagger_{m}
                                  \hat{a}_{m''}.
  \end{split}
\end{equation}
Equation~\eqref{Eq:aa} contains coupling between $\hat{b}_{\sigma
\mathbf{q}}$ and $\hat{a}_m^\dagger \hat{a}_{m'}$ and therefore
the solution of the operator equation system of Eqs.~\eqref{Eq:b}
and \eqref{Eq:aa} requires additional approximations.

\subsection{Classical radiation
contribution}\label{Sec:radiative_damping_classical}

The system of Eqs.~\eqref{Eq:b} and \eqref{Eq:aa} can be
simplified to an ordinary differential equation system by making
use of the semiclassical approximation. In the framework of this
approximation we allow only for classical (coherent) electric
fields. This is equivalent to the following factorization of the
mean value of the operator $\hat{b}_{\sigma\mathbf{q}}
\hat{a}^\dagger_m\hat{a}_{m'}$:
\begin{equation}\label{Eq:decompos_semiclass}
   \langle \hat{b}_{\sigma\mathbf{q}} \hat{a}^\dagger_m\hat{a}_{m'} \rangle=
   \langle \hat{b}_{\sigma\mathbf{q}}\rangle \langle \hat{a}^\dagger_m\hat{a}_{m'}
   \rangle\;.
\end{equation}
We take the mean values of Eqs.~\eqref{Eq:b} and \eqref{Eq:aa},
introduce simplified notations $b_{\sigma\mathbf{q}}=\langle
\hat{b}_{\sigma\mathbf{q}}\rangle$ and
$\rho_{mm'}=\langle\hat{a}^\dagger_m\hat{a}_{m'}\rangle$, write an
implicit solution of Eq.~\eqref{Eq:b} as
\begin{equation}\label{Eq:b_integrated}
 b_{\sigma\mathbf{q}}(t)=\frac{1}{\hslash}\int_0^t\!\! {\rm d}t'\;
 {\rm e}^{i\omega_q(t'-t)}\sum_{mm'}{\cal
 E}_q^*d_{\sigma,m'm}^*(\mathbf{q})\rho_{mm'}(t')
\end{equation}
and insert it into Eq.~\eqref{Eq:aa}. Here we have assumed that
before the time moment at $t=0$ the electronic system is unexcited
and there is no coherent light field.  The resulting
integrodifferential equation fully describes the radiative damping
dynamics in the case of classical fields. This equation can be
further simplified taking into account that the decay of the
density matrix induced by the radiative damping takes place on
considerably longer time scales than the characteristic electronic
time scale of the QR  given by the ballistic time $\tau_{_\mathrm{F}}$.
This leads to a so-called adiabatic approximation
\cite{Lehmberg1969,Milonni1975,Khoo1976}. Rewriting
Eq.~\eqref{Eq:b_integrated} in the framework of this
approximation, inserting it into Eq.~\eqref{Eq:aa}, neglecting
small renormalizations of the single-particle energies
\cite{Rossi_Kuhn}, and producing the summation over $\mathbf{q}$
and $\sigma$ in the dipole approximation, we obtain a system of
ordinary differential equations for the density matrix components:
\begin{equation}\label{Eq:radiative_damping_full}
  \begin{split}
    \frac{{\rm d}\rho_{mm'}}{{\rm d} t}\!=&-\frac{i}{\hslash}(\varepsilon_{m'}-\varepsilon_m)\rho_{mm'}\\
    &-s(\rho_{m-1m'}\!-\rho_{mm'+1})\sum_\nu(\varepsilon_\nu\!-\varepsilon_{\nu-1})^3\rho_{\nu\nu-1}\\
    &-s(\rho_{mm'-1}\!-\rho_{mm'-1})\sum_\nu(\varepsilon_\nu\!-\varepsilon_{\nu-1})^3\rho_{\nu-1\nu}\;,
  \end{split}
\end{equation}
where
\begin{equation}\label{Eq:s}
    s=\frac{1}{3}\sqrt{\kappa}\frac{e^2r_0^2}{\hslash^4c^3}\;
\end{equation}
and $r_0$ is the QR radius (see Fig.~\ref{Fig:ring}).

It can be seen that the diagonal density matrix components, which
determine the charge current $I$ flowing in the QR, do not possess
an own decay dynamics, as expected for the case of the classical
radiation. They are driven only by the near-diagonal components,
which dictate the charge polarization $\boldsymbol{\mu}$ of the
QR. The decay rate of $\boldsymbol{\mu}$ can be obtained from
Eq.~\eqref{Eq:radiative_damping_full} if we assume that the
density matrix is only slightly disturbed from the equilibrium
diagonal Fermi-Dirac distribution, the number of electrons $N$ in
the ring is large, and the temperature $T$ is much lower than the
electron energy at the Fermi level. In this case we can calculate
the decay rate of the charge polarization as
\begin{equation}\label{Eq:gamma}
    \gamma=\frac{1}{6}\sqrt{\kappa}\frac{e^2\omega_{\mathrm{F}}^2}{m^*c^3}N\;,
\end{equation}
where $\omega_{_\mathrm{F}}=2\pi/\tau_{_\mathrm{F}}$. The value of $\gamma$
was estimated for
electrons in GaAs with $m^*=0.067m_0$ and $\kappa=12.5$
\cite{Moskalenko_PRA2008}. The estimation gives
$\gamma=1.5\times10^2~{\rm s}^{-1}$ for $r_0=1.35~\mu$m, $N=400$
and  $\gamma=0.4\times 10^4~{\rm s}^{-1}$ for $r_0=0.3~\mu$m,
$N=160$.

\subsection{Spontaneous emission contribution}\label{Sec:radiative_damping_spontaneous}
In order to describe contributions to the radiative damping beyond
the semiclassical approximation, correlations between the electron
and field operators should be taken into account. We need to consider
the correction to Eq.~\eqref{Eq:decompos_semiclass}
\begin{equation}\label{Eq:baa_def}
   t_{\sigma\mathbf{q}}^{mm'}\equiv\Delta\langle \hat{b}_{\sigma\mathbf{q}} \hat{a}^\dagger_m\hat{a}_{m'}
   \rangle\equiv \langle \hat{b}_{\sigma\mathbf{q}} \hat{a}^\dagger_m\hat{a}_{m'}
   \rangle -
   \langle \hat{b}_{\sigma\mathbf{q}}\rangle \langle \hat{a}^\dagger_m\hat{a}_{m'}
   \rangle\;.
\end{equation}
Then we write the equation of motion for
$\hat{b}_{\sigma\mathbf{q}} \hat{a}^\dagger_m\hat{a}_{m'}$ and
calculate the mean value of it. After this we utilize the
Hartree-Fock approximation for the electrons writing
$\langle\hat{a}_{m_1}^\dagger \hat{a}_{m_2}^\dagger \hat{a}_{m_3}
\hat{a}_{m_4} \rangle=\rho_{m_1 m_4}\rho_{m_2 m_3}-\rho_{m_1
m_3}\rho_{m_2 m_4}$, take into account Eqs.~\eqref{Eq:b} and
\eqref{Eq:aa}, and neglect correlations including two photon
operators (neglecting in this way the influence of the incoherent
photon density and two-photon processes). In that way we arrive at
\begin{equation}\label{Eq:baa_dyn}
  \begin{split}
    i\hslash \frac{\partial}{\partial t}t_{\sigma\mathbf{q}}^{mm'}
    =&(\varepsilon_{m'}-\varepsilon_m+\hslash\omega_q)t_{\sigma\mathbf{q}}^{mm'} \\
    &+\sum_{\nu\nu'}i{\cal E}_{q}d_{\sigma,\nu \nu'}^*(\mathbf{q})\rho_{m\nu}\bar{\rho}_{\nu'm'},
 \end{split}
\end{equation}
where we have introduced the notation
$\bar{\rho}_{mm'}=\delta_{mm'}-\rho_{mm'}$. As for the case of the
classical radiation contribution, we solve Eq.~\eqref{Eq:baa_dyn}
in the adiabatic approximation assuming vanishing
$t_{\sigma\mathbf{q}}^{mm'}$ before the excitation,
insert the solution to the mean value of Eq.~\eqref{Eq:aa}, take
the sum over $\mathbf{q}\;$ and $\sigma$ in the dipole
approximation, and obtain finally the following correction to the
rhs of Eq.~\eqref{Eq:radiative_damping_full}:
\begin{equation}\label{Eq:rho_sp_general}
  \begin{split}
     \frac{{\rm d}\rho_{mm'}}{{\rm d} t}\Big|_{\rm sp}
     \!=\sum_{\nu}\!\!\sum_{j=\pm 1}W_{\nu+j,\nu}&
     \left(\bar{\rho}_{\nu,m'}\rho_{m+j,\nu+j}+\bar{\rho}_{m,\nu}\rho_{\nu+j,m'+j}\right.\\
     &\hspace{-0.3cm}\left.-\bar{\rho}_{m-j,\nu}\rho_{\nu+j,m'}-\bar{\rho}_{\nu,m'-j}\rho_{m,\nu+j}\right)\!,
 \end{split}
\end{equation}
where
\begin{equation}\label{Eq:W_def}
  \begin{split}
    W_{\nu\nu'}=\frac{s}{2}
    (\varepsilon_\nu-\varepsilon_{\nu'})^3
    \Theta(\varepsilon_\nu-\varepsilon_{\nu'})\;,
   \end{split}
\end{equation}
$s$ is determined by Eq.~\eqref{Eq:s} and $\Theta(\varepsilon)$
denotes the Heaviside step function.

Equation system following from the inclusion of \eqref{Eq:rho_sp_general} is
nonlinear and, in general, should be solved numerically. In the
weak excitation regime, we may represent the density matrix as
\begin{equation}\label{Eq:rho_decomposition}
    \rho_{mm'}=f^0_{m}\delta_{mm'}+\tilde{\rho}_{mm'},
\end{equation}
where $f^0_{m}$ is the equilibrium distribution and
$\tilde{\rho}_{mm'}$ is a small correction
caused by the excitation. If only the electron-photon interaction
is taken into account the equilibrium distribution corresponds to
the zero-temperature Fermi-Dirac distribution. More correctly,
it should be determined by the equilibrium with the phonon bath having
temperature $T$ (see Appendix \ref{Sec:relaxation_phonons}) and
obeys  the corresponding finite temperature Fermi-Dirac
distribution. Inserting Eq.~\eqref{Eq:rho_decomposition} into
Eq.~\eqref{Eq:rho_sp_general} and keeping only first order terms
in $\tilde{\rho}_{mm'}$, we find
\begin{equation}\label{Eq:rho_sp_result}
  \begin{split}
   \frac{{\rm d}\tilde{\rho}_{mm'}}{{\rm d}t}\!\Big|_{\rm
   sp}=
   &-s\!\sum_{j=\pm
   1}\!(Q_m^{m+j}+Q_{m'}^{m'+j})\tilde{\rho}_{mm'}\\
   &+s\!\sum_{j=\pm
   1}\!(Q_{m+j}^m+Q_{m'+j}^{m'})\tilde{\rho}_{{m+j}\;{m'+j}}\;,
 \end{split}
\end{equation}
where the coefficients $Q^m_{m'}$ are given by
\begin{equation}\label{Eq:Q}
  \begin{split}
   Q^m_{m'}=& (\varepsilon_m-\varepsilon_{m'})^3f_m^0\; , \hspace{1.1cm} \varepsilon_m \geq \varepsilon_{m'}; \\
   Q^m_{m'}=& (\varepsilon_m'-\varepsilon_{m})^3(1-f_m^0)\; , \ \ \varepsilon_m < \varepsilon_{m'}\;.
  \end{split}
\end{equation}

Equation \eqref{Eq:rho_sp_result} can be used directly for a
numerical simulation of the density matrix relaxation dynamics
associated with the spontaneous emission of radiation. However,
for the non-diagonal density matrix elements, which induce the
coherent part of radiation, a further approximation is possible
under certain conditions. If we consider the density matrix
components in the vicinity of the Fermi level, where $m\approx
m_{_\mathrm{F}}$, one can notice that the dynamics of the terms in the
second line of Eq.~\eqref{Eq:rho_sp_result} is dominated by an
oscillation with a frequency which is different from the frequency
$(\varepsilon_m-\varepsilon_{m'})/\hslash$ of the dominant
oscillation of $\rho_{mm'}$ by around
$\omega_{_\mathrm{F}}|m-m'|/m_{_\mathrm{F}}$. Therefore these terms are averaged out to zero on the time
scale exceeding $\tau_{_\mathrm{F}}m_{_\mathrm{F}}/|m-m'|$ if it is still
shorter than the relaxation time scale. We should only analyze the negative coefficients in the diagonal
term in the first line on the rhs of
Eq.~\eqref{Eq:rho_sp_result}, denoting
\begin{equation}\label{Eq:gamma_sp_def}
  \begin{split}
   \gamma^{\rm sp}_{mm'}\equiv s\sum_{j=\pm
   1}\!(Q_m^{m+j}+Q_{m'}^{m'+j}).
 \end{split}
\end{equation}
From this equation, we can approximately calculate the effective
relaxation rate of the dipole moment due to the spontaneous
emission $\gamma^{\rm sp}_\mathrm{eff}$. This rate is determined by
the decay of the near-diagonal elements $\rho_{m,m-1}$ of the
density matrix close to the Fermi level. Using Eq.~\eqref{Eq:Q} in
Eq.~\eqref{Eq:gamma_sp_def}, setting $m'=m-1$, and summing over
$m$, we find
\begin{equation}\label{Eq:rho_sp_result_rate}
  \begin{split}
   \gamma^{\rm sp}_\mathrm{eff}=\gamma \sum_{m>0}&
   (f^0_{m-1}-f^0_{m})\\
   &\times(2+f^0_{m+1}+f^0_{m}-f^0_{m-1}-f^0_{m-2})\;,
 \end{split}
\end{equation}
where $\gamma$ is given by Eq.~\eqref{Eq:gamma}. One can see that
for $T\gg \hslash\omega_{_\mathrm{F}}/k_{\rm B}$ the effective
spontaneous decoherence rate $\gamma^{\rm sp}_\mathrm{eff}$
reaches $2\gamma$. As it follows from
Eq.~\eqref{Eq:rho_sp_result_rate}, with decrease of $T$ below
$\hslash\omega_{_\mathrm{F}}/k_{\rm B}$ this decoherence rate decreases
rapidly towards zero if the Fermi level is located between the QR
levels. It decreases towards $\gamma/2$ if the Fermi level
coincides with one of the single-particle levels so that the
latter is half-filled. Most importantly, in all cases it does not
exceed the value $2\gamma$.

\section{Relaxation by interaction with phonons in semiconductor quantum rings}\label{Sec:relaxation_phonons}
In the case of a semiconductor QR, the electron-phonon interaction
Hamiltonian has the form
\begin{equation}\label{Eq:H_P}
    \hat{H}_{\rm P}=\sum_{\mathbf{q},m,m'}{\cal G}_{\sigma\mathbf{q}}^{m,m'}\hat{c}_{\sigma\mathbf{q}}
    \hat{a}_m^\dagger \hat{a}_{m'}+{\rm H.c.}\
\end{equation}
and the free-phonon Hamiltonian $\hat{H}_0^{\rm phon}$ is given by
\begin{equation}\label{Eq:H0_phon}
  \hat{H}_0^{\rm phon}
  = \sum_{\sigma\mathbf{q}} \hslash \omega^{\rm s}_{\sigma\mathbf{q}}
  \left( \hat{c}^{\dagger}_{\sigma\mathbf{q}} \hat{c}_{\sigma\mathbf{q}}
  + \frac{1}{2} \right),
\end{equation}
where $\;\omega^{\rm s}_{\sigma\mathbf{q}}\;$ is the frequency of
a phonon with momentum $\mathbf{q}$ and polarization $\sigma$;
$\hat{c}_{\sigma\mathbf{q}}^\dagger$ and
$\hat{c}_{\sigma\mathbf{q}}$ are the phonon creation and
annihilation operators, respectively. The electron-phonon coupling
constant ${\cal G}_{\sigma\mathbf{q}}^{m,m'}$ for electrons
confined in the ring is given by
\begin{equation}\label{Eq:electron_phonon_coupling_constant}
  \begin{split}
    {\cal G}_{\sigma\mathbf{q}}^{m,m'}={\cal G}_{\sigma\mathbf{q}}^{\rm bulk}\int\!\!\mbox{d}^3r
    \ \psi_{1,m}^*(\mathbf{r})\mbox{e}^{i\mathbf{q}\mathbf{r}}\psi_{1,m'}(\mathbf{r})
  \end{split}\;,
\end{equation}
where $\psi_{1,m}(\mathbf{r})$ is determined by
Eq.~\eqref{Eq:ring_wave_function}. The expression for the bulk
electron-phonon coupling constant ${\cal
G}_{\sigma\mathbf{q}}^{\rm bulk}$ depends on the type of phonons.
In the case of a relatively weak excitation, which is considered
here, optical phonons can not provide a source for the relaxation
process because their energy is larger than the electron
excitation energies.
Here we consider only the relaxation of the excited electronic
states in QRs due to the scattering by longitudinal acoustic (LA)
phonons and omit the index $\sigma$.

We find that ${\cal G}_{\mathbf{q}}^{m,m'}$ depends only on $m-m'$
so that we can use the notation ${\cal G}_{\mathbf{q}}^{m-m'}$ in
place of ${\cal G}_{\mathbf{q}}^{m,m'}$ and recast the
carrier-phonon interaction Hamiltonian as
\begin{equation}\label{Eq:H_P_rewritten}
    \hat{H}_{\rm P}=\sum_{\mathbf{q},m,m'}{\cal G}_{\mathbf{q}}^{m^\prime}\hat{b}_{\mathbf{q}}
    \hat{a}_m^\dagger \hat{a}_{m-m'}+{\rm H.c.}\ .
\end{equation}

\subsection{Coherent wave contribution}\label{Sec:phonon_damping_coherent}

Similar steps and arguments as in the case of the radiative
damping, discussed in Appendix \ref{Sec:radiative_damping}, lead
to the equation
\begin{equation}\label{Eq:coh_phon_dyn_dens_mat}
  \begin{split}
    \frac{{\mathrm d}{\rho}_{mm'}}{{\mathrm d}t}=-\frac{i}{\hslash}&(\varepsilon_{m'}-\varepsilon_m)\rho_{mm'}\\
    -\sum_{\nu\nu'}\Big\{ & \big[M_{\nu'-\nu}(\varepsilon_\nu-\varepsilon_{\nu'})-M_{\nu'-\nu}(\varepsilon_{\nu'}-\varepsilon_\nu)\big]\\
    &\times \rho_{\nu\nu'} \left(\rho_{m+\nu'-\nu,m'}-\rho_{m,m'+\nu-\nu'}\right)\Big\}\;,
  \end{split}
\end{equation}
where the coupling coefficient $M_{m}(\varepsilon)$ is given by
\begin{equation}\label{Eq:M_elements}
   M_{m}(\varepsilon)=\frac{\pi}{\hslash}\sum_{\mathbf{q}}\left|{\cal G}_{\mathbf{q}}^{m}\right|^2
   \delta(\varepsilon-\hslash \omega^{\rm s}_{\mathbf{q}})\;.
\end{equation}
In this case the summation over phonon wave vectors $\mathbf{q}$
is more cumbersome than in the case of the radiation damping. In
contrast to photons, which couple only neighboring electronic
states, acoustic phonons can couple electronic states occuring in the
range covered by the largest possible phonon energy. The coupling
coefficient $M_{m}(\varepsilon)$ was analyzed under certain
assumptions and used in the calculations of
Refs.~\cite{Moskalenko_PRB2006,Moskalenko_PRA2008}. In particular, the Debye
model for the acoustic phonon spectrum with an isotropic linear
dispersion relation and the radial wave functions $R_l(r)$
corresponding to a quantum well with infinite energy barriers were
assumed. Only transitions between the
electronic states in the neighborhood of the Fermi level are relevant.
In the resulting model and for typical parameters of semiconductor
QRs, the coefficient $M_m(\varepsilon)\equiv M(\varepsilon)$ does
not depend on the quantum number $m$, i.e. on the transferred
angular momentum, if both involved electronic states belong to the
same energy branch (have the same sign of the angular quantum
number). Under such conditions one can find \cite{Moskalenko_PRB2006,Moskalenko_PRA2008}
\begin{equation}\label{Eq:typical_integrals6}
  \begin{split}
    M(\varepsilon)&=\frac{1}{\tau_{\rm LA}}
    F\!\left(\frac{\varepsilon d}{\hslash c_{\mathrm LA}}\right)\Theta(\varepsilon)\Theta(\hslash\omega_{_\mathrm D}-\varepsilon)\; ,
  \end{split}
\end{equation}
where $\Theta(\varepsilon)$ denotes the Heaviside step function,
$c_{{\mathrm LA}}$ is the LA-velocity of sound, the function
$F(x)$ is defined as
\begin{equation}\label{Eq:CI_def}
  F(x)=8\pi^2x\int_0^{x}\!\!{\rm d}x'\;
    \frac{1}{\sqrt{1-x'^2/x^2}}\;
    \frac{\sin^2\!(x'/2)}{x'^2\left[x'^2-(2\pi)^2\right]^2}\;\;,
\end{equation}
and the time constant $\tau_{\rm LA}$ is given by
\cite{Moskalenko_PRB2006,Moskalenko_PRA2008}
\begin{equation}\label{Eq:tau_LA}
  \tau_{\rm LA}=\frac{\hslash c_{\rm LA}^2\rho_{\mathrm s} d^2
  r_0}{|D|^2}\; .
\end{equation}
Here $\omega_{_{\mathrm D}}$ is the Debye frequency, $D$ is the
deformation constant, $\rho_{\mathrm s}$ is the lattice density
and $d$ is the QR radius (see Fig.~\ref{Fig:ring}).
$M(\varepsilon)$ vanishes for the states from different energy
branches.

In the weak excitation regime, only diagonal and near-diagonal
density matrix elements in the neighborhood of the Fermi level are
important and we can proceed similarly as in the case of the
damping by coherent radiation. At low temperatures the damping
constant of the charge polarization in a semiconductor QR due to
emission of coherent phonon waves is then approximately given by
\begin{equation}\label{Eq:gamma_phon}
 \gamma^{\rm s}=\frac{1}{\tau_{\rm LA}}F\!\left(\frac{\omega_{\mathrm F}d}{c_{\rm
 LA}}\right).
\end{equation}
Here the condition $\omega_{_\mathrm F}<\omega_{_{\rm D}}$ must be
fulfilled that is valid for typical semiconductor QRs under
consideration. For a GaAs semiconductor QR (for the used material
parameters see Refs.~\cite{Moskalenko_PRB2006,Moskalenko_PRA2008}) with
$r_0=1.35~\mu$m, $d=50$~nm, and $N=400$, Eq.~\eqref{Eq:gamma_phon}
gives $\gamma^{\rm s}=0.8\times 10^6~{\rm s}^{-1}$, whereas for
$r_0=0.3~\mu$m, $d=20$~nm,  and $N=160$ it leads to $\gamma^{\rm
s}=2.1\times 10^8~{\rm s}^{-1}$ \cite{Moskalenko_PRA2008}.

\subsection{Scattering by incoherent phonons}
In case of the scattering of excited electronic states by
incoherent phonons, performing the same steps like when
considering the emission of incoherent light [leading to
Eq.~\eqref{Eq:rho_sp_result}] and calculating the phonon coupling
coefficient $M_m(\varepsilon)$ as in the previous section, in the
weak excitation regime we infer \cite{Moskalenko_PRB2006}
\begin{equation}\label{Eq:ac_phon_scatt_answer2}
  \begin{split}
     \frac{{\rm d}\tilde{\rho}_{mm'}}{{\rm d}t}\!\Big|_{\rm sp}^{\rm phon}\!=&
      -\!\frac{1}{\tau_{_{\rm LA}}}\sum_{\nu}
     (R^{m+\nu}_{m}\!+R^{m'+\nu}_{m'})\tilde{\rho}_{mm'}\\
     &+\!\frac{1}{\tau_{_{\rm LA}}}\sum_{\nu}
     (R^{m}_{m+\nu}\!+R^{m'}_{m'+\nu})\tilde{\rho}_{m+\nu\: m'+\nu}\;,
   \end{split}
\end{equation}
where
\begin{equation}\label{Eq:R}
  \begin{split}
   R^{m}_{m'}\!\!=& F(q^{m}_{m'} d)\chi^m_{m'}\:[n^0(
     q^{m}_{m'})+f^0_{m}],\ \ \ \ q^{m}_{m'}\!
     \in\!\!\Big(0,\frac{\omega_{_{\rm D}}}{c_{_{\rm LA}}}\Big);\\
     R^{m}_{m'}\!\!=& F(q^{m}_{m'} d)\chi^m_{m'}\:[n^0(
     q^{m}_{m'})\!+\!1\!-\!f^0_{m}],\ q^{m}_{m'}\!\in\!\!\Big(\!\!-\!\frac{\omega_{_{\rm D}}}
     {c_{_{\rm LA}}},\!0\Big);\\
      R^{m}_{m'}\!\!=&0,\hspace{4.1cm} \mbox{otherwise}.
  \end{split}
\end{equation}
Here $q^{m}_{m'}=(\varepsilon_m-\varepsilon_{m'})/(\hslash
c_{_{\rm
  LA}})$;
$\chi^m_{m'}=1$ if ${\rm sgn}(m)={\rm sgn}(m')$ and
$|q^{m}_{m'}|<\frac{\omega_{_{\rm D}}}{c_{_{\rm LA}}}$,
$\chi^m_{m'}=0$ otherwise.
The influence of the phonon bath was included via its equilibrium
distribution $n^0(q)$ having the same temperature $T$ as the
equilibrium electron temperature.

\begin{figure*}[t]
  \centering
  \includegraphics[width=7cm]{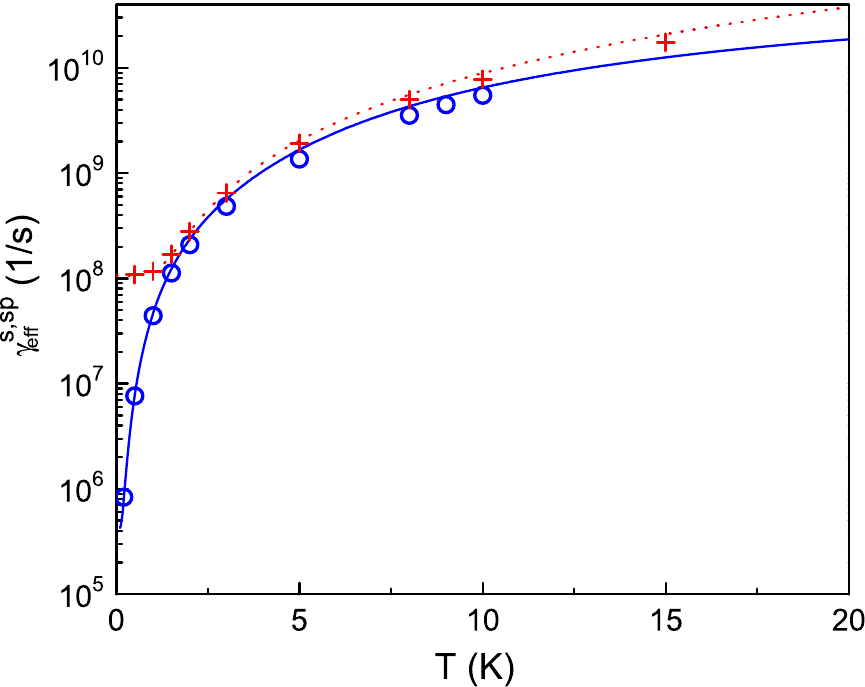}
  \caption{\label{Fig:eff_rate} Temperature dependence of the effective decoherence
  rate of the QR dipole moment induced by the application
  of a single weak HCP. The calculation result after Eq.~\eqref{Eq:gamma_sp_phon_eff} for a GaAs QR with
  $r_0=1.35~\mu$m, $d=50$~nm, and $N=400$ ($r_0=0.3~\mu$m, $d=20$~nm, and $N=160$) is shown by full blue
  line (dotted red line), whereas the results of the numerical estimation using Eq.~\eqref{Eq:ac_phon_scatt_answer2}
  for the same parameters of the QR and varied temperature are shown by empty circles (crosses).
  The material parameters are as in Fig.~\ref{Fig:revivals_PRB}.}
\end{figure*}

In analogy to the derivation of Eq.~\eqref{Eq:gamma_sp_def}, we
can neglect under the mentioned conditions the second line of
Eq.~\eqref{Eq:ac_phon_scatt_answer2} and derive from the first
line on the rhs of
Eq.~\eqref{Eq:ac_phon_scatt_answer2} the coefficients
\begin{equation}\label{Eq:gamma_sp_phon_def}
  \begin{split}
   \gamma^{\rm s,sp}_{mm'}\equiv \frac{1}{\tau_{_{\rm LA}}}\sum_{\nu}\!(R_m^{m+\nu}+R_{m'}^{m'+\nu})
 \end{split}
\end{equation}
determining the relaxation of the non-diagonal density matrix
elements $\rho_{mm'}$. The effective decoherence rate of the
dipole moment created by relatively weak ultrashort pulses for not too high temperatures
is then approximately given by
\begin{equation}\label{Eq:gamma_sp_phon_eff}
  \begin{split}
   \gamma^{\rm s,sp}_\mathrm{eff}=\sum_{m>0}&
   (f^0_{m-1}-f^0_{m})\gamma^{\rm s,sp}_{m\:m-1}\;.
 \end{split}
\end{equation}

We have calculated the temperature dependence of the effective
decoherence rate of the dipole moment using
Eq.~\eqref{Eq:gamma_sp_phon_eff} for two parameter sets. We have
also estimated this rate by simulating the dynamics of the dipole
moment, induced by the application of a single weak HCP to a GaAs
QR, numerically using Eq.~\eqref{Eq:ac_phon_scatt_answer2}. In the
latter approach the decoherence rate is estimated from the
relation between the amplitudes of the zero and first revivals of
the dipole moment dynamics \cite{Moskalenko_PRB2006}. The results
are shown in Fig.~\ref{Fig:eff_rate}. We see that the outcome of
the numerical estimation is in good agreement with the
dependencies following from Eq.~\eqref{Eq:gamma_sp_phon_eff}. Small
deviations are caused by the off-resonant terms [second line of
Eq.~\eqref{Eq:ac_phon_scatt_answer2}] leading to the decay
dynamics of the dipole moment being actually different from a
single-constant decay.

\section{Emission intensity and spectrum}\label{Sec:intensity_spectrum_def}
\subsection{Time-integrated spectra}
Let us consider a quantum system containing confined
non-relativistic electrons. We are interested in the radiation
from this system in the far-field region, i.e. at some distance
$R$ which is much larger than the system size.
Then the radiation intensity, i.e. power per unit solid angle,
is given by \cite{Landau_Lifshitz_second,Eberly1977,Raymer1995}
\begin{equation}\label{Eq:Intensity_per_solid_angle}
    I(\mathbf{n},t)=\frac{cR^2}{4\pi}\langle
\hat{\mathbf{E}}^\dagger(\mathbf{R},t)\cdot\hat{\mathbf{E}}(\mathbf{R},t)\rangle\;,
\end{equation}
where $\hat{\mathbf{E}}(\mathbf{R},t)$ is the operator of the
quantized electric field, $\mathbf{n}=\mathbf{R}/R$ is the unit vector determining the emission direction, $\langle ... \rangle$ denotes the
expectation value and $c$ is the speed of light in vacuum.
If we are interested only in the coherent emission contribution,
then $\langle
\hat{\mathbf{E}}^\dagger(\mathbf{R},t)\cdot\hat{\mathbf{E}}(\mathbf{R},t)\rangle$
simplifies to $|E(\mathbf{R},t)|^2$, where
$\mathbf{E}(\mathbf{R},t)=\langle \hat{\mathbf{E}}(\mathbf{R},t)
\rangle$ is the classical electric field (cf. Appendix
\ref{Sec:radiative_damping}).  We assume that
the emitted electric field vanishes quick enough for $t\rightarrow
\pm\infty$. Then using the Parseval's theorem we can write
\begin{equation}\label{Eq:Parseval}
    \int_{-\infty}^{\infty}I(\mathbf{n},t)\mathrm{d}t=2
    \int_{0}^{\infty}I_\mathrm{s}(\mathbf{n},\omega)\frac{\mathrm{d}\omega}{2\pi}\;,
\end{equation}
where the power spectrum for a particular emission direction $\mathbf{n}$ is given by
\begin{equation}\label{Eq:Intensity_per_solid_angle_per_frequency}
    I_{\mathrm{s}}(\mathbf{n},\omega)=\frac{cR^2}{4\pi}|E_\omega(\mathbf{R})|^2\;
\end{equation}
with
\begin{equation}\label{Eq:E_Fourier}
    \mathbf{E}_\omega(\mathbf{R})=\int_{-\infty}^{\infty}\mathbf{E}(\mathbf{R},t)e^{i\omega
    t}\mathrm{d}t\;.
\end{equation}

Using the solution of the inhomogeneous wave equation and making
the standard multipole expansion in the far-field region up to the
quadrupole term, leads to \cite{Landau_Lifshitz_second}
\begin{equation}\label{Eq:E_multipole}
      \mathbf{E}(\mathbf{R},t)=\frac{1}{c^2R}\left\{\mathbf{n}\!\times\!\big[\mathbf{n}\times\ddot{\boldsymbol{\mu}}(t-t_0)\big]
      +\mathbf{n}\!\times\!\ddot{\mathbf{M}}(t-t_0)
      +\frac{1}{6c}\mathbf{n}\!\times\!\big[\mathbf{n}\!\times\!\dddot{\mathbf{D}}(t-t_0)\big]\right\}\;,
\end{equation}
where $t_0=R/c$ and the dots over the variables denote the
corresponding time derivatives. The terms on the rhs
of Eq.~\eqref{Eq:E_multipole} describe the electric dipole,
magnetic dipole, and quadrupole contributions, respectively. For
the cases which we are going to consider here the electric dipole
contribution is the dominating one, and the next contributions are
by orders of magnitude weaker. Then from
Eqs.~\eqref{Eq:E_multipole} and
\eqref{Eq:Intensity_per_solid_angle_per_frequency} follows
\begin{equation}\label{Eq:Intensity_per_solid_angle_per_frequency_mu}
    I_{\mathrm{s}}(\mathbf{n},\omega)=\frac{\omega^4}{4\pi
    c^3}|\mathbf{n}\times\boldsymbol{\mu}_{\omega}|^2\;.
\end{equation}
If the dynamics of the dipole moment is confined to a plane,
selecting the coordinate system so that it is the $(x,y)$-plane,
we can write
\begin{equation}\label{Eq:Intensity_theta0_per_frequency_mu}
    I_{\mathrm{s}}(\vartheta=0,\omega)=\frac{1}{4\pi
    c^3}\omega^4|\mu_{\omega}|^2\;,
\end{equation}
where $\vartheta$ is the angle between $\mathbf{n}$ and the
$z$-axis and $\mu_{\omega}$ is the Fourier component of the dipole
moment [cf. Eq.~\eqref{Eq:E_Fourier}]. Collecting the emission in
a cone limited by the condition $\vartheta<\bar{\vartheta}$ leads
to
\begin{equation}\label{Eq:Intensity_collected_per_frequency_mu}
    I_{\mathrm{s}}(\vartheta<\bar{\vartheta},\omega)=I_{\mathrm{s}}(\vartheta=0,\omega)f(\bar{\vartheta})\;,
\end{equation}
where
\begin{equation}\label{Eq:f_dipole}
    f(\bar{\vartheta})=\pi
    \left(\frac{4}{3}-\cos\bar{\theta}-\frac{1}{3}\cos^3\bar{\theta}\right).
\end{equation}

\subsection{Time-resolved spectra}
In order to determine simultaneously the time and frequency
dependence of the intensity of the emitted radiation on ultrashort
time scales we can make use of the theory of the time-dependent
spectrum of radiation \cite{Eberly1977,Raymer1995}. Then the
time-resolved emission spectrum detected at the time moment $t$ in
a particular emission direction $\mathbf{n}=\mathbf{R}/R$, corresponding to the position $\mathbf{R}$ of the
detector with respect to the emitting system, in the far-field
region is given by
\begin{equation}\label{Eq:S0_start}
     S_0(\omega,t;\mathbf{n})=\frac{c R^2}{4\pi} \langle
     (\hat{\mathbf{E}}^\dagger)_\mathrm{d}(-\omega,t+t_{\rm d};\mathbf{R})\cdot(\hat{\mathbf{E}})_\mathrm{d}(\omega,t+t_{\rm d};\mathbf{R})\rangle\;,\\
\end{equation}
where $\hat{\mathbf{E}}(\mathbf{R},t)$ is the operator of the
quantized electric field, $\langle ... \rangle$ denotes the
expectation value, $c$ is the speed of light, $t_{\rm d}$ is a
delay time controlled by the detector, and
$(f)_\mathrm{d}(\omega,t)$ denotes the convolution
\begin{equation}\label{Eq:convolution}
     (f)_\mathrm{d}(\omega,t)\equiv \int_{-\infty}^{\infty} f(t') G(t'-t) {\rm
    e}^{i\omega t'} {\rm d} t'\;
\end{equation}
for a time-dependent quantity $f(t)$. Here $G(t)$ is a detector
function. We can select $t_{\rm d}=R/c$ and use \cite{Raymer1995}
\begin{equation}\label{Eq:detector_function}
     G(t)=\left(\frac{2}{\pi}\right)^{\!\!\!1/4}\!\!\frac{1}{\sqrt{\Delta T}}\:{\rm e}^{-t^2/\Delta T^2}\; ,
\end{equation}
where $\Delta T$ is the duration of the local oscillator pulse.
Notice that $S_0(\omega,t;\mathbf{n})$ has the dimension of power per
unit solid angle and per unit angular frequency interval.
Integrated  {over} certain solid angle, angular frequency and time
intervals, it gives the corresponding energy of the
electromagnetic field collected by the detector. Taking into
account only the coherent electric dipole contribution, assuming
that the dynamics of the dipole moment $\boldsymbol{\mu}(t)$ is
confined to the $(x,y)$-plane and that the emission is collected
in a cone limited by the angle $\bar{\vartheta}$ from the $z$-axis
(see  {Appendix~\ref{Sec:intensity_spectrum_def}.1}), we get
\begin{equation}\label{Eq:S0_angle_limited}
     S_0(\omega,t)=S_0^\bot(\omega,t) f(\bar{\vartheta})\;.
\end{equation}
Here the time-dependent emission spectrum in the direction
perpendicular to the plane is given by
\begin{equation}\label{Eq:S0_bot}
     S_0^\bot(\omega,t)=\frac{1}{4\pi
     c^3}
     \Big[\big|(\ddot{\mu}_x)_\mathrm{d}
     (\omega,t)\big|^2+\big|(\ddot{\mu}_y)_\mathrm{d}(\omega,t)\big|^2\Big]\;
\end{equation}
and the angular function $f(\bar{\vartheta})$ is determined by
Eq.~\eqref{Eq:f_dipole}. The double dots in Eq.~\eqref{Eq:S0_bot}
denote the second time derivatives.

\section{Time-dependent Stokes parameters and degree
of circular polarization}\label{Sec:Stokes}
Generally, the polarization properties of light are characterized
by the Stokes parameters $S_0$, $S_1$, $S_2$, and $S_3$
\cite{Schmidt_synchrotron,Born_Wolf_book}. $S_0$ is just the
intensity of radiation, $S_1$ and $S_2$ determine the part of the
intensity corresponding to the linearly polarized light, $S_3$
provides the contribution to the intensity from the circular
polarized light. The quantity $P=\sqrt{S_1^2+S_2^2+S_3^2}/S_0$
gives then the total degree of light polarization. The Stokes
parameters $S_1$, $S_2$, and $S_3$ are measured by using
polarization sensitive detectors. If we want to study the time
dependence of these quantities on ultrashort time scales, i.e. the
time scales comparable with the reciprocal frequencies of the
emitted radiation, it is necessary to define them appropriately
for such a situation. This is immediately evident in the case of
the parameter $S_3$, which obviously can not be extracted from the
information contained in an electromagnetic wave at fixed position
and time. In Ref.~\cite{Moskalenko_PRA2008} it was shown
that the time-dependent Stokes parameters can be defined on the
basis of the theory of the time-dependent emission spectrum,
discussed in Appendix \ref{Sec:intensity_spectrum_def}. Generalizing
Eq.~\eqref{Eq:S0_start}, we can introduce the
polarization-resolved time-dependent emission spectrum
\begin{equation}\label{Eq:S0_start_polarized}
     S^{(\alpha)}(\omega,t;\mathbf{n})=\frac{c R^2}{4\pi} \langle
     (\mathbf{e}_{\alpha}\cdot\hat{\mathbf{E}}^\dagger)_\mathrm{d}(-\omega,t+t_{\rm d};\mathbf{R})
     (\mathbf{e}_{\alpha}^{\:*}\!\cdot\hat{\mathbf{E}})_\mathrm{d}(\omega,t+t_{\rm d};\mathbf{R})\rangle\;,\\
\end{equation}
where the polarization vectors $\mathbf{e}_{\sigma}$ and
$\mathbf{e}_{\pi}$ are perpendicular to the propagation direction
$\mathbf{n}=\mathbf{R}/R$ and to each other, $\mathbf{e}_{\pm
45^\circ}=\frac{1}{\sqrt{2}}(\mathbf{e}_{\sigma}\pm\mathbf{e}_{\pi})$,
and $\mathbf{e}_{\pm}=\frac{1}{\sqrt{2}}(\mathbf{e}_{\sigma}\pm i
\mathbf{e}_{\pi})$. Clearly,
$S_0(\omega,t;\mathbf{n})=S^{(\sigma)}(\omega,t;\mathbf{n})+S^{(\pi)}(\omega,t;\mathbf{n})=S^{(45^\circ)}(\omega,t;\mathbf{n})+S^{(-45^\circ)}(\omega,t;\mathbf{n})
=S^{(+)}(\omega,t;\mathbf{n})+S^{(-)}(\omega,t;\mathbf{n})$ gives the
time-dependent emission spectrum \eqref{Eq:S0_start}, i.e. the
frequency-, time-, and angle-resolved Stokes parameter $S_0$. Then
the frequency-, time-, and angle-resolved Stokes parameters $S_1$,
$S_2$ and $S_3$ are defined by
\begin{equation}\label{Eq:S1_omega_t_theta_phi}
     S_{1(2,3)}(\omega,t;\mathbf{n})=
     S_{\sigma(45^\circ,+)}(\omega,t;\mathbf{n})-S_{\pi(-45^\circ,-)}(\omega,t;\mathbf{n})\;.
 \end{equation}
The frequency-integrated Stokes parameters are given by
\begin{equation}\label{Eq:S3_i}
  S_i(t;\mathbf{n})=2\int_0^\infty\! \frac{{\rm d}\omega}{2\pi}\;
S_i(\omega,t;\mathbf{n})\;,
\end{equation}
where $i=0,1,2,3$. Notice that the integration here is performed
over all frequencies. It can be, however, also restricted to a
particularly selected frequency range determined, e.g., by the sensitivity of the detection setup. A slightly different
realization of the frequency integration of the Stokes parameters
is proposed in Ref.~\cite{Agarwal2003}. However, it leads to
the same qualitative interpretation of the polarization
properties.

Let us concentrate on the degree of circular polarization and consider
the parameters $S_0$ and $S_3$.  Proceeding as described in Appendix~\ref{Sec:intensity_spectrum_def}, we get for the time-dependent Stokes
parameter $S_0(t)$:
\begin{equation}\label{Eq:S0_t}
     S_0(t)=S_0^\bot(t)f(\bar{\vartheta})\;,
\end{equation}
where
\begin{equation}\label{Eq:S0_bot_t}
     S_0^\bot(t)=\frac{1}{2\pi
     c^3}
     \int_0^\infty\! \frac{{\rm d}\omega}{2\pi}\; \Big[\big|(\ddot{\mu}_x)_\mathrm{d}
     (\omega,t)\big|^2+\big|(\ddot{\mu}_y)_\mathrm{d}(\omega,t)\big|^2\Big]\;
\end{equation}
 and the angular
function $f(\bar{\vartheta})$ is determined by
Eq.~\eqref{Eq:f_dipole}. $S_0(t)$ has the meaning of the
time-dependent detected power.
For $S_3(t)$ we obtain
\begin{equation}\label{Eq:S3_t}
     S_3(t)=S_3^\bot(t) g(\bar{\vartheta})\;,
\end{equation}
where
\begin{equation}\label{Eq:S3_bot_t}
     S_3^\bot(t)=\frac{1}{\pi
     c^3}\int_0^\infty\! \frac{{\rm d}\omega}{2\pi}\;
     \mathrm{Im}\big[(\ddot{\mu}_y)_\mathrm{d}(-\omega,t)(\ddot{\mu}_x)_\mathrm{d}(\omega,t)\big]
\end{equation}
and the angular function $g(\bar{\vartheta})$ reads
\begin{equation}\label{Eq:g_angular_function}
   g(\bar{\vartheta})=\pi\sin^2\bar{\vartheta}\;.
\end{equation}
The signed degree of circular polarization is given by
\begin{equation}\label{Eq:S_3_collected}
    P_{\rm circ}(t)=\frac{S_3^\bot(t)}{S_0^\bot(t)} \frac{g(\bar{\vartheta})}{f(\bar{\vartheta})}\;.
\end{equation}
The factor $g(\bar{\vartheta})/f(\bar{\vartheta})$ determines the
attenuation of the degree of circular polarization by increasing
the collection angle $\bar{\vartheta}$. It is equal to 1 at
$\bar{\vartheta}=0$, i.e. $P_{\rm
circ}^\bot(t)=S_3^\bot(t)/S_0^\bot(t)$, and decreases
monotonically to 0 as $\bar{\vartheta}$ is increased to
$\bar{\vartheta}=\pi$.

\paragraph*{Rotating dipole.} Let us consider a simple example
with $\mu_x(t)=\mu_0\cos(\omega_0 t)$ and
$\mu_y(t)=-\mu_0\sin(\omega_0 t)$. Such a dipole moment dynamics
(together with a circulating charge current) can be
realized in a semiconductor QR excited by two delayed mutually
perpendicular HCPs (see
Fig.~\ref{Fig:current_generation_mechanism}) at low temperature.
Then we calculate from Eqs.~\eqref{Eq:S0_bot_t} and
\eqref{Eq:S3_bot_t}
\begin{eqnarray}
     S_0^\bot(t)&=&\frac{\mu_0^2\omega_0^4}{4\pi
     c^3}, \label{Eq:rot_dipole_S0}\\
      S_3^\bot(t)&=&\frac{\mu_0^2\omega_0^4}{4\pi
     c^3}\:{\rm erf}\!\left[\frac{\omega_0\Delta T}{\sqrt{2}}\right].
\end{eqnarray}
Consequently, the degree of circular polarization is given by
\begin{equation}\label{Eq:rot_dipole_circ}
    P_{\rm circ}(t)={\rm erf}\!\left[\frac{\omega_0\Delta T}{\sqrt{2}}\right]
    \frac{g(\bar{\vartheta})}{f(\bar{\vartheta})}\;,
\end{equation}
where ${\rm erf}(x)$ denotes the error function. The plus (minus)
sign corresponds to the clockwise (anticlockwise) sense of
rotation of the dipole moment. If $\omega_0\Delta T>\pi$ (meaning
that the detector time is larger than the half of the rotation
period) than the value of the error function in
Eq.~\eqref{Eq:rot_dipole_circ} is larger than 0.995. Thus the
circular polarization is well defined and detectable in this
case for not too large collection angles $\bar{\vartheta}$.

\section{List of abbreviations}
{
\renewcommand{\arraystretch}{1}
 \flushleft
\begin{longtable}[l]{ll}
  1D \hspace{1.5cm} & one-dimensional \\
  CW & continuous wave \\
  DC & direct current\\
  DQD & double quantum dot\\
  DQW & double quantum wells\\
  FME & first term in the Magnus expansion\\
  HCP  & half-cycle pulse \\
  IA & impulsive approximation \\
  LA & longitudinal acoustic \\
  NS & numerical solution\\
  PME & unitary perturbation theory based on the Magnus expansion\\
  QR & quantum ring\\
  rhs & right hand side\\
  RWA & rotating wave approximation\\
  SFET & spin field-effect transistor\\
  SL & semiconductor superlattice\\
  SOI & spin-orbit interaction\\
  SVS & short but very strong (interaction)\\
  TDPT & time-dependent perturbation theory\\
  TDSE & time-dependent Schr\"{o}dinger equation\\
  TLS & two-level system\\
  TLSA & two-level system approximation
\end{longtable}
}


\bibliographystyle{elsarticle-num}
\bibliography{report_v3}

\end{document}